\pdfoutput=1
\documentclass[modern]{aastex61}
\usepackage{graphicx}
\usepackage{amsmath}
\usepackage{hyperref}
\usepackage[utf8]{inputenc}
\usepackage{natbib}
\usepackage{cmap}
\usepackage{amsfonts}
\usepackage{booktabs}

\newcommand{\diff}{\mathrm{d}}
\let\vec\mathbf

\begin{document}

\title{COBAIN: generalized 3D radiative transfer code for contact binary atmospheres}

\author{Angela Kochoska}
\affiliation{Villanova University, Dept.~of Astrophysics and Planetary Sciences, 800 E.\ Lancaster Ave, Villanova, PA 19085, USA}
\affiliation{University of Ljubljana, Dept.~of Physics, Jadranska 19, SI-1000 Ljubljana, Slovenia}

\author{Andrej Pr\v sa}
\affiliation{Villanova University, Dept.~of Astrophysics and Planetary Sciences, 800 E.\ Lancaster Ave, Villanova, PA 19085, USA}

\author{Toma\v{z} Zwitter}
\affiliation{University of Ljubljana, Dept.~of Physics, Jadranska 19, SI-1000 Ljubljana, Slovenia}

\author{Martin Horvat}
\affiliation{University of Ljubljana, Dept.~of Physics, Jadranska 19, SI-1000 Ljubljana, Slovenia}
\affiliation{Villanova University, Dept.~of Astrophysics and Planetary Sciences, 800 E.\ Lancaster Ave, Villanova, PA 19085, USA}

\author{Kyle E.~Conroy}
\affiliation{Vanderbilt University, Dept.~of Physics and Astronomy, 6301 Stevenson Center Ln, Nashville TN, 37235, USA}
\affiliation{Villanova University, Dept.~of Astrophysics and Planetary Sciences, 800 E.\ Lancaster Ave, Villanova, PA 19085, USA}


\begin{abstract}
Contact binary stars have been known to have a peculiar and somewhat mysterious hydro- and thermodynamical structure since their discovery, which directly affects the radiation distribution in their atmospheres.Over the past several decades, however, observational data of contact binaries have been modeled through a simplified approach, involving the artificial concatenation of the two components of the contact envelope and populating their respective surfaces with either blackbody atmospheres or plane-parallel model atmospheres of single stars. We show the implications this approach has on the reliability of the system parameter values and propose a method to overcome these issues with a new generalized radiative transfer code, COBAIN (COntact Binary Atmospheres with INterpolation). The basic principles of COBAIN are outlined and their application to different geometries and polytropic stellar structures is discussed. We present initial tests on single non-rotating, uniformly rotating and differentially rotating stars, as well as on simplified polytropic structural models of contact binaries. We briefly discuss the final goal of this ambitious project, which is the computation of model atmosphere tables under the correct assumptions for contact binary stars, to be used in modern binary star analysis codes. 

\end{abstract}

\section{Introduction}\label{sect1}

Contact binary stars are the consequence of a peculiar stellar evolutionary path in which binary star components have reached, and spend most of their lifetimes in, physical contact. They have very short orbital periods of 0.3 - 1 day \citep{webbink2003}, that makes them easy to detect in photometric studies due to their characteristic light curves. Thus, they comprise about $\sim 10-20\%$ of all binaries in current photometric sky-surveys \citep{kirk2016, sozynsky2016} and this number is expected to increase with new data from large scale sky surveys (e.g., {\sl Gaia} \citep{gaia2}, {\sl LSST} \citep{lsst}, {\sl TESS} \citep{tess}, {\sl PLATO} \citep{plato}) that favor shorter period binaries with prominent light curve variation \citep{kochoska2017}. W UMa stars, the most prominent subgroup of eclipsing contact binaries, are characterized by light curves exhibiting eclipses of almost equal depth and little color variation, suggesting similar surface temperatures of the two components. As their evolutionary phase and effective temperatures are similar, one would expect the two components to also have similar masses, but spectroscopic analyses have shown that the majority of contact binaries have mass ratios ($q$) that deviate from unity, even by a large amount -- a typical value for W UMa systems is $q = 0.5$, but can go as low as 0.07-0.08 \citep{rucinski2001,arbutina2009}. This in turn requires elaborate modeling of the mass and energy transfer inside the common envelope to reproduce the observed effects. 

The hydro- and thermodynamical structure of contact binaries has been studied in great detail over the last few decades and has shown to be rather complex and does not necessarily fit the predictions of simple hydrodynamical models. In all cases, a mechanism of stable mass and heat transfer is required, which in many models leads to problems such as inconsistency with observations \citep{lucy1961}, cycles between contact and detached phases \citep{flannery1976}, a representative sample of which has not been identified observationally \citep{lucy1976,yakut2005}, or complexity that surpasses the current computational capabilities, thus imposing many a-priori assumptions \citep{kahler2002b,kahler2002a,kahler2003,stepien2009}. 

However, the sheer abundance of these objects has led to frequent analysis of contact binary data and published models of the determined system parameters. This analysis is usually done with binary star modeling software, like Wilson-Devinney \citep{wd1971,wilson1976, vanhamme2003} or PHOEBE \citep{prsazwitter05b,phoebe2paper} in which the contact envelope surface is populated with values derived from radiative transfer models of single stars, with underlying assumptions that do not hold in contact binaries. Because of this, system parameters computed with these models are not necessarily accurate and reliable and can produce unphysical jumps in the surface temperature distribution of the model, if the temperature ratio of the two components is not unity. This paper addresses the current issues and deficiencies of the state of modeling contact binary atmospheres and outlines the foundations of the new radiative transfer code COBAIN (COntact Binary Atmospheres with INterpolation). COBAIN is being developed with the main goal to compute contact binary atmospheres that would replace the current single-star atmosphere approximations used in eclipsing binary codes, in particular PHOEBE 2. 

The paper is structured as follows: Section~\ref{sect2} is devoted to a brief overview of problems with current modeling of contact binary atmospheres in Wilson-Devinney (WD) and PHOEBE. Section~\ref{sect3} provides a detailed overview of the radiative transfer code COBAIN. Section~\ref{sect4} demonstrates the code performance on non-rotating and rotating single stars, while Section~\ref{sect5} showcases the results of gray radiative transfer simulations on a set of various contact binary system geometries.

\section{The problem with current modeling}\label{sect2}

Stars in WD and PHOEBE are represented by their surface meshes, where each surface point is populated with radiative properties computed through analytical formulae (limb- and gravity darkening laws) and interpolated from model atmosphere tables. Meshes in WD and PHOEBE 1 are discretized in a set of spherical angles, with corresponding radii computed from the Roche potential value of the surface \citep{wd1971,prsazwitter05b}. Each mesh point is the center of a trapezoid that spans a portion of the surface of the star. Meshes in PHOEBE 2 are triangulated grids computed with the marching algorithm \citep{hartmann}, where grid points are represented by the vertices of the triangles, while the radiative properties of all vertices are averaged to yield the emergent intensity and flux of each triangle. Unlike detached stars, the meshing of contact binaries is handled differently by the two approaches: in the trapezoidal approach, each component is created separately and populated with radiative properties corresponding to its respective polar temperature. The two sides of the contact binary envelope are then joined in the neck, which often leaves gaps in the neck region and produces systematic effects in simulated light- and radial velocity curves. The marching algorithm used by PHOEBE 2 produces the entire surface of the contact envelope as one closed triangularized mesh, which efficiently avoids the gap problem of the trapezoidal approach. However, the current computation of surface radiative properties of the triangularized mesh retains the approach introduced in trapezoidal meshes: both components receive their own polar temperature which dictates the radiative distribution of each component individually. 

\begin{figure}[ht]
\centering
\includegraphics[width=0.49\textwidth]{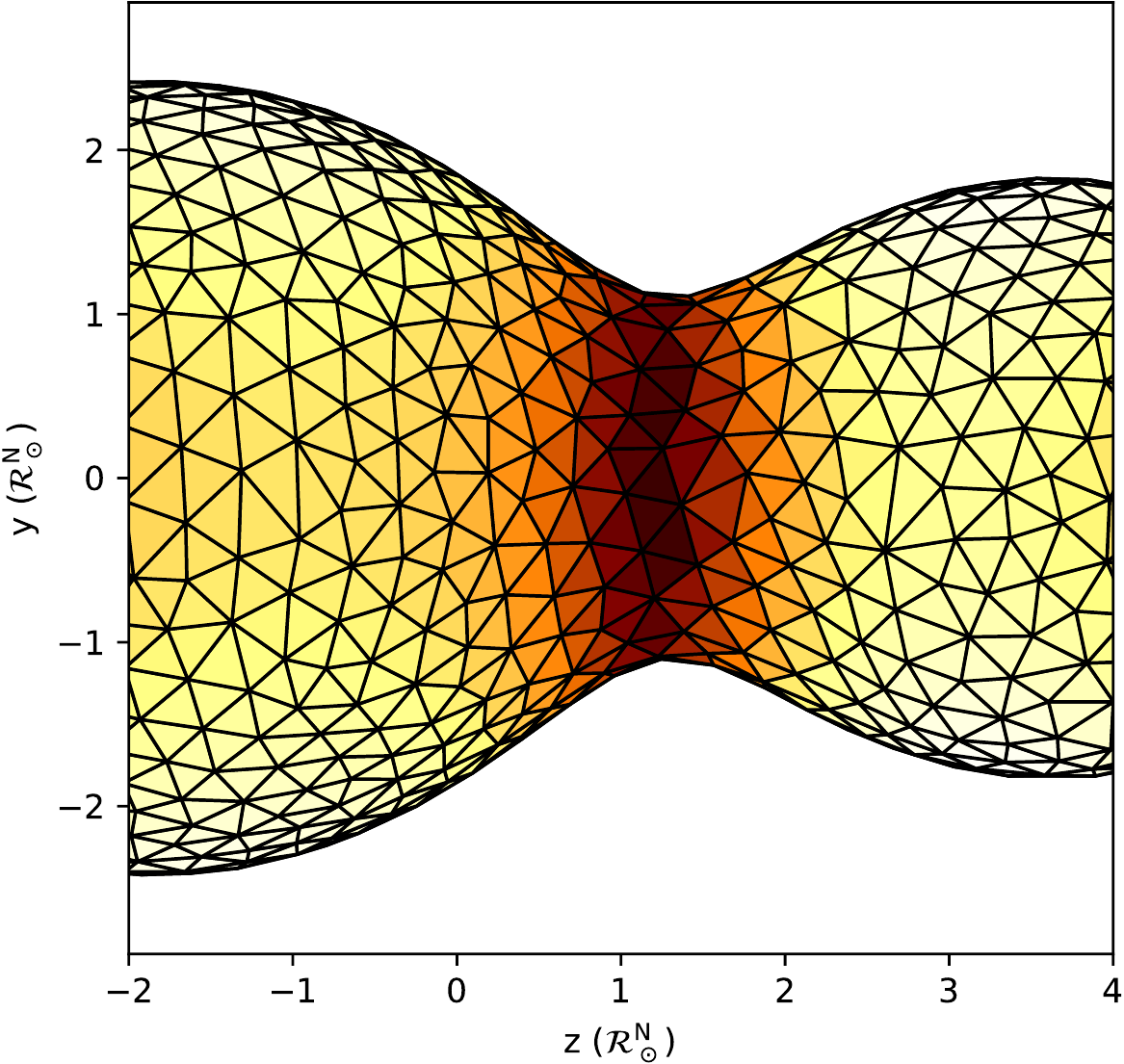}
\includegraphics[width=0.49\textwidth]{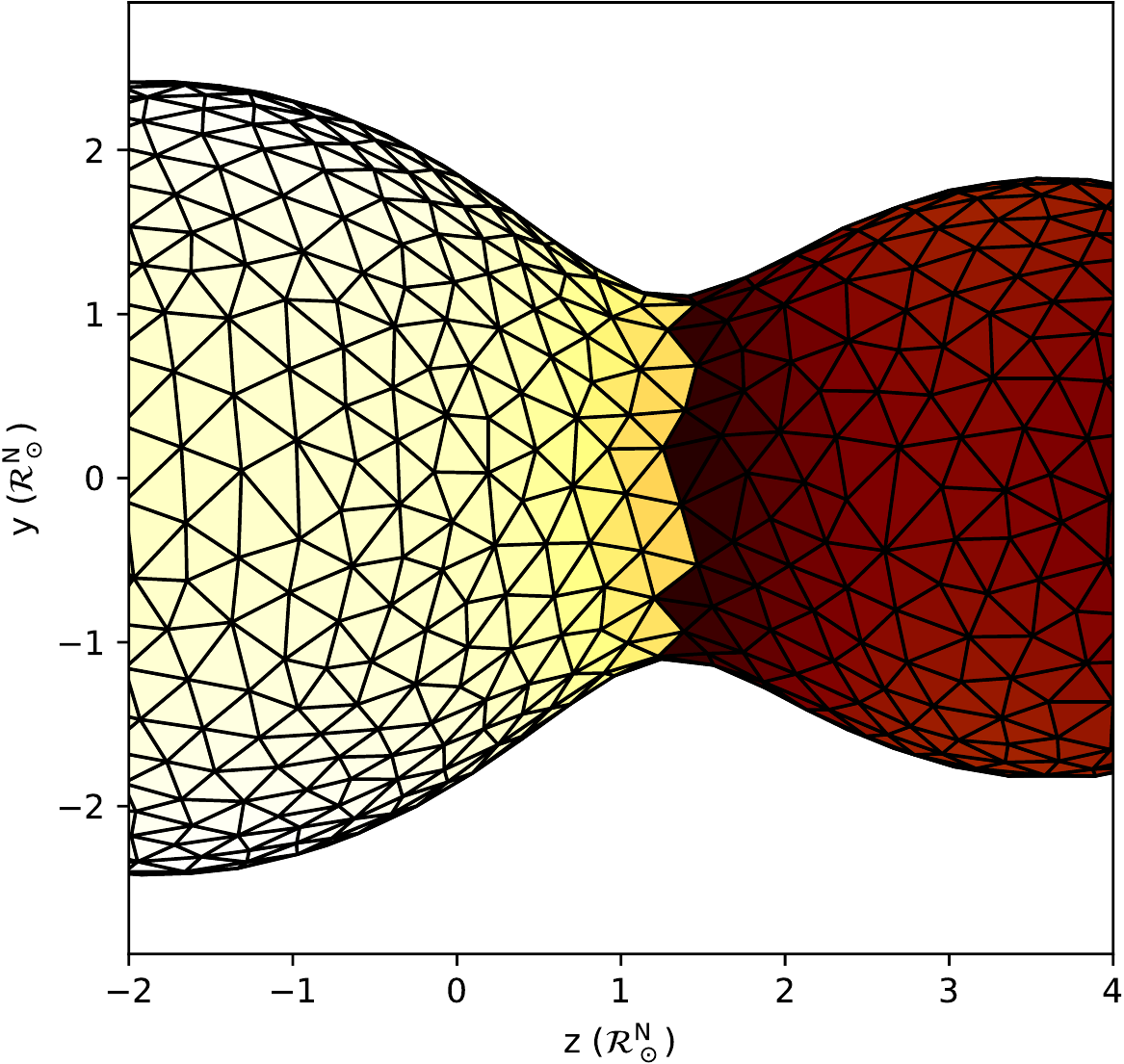}
\caption{Temperature distributions over the marching mesh of a contact binary in PHOEBE 2.0. Left: contact binary with equal component temperatures. Right: contact binary with different component temperatures.}
\label{oc_meshes_teffs}
\end{figure}

WD and PHOEBE~1 split the two components by computing a boundary plane, located at the position of minimum radius of the neck \citep{wilson1976}. In PHOEBE~2, the neck can be composed of triangles whose vertices are split between the two components. To retain the completely separate treatment of the two components as implemented in WD and PHOEBE 1, PHOEBE~2 currently assigns a weight to each triangle that intersects the neck, based on the number of vertices that belong to the primary and secondary component. If at least two of the vertices belong to one component, the whole triangle is populated with radiative properties determined by the effective temperature of that component. In this way, when the quantities over all vertices are averaged, there is no mixing of the two components (Figure~\ref{oc_meshes_teffs}). This is clearly unphysical as it produces temperature jumps in the neck of the contact systems out of thermal equilibrium, but it closely matches the current wide-spread implementation of contact binary modeling in state-of-the-art software. We use it primarily to demonstrate its deficiencies.

To showcase the implications of the unphysical modeling of contact binaries with unequal component temperatures, we have implemented an ad-hoc smoothing of the temperature distribution and study its effects on the synthetic light curves of several contact binary systems with published model solutions. 

The smoothing algorithm takes several reference points from the WD temperature distribution: the pole, the two equatorial points in the $+y$ and $-y$ directions, respectively, and the closest point to the neck of each component. As the $z$ coordinate is a parametric function of $x$ and $y$ described by the Roche potential, the temperature distribution is fit as a function of the coordinates $x$ and $y$ of the reference points in the $x=0$ and $x=a$ planes of the stellar poles and the $y=0$ plane passing through the stellar centers. If the neck temperatures are not equal, a weighted average value of the two is computed and used in the new temperature distribution. This ensures a smooth temperature transition over the neck of the contact binary.

The two-dimensional Gaussian functions used to recompute the neck region temperatures are given by:
\begin{equation}
T(x,y)_i =  T_{pole,i} \exp\left(-\frac{(x - x_{0,i}) ^ 2}{2 \sigma_{x,i} ^ 2} - \frac{(y - y_{0,i}) ^ 2}{2\sigma_{y,i} ^ 2}\right) + \mathrm{offset} \; ,
\end{equation}
where  $T(x,y)_i$ is the temperature at coordinates $(x,y)$ for component $i = 1,2$, $T_{pole,i}$ is the corresponding polar temperature that is used to define the amplitude of the Gaussians, and $x_{0,i}$, $ y_{0,i}$, $\sigma_{x,i}$, $\sigma_{y,i}$ are the Gaussian means and variances in the $x$ and $y$ direction, respectively. The back hemisphere of each component retains its original temperature distribution, obtained from the gravity darkening law, while the two hemispheres facing the neck are repopulated with the values from their corresponding Gaussian functions. The points used for fitting the two Gaussian distributions are summarized in Table~\ref{gaussianfitpoints}.

The weighted average temperature $T_{avg}$ is computed via:
\begin{equation}
T_{avg} = w \;T_{neck,1} + (1-w) \; T_{neck,2} \; ,
\end{equation}
where $w$ is a user-defined weight that determines the component with a higher contribution to the neck temperature. In physical systems, the value of $w$ and the shape of the temperature distribution of the neck would the depend on the hydrodynamical model of the contact binary interior structure, as well as on any potential presence and direction of mass and heat flows inside the envelope.  

\begin{table}[ht]
\centering
\caption{Contact binary points used for the Gaussian fit of the temperature distribution.}
\label{gaussianfitpoints}
{\def\arraystretch{1}\tabcolsep=10pt
\begin{tabular}{@{}lllll@{}}
\toprule
\multicolumn{2}{l}{\textit{primary component}} & \multicolumn{2}{l}{\textit{secondary component}} &                                     \\ 
\textbf{$(x,y)$}         & \textbf{$T$}        & \textbf{$(x,y)$}          & \textbf{$T$}         & \textit{\textbf{point description}} \\ \midrule
$(0,0)$                  & $T_{pole,1}$        & $(a,0)$                   & $T_{pole,2}$         & pole of the star                    \\
$(0,r_{eq,1})$           & $T_{eq,1}$          & $(a,r_{eq,2})$            & $T_{eq,2}$           & equatorial point in $+y$            \\
$(0,-r_{eq,1})$           & $T_{eq,1}$          & $(a,-r_{eq,2})$            & $T_{eq,2}$           & equatorial point in $-y$            \\
$(x_{neck,1},0)$         & $T_{avg}$           & $(x_{neck,2},0)$          & $T_{avg}$            & closest point to the neck           \\ \bottomrule
\end{tabular}}
\end{table}

Note again that this temperature smoothing is clearly ad-hoc and should not be used as a replacement for any physical mechanism of mixing and temperature distribution smoothing in contact binaries. It has been used here only to demonstrate the potential effects that mixing and smoothing of the temperature distribution would have on the synthetic light curves and model solutions obtained through current binary star modeling software. For this purpose, we have computed the light curves of two contact binary systems with published system parameters with unequal component temperatures: BL Eri  \citep{BLEripaper} and BL And \citep{BLAndpaper}. The published model parameters of the two systems are given in Table~\ref{calebmodelparams}.

\begin{table}[h!]
\centering
\caption{Model parameters for BL Eri and BL And. Adapted from the CALEB catalog, with model solutions published in \citet{BLEripaper} and \citet{BLAndpaper}, respectively. The parameters with uncertainties have been fitted to the light curve, while the ones without uncertainties have been kept fixed or error estimates are not given in the cited sources.}
\label{calebmodelparams}
{\def\arraystretch{1}\tabcolsep=10pt
\begin{tabular}{@{}lll@{}}
\toprule
\textbf{Parameters}            & \textbf{BL Eri}            & \textbf{BL And}            \\ \midrule 
$q$          & 0.546 $\pm$ 0.002            & 0.311  $\pm$ 0.008       \\
$\Omega_1$  =   $\Omega_2$         & 2.9108 $\pm$ 0.0030         & 2.394 $\pm$ 0.019         \\
T$_{\mathrm{eff,1}}$       & 5980     & 7500     \\
T$_{\mathrm{eff,2}}$       & 5603 $\pm$ 7      & 5370 $\pm$ 100       \\
$\beta_1$           & 0.32          & 0.32          \\
$\beta_2$           & 0.32          & 0.32          \\
$x_1^{ld,1}$, $x_2^{ld,1}$               & 0.55          & 0.632          \\
$x_1^{ld,2}$, $x_2^{ld,2}$              & 0.55          & 0.824          \\
$A_1$        & 0.5          & 1.0          \\
$A_2 $       & 0.5          & 0.5          \\
$l_3$         & 0.0          & 0.0          \\
$i$          & 89.8 $\pm$ 0.8         & 88.0 $\pm$ 1.7         \\
\bottomrule
\end{tabular}
}
\end{table}

\begin{figure}[ht]
\centering
\includegraphics[width=0.49\textwidth]{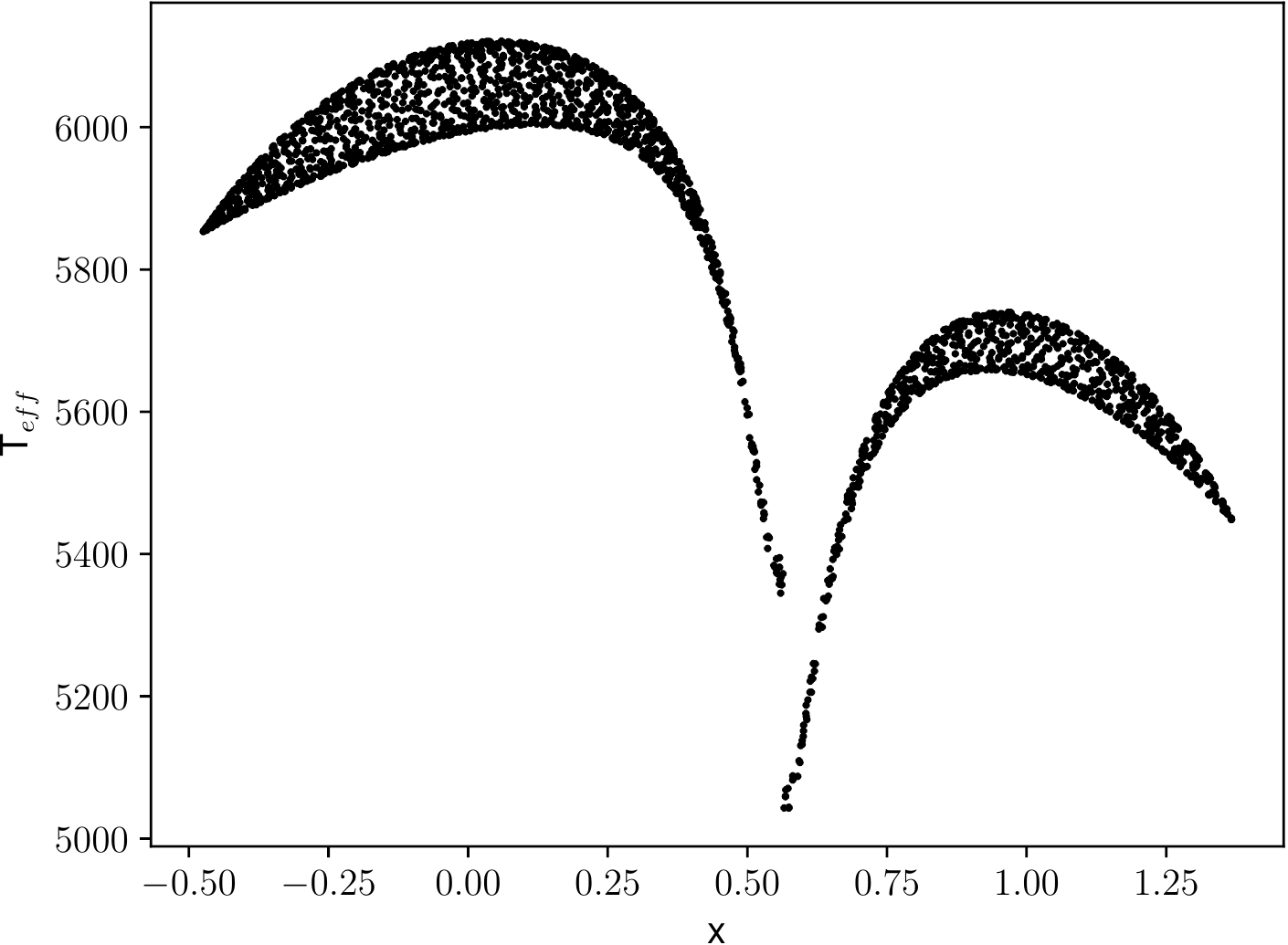}
\includegraphics[width=0.49\textwidth]{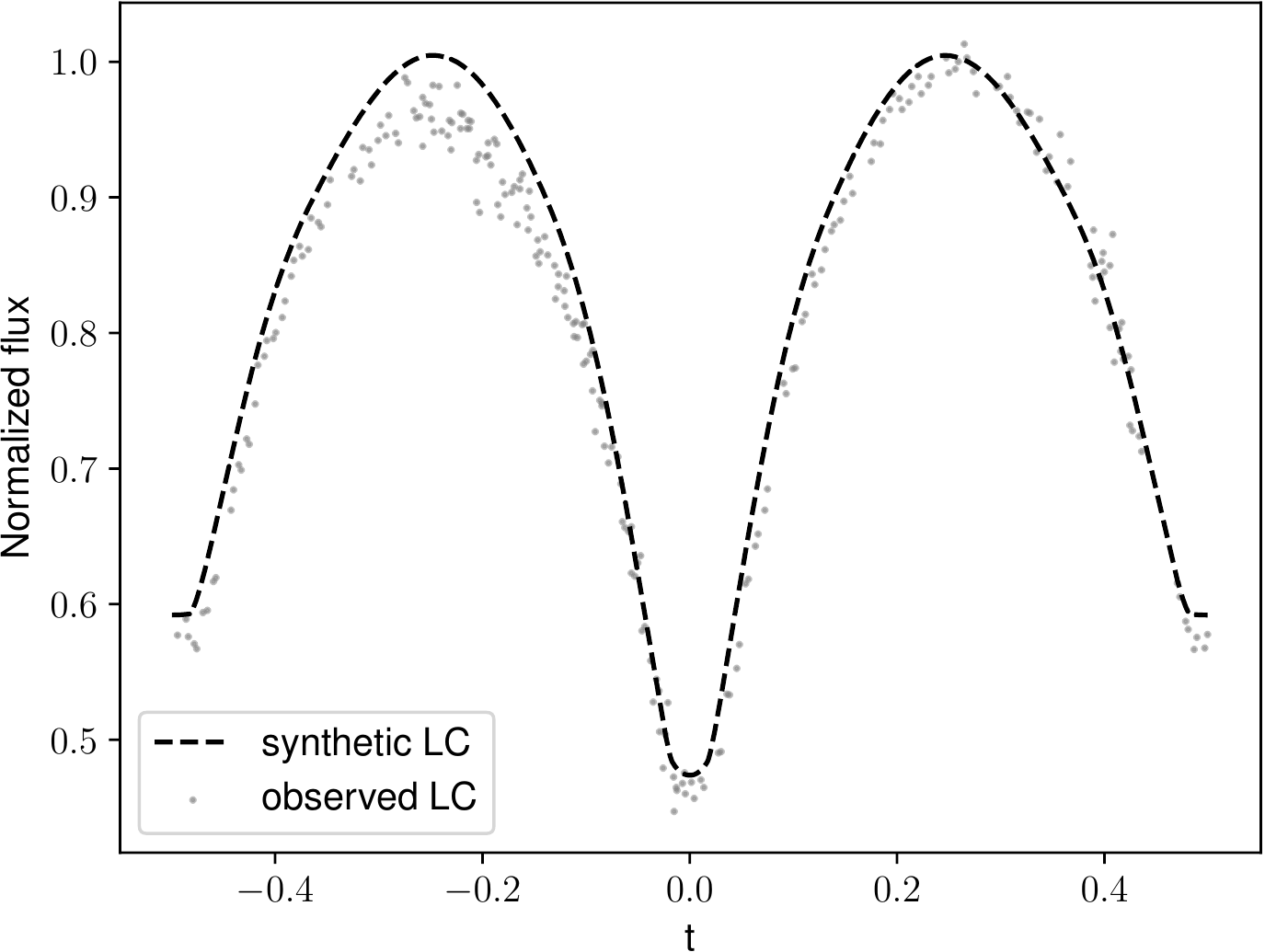}
\caption{Left: the original WD temperature distribution of BL Eri along the $x$ axis of the Roche coordinate system. Right: the corresponding observed (gray dots) and synthetic (dashed line) light curves.}
\label{BLEri_lc}
\end{figure}

\begin{figure}[ht]
\centering
\includegraphics[width=0.49\textwidth]{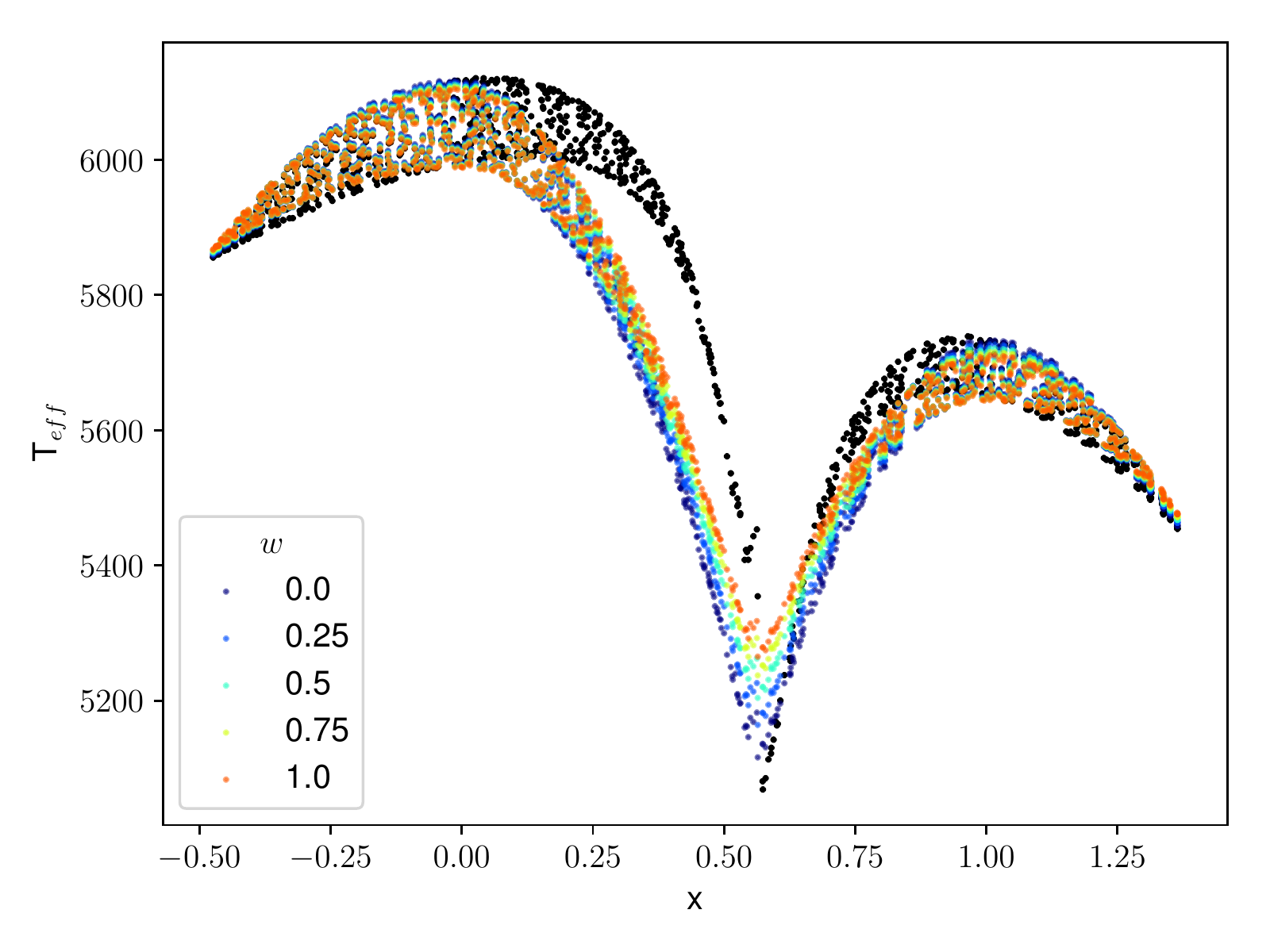}
\includegraphics[width=0.49\textwidth]{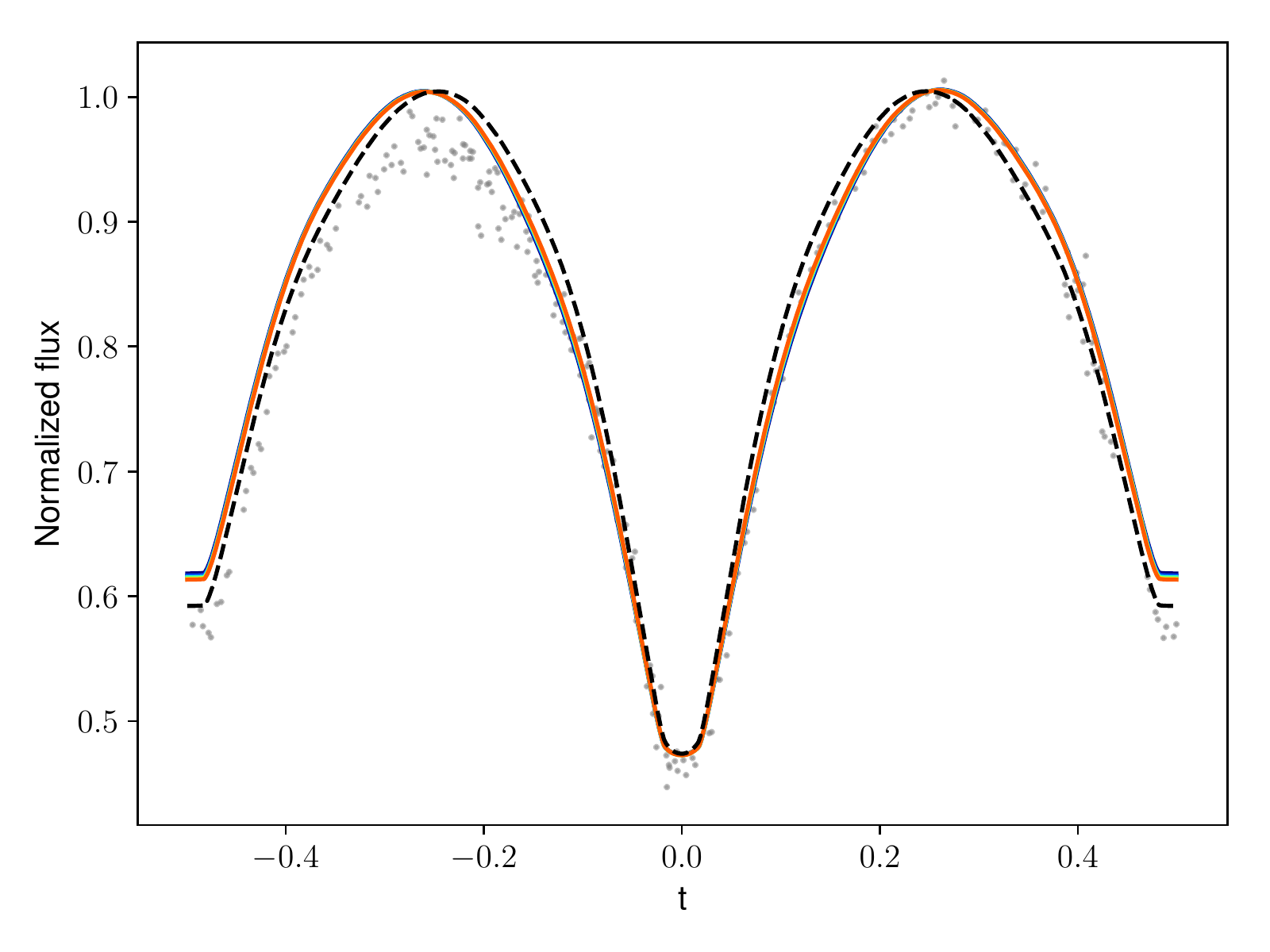}
\caption{Modified temperature distributions with different primary component weights of the average neck temperature (left) and their corresponding light curves (right) for BL Eri. The modified region spans the whole contact binary surface between the poles of the two components.}
\label{BLEri_teff_lc}
\end{figure}

\begin{figure}[ht]
\centering
\includegraphics[width=0.49\textwidth]{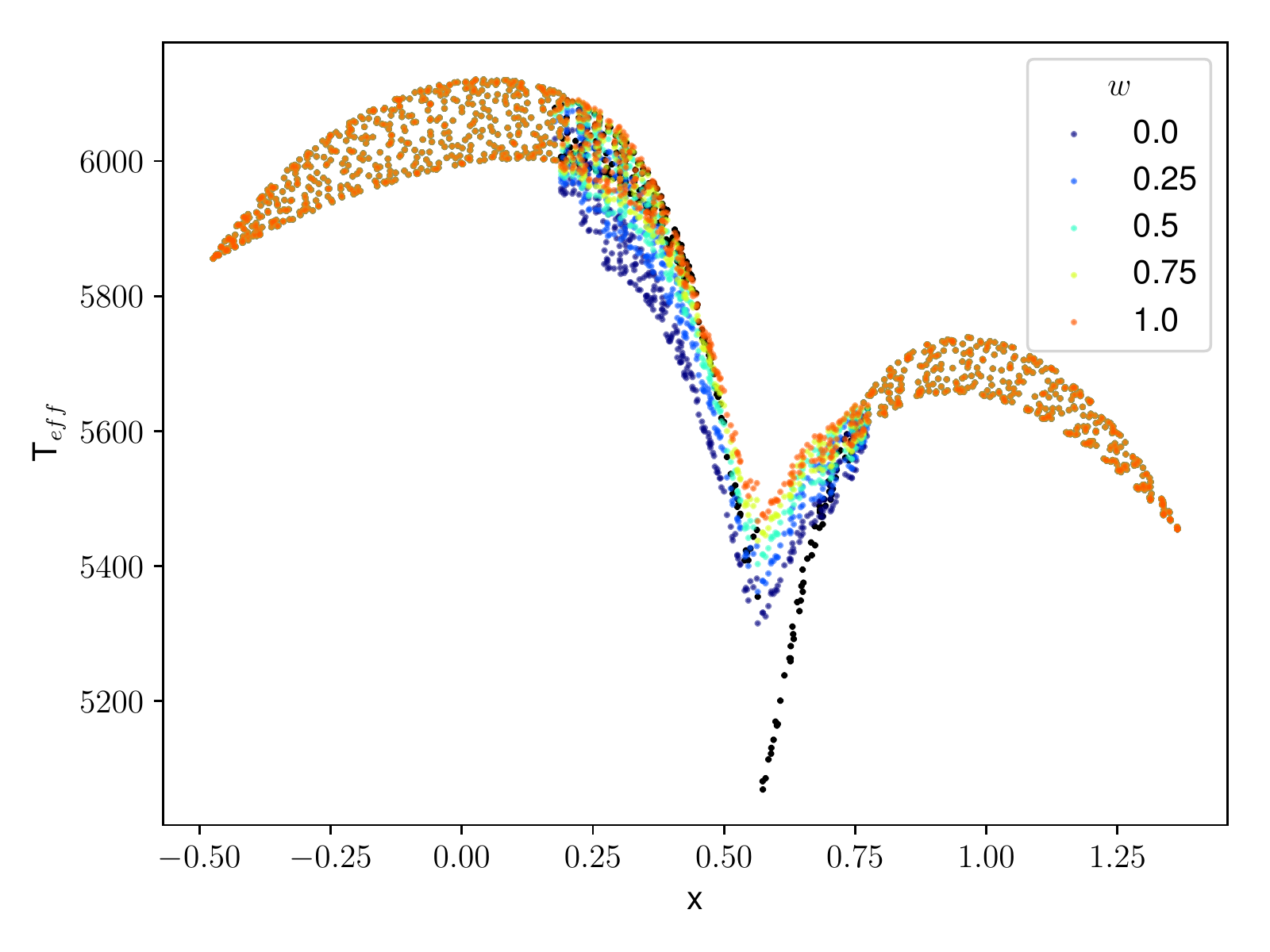}
\includegraphics[width=0.49\textwidth]{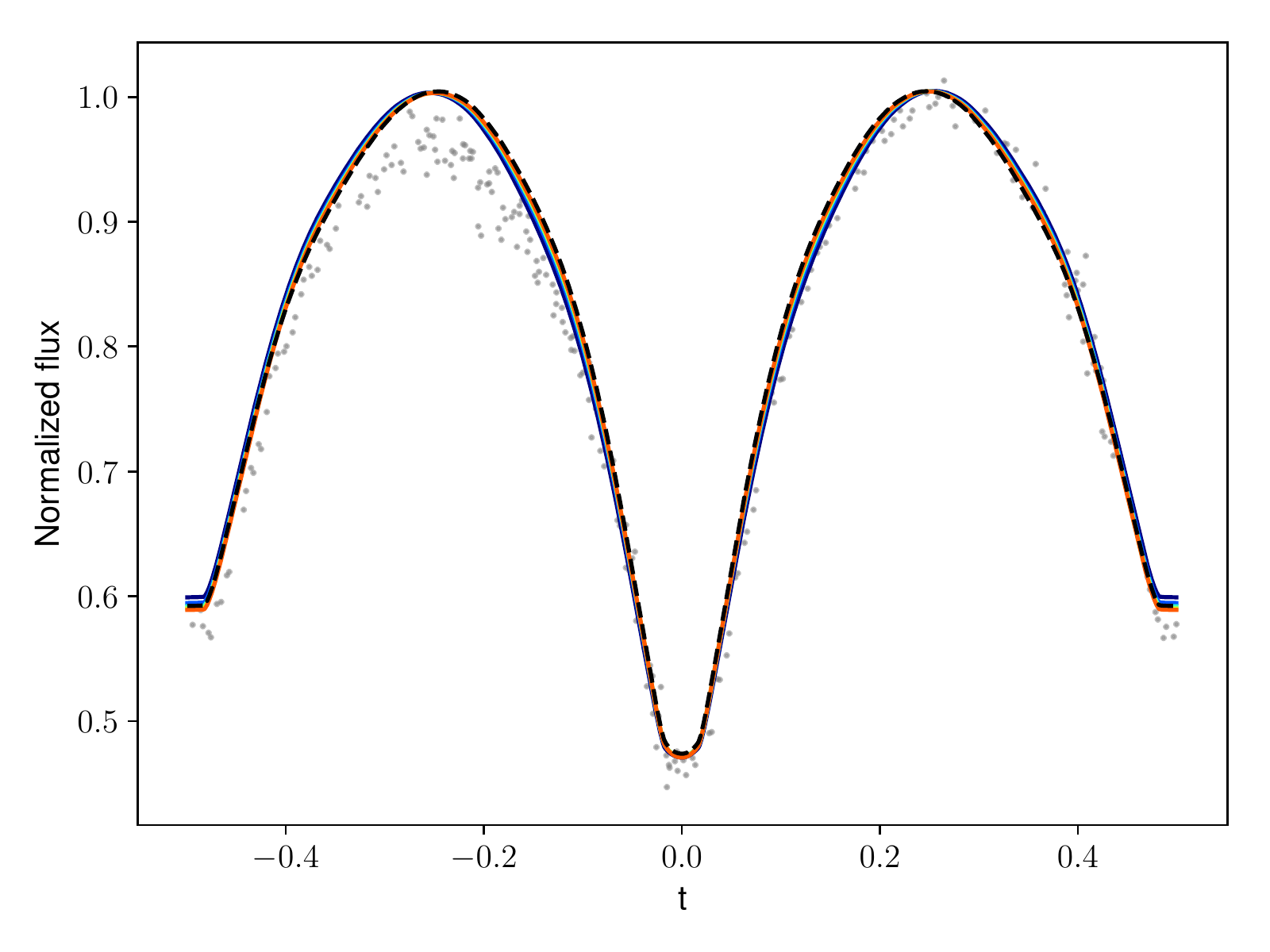}
\caption{Modified temperature distributions with different primary component weights of the average neck temperature (left) and their corresponding light curves (right) for BL Eri.  The modified region spans the stellar surface in the range $(x_{neck}-0.4, x_{neck}+0.4)$.}
\label{BLEri_neck}
\end{figure}

\begin{figure}[ht]
\centering
\includegraphics[width=0.49\textwidth]{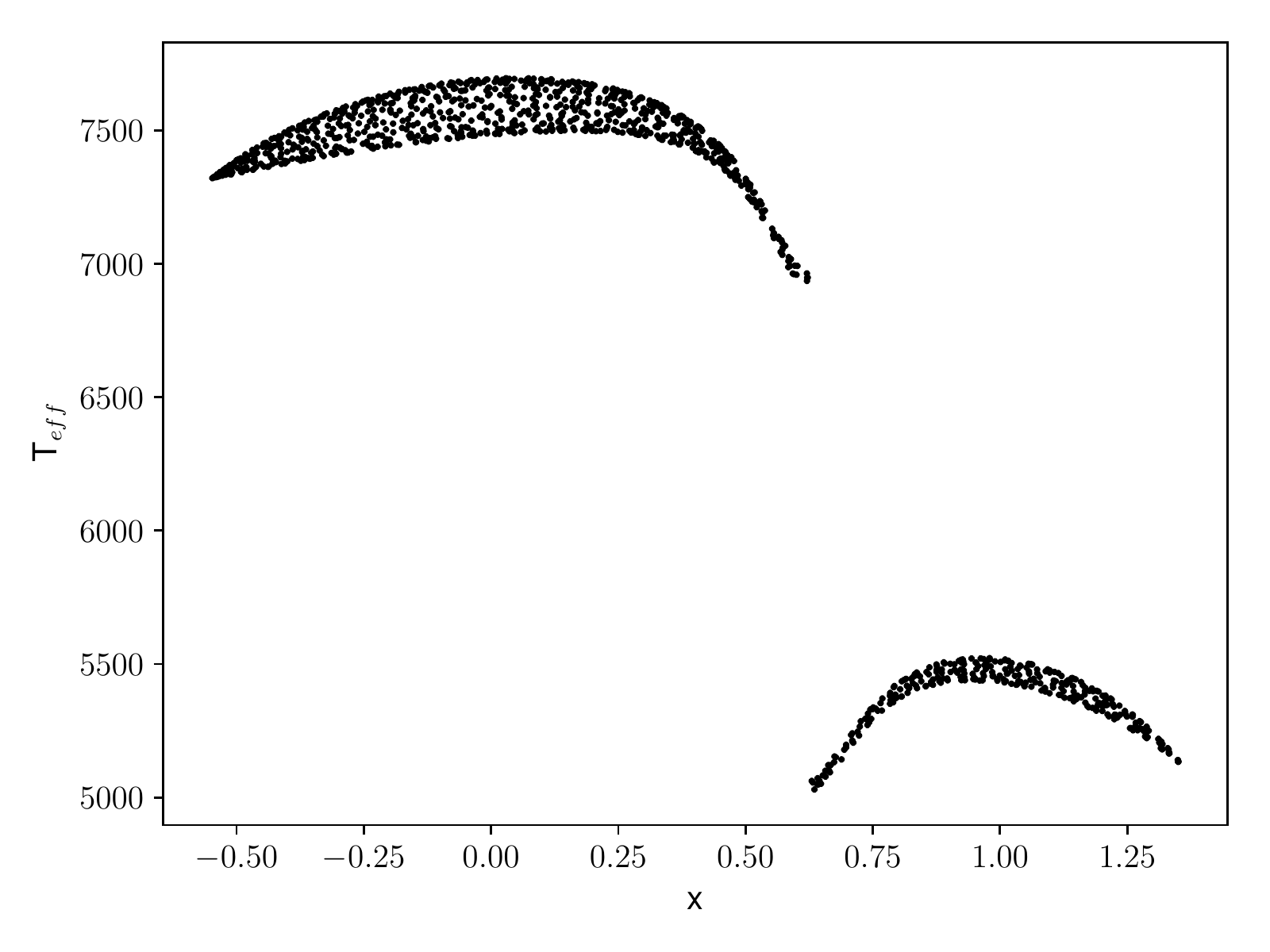}
\includegraphics[width=0.49\textwidth]{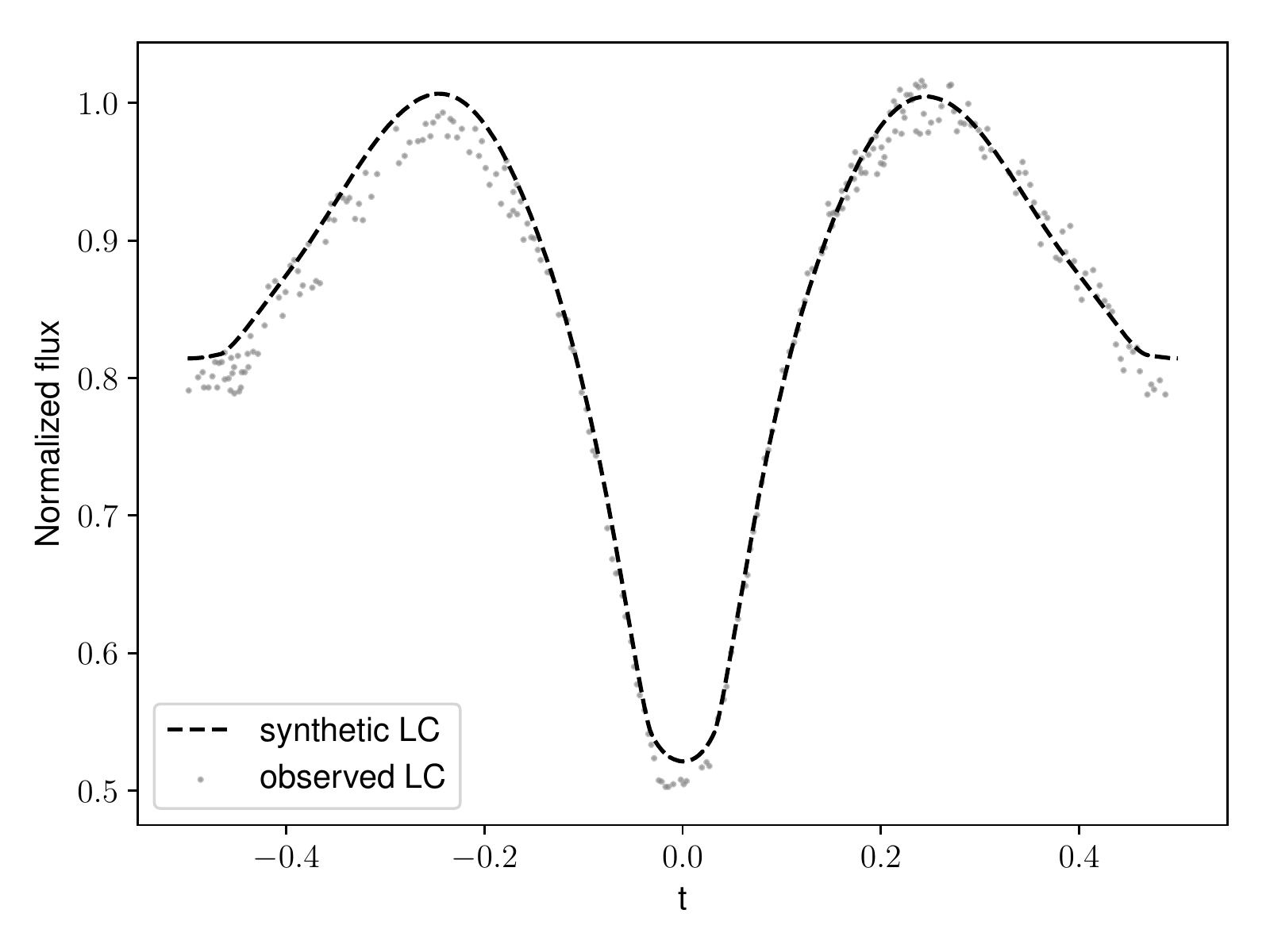}
\caption{Left: the original WD temperature distribution of BL And along the $x$ axis of the Roche coordinate system. Right: the corresponding observed (gray dots) and synthetic (dashed line) light curves.}
\label{BLAnd_lc}
\end{figure}

\begin{figure}[ht]
\centering
\includegraphics[width=0.49\textwidth]{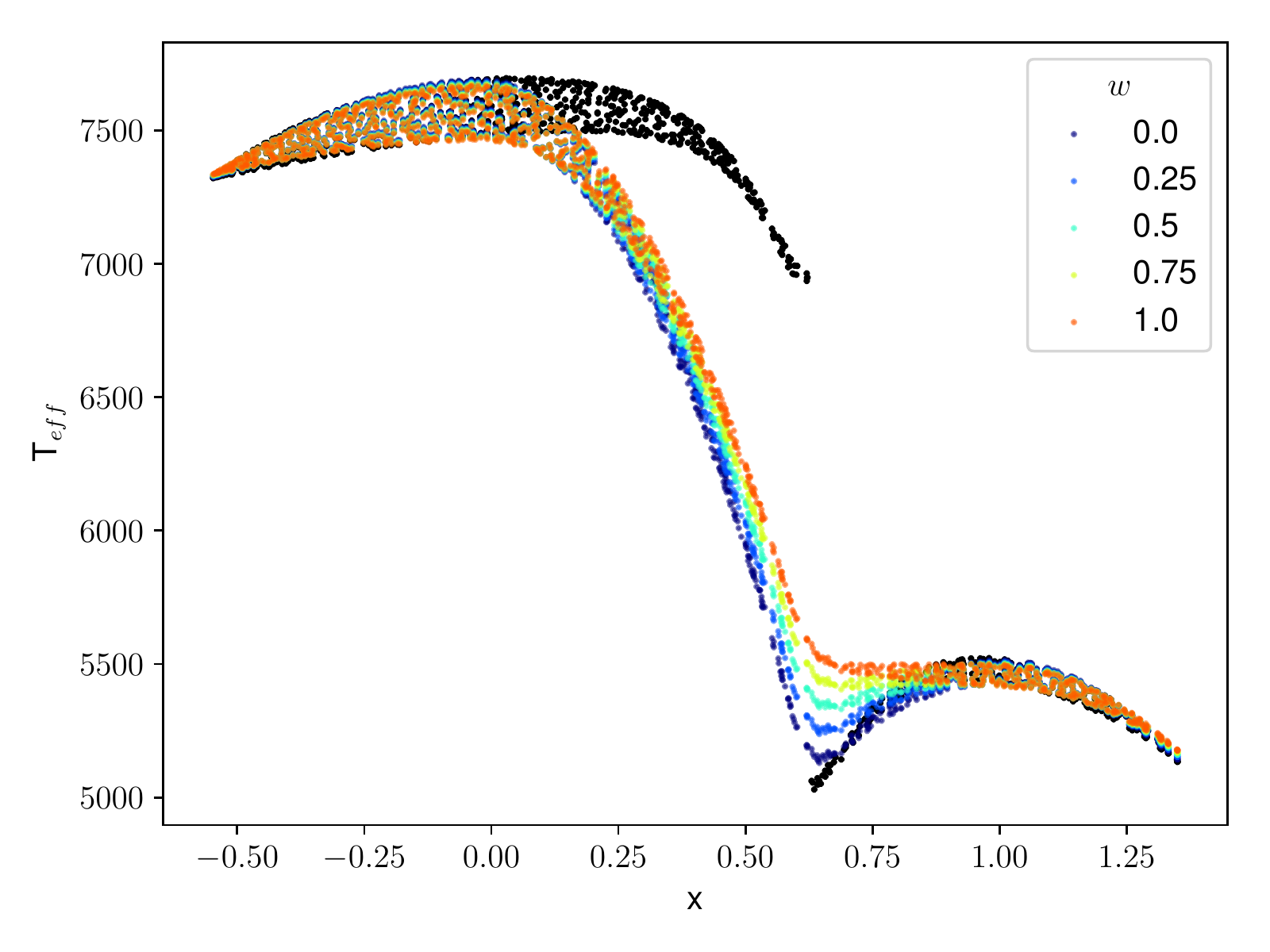}
\includegraphics[width=0.49\textwidth]{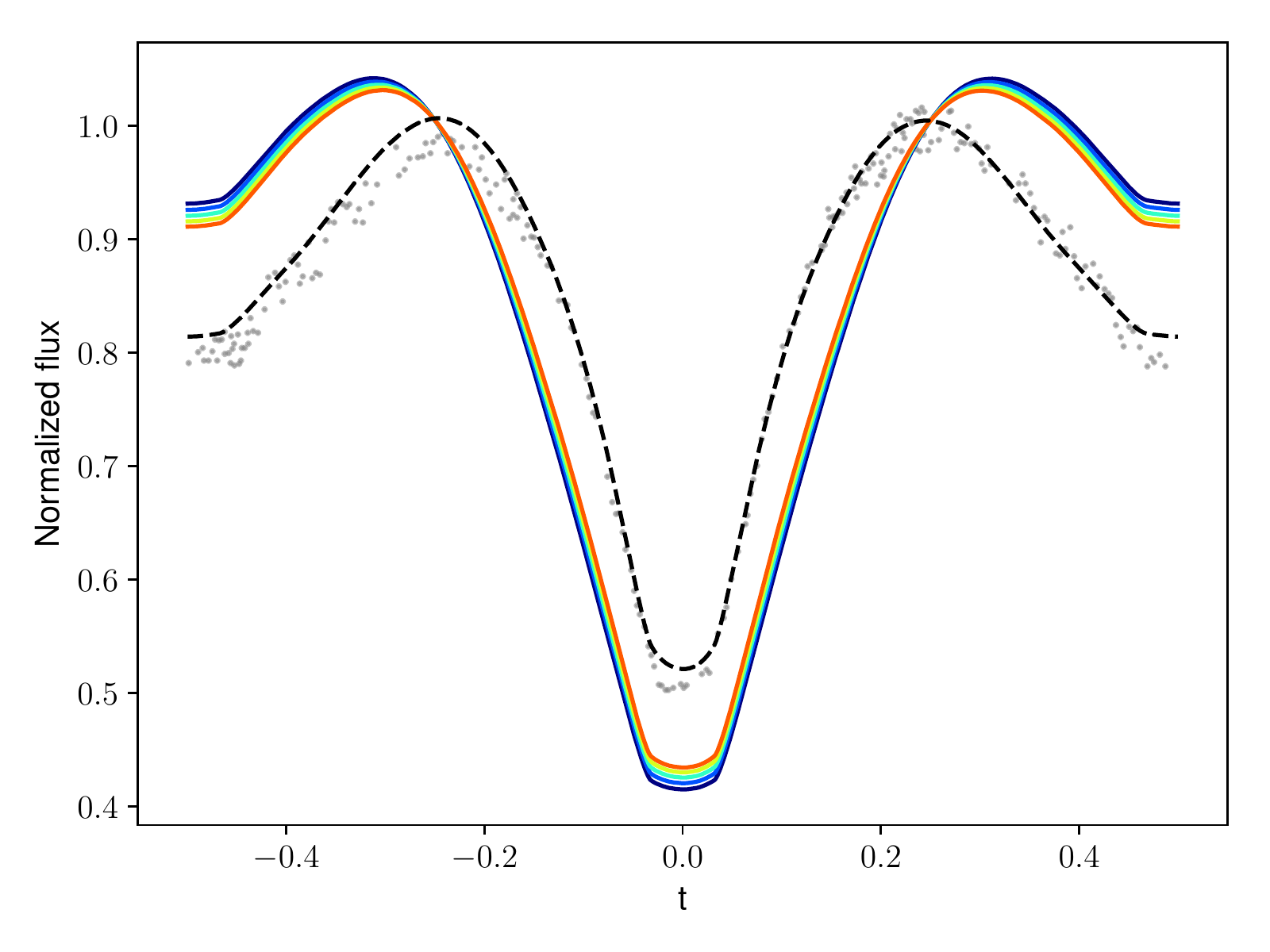}
\caption{Modified temperature distributions with different primary component weights of the average neck temperature (left) and their corresponding light curves (right) for BL And. The modified region spans the whole contact binary surface between the poles of the two components.}
\label{BLAnd_teff_lc}
\end{figure}

\begin{figure}[ht]
\centering
\includegraphics[width=0.49\textwidth]{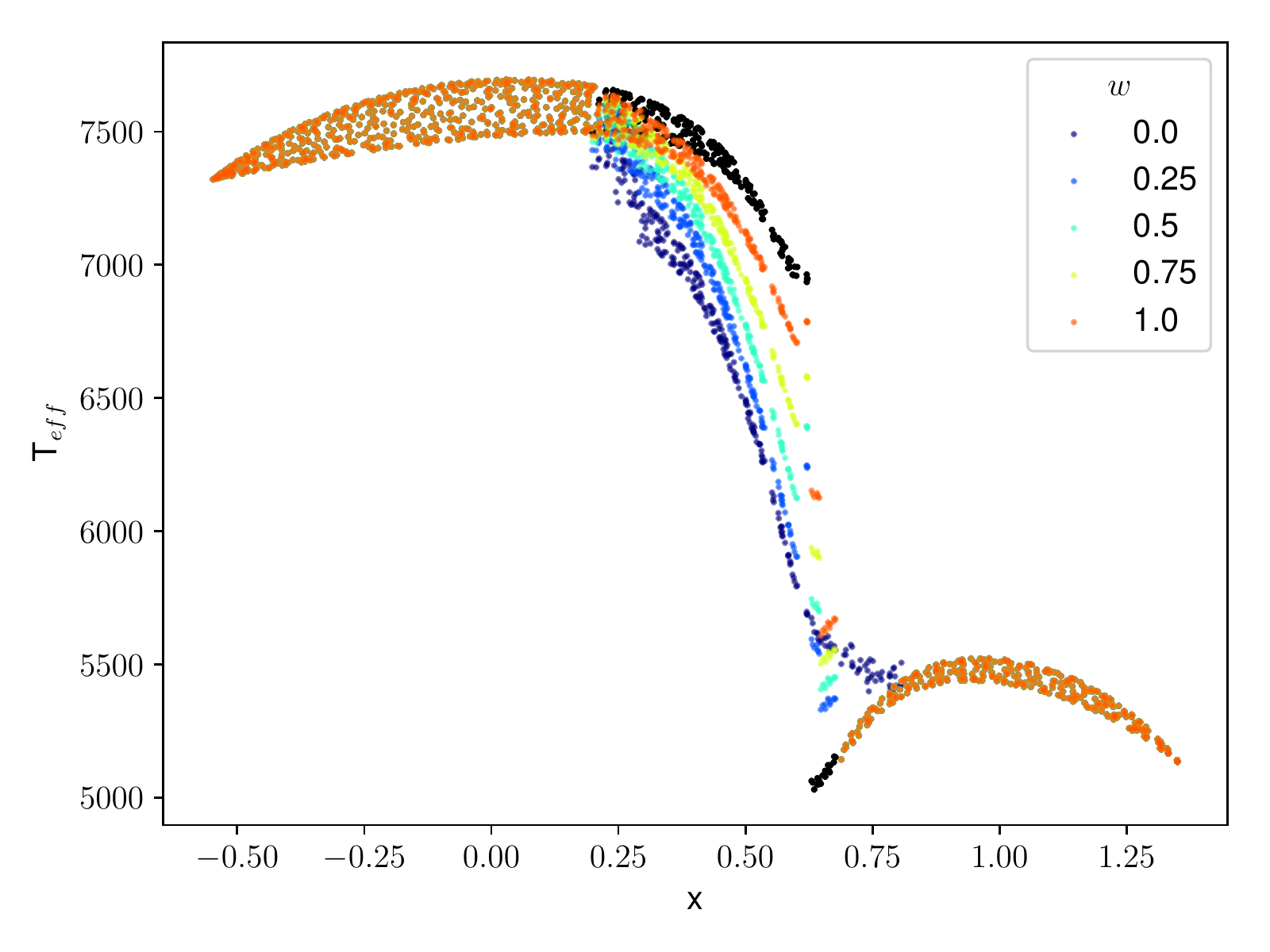}
\includegraphics[width=0.49\textwidth]{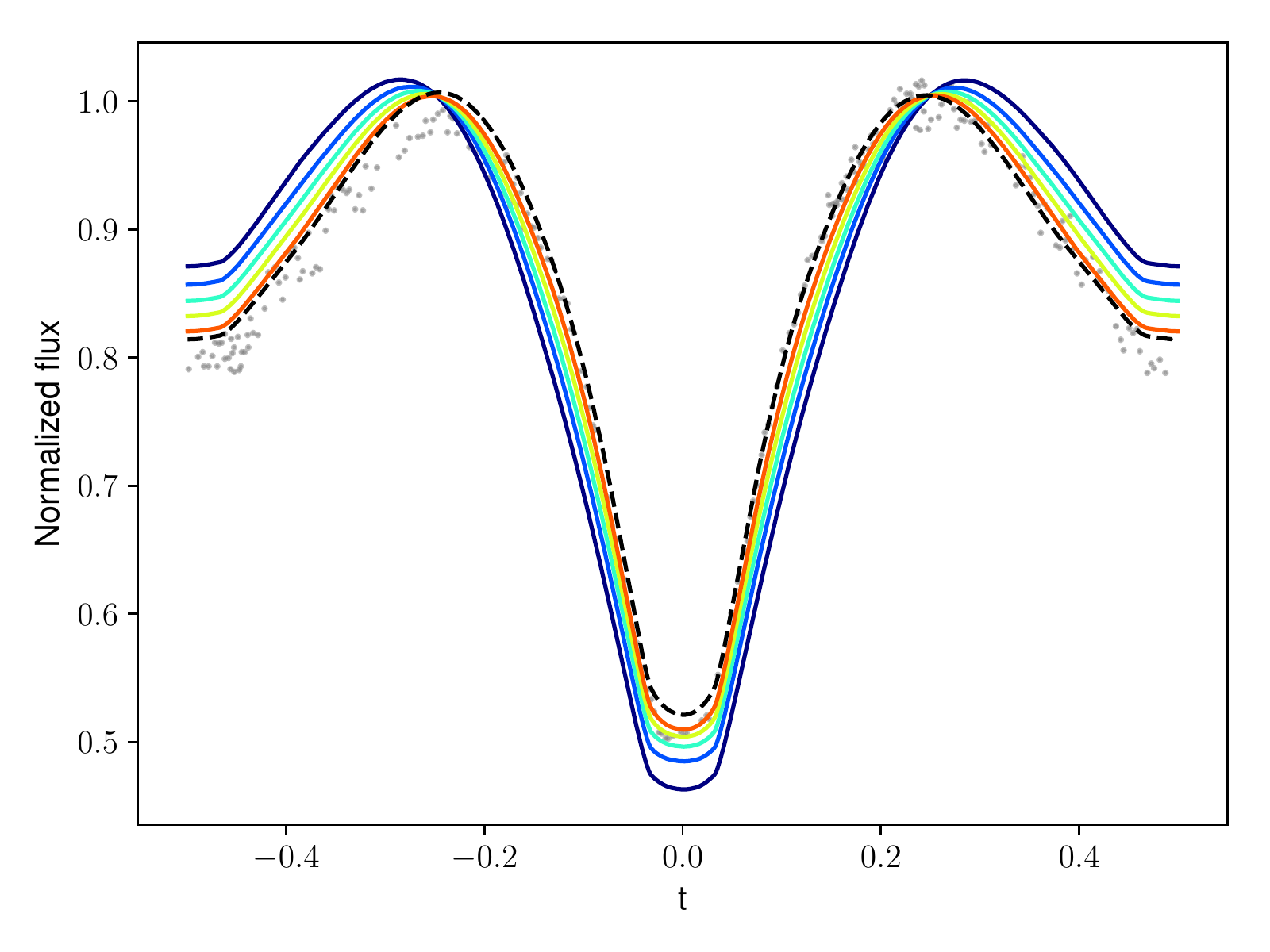}
\caption{Modified temperature distributions with different primary component weights of the average neck temperature (left) and their corresponding light curves (right) for BL And. The modified region spans the stellar surface in the range $(x_{neck}-0.4, x_{neck}+0.4)$.}
\label{BLAnd_neck}
\end{figure}

\begin{itemize}
\item \textbf{BL Eri}

A photometric model solution of the contact binary BL Eri has been published by \citet{BLEripaper}, with reported component temperatures that differ by about $\sim400$ K. This difference manifests itself in the model surface temperature distribution as a discontinuous temperature jump (notable at $x~\sim 0.55$ on the left panel of Figure~\ref{BLEri_lc}). A synthetic light curve is produced with PHOEBE~2.0 by adopting the published model parameters and neglecting the presence of spots. The right panel of Figure~\ref{BLEri_lc} shows the synthetic light curve that is obtained with this discontinuous temperature distribution model, which agrees reasonably well with the observed light curve. However, the synthetic light curves produced with smoothed temperature distribution (Figure~\ref{BLEri_teff_lc}) clearly do not agree with the observations well. The largest difference is notable in the secondary depth, which is significantly lower in the smoothed temperature model. A more subtle temperature modification can be introduced by recomputing the temperature distribution in a small area around the neck (Figure~\ref{BLEri_neck}), where the difference in eclipse depth becomes less notable.

The ratio of eclipse depths is proportional to the flux ratio of the two components, therefore neglecting the physical requirement for some smoothing of the properties at the boundary can lead to erroneous estimates of the flux contribution of each component. On the other hand, contact binaries cannot really be thought of in terms of components since we are dealing with only one surface, thus a unique and smooth temperature distribution over the surface mesh would be preferred for accurate modeling of these systems.

\item \textbf{BL And}

A more extreme case of a published model of a contact binary with temperature discontinuity is BL And \citep{BLAndpaper}. The reported model temperatures differ by $\sim 2000$ K, which causes a large unphysical jump in the neck temperatures (left panel of Figure~\ref{BLAnd_lc}). The published model with this discontinuity agrees well with the observations, but the temperature-smoothed models change the synthetic light curves drastically (Figure~\ref{BLAnd_teff_lc}). The less pronounced temperature modification localized in an area around the neck (Figure~\ref{BLAnd_neck}) still produces light curves that differ greatly from the one computed with the published model. The amplitude of the maxima and minima changes in all cases, as well as the width of the eclipses. This in turn influences the values of all reported system parameters: mass ratio, surface potential, temperature ratio and inclination. If radial velocity curves for the system are also available, these quantities can be related to the physical parameters of the two stars, like masses and radii, which are the key point of interest of modeling binary stars. This unphysical model solution would without a doubt result in stellar parameters that differ substantially from the true system parameters. Considering the large temperature difference and the drastic effects even localized smoothing has on the light curves, this system is more likely a very close detached or semi-detached binary and not a contact binary at all. The fact that it is possible to get a contact binary light curve that is a good fit to the observations of a system like this just further reinforces the need for better modeling of contact binaries. The binary star model has inherent parameter degeneracies that render the task of obtaining a perfect fit difficult in itself, however, improving the modeling of contact binaries in binary star analysis codes would ensure that these solutions are obtained through physically consistent underlying models.

\end{itemize}

In practice, this problem is not as pronounced for the majority of contact binaries because of the similar surface temperatures of the components. This can be seen by comparing the results on the smoothing of BL Eri, where $\Delta T \sim 400$ K, and BL And with $\Delta T \sim 2000$ K. The light curves produced with smoothed temperatures in a localized neck area fall within the observational error for BL Eri and their effect is small, while their disagreement with the observations is still excessive for BL And.

Thus, we can conclude that the smaller the difference $\Delta T$ between the two component temperatures, the smaller the effect of the smoothing on the light curve. Nonetheless, this does not mean that the temperatures used to populate the component surfaces are correct if they do not have a neck discontinuity. As mentioned before, the computation of the light curves in binary star analysis software available today uses atmosphere tables computed for single stars, under assumptions valid for single stars, very often in local thermodynamic equilibrium, many of which do not apply to contact binary stars. This in turn affects the determination of model parameter values, which are directly related to the physical properties of the system. This fact alone, combined with the inflow of large quantities of high-quality data that we have on hand and still expect from large scale surveys, calls for an imminent action on developing a more consistent approach to modeling of contact binary atmospheres using hydrodynamical models of their peculiar structure. The COBAIN (COntact Binary Atmospheres with INterpolation) code is the first step towards the accomplishment of this ambitious goal. 

\clearpage

\section{The COBAIN (COntact Binary Atmospheres with INterpolation) radiative transfer code}\label{sect3}

COBAIN is a free and open source code that currently supports radiative transfer computation in gray atmospheres of spherical stars, rotating stars, differentially rotating stars and contact binaries whose interior structure is approximated by polytropes. The radiative transfer computation alone is completely general and can be used with any chosen structure model of stellar objects, preferably easily parametrized in spherical or cartesian coordinates to ensure fast interpolation and convergence, while other geometries can greatly affect its speed. In the current version of the code, the simulations are carried out in nested-equipotential grids which allow for fast 3D interpolation and each simulation is run on average 5-10 iterations. This Section is devoted to a detailed overview of the capabilities and structure of the code, while initial results follow in the subsequent Sections.

\subsection{The nested-equipotential grid}

In large optical depths in the stellar interior and deeper layers of the atmosphere the radiation is in near-perfect thermodynamical equilibrium and closely follows Planck's law. However, as the optical depth decreases in the outermost layers of the atmosphere, the mean free path of photons becomes larger and the effects of absorption and scattering become more pronounced. Therefore, the points of interest in radiative transfer computations are the ones located in a thin layer close to the stellar surface. To account for this, the grids in COBAIN span only these atmosphere layers, while the radiative properties of points deeper in the star are computed analytically through the blackbody approximation. 

\begin{figure}[ht]
\begin{center}
\includegraphics[height=5.55cm]{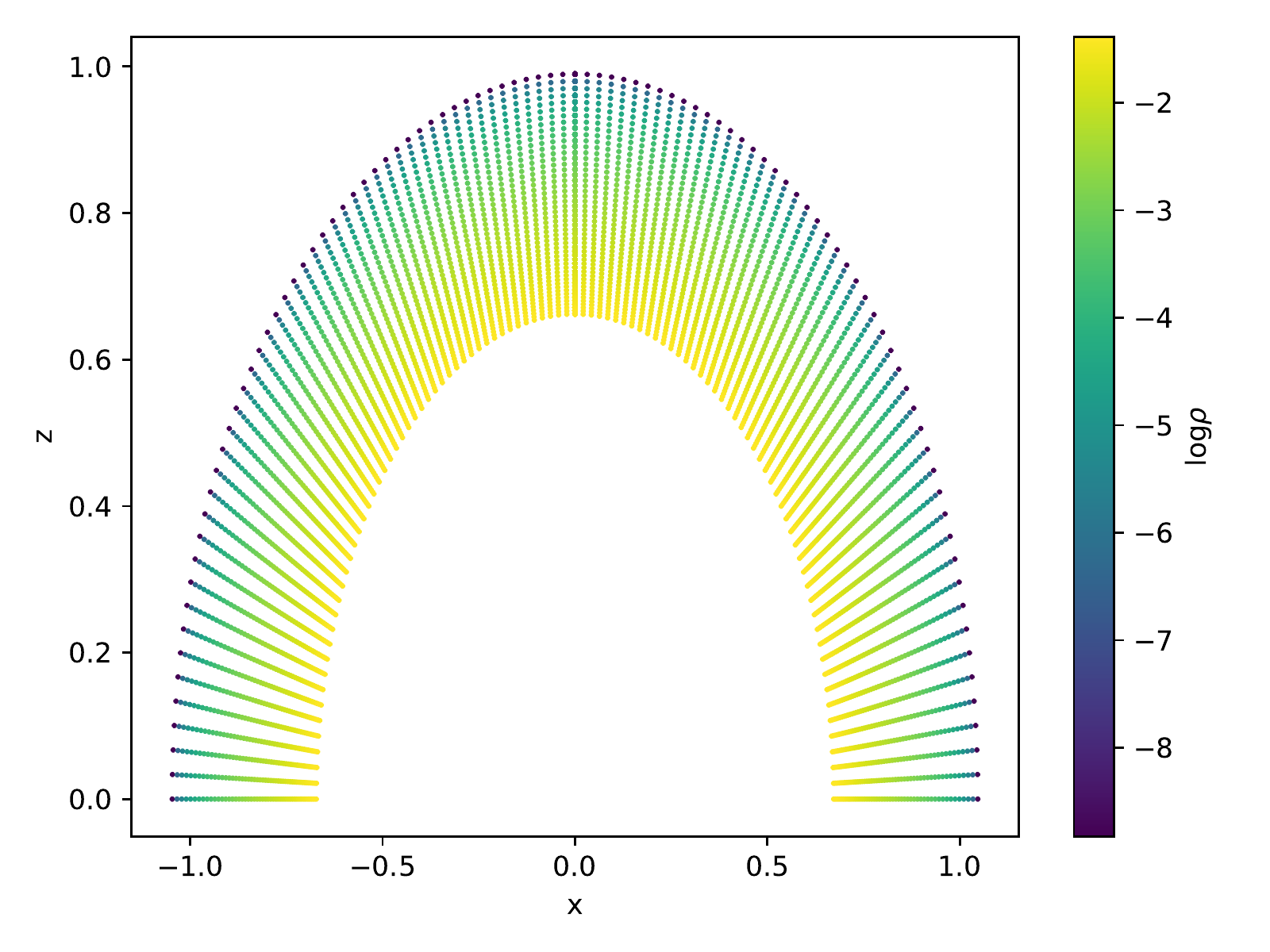}
\includegraphics[height=5.55cm]{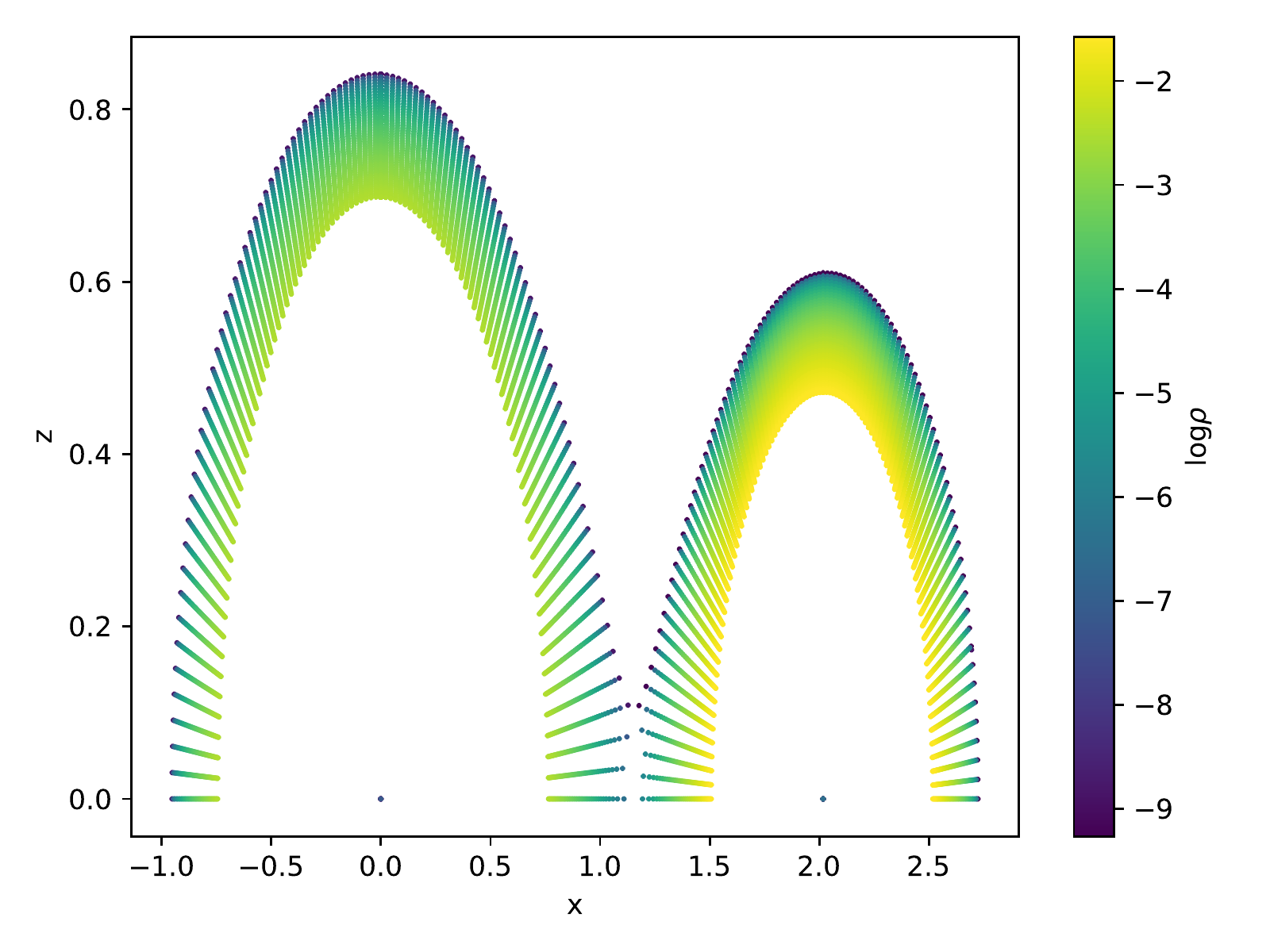}
\end{center}
\caption{The $xz$ intersection of the nested equipotential grids of a uniformly rotating single star (left) and a contact binary with mass ratio $q=0.5$ and fillout factor $FF=0.05$. The potential range of both is $\Delta\Omega = 0.5$ and the grid size per star is $(N_{\Omega},N_{\theta},N_{\varphi}) =  (50,50,50)$. The color bars represent the logarithm of the density in solar units.}
\label{fig:neq_grid}
\end{figure}

The volume grid (Figure~\ref{fig:neq_grid}) of a star or a binary component is computed in a set of equidistant potentials ($\Omega$) discretized in equidistant colatitude ($\theta$) and longitude ($\phi$) values. Currently supported potential functions are the Roche model for binary stars and gravitational and rotating potentials of single stars (Table~\ref{tab:potentials}). All potentials are dimensionless - the lengths in the Roche potential scale with the semi-major axis, while in the gravitational and rotational potential for single stars they scale with the radius of the star. The forms of the Roche potentials are adapted from \citet{wilson79}, while the rotational form is adapted from \citet{mohan1991}. $q$ is the mass ratio, $r$ the radial distance, $\lambda=\sin\theta\cos\phi$ and $\nu=\cos\theta$ are the direction cosines, $d$ is the instantaneous distance between the components of an eccentric binary and $F$ is the synchronicity parameter in the case of asynchronously rotating components. In the rotational potential, $s$ is the distance from the axis of rotation and $(b_0,b_1,b_2)$ are the differential rotation model parameters.

\begin{table}[htb]
\centering
\caption{Currently supported potentials for the nested-equipotential grid in COBAIN.}
\label{tab:potentials}
\begin{tabular}{@{}ll@{}}
\toprule
Potential & Equation \\ \midrule
Roche (circular, synchronous) & $\Omega (\vec{r},q) = \frac{1}{r} + q \left(\frac{1}{\sqrt{1-2\lambda r + r^2}} - \lambda r\right) + \frac{1}{2}(q+1)r^2(1-\nu^2)$ \\
Roche (eccentric, asynchronous) & $\Omega(\vec{r},q,d) = \frac{1}{r} + q \left(\frac{1}{\sqrt{d^2-2\lambda r d + r^2}} - \frac{\lambda r}{d^2}\right) + \frac{1}{2}F^2(q+1)r^2(1-\nu^2)$ \\
Gravitational + rotational & $\Omega (\vec{r},\omega) = \frac{1}{r} + \frac{1}{2} s^2\omega^2 = \frac{1}{r} + \frac{1}{2}s^2\left(b_0+ \frac{1}{2} b_1 s^2 + \frac{1}{3} b_2 s^4\right)$ \\ \bottomrule
\end{tabular}
\end{table}

For each $(\Omega,\theta,\varphi)$ combination, the corresponding radius vector is computed iteratively with the Newton-Raphson method \citep{wd1971}. The resulting nested grid is thus not regular in $(x,y,z)$ nor $(r,\theta,\varphi)$ but rather in $(\Omega, \theta, \varphi)$ and can be used in fast and efficient multidimensional interpolation of all structural quantities in multiple iterations. A disadvantage of this grid is the size required to cover the atmosphere, which greatly cripples its resolution - grids of about $(N_{\Omega}, N_{\theta}, N_{\varphi}) = (50,50,50)$ per star/component are being used at the moment to avoid the creation of large grid files and long CPU times for one iteration of the radiative transfer code. To further reduce the computation time, a symmetry with respect to the $xy$ and $xz$ planes is assumed:
\begin{equation}\label{gridsymmetry}
f(\theta) = f(\pi - \theta) \;\;\; \mathrm{and} \;\;\; f(\varphi) = f(-\varphi) 
\end{equation}
and only a quarter of the mesh is built in the range $\theta \in [0, \pi/2]$ and $\varphi \in [0, \pi]$. The remaining values are populated utilizing the assumed symmetry of the grid. To ensure a more accurate interpolation, the differences between the values of the regular grid points need to be sufficiently small. As the range of spherical angles is already predetermined, the toll on the size is paid in the range of possible potential values, thus confining this grid to the outermost layers of the star. 

The points corresponding to the $(\theta,\varphi)$ values of the neck in contact binaries cannot be covered because of their diverging Newton-Raphson solutions. They are nonetheless kept in the contact binary mesh with a zero value of the radius, to ensure the regularity of the grid for interpolation (however, there is no interpolation in the diverging points, all radiative properties there are computed analytically through the black body approximation instead). 

\subsection{Stellar structure}

In order to compute the radiative transfer, the mesh first needs to be populated with structural quantities (density, pressure, temperature, opacities, etc). As this step is independent from the radiative transfer, the structure can be based on any desired model that can be easily parametrized on a regular grid, which provides the liberty to explore all available hydrodynamical models and test new models of contact binaries. Due to their simple parametrization, polytropic solutions of differential rotation models \citep{mohan1991} and tidally distorted stars \citep{mohan1983} are being used in this work. 

The amount and complexity of the contact binary structure models make them impractical to use at this point of the development of the code, and for that purpose a simplified treatment of the structure of contact binaries is retained in the pilot phase, using  tidally distorted polytropic solutions \citep{mohan1983}. This treatment will be replaced by the results from hydrodynamical models of the contact binary structure \cite{lucy1961,flannery1976,kahler2002a,kahler2002b,kahler2003,stepien2009}. The mass of the primary star and mass ratio of the system are used to compute the corresponding undistorted radii and effective temperatures under the assumption of main-sequence stars:
\begin{equation}
  \begin{split}
\frac{R}{R_{\odot}} & = \left(\frac{M}{M_{\odot}}\right)^{\zeta} \\
\frac{T}{T_{\odot}} & = \left(\frac{M}{M_{\odot}}\right)^{\frac{1}{4}\left(\eta + \frac{1}{2\zeta}\right)}
  \end{split}
 \end{equation}
where
\begin{equation}
  \begin{split}
  \zeta & = 0.8 \;,  \; \; M < M_{\odot} \\
  \zeta & = 0.57\;,  \; \; M > M_{\odot}
  \end{split}
 \end{equation}
and
\begin{equation}
  \begin{split}
  \eta & = 2.5 \;, \; \; M < 0.25 M_{\odot} \\
  \eta & = 4 \;,  \; \;  0.25 M_{\odot} < M < 10 M_{\odot} \\
  \eta & = 3 \;, \; \; M > 10 M_{\odot}
  \end{split}
 \end{equation}
The structure of each star is then populated with the corresponding tidally distorted polytropic solutions via:
\begin{equation}\label{eq:polytrope_fs}
\begin{split}
P_{\psi}  & = P_c \theta_{\psi}^{n+1}  \\
\rho_{\psi}  & = \rho_c \theta_{\psi}^n \\
T_{\psi}  & = T_c \theta_{\psi} 
\end{split}
\end{equation}
where $P_c$, $\rho_c$ and $T_c$ are the central pressure, density and temperature, respectively, while $P_{\psi}$, $\rho_{\psi}$, $T_{\psi}$, $g_{\psi}$ are the mean values of the pressure, density, temperature and surface gravity averaged over an equipotential. $\theta_{\psi}$ is the dimensionless Lane-Emden variable in the polytropic solution. Currently, the structure of the two components of a contact binary is populated separately based on the given stellar mass. Additional smoothing in the neck region or weighted averaging of the structural values in the equipotentials above the critical one can be introduced to account for the expected mixing of the material. However, populating the structure of contact binaries with tidally distorted polytropes is an over-simplified treatment and used for pilot purposes and proof of concept only. Part of the main focus of future work will be replacing it with a more complex hydrodynamical model of contact binary stars.

Once the grid is populated with a hydrodynamical model of the stellar interior, the opacity of each grid point is computed with the use of OPAL \citep{iglesias1996} opacity tables. COBAIN currently uses interpolation in the solar-type fixed chemical composition tables. 

\subsection{Generalized radiative transfer in an arbitrary grid and stellar structure}

When the assumptions of plane-parallel, one- or multi-dimensional spherically symmetric atmospheres are abandoned, all simplifications and analytical approximations of the variation of the stellar parameters across the grid no longer hold and their true distributions across the grid need to be taken into account in the propagation of radiation. The core of COBAIN lies in the short characteristics method \citep{VanNoort_Hubeny}, where the intensity in each grid point is computed by propagating discretized rays in a predefined set of directions spanning the whole unit sphere around each grid point and numerically integrating the formal solution of the radiative transfer equation:

\begin{equation}
I (s) = I(s_0) e^{-\Delta\tau} + \int_{0}^{\Delta\tau} S(t) e^{-(\Delta\tau - t)} \diff t \; .
\label{eq:rt_fs5}
\end{equation}

The short characteristics method propagates a ray until it intersects a side of a grid cell. The values of the relevant properties in the intersecting points, like opacity, source function and intensity, are interpolated from the neighbouring points of the grid by assuming an analytical form of the property variation across the grid cell (most typically quadratic, but more advanced forms are being used as well \citep{ibgui_sc}). The formal solution is then integrated using three points (central point, in which the intensity is being computed, downwind point, which is the intersection with the cell the radiation propagates from, and upwind point, which is the intersection with the cell where the radiation propagates to) and assuming an analytical form of the source function variation with optical depth. This method works reasonably well on single star atmospheres in simple geometries (cylindrical, spherical, 2D and 3D plane-parallel atmospheres), but is insufficient in the case of contact binaries due to the complex geometry and potentially complex hydrodynamical models of their structure. To overcome these issues, \textsc{COBAIN} chooses the propagation end-points adaptively, instead of confining them to the grid cells, while the structural quantities are interpolated from the entire grid instead of only the neighboring points, in about $\sim1000$ points along the ray, instead of only three. This avoids the need to assume an analytical form of the variation of the structural quantities along the ray, which can introduce significant deviations from their true values. Numerical deviations can still be present based on the choice of the grid interpolating function, but, in this case, they can be easily decreased with a finer grid or controlled by a rescaling function (Sect.~\ref{sect:parallelization}). 



The numerical integration of the formal radiative transfer solution for each point $s$ requires the knowledge of the intensity value in a starting point $s_0$ and the values of the source function in all discretized points along the ray from $s_0$ to $s$. The initial values of these quantities are computed through the gray approximation: $ I = S = \sigma T^4/\pi$. 

In each subsequent iteration, the values of the previous iteration are used in the integration of the formal solution. The solution of the intensity at point $s$ in iteration $i$ is then:
\begin{equation}\label{eq:formal_solution}
I_i(s) = I_{i-1}(s_0) + \int_{0}^{\Delta\tau_{i-1}} S_{i-1}(t) e^{-(\Delta\tau_{i-1} - t)} \diff t \; ,
\end{equation}
where
\begin{equation}\label{eq:optical_depth}
\Delta\tau_{i} = \int_{s_0}^{s} \rho_{i-1}(s') \kappa_{i-1} (s') \diff s' \; .
\end{equation}

If we assume that the total energy flux is equal to the radiative flux, at each point inside the star all emitted energy must equal all extinct energy flux (obtained through the mean over all directions of intensity propagation). 
%
%
In gray atmospheres, where the dependence on frequency of the quantities is replaced with a mean value, the integral reduces to $J=S$. Therefore, in the current gray treatment of atmospheres in \textsc{COBAIN}, the source function after each iteration is computed through the mean intensity value in each point:
\begin{equation}\label{eq:re_cobain}
S_{i} = J_{i} \; .
\end{equation}
The radiative transfer computation is iterated until $|\overline{J_i - J_{i-1}}|/|\overline{J_i}| < \varepsilon$ in each grid point, with the user-defined convergence threshold $\varepsilon$.


The mean intensity and flux are computed through integration of the specific intensity over the solid angle using Lebedev quadratures \citep{lebedev}. The choice of Lebedev quadratures is not coincidental - they offer the most uniformly distributed set of directions over the unit sphere, 
which is of major importance when the function that needs to be integrated can vary greatly in different directions. In stellar atmospheres this is especially true for points near the surface where the intensity drops quickly in directions facing away from the surface normal. The uniform distribution of quadrature directions is thus essential for consistent weighting of the significant contributions to the mean intensity and flux integrals.
 
To further ensure that there is an equal number of directions facing inward and outward, the quadrature directions are rotated with respect to the equipotential normal $\hat{n}$ in each grid point.  The simplest way to achieve this is through the computation of the transformation matrix that rotates the coordinate system of the quadrature directions into the orthonormal coordinate system spanned by the normal and tangential plane of the equipotential surface at point $\vec{r}$.
 
The normal of the equipotential $\Omega$ at point $\vec{r} = (x,y,z)$ is given by:
 \begin{equation}
 \vec{n} = (n_x, n_y, n_z) = \frac{1}{|\nabla \Omega|}\left(\frac{\diff\Omega}{\diff x}, \frac{\diff\Omega}{\diff y}, \frac{\diff\Omega}{\diff z} \right) \; .
 \end{equation}
 The tangential plane is defined by two vectors orthogonal to the normal and to each other:
 \begin{equation}
 \vec{t_1} = (t_{1x}, t_{1y}, t_{1z}) =  (n_y, -n_x, 0) \;\;\;\;\;\; \mathrm{and} \;\;\;\;\;\; \vec{t_2} = (t_{2x}, t_{2y}, t_{2z})= \vec{n} \times \vec{t_1} \; .
 \end{equation}
 
The quadrature coordinate system is spun by the unit axis vectors $(\hat{i},\hat{j},\hat{k})$:
 
The transformation matrix from the quadrature to the normal equipotential coordinate system is computed as:
\begin{center} 
\(
R_{\Omega} = 
\begin{bmatrix}
    \hat{i}\cdot\vec{t_1} & \;\; \hat{i}\cdot\vec{t_2} & \;\;  \hat{i}\cdot\vec{n} \\
    \hat{j}\cdot\vec{t_1} & \;\;   \hat{j}\cdot\vec{t_2} & \;\;  \hat{j}\cdot\vec{n} \\
    \hat{k}\cdot\vec{t_1} &  \;\; \hat{k}\cdot\vec{t_2} &  \;\; \hat{k}\cdot\vec{n}  
\end{bmatrix}
\)
\end{center}
Each quadrature direction $\vec{d}_Q$ is then rotated into the coordinate system of the equipotential surface:
 \begin{equation}
 \vec{d}_{\Omega} = R_{\Omega} \vec{d}_Q \; .
 \end{equation}
 
This ensures that all directions with latitude $0 < \theta < \pi/2$ in the original quadrature coordinate system are pointing outwards with respect to the equipotential surface, while all directions with $\pi/2 <\theta < \pi$ are pointing inwards.
 
The integrals of the mean intensity and flux are computed as:
 \begin{equation}
 J = \frac{1}{4 \pi} 4\pi \sum_{i=0}^{N-1} w_i I_i = \sum_{i=0}^{N-1} w_i I_i 
 \end{equation}
 \begin{equation}
 F = 4 \pi \left(\sum_{i=0}^{N/2} w_i^{out} I_i^{out} - \sum_{i=N/2}^{N-1} w_i^{in} I_i^{in}\right)
 \end{equation}
where $w_i^{out}$ and $I_i^{out}$ correspond to the outward directions with colatitude $0 < \theta < \pi/2$, while $w_i^{in}$ and $I_i^{in}$ correspond to the inward directions with colatitude $\pi/2 <\theta < \pi$.
 
The blackbody temperature in each point of the grid is then recomputed as $T = (\pi J/\sigma)^{1/4}$ and new OPAL opacities are interpolated from the value of density (which is kept unchanged) and the new values of grid temperatures.
 
\begin{figure}[ht]
\begin{center}
\includegraphics[width=\hsize]{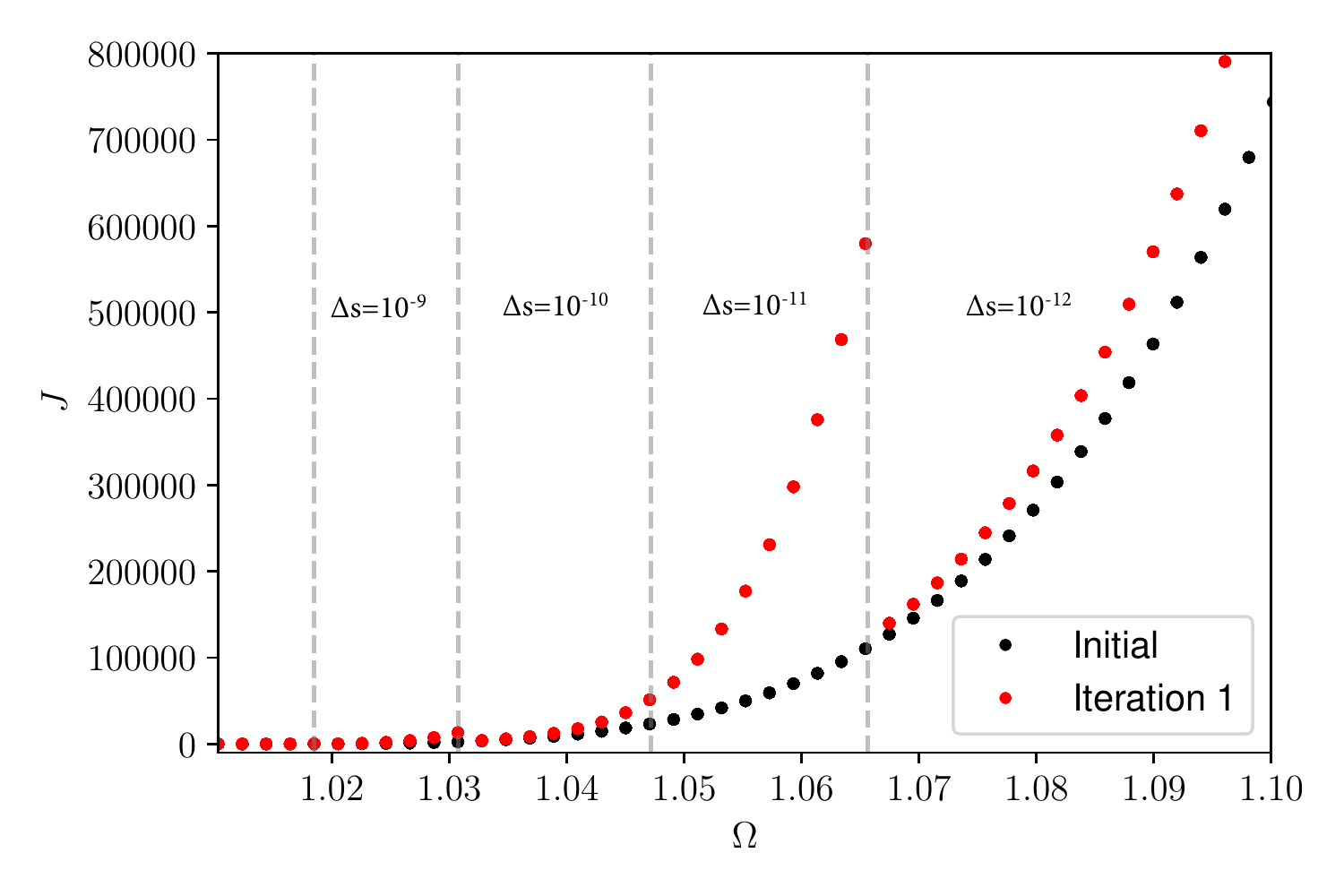}
\end{center}
\caption{The effect of fixing the step size within different regions of the star: the more the mean free path decreases (towards higher potentials deeper in the atmosphere), the larger the numerical errors that arise due to the insufficient discretization of the ray because of the deviations from the assumed local linearity of the absorption coefficient and source function. Gray dashed lines mark the regions where the step size value decreases by one order of magnitude.}
\label{stepsize}
\end{figure}

Once the quadrature directions are rotated into the normal plane of the equipotential surface at a given grid point $\vec{r}$, the rays are discretized in a set of points in all directions. A step size $\Delta s$ is computed and the discretized points are obtained via:
\begin{equation}
\vec{p}_i = \vec{r} - i \;\Delta s\; \vec{d}_{\Omega}  \;\;\;\;\; \mathrm{for} \;\; i = 0,1,2,3,...,N
\end{equation}
The values of  $\Delta s$ and $N$ are computed based on the position of the grid point inside the star. Uniform or pre-determined step size can result in large numerical errors due to deviations from the assumed local linearity of the structural quantities along the ray (Figure~\ref{stepsize}).

The variation of all structural quantities near the atmosphere is very rapid and the discretization needs to be optimized to be roughly equal in all points and all directions. This is achieved through an empirical search of the step size equivalent to a difference in optical depth of 1 in the following way:
\begin{itemize}
\item an array of log-equidistant step sizes is created as $ \{s | s \in {10^{range(-10,-4,1000)}}\}$, where $range(-10,-4,1000)$ spans the array of exponential factors in 1000 points between $-10$ and $-4$;
\item the values of density and opacity for each resulting radius vector $\vec{p}_s = \vec{r} - s \; \vec{d}_{\Omega}$ are interpolated from the grid;
\item assuming linear variation of the absorption coefficient $\chi(s') = \rho(s') \kappa(s')$ from $\vec{r}$ to $\vec{p}_s$, an estimate of the optical depth is computed as $\Delta\tau_s = (1/2) \chi_s s$;
\item the step size resulting in a value of the optical depth closest to 1 is chosen for the ray propagation.
\end{itemize}
In some cases the variation of the absorption coefficient is extremely non-linear and the resulting step size is overestimated. In these cases, the step size value is recomputed after the first run in a way that achieves discretization in $N = 1000$ points. This is done by multiplying the initial step size by the ratio of the number of discretization steps achieved with the initial step size and 1000:
\begin{equation}
    \Delta s_{\mathrm{new}} = \Delta s_{\mathrm{old}} \frac{N_{\mathrm{old}}}{1000} .
\end{equation}
The computation is then rerun with $\Delta s_{\mathrm{new}}$.

The absorption coefficient and source function values in each ray point: 
\begin{equation}
\vec{p}_i = r_i (\sin\theta_i\cos\varphi_i,\sin\theta_i\sin\varphi_i,\cos\theta_i),
\end{equation}
can be obtained in several different regimes based on the corresponding value of the Roche potential $\Omega_i$:
\begin{itemize}
\item If $\Omega_i < \Omega_{S}$, where $\Omega_S$ is the Roche potential value at the surface, $\vec{p}_i$ lies outside of the star and $\chi_i = S_i = 0$.
\item If  $\Omega_{S} < \Omega_i < \Omega_{max}$, where $\Omega_{max}$ is the Roche potential value of the innermost equipotential of the grid, $\chi(\vec{p}_i)$ and $S(\vec{p}_i)$ are interpolated in $(\Omega_i,\theta_i,\varphi_i)$.
\item If $\Omega_i > \Omega_{max}$ or the point lies in the neck region of a contact binary not covered by the mesh, the values of density and temperature are computed from the Lane-Emden solution and used to interpolate the opacity. The absorption coefficient is computed as $\chi_i = \rho_i \kappa_i$ and the source function is computed through the gray approximation as $S_i = \sigma T_i^4/\pi$.
\end{itemize}

Numerical integration of the absorption coefficient and source function integrals in Eqs.~\ref{eq:formal_solution} and \ref{eq:optical_depth} would involve assumptions on the variation of the absorption coefficient and source function between successive ray points, which, in the case of large values of these quantities, is a source of large numerical errors. To avoid this problem, a spline function is fit through the points $\chi (s)$ and $S(\tau) e^{-\tau}$ and analytically integrated to produce the corresponding integrals. Since the ray is propagated from the point where we compute the specific intensity, the integral $ \int_{0}^{\Delta\tau} S(t) e^{-(\Delta\tau - t)} \diff t$ is replaced with the local integral $ \int_{0}^{\tau_i} S(t) e^{-t} \diff t$. The final boundaries are also inverted: $\tau=0$ at the point of interest $\vec{p}_0=\vec{r}$, while $\tau=\Delta\tau$ at the point $\vec{p}_N = \vec{r} - N \;\Delta s\; \vec{d}_{\Omega}$. 

The ray is propagated until the exponential term $e^{-\tau}$ becomes numerically indistinguishable from 0. Double-precision floating points are used and the computation is cut-off at the point where the cumulative optical depth along a ray reaches $\Delta\tau > 740$. $N$ is then the number of points between $e^{-\tau} = 1$ and $e^{-\tau} = 0$ and, depending on the step size, it ranges between $200$ and $\sim1000$.

The resulting intensity in point $\vec{r}$ and direction $\vec{d}_{\Omega}$ can then be written as:
\begin{equation}
I (\vec{r},\vec{d}_{\Omega}) = I (\vec{p}_N) e^{-\tau_N} + \int_{0}^{\tau_N} S(t) e^{-t} \diff t .
\end{equation}

The time for the specific intensity computation in one direction spans $\sim10^9$ CPU cycles due to the large number of points that need to be interpolated along the ray and scales with $\sim\mathcal{O}(N\times N_d)$, where $N$ is the number of points in the grid and $N_d$ is the number of directions. The computation time can be decreased with the choice of a larger step size, but that may significantly affect the accuracy of the integration, as demonstrated in Figure~\ref{stepsize}, thus we retain the current choice of step size computation and optimize the computation time with code parallelization.

\subsubsection{Parallelization}\label{sect:parallelization}

As the intensities in each iteration are computed using the structural values of the grid from the previous iteration, there is no inter-grid communication and the grid points can be easily split for parallelization. The main function of the code that handles radiative transfer computations takes as input an array of arguments of the grid points and reads the stored grid quantities from the previous iteration for interpolation. After the specific intensity, mean intensity and flux are computed in all points per process for a given iteration, the values of the structural arrays are stored in new files, which can be used for interpolation in the next iteration. The linear interpolation in the grid with the choice of a $\Delta\tau\approx1$ stepsize introduces a roughly constant numerical error that causes an increase of $\sim8\%$ of the values in the deeper layers where the solutions approach the black body approximation. Rescaling the values after each iteration so that the structural quantities in the near-black body regime stay approximately the same ensures that the intensities do not artificially increase after each iteration due to this numerical error. 

\section{COBAIN gray atmosphere simulations on single stars}\label{sect4}

\begin{figure}[ht]
\begin{center}
\includegraphics[width=0.495\hsize]{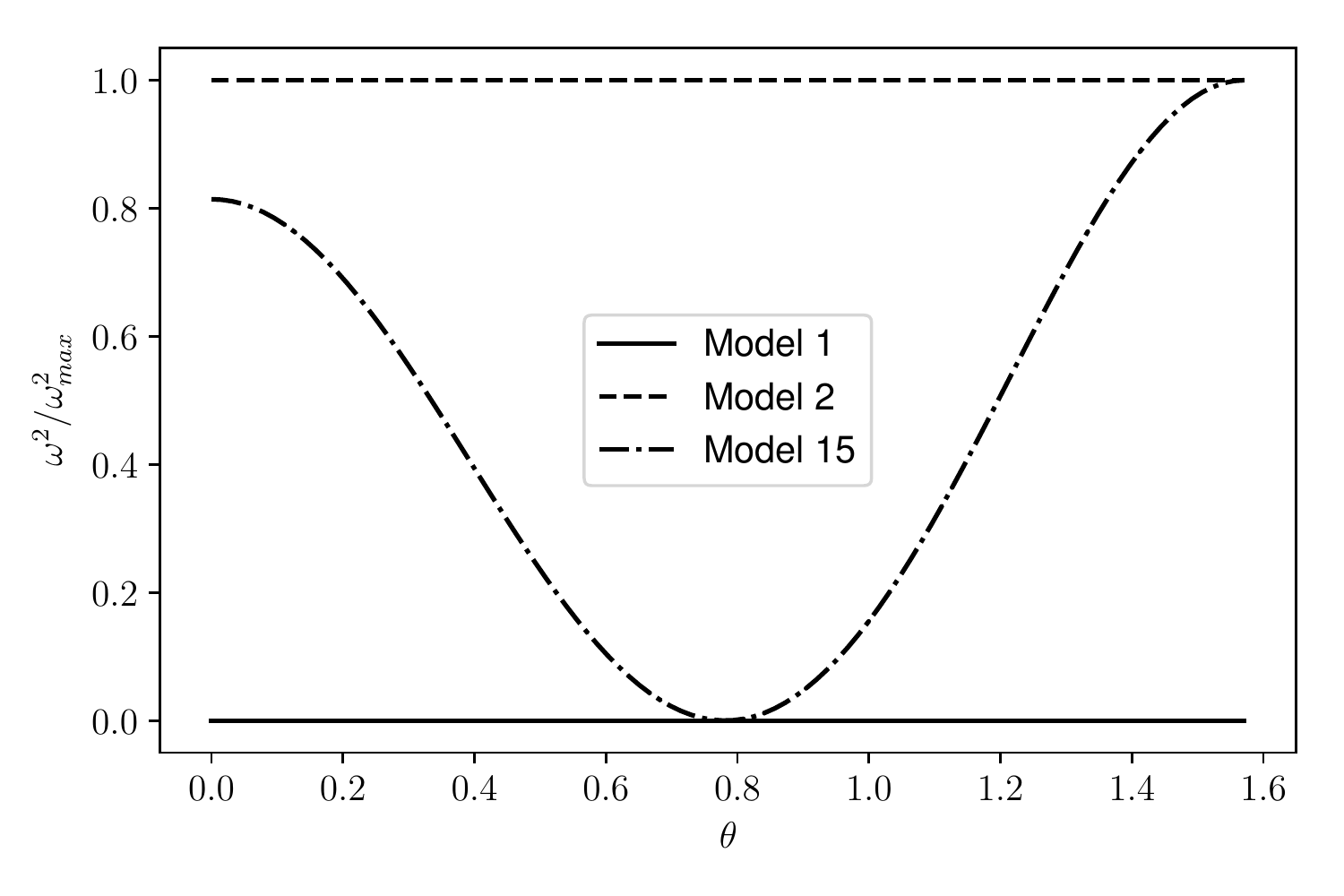}
\includegraphics[width=0.495\hsize]{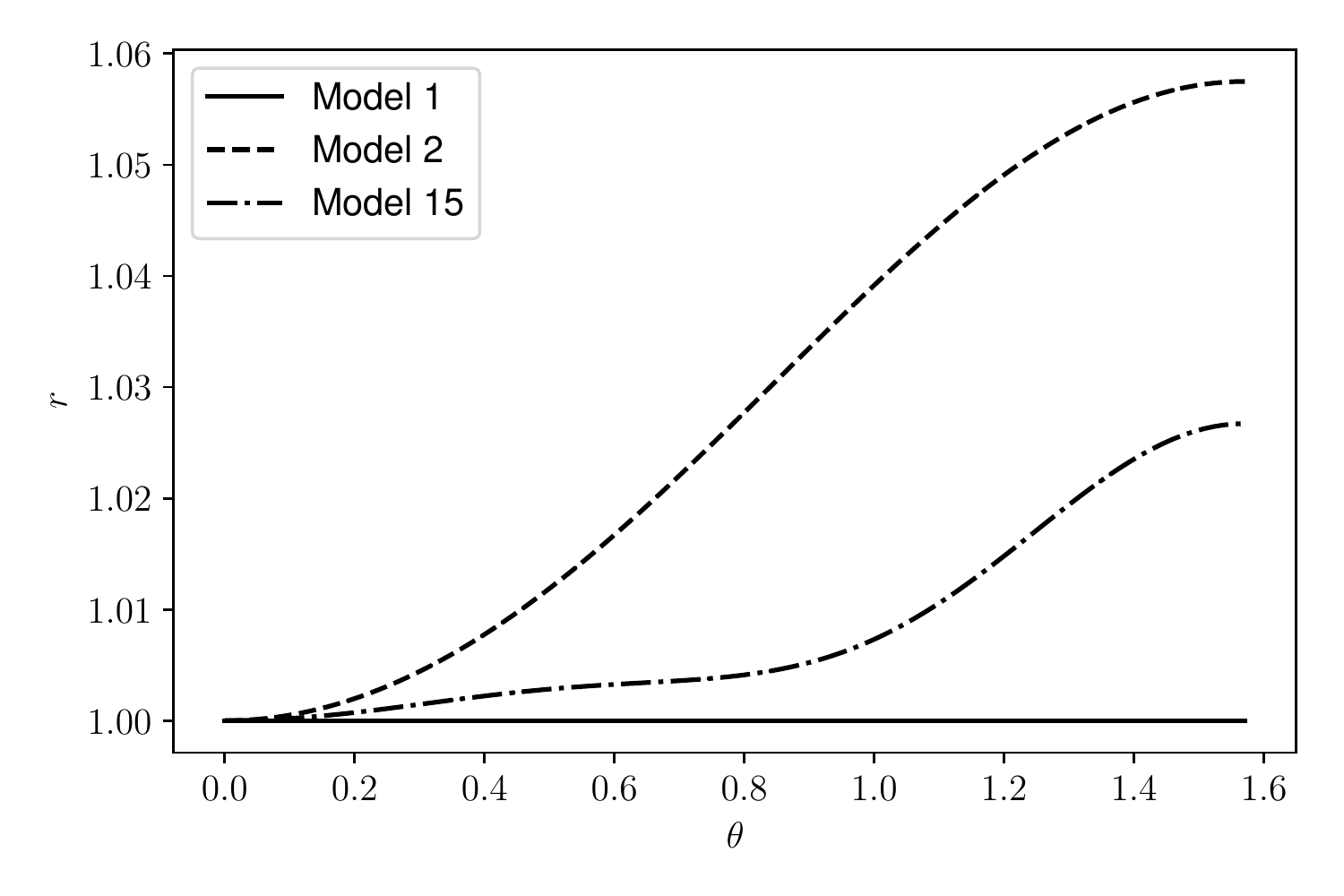}
\end{center}
\caption{The normalized angular velocity (left) and radius (right) as functions of colatitude in a non-rotating (solid line), uniformly rotating (dashed line) and an unstable (dot-dashed line) differential rotation model.}
\label{fig:diffrot_r_w}
\end{figure}

All single stellar models are computed for a solar-type star of $M = 1 \;M_{\odot}$, $R= 1 \;R_{\odot}$, and solar mean molecular mass $\mu = 0.6$. The polytropic solutions are computed for a radiative atmosphere with a polytropic index $n=3$ and model parameters adapted from \citet{mohan1991} (cf. Table~\ref{tab:diffrot}). The resulting angular velocity functions and geometrical configurations of the stellar surface are depicted in Figure~\ref{fig:diffrot_r_w}. The nested equipotential grids of the stars are constructed in $(N_{\Omega,i},N_{\theta,i},N_{\varphi,i}) = (100,50,50)$, and the range of potentials covered with the grid is $\Delta\Omega = 0.01$ regardless of the surface potential value.

\begin{table}[ht]
\centering
\caption{Differential rotation models of a non-rotating body, a solid body rotator and an unstable differential rotator. Adapted from \citep{mohan1991} with the same model numbers as in Table 1 of the paper.}
\label{tab:diffrot}
\begin{tabular}{@{}ccllcc@{}}
\toprule
Model No. & \multicolumn{3}{c}{$(b_0,b_1,b_2)$}     & Model description                                                                                  & Model stability \\ \midrule
1         & \multicolumn{3}{c}{$(0.0, 0.0, 0.0)$}   & Non-rotating model                                                                                 & Stable          \\
2         & \multicolumn{3}{c}{$(0.1,0.0,0.0)$}     & \begin{tabular}[c]{@{}c@{}}Solid body rotation \\ ($\omega^2=0.1$)\end{tabular}                    & Stable          \\
15        & \multicolumn{3}{c}{$(0.04,-0.16,0.16)$} & \begin{tabular}[c]{@{}c@{}}Differential rotation model\\  ($0.0 \leq \omega^2 \leq 0.04$)\end{tabular} & Unstable        \\ \bottomrule
\end{tabular}
\end{table}

We have chosen these particular polytropes to test whether the results of the radiative transfer simulation match the expected outcome for the stable models - a uniform normal emergent intensity and flux distribution for the non-rotating model and a gravity-darkened distribution for the uniformly rotating star. The last model (Model 15 in Table~\ref{tab:diffrot}) is constructed with a choice of model parameters that yield an unstable differential rotator with a peculiar geometry - the angular velocity, and, consequently, the radius, decrease half-way between the pole and the equator. This model was used to test how well the radiative transfer code follows the geometry of the grid. Results of the radiative transfer simulations are presented below.

\subsection{Non-rotator}

\begin{figure}[htb]
\begin{center}
\includegraphics[width = 0.48\hsize]{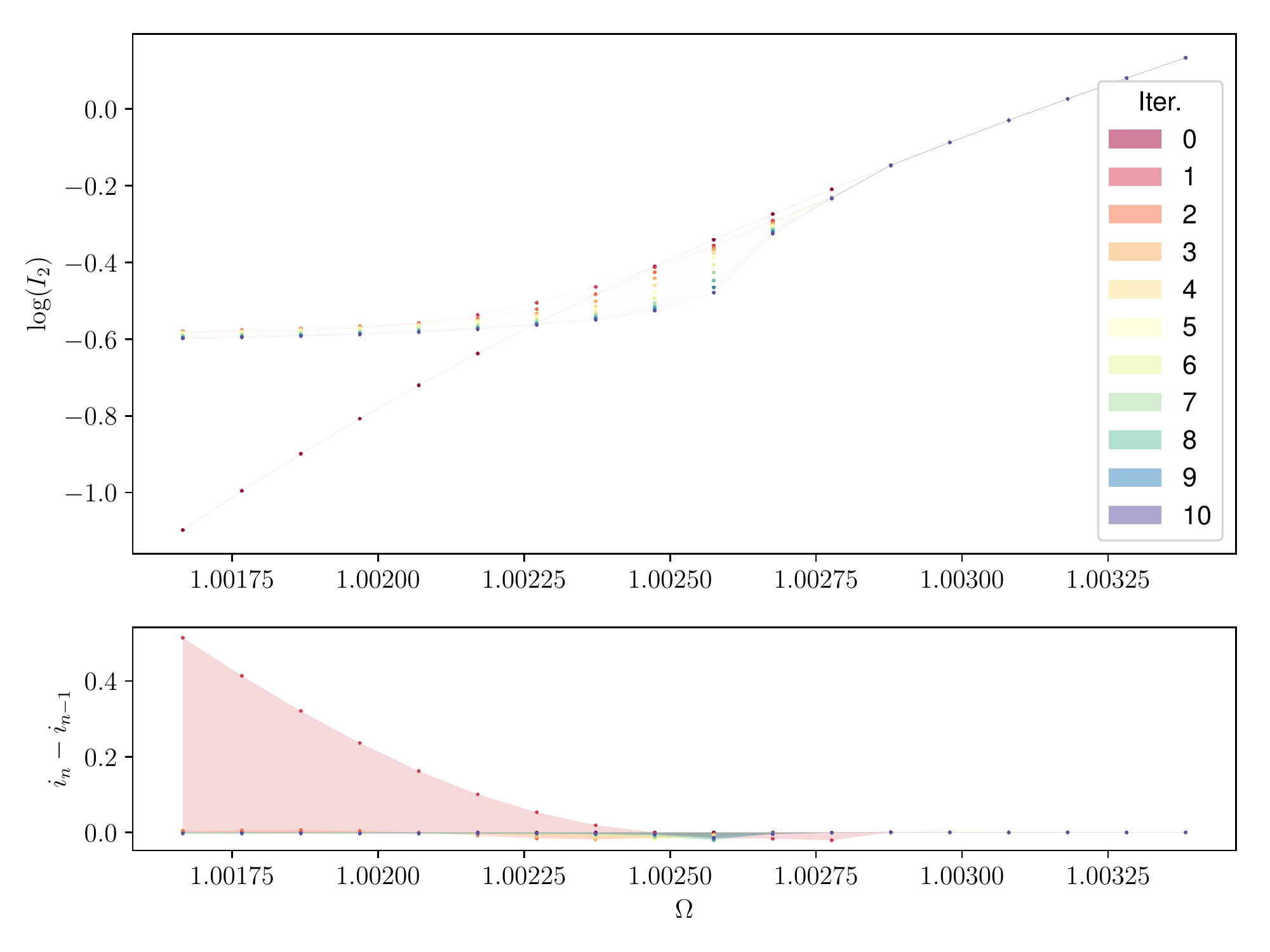}
\includegraphics[width = 0.48\hsize]{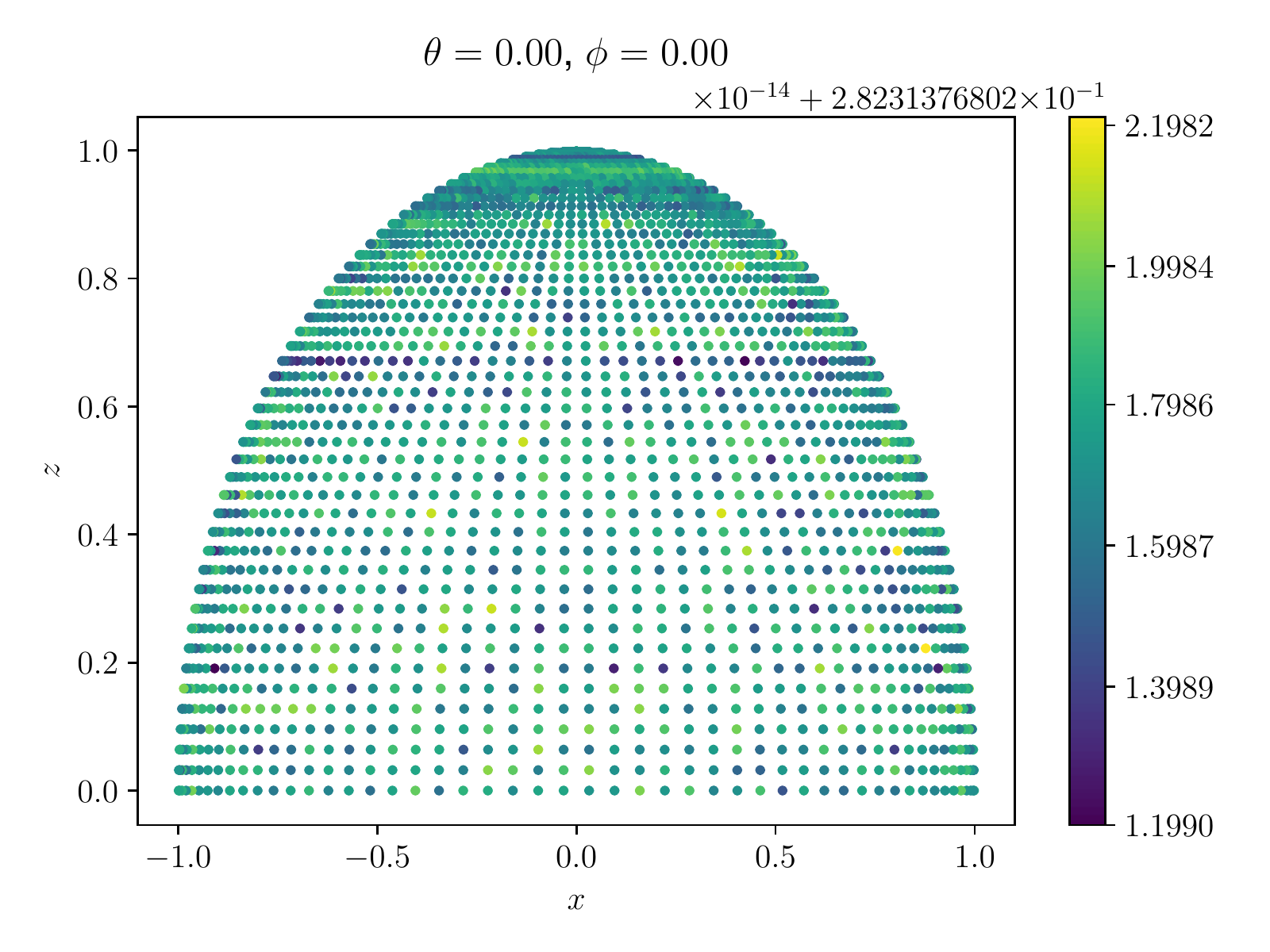}
\end{center}
\caption{The normal outward intensity as a function of the potential (top left) with differences between successive iterations (bottom left) and its surface distribution (right) for a non-rotating radiative polytrope.}
\label{fig:m1_I2}
\end{figure}

\begin{figure}[htb]
\begin{center}
\includegraphics[width = 0.48\hsize]{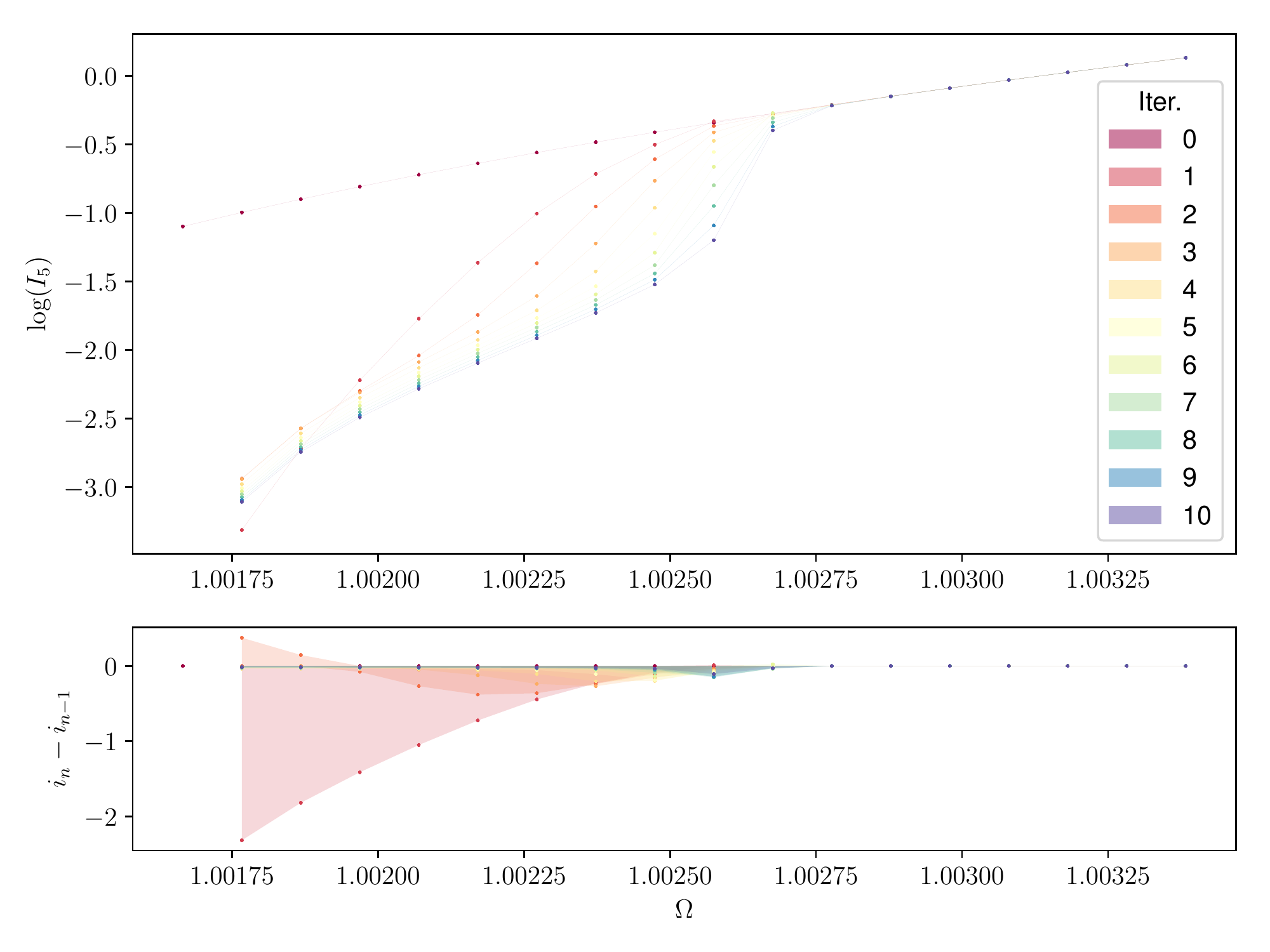}
\includegraphics[width = 0.48\hsize]{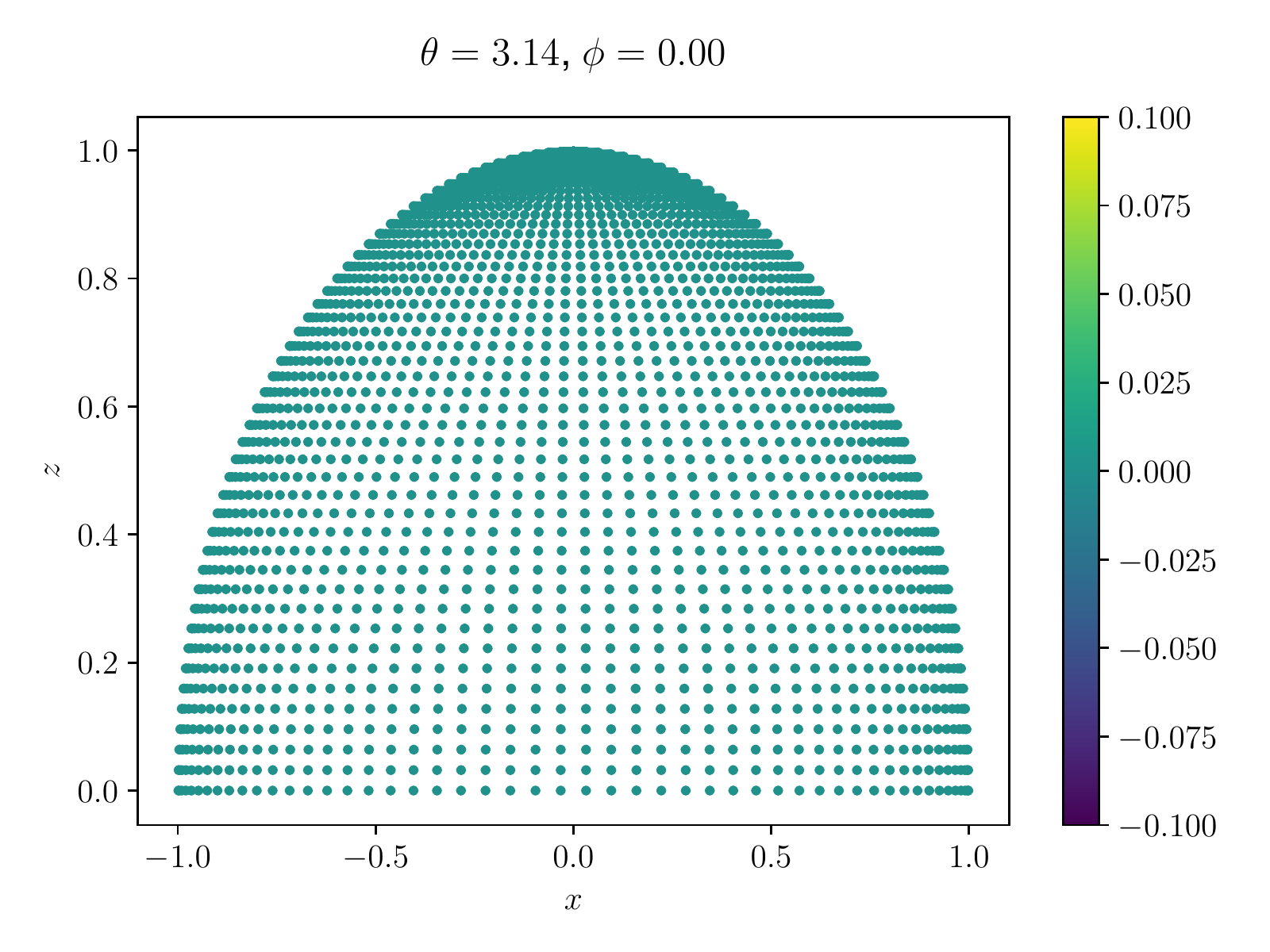}
\end{center}
\caption{The normal inward intensity as a function of the potential (top left) with differences between successive iterations (bottom left) and its surface distribution (right) for a non-rotating radiative polytrope.}
\label{fig:m1_I5}
\end{figure}

\begin{figure}[htb]
\begin{center}
\includegraphics[width = 0.48\hsize]{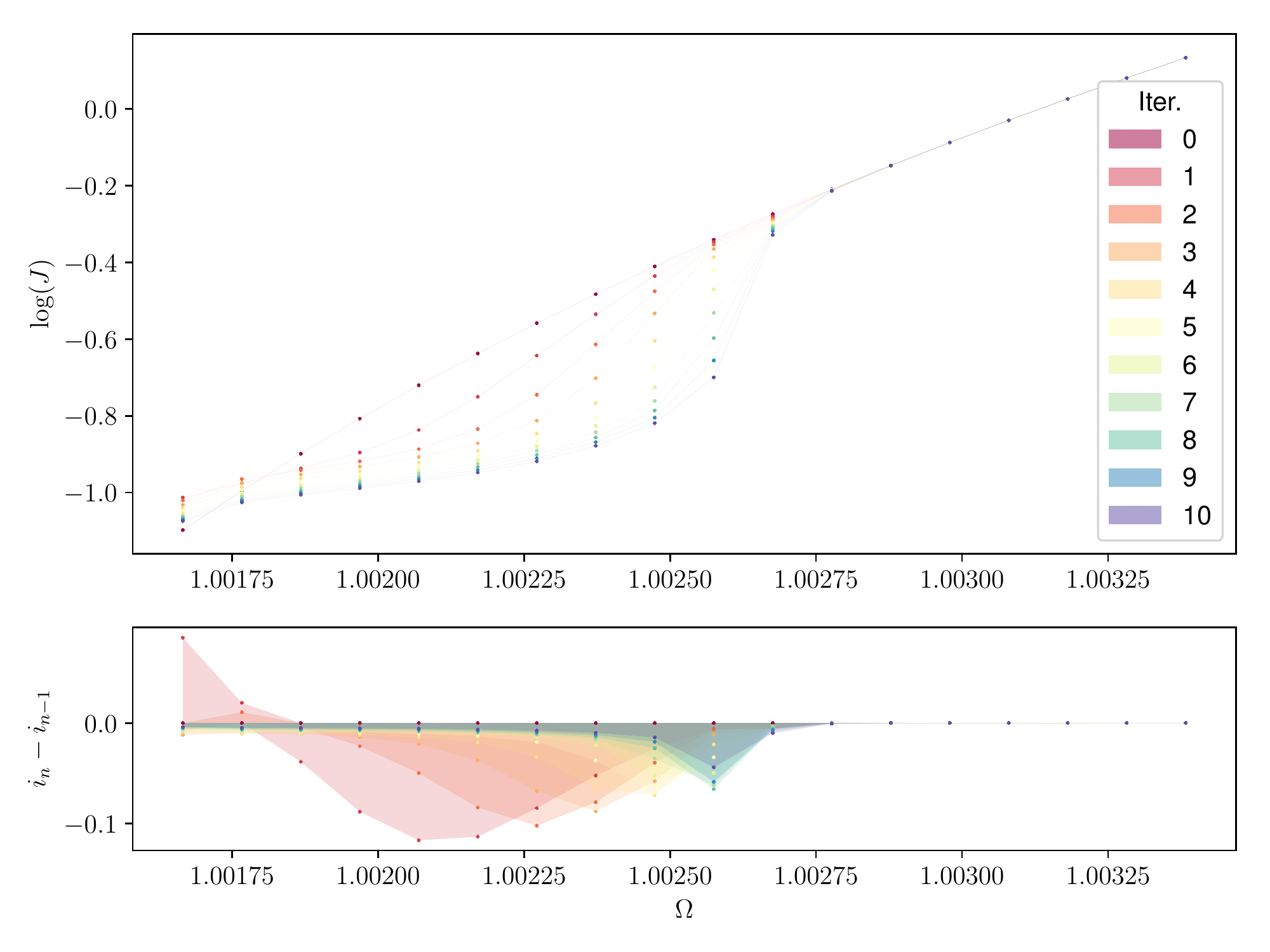}
\includegraphics[width = 0.48\hsize]{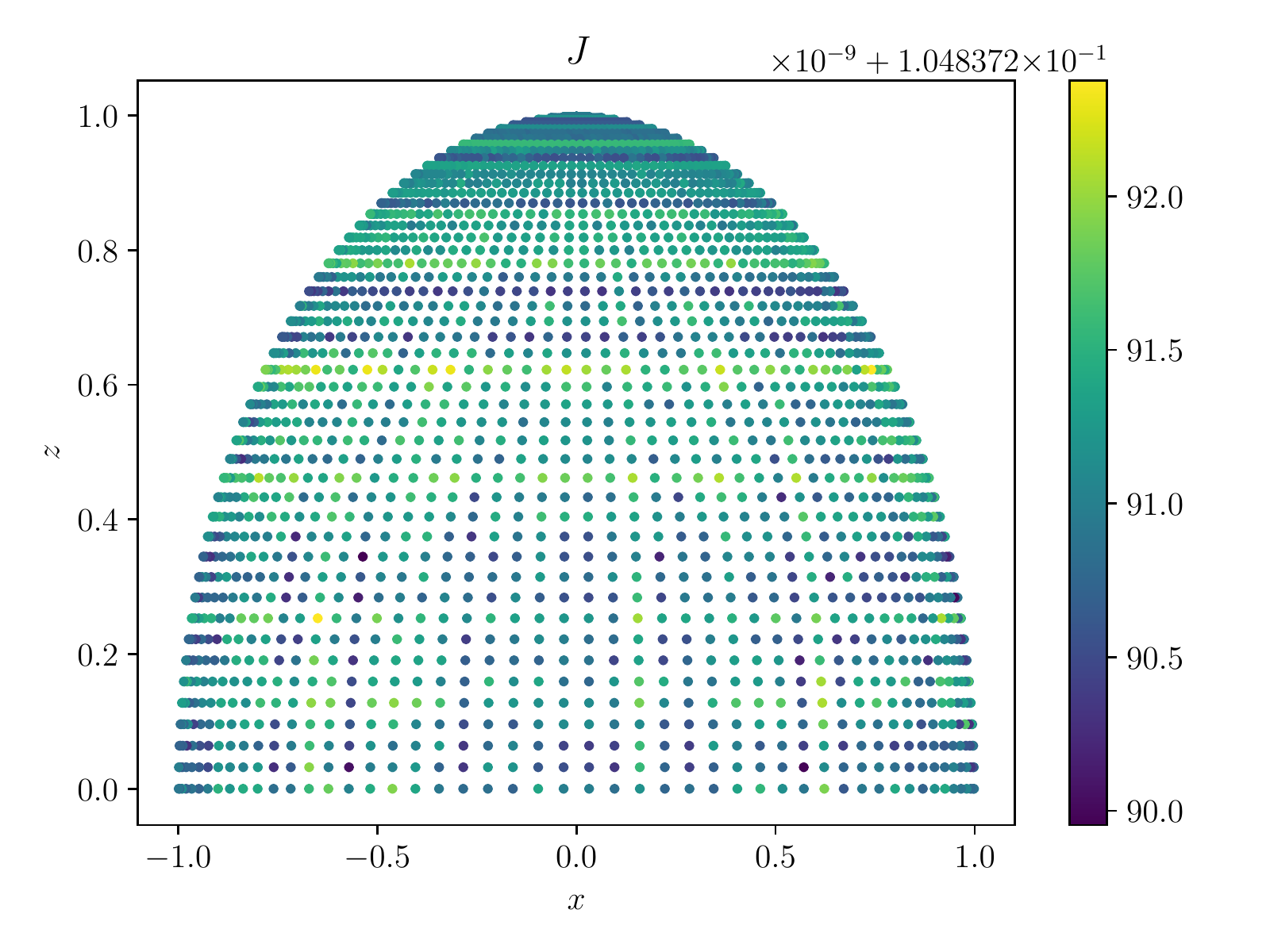}
\end{center}
\caption{The mean intensity as a function of the potential (top left) with differences between successive iterations (bottom left) and its surface distribution (right) for a non-rotating radiative polytrope.}
\label{fig:m1_J}
\end{figure}

For a well-performing code the non-rotator is expected to yield a uniform distribution of the normal emergent intensity and its integrals. The right panels of Figures \ref{fig:m1_I2} - \ref{fig:m1_J} show the normal outward, inward and mean intensity distributions over the surface of the star. The differences are of the order of $\sim 10^{-14}$, hence purely numerical, and the surface distributions are essentially uniform. The Left panelss of Figures \ref{fig:m1_I2} - \ref{fig:m1_J} show the normal outward, inward and mean intensity distributions as functions of the potential. As expected, deeper into the atmosphere, the solution approaches the initial blackbody conditions, while in the upper layers of the atmosphere there is a clear deviation from the blackbody solution. The initial decrease of intensity with respect to blackbody is due to the break of symmetry of incoming radiation (less propagates inward than outward), which can also be seen in Figure~\ref{fig:m1_I5}, where there is a steep decrease of inward intensity and reaches zero at the surface. On the other hand, in the layers close to the surface, the outward intensity reaches an almost constant value due to the decrease in density and temperature that decrease the optical depths and increase the mean free path of the rays. This behavior is visible in all radiative transfer models computed with COBAIN, while differences arise in the ranges of intensity values that are determined by the particular geometry of the model equipotentials.



\subsection{Rigid rotator}

\begin{figure}[htb]
\begin{center}
\includegraphics[width = 0.48\hsize]{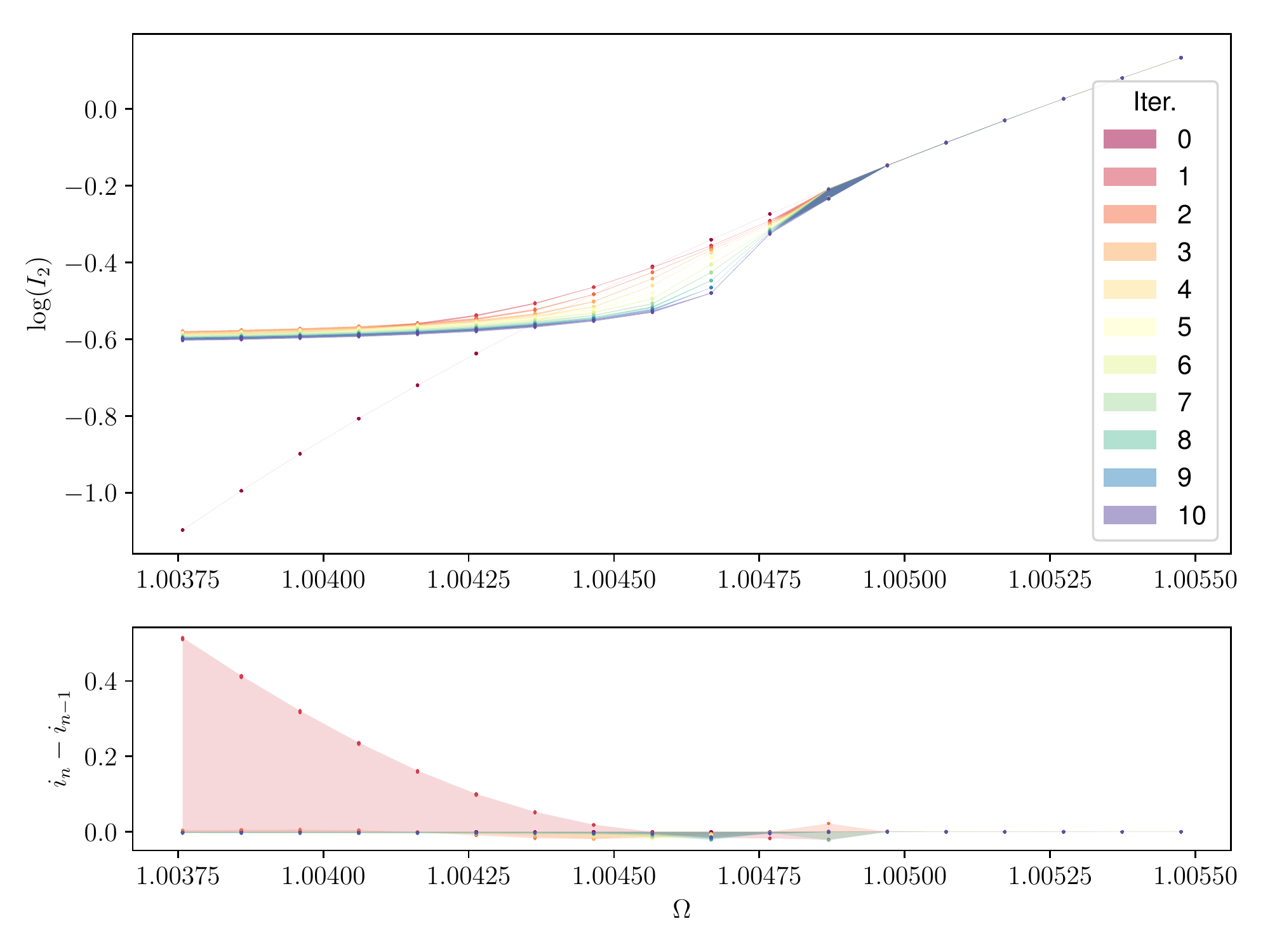}
\includegraphics[width = 0.48\hsize]{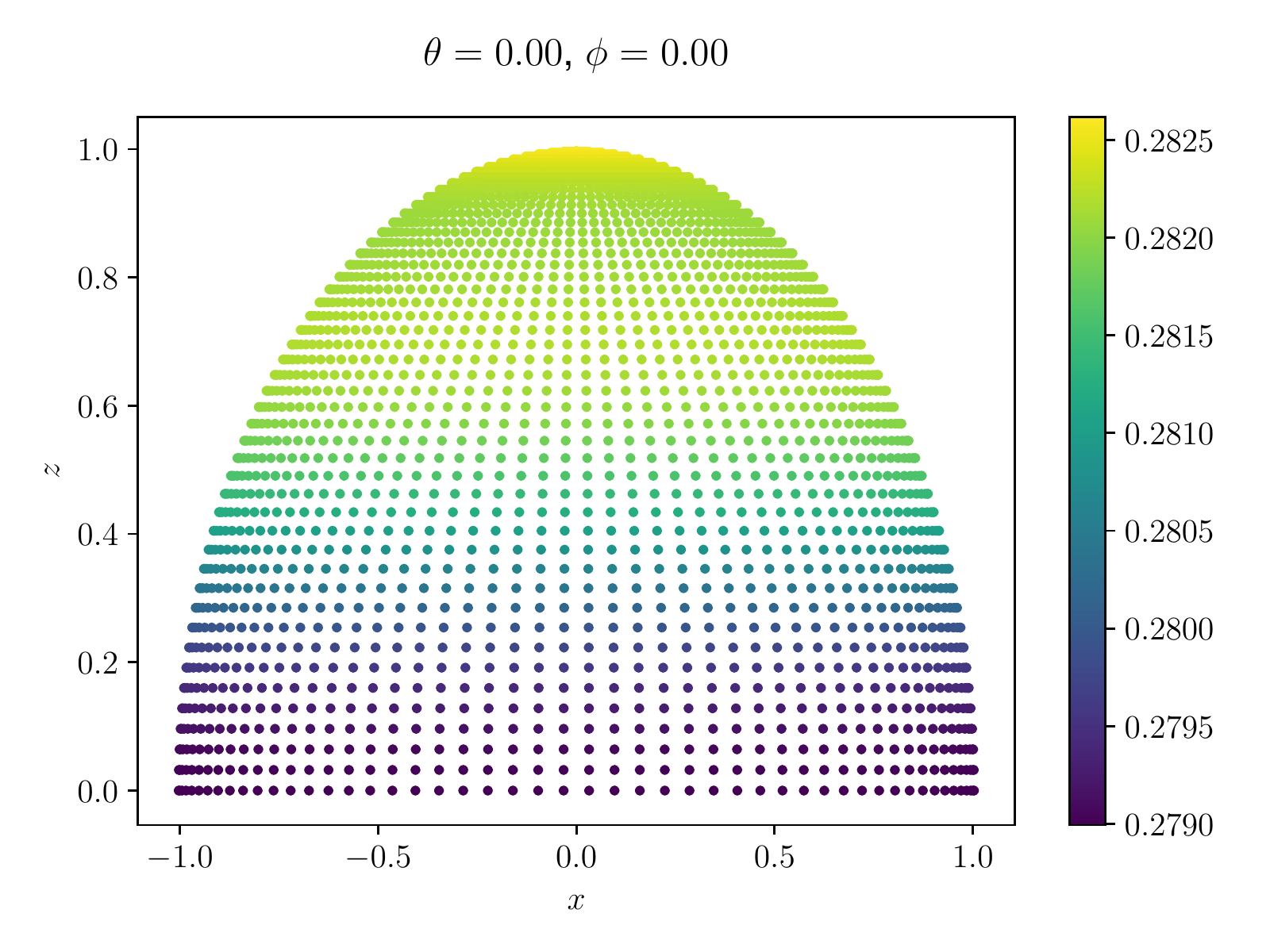}
\end{center}
\caption{The normal outward intensity as a function of the potential (top left) with differences between successive iterations (bottom left) and its surface distribution (right) for a uniformly rotating radiative polytrope.}
\label{fig:m2_I2}
\end{figure}

\begin{figure}[htb]
\begin{center}
\includegraphics[width = 0.48\hsize]{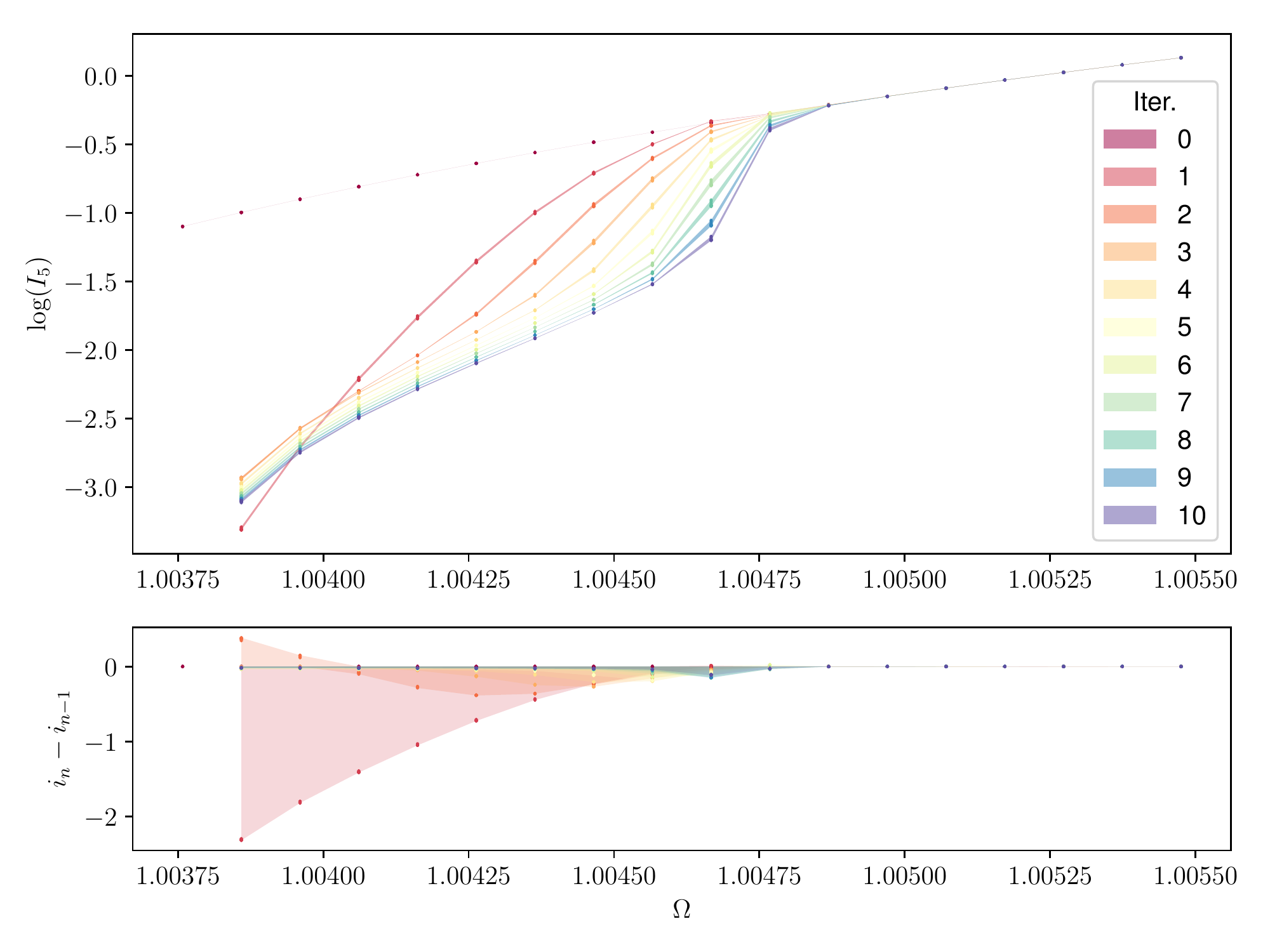}
\includegraphics[width = 0.48\hsize]{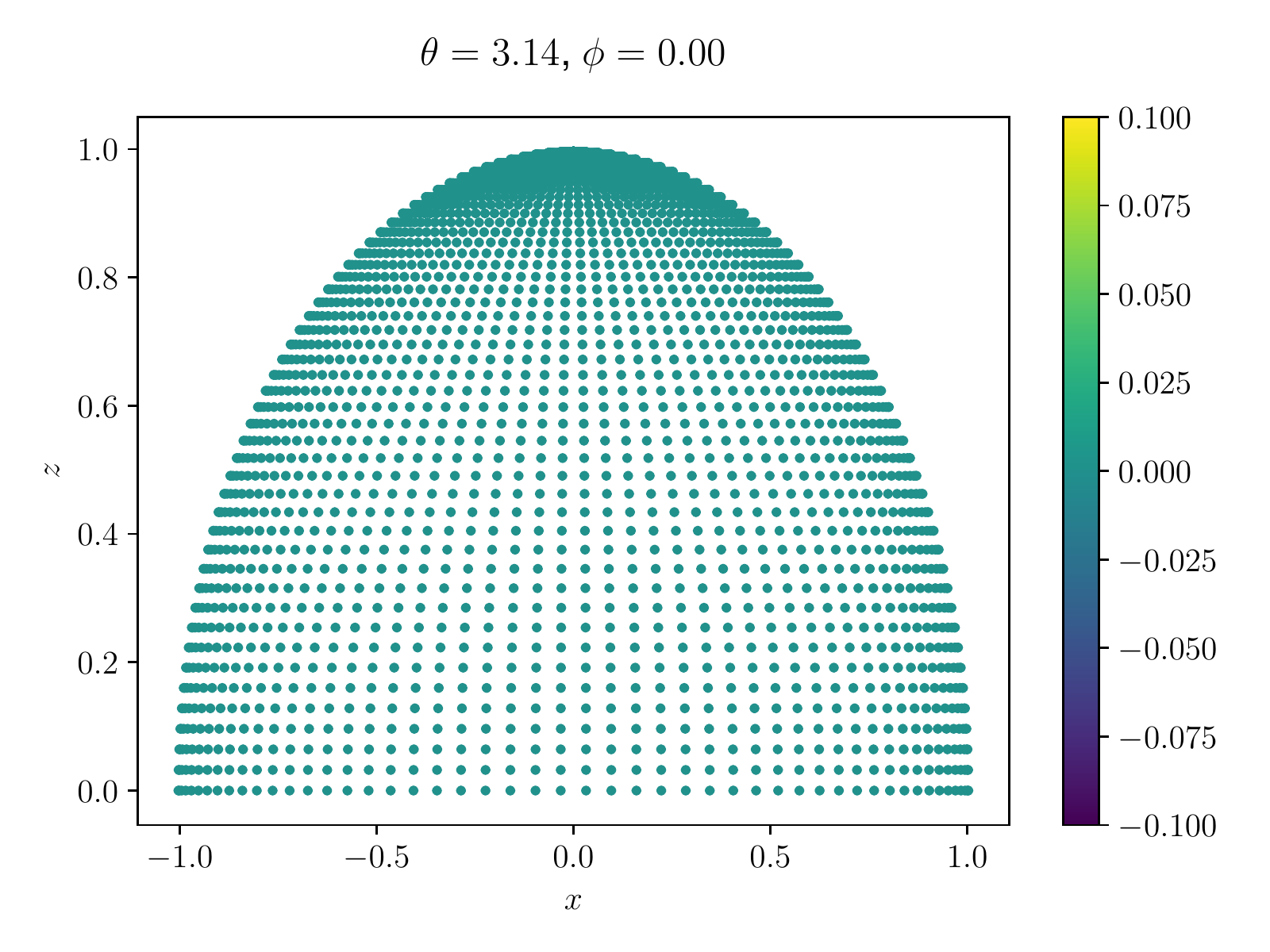}
\end{center}
\caption{The normal inward intensity as a function of the potential (top left) with differences between successive iterations (bottom left) and its surface distribution (right) for a uniformly rotating radiative polytrope.}
\label{fig:m2_I5}
\end{figure}

\begin{figure}[htb]
\begin{center}
\includegraphics[width = 0.48\hsize]{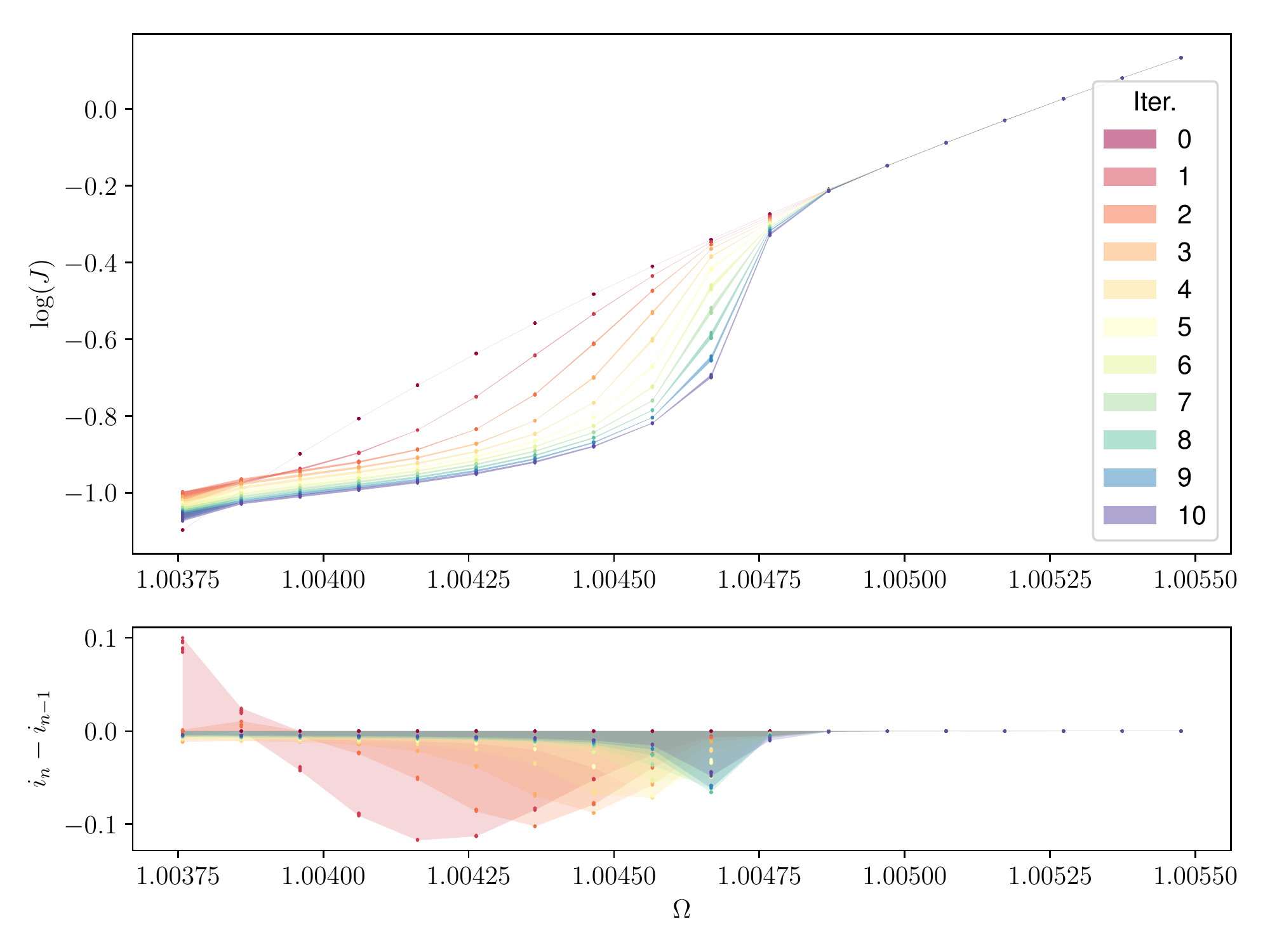}
\includegraphics[width = 0.48\hsize]{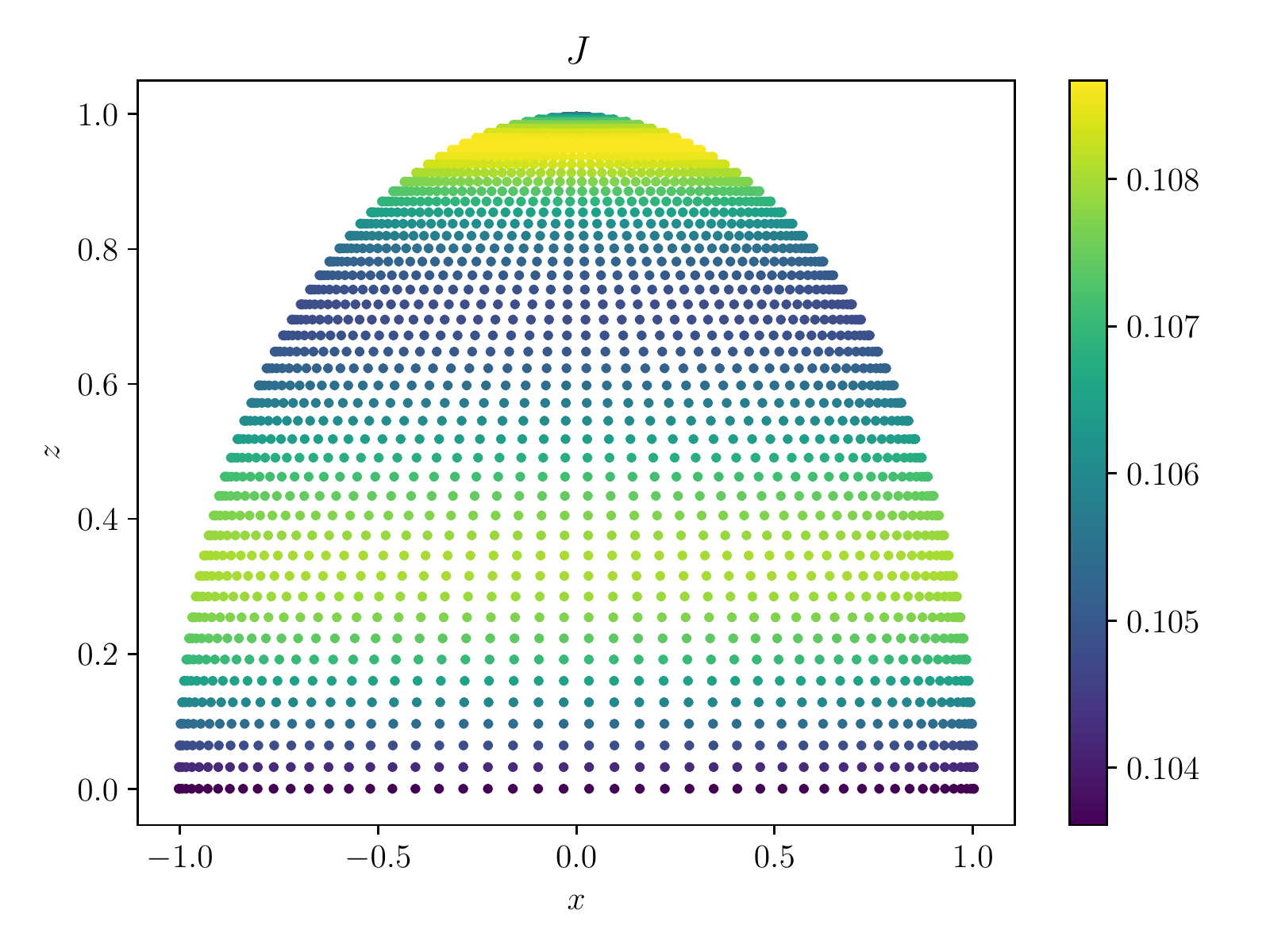}
\end{center}
\caption{The mean intensity as a function of the potential (top left) with differences between successive iterations (bottom left) and its surface distribution (right) for a uniformly rotating radiative polytrope.}
\label{fig:m2_J}
\end{figure}

The shape of a uniformly rotating star is ellipsoidal, with a larger radius at the equator than the pole, and consequently a gradient of the normal emergent intensity is expected, from higher values near the pole, which is closer to the center, to lowest values near the equator, where the same difference of optical depth corresponds to smaller geometrical path lengths. The intensity distribution obtained for a rigid rotating model with $\omega^2 = 0.1$ with \textsc{COBAIN} shows precisely that. The gradient of the surface distributions is evident from the right panels of Figures \ref{fig:m2_I2} - \ref{fig:m2_J}, while the gradual convergence to blackbody values towards the deeper layers of the atmosphere is visible in the functional dependence of the radiative properties with increasing potential, given in the Left panelss of Figures \ref{fig:m2_I2} - \ref{fig:m2_J}.



\subsection{Unstable differential rotator}

\begin{figure}[htb]
\begin{center}
\includegraphics[width = 0.48\hsize]{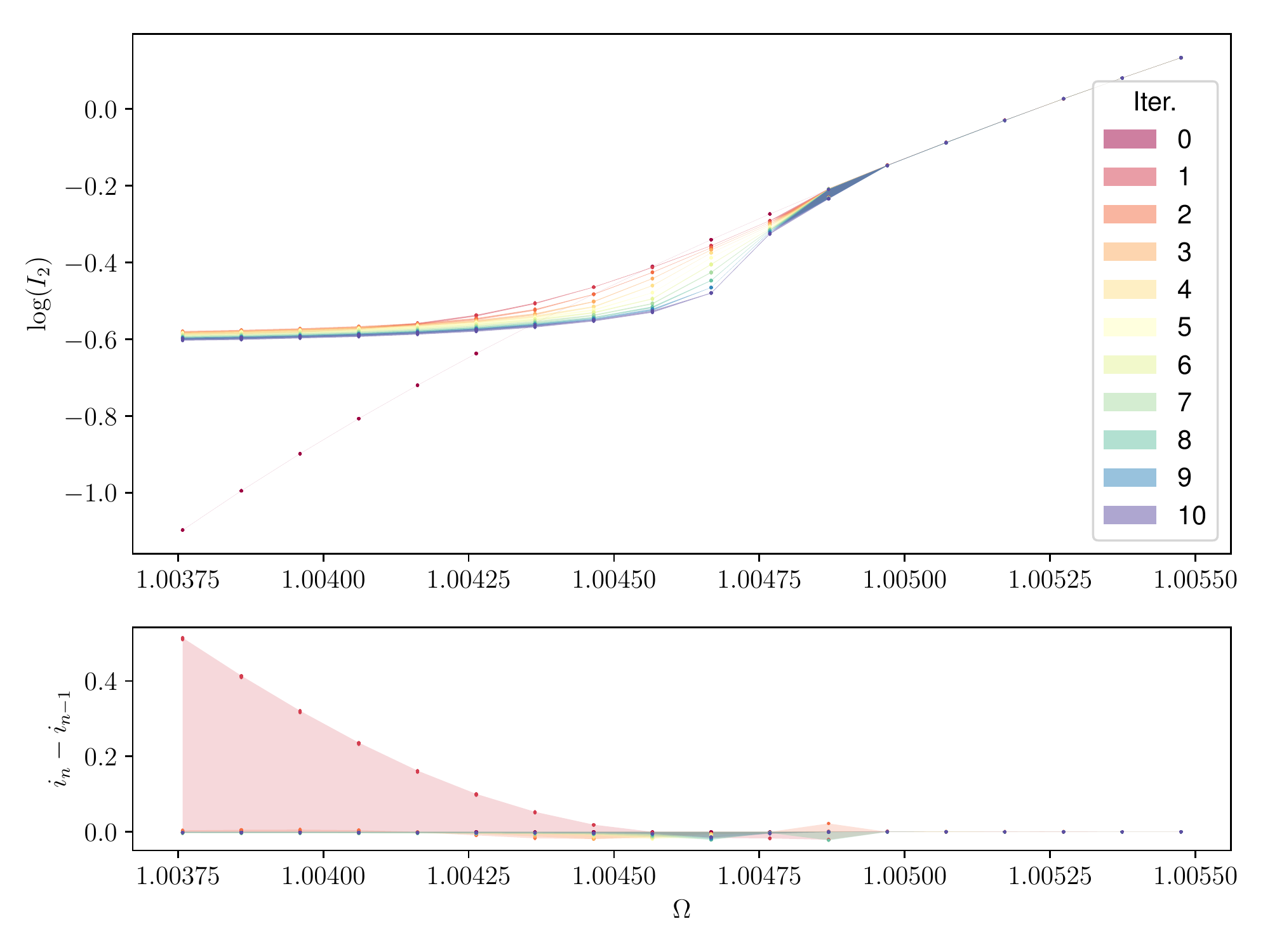}
\includegraphics[width = 0.48\hsize]{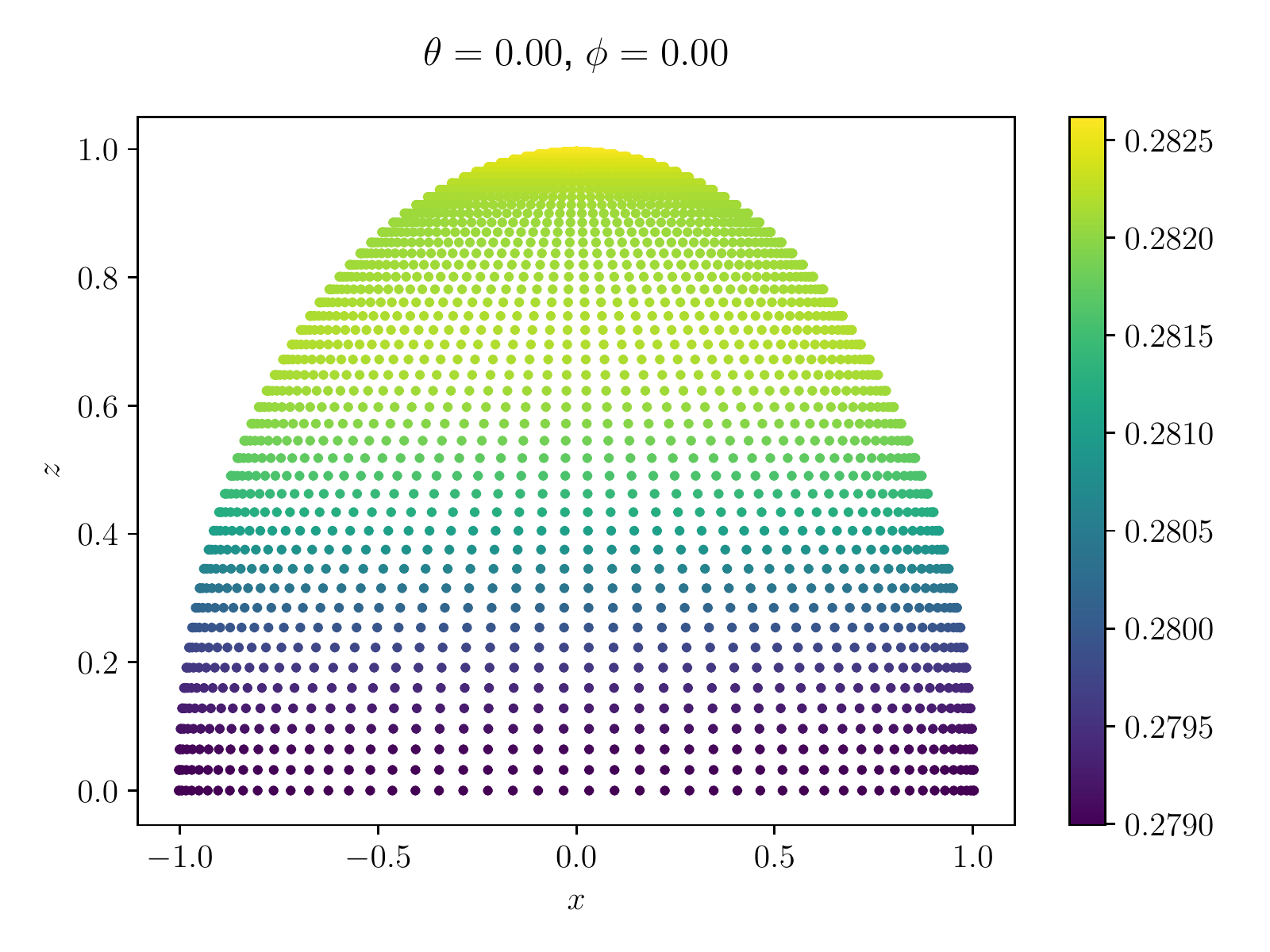}
\end{center}
\caption{The normal outward intensity as a function of the potential (top left) with differences between successive iterations (bottom left) and its surface distribution (right) for an unstable differentially rotating radiative polytrope.}
\label{fig:m15_I2}
\end{figure}

\begin{figure}[htb]
\begin{center}
\includegraphics[width = 0.48\hsize]{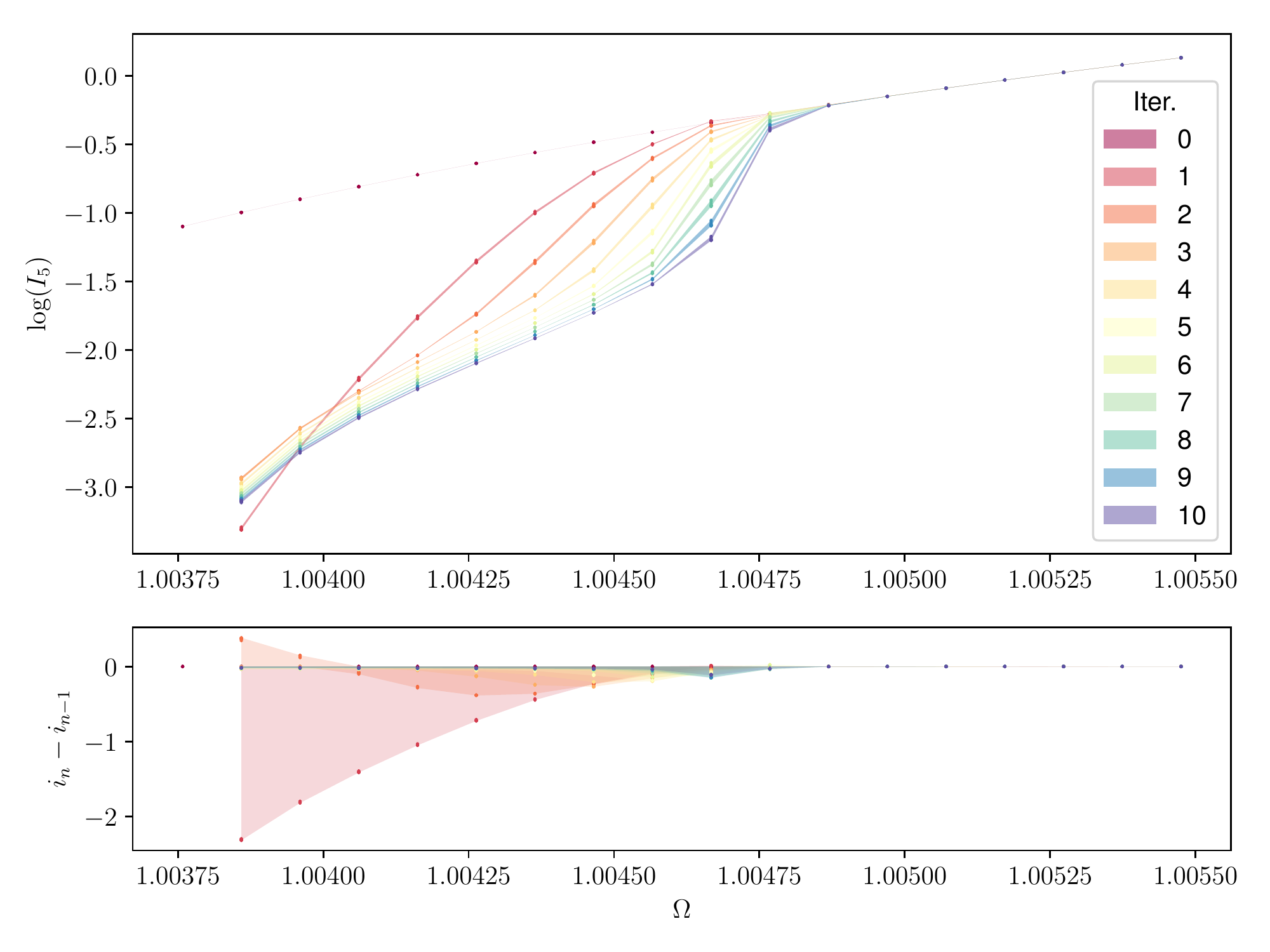}
\includegraphics[width = 0.48\hsize]{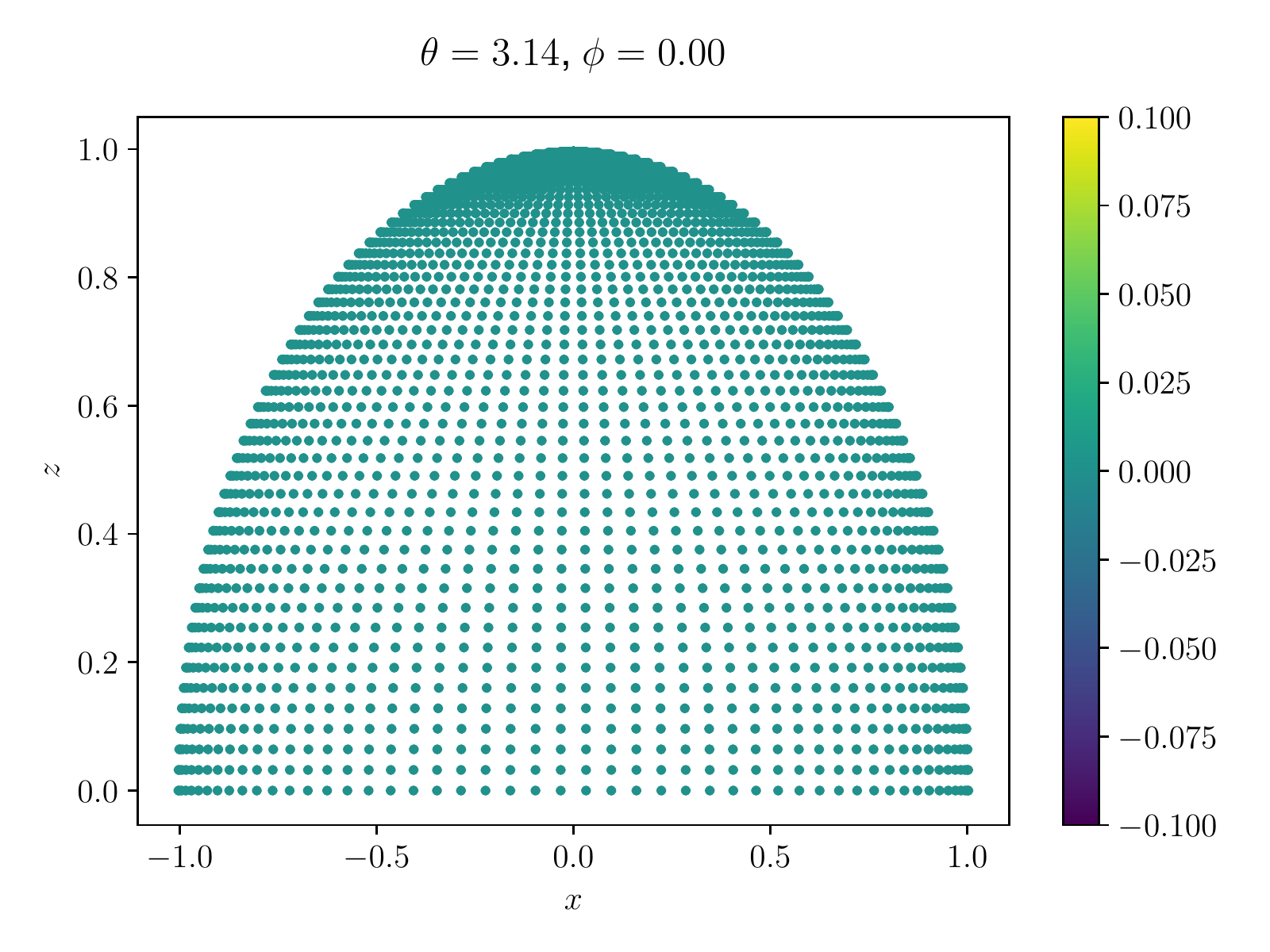}
\end{center}
\caption{The normal inward intensity as a function of the potential (top left) with differences between successive iterations (bottom left) and its surface distribution (right) for an unstable differentially rotating radiative polytrope.}
\label{fig:m15_I5}
\end{figure}

\begin{figure}[htb]
\begin{center}
\includegraphics[width = 0.48\hsize]{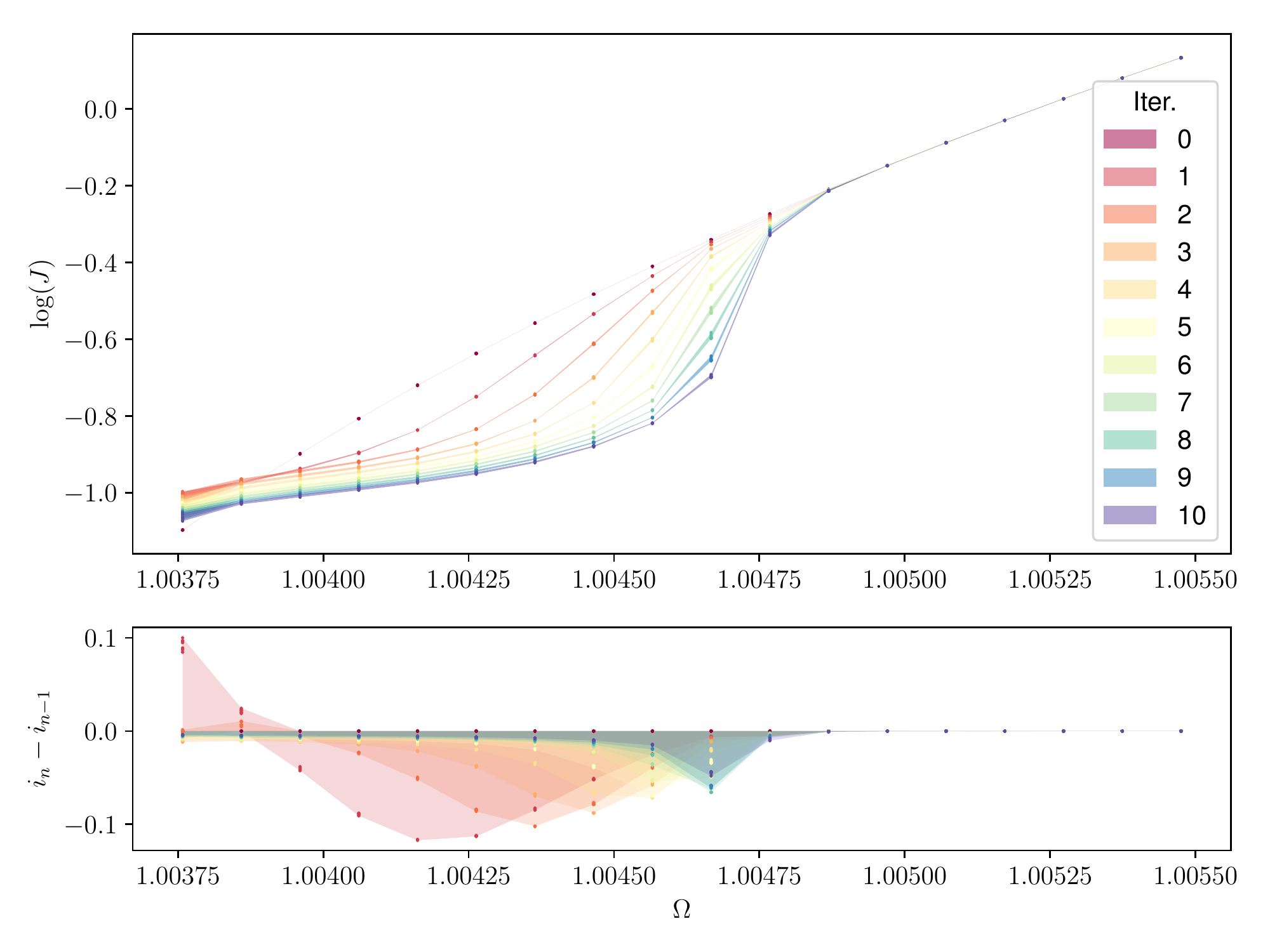}
\includegraphics[width = 0.48\hsize]{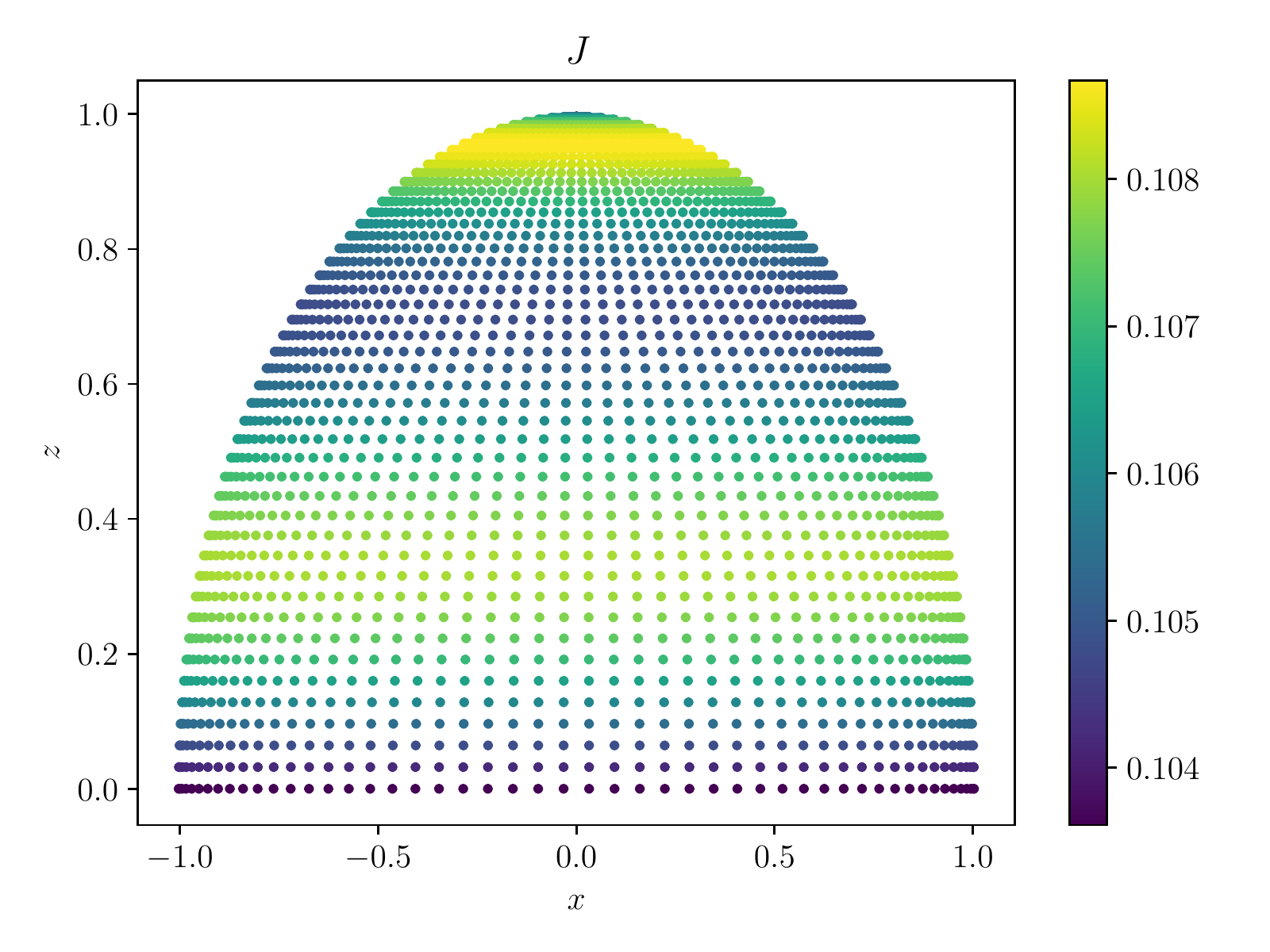}
\end{center}
\caption{The mean intensity as a function of the potential (top left) with differences between successive iterations (bottom left) and its surface distribution (right) for an unstable differentially rotating radiative polytrope.}
\label{fig:m15_J}
\end{figure}

The radiative transfer in the unstable differential rotator simulated with \textsc{COBAIN} clearly demonstrates the performance of the code in different geometries and how closely it follows the structure of the input model. The peculiar pattern that can be seen on the surface distributions in Figures \ref{fig:m15_I2} - \ref{fig:m15_J}, especially in the mean intensity distribution in Figure~\ref{fig:m15_J}, comes directly from the variation of the radius length from the pole to the equator (Figure~\ref{fig:diffrot_r_w}). This result, although of limited physical significance, confirms the reliability of the model to perform well in various structures and geometries, without any substantial loss of accuracy. This puts it in a prime position to compute radiative transfer in complex structural models of objects like contact binaries with differential rotation, convection, flows, etc.

\clearpage

\section{COBAIN gray radiative transfer simulations of contact binaries}\label{sect5}

To demonstrate the performance of COBAIN on the peculiar geometry of contact binary stars, we did initial tests on the simplified structural models with tidally distorted polytropes. The default test system has the following parameter values: $M_{1} = 1 M_{\odot}$, $q=1$, $FF = 0.05$, $(N_{\Omega}, N_{\theta}, N_{\phi}) = (50, 50, 50)$. The fillout factor $FF$ determines the degree of contact and is used to compute the surface potential of the contact envelope: 
\begin{equation}
    \Omega = \Omega_{L1} - FF (\Omega_{L1}-\Omega_{L2}) \; ,
\end{equation}
where $\Omega_{L1}$ and $\Omega_{L2}$ denote the values of the first and second critical potential of the binary.
The polytropic models all have a polytropic index $n=3$, the mean molecular mass is $\mu = 0.6$ and the range of potentials covered with the grid is $\Delta\Omega = 0.01$ regardless of the surface potential value. Twelve system configurations were built for three different values of each of these parameters: $q \in \{0.5, 0.75, 0.9\}$, $M_1 \in \{0.75 M_{\odot}, 1 M_{\odot}, 5 M_{\odot}\} $, $FF \in \{0.1, 0.45, 0.9\}$ and $(N_{\Omega}, N_{\theta}, N_{\phi}) \in \{(50,25,25), (35, 75, 75), (75, 50, 50)\}$. In each different system, only the respective parameter value is changed, while others are kept at their default values. A convergence test is performed by computing the mean difference between each two successive iterations and the threshold $|\overline{\Delta J}/\overline{J}| = 0.01\%$ is achieved after 5 iterations in most cases. Results of radiative transfer simulations and the effects of various parameter values are discussed below.

\subsection{Grid dimensions}

The default test system ($M_{1} = 1 M_{\odot}$, $q=1$, $FF = 0.05$) was build in three different grid dimension configurations, besides the default one, to test the performance of grid interpolation in different regimes:
\begin{itemize}
    \item $50\times25\times25$ - default potential sampling, lower spherical angle resolution
    \item $35\times75\times75$ - lower potential resolution, higher spherical angle resolution
    \item $75\times50\times50$ - higher potential resolution, default spherical angle sampling
\end{itemize}
A system in the other limit of the fillout factor: $FF=0.95$, was also built in all grid configurations, to test the potential differences between different geometries in combination with different sampling.

\begin{figure}[h]
    \centering
    \includegraphics[width=\hsize]{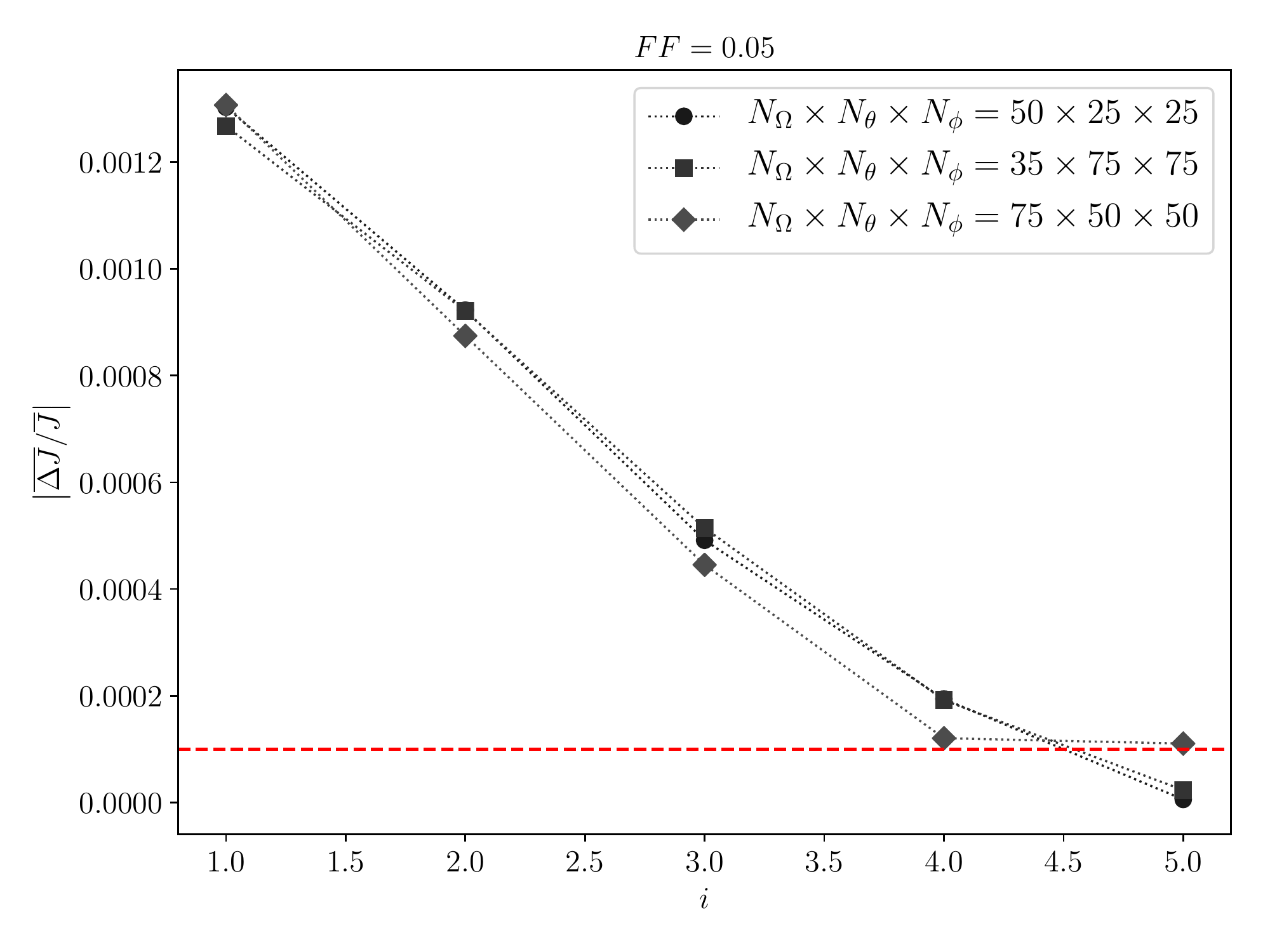}
    \includegraphics[width=\hsize]{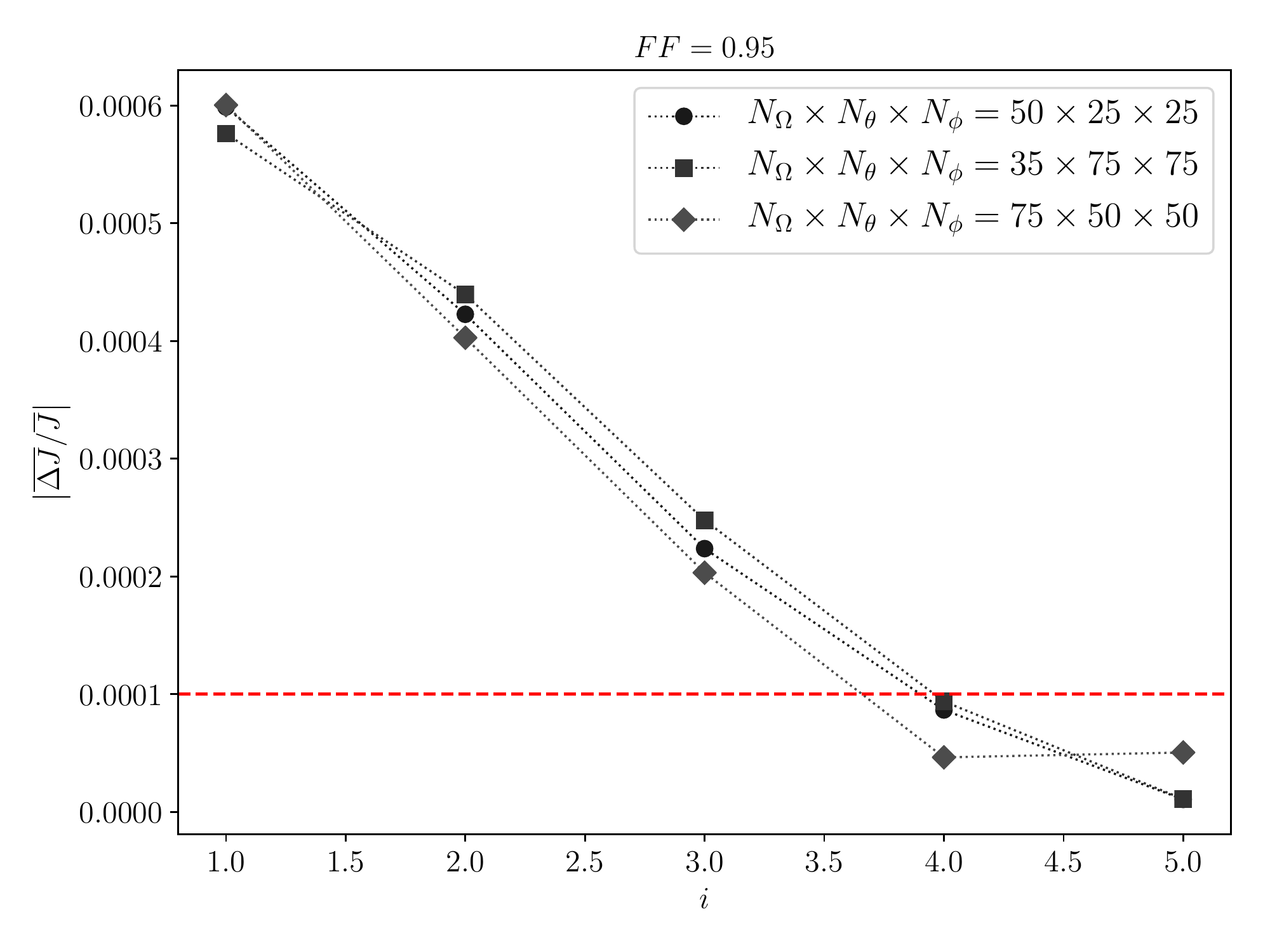}
    \caption{Convergence of the mean intensity in test systems with different grid dimensions, for two different fillout factors. Top: $FF=0.05$, bottom: $FF=0.95$.}
    \label{fig:iters_dims}
\end{figure}

\begin{figure}[h]
    \centering
    \includegraphics[width=0.495\hsize]{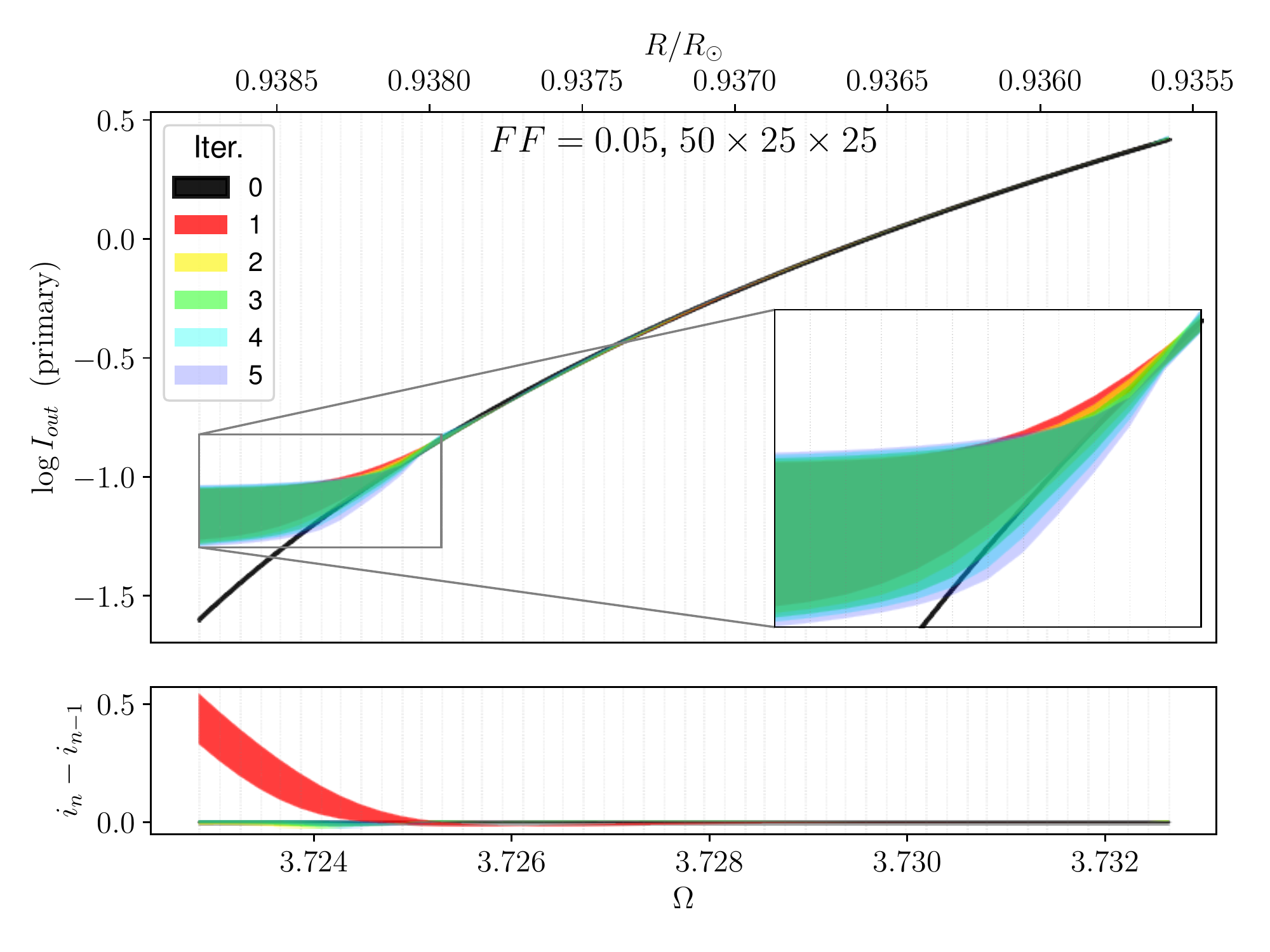}
    \includegraphics[width=0.495\hsize]{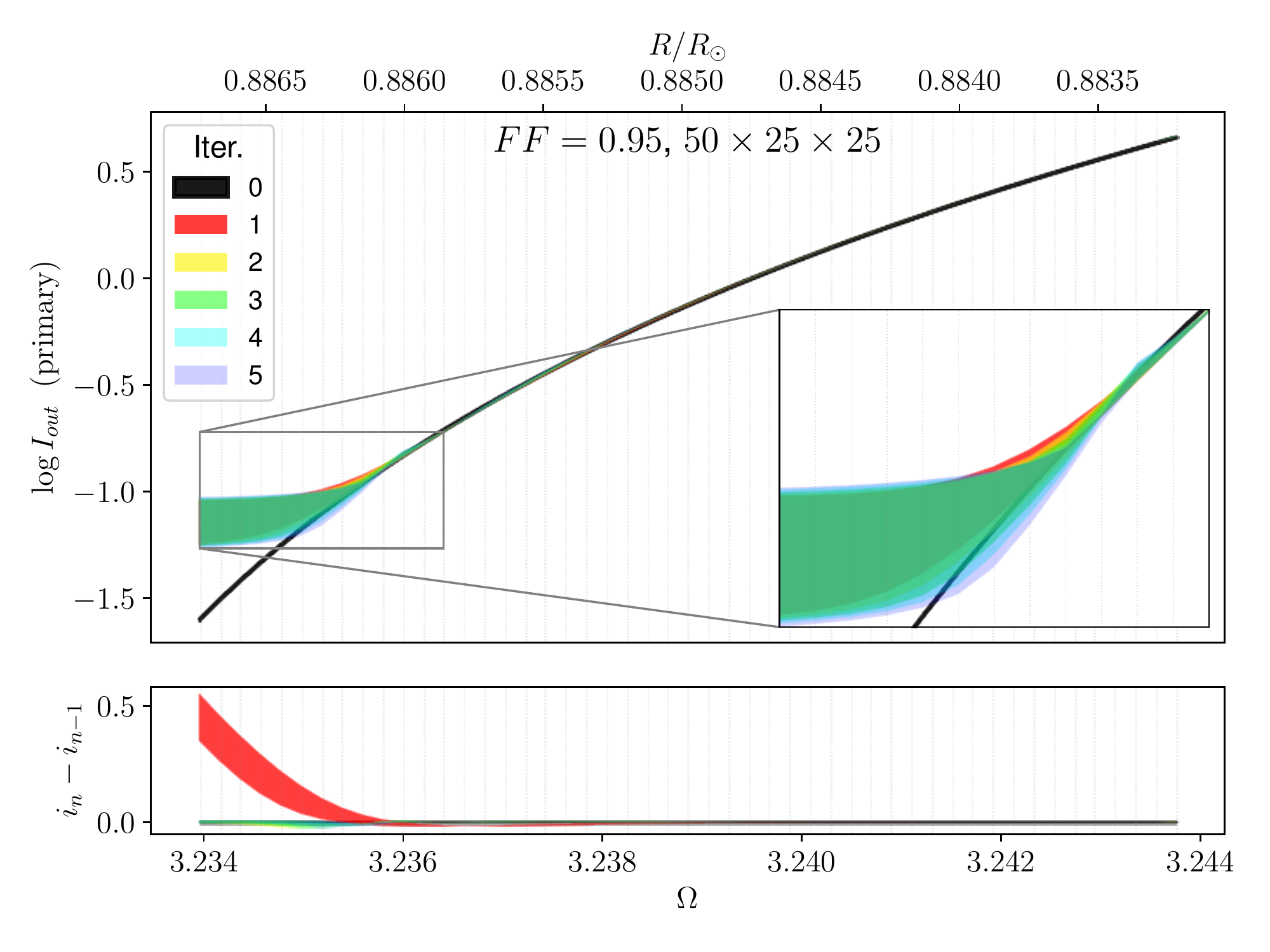}
    
    \includegraphics[width=0.495\hsize]{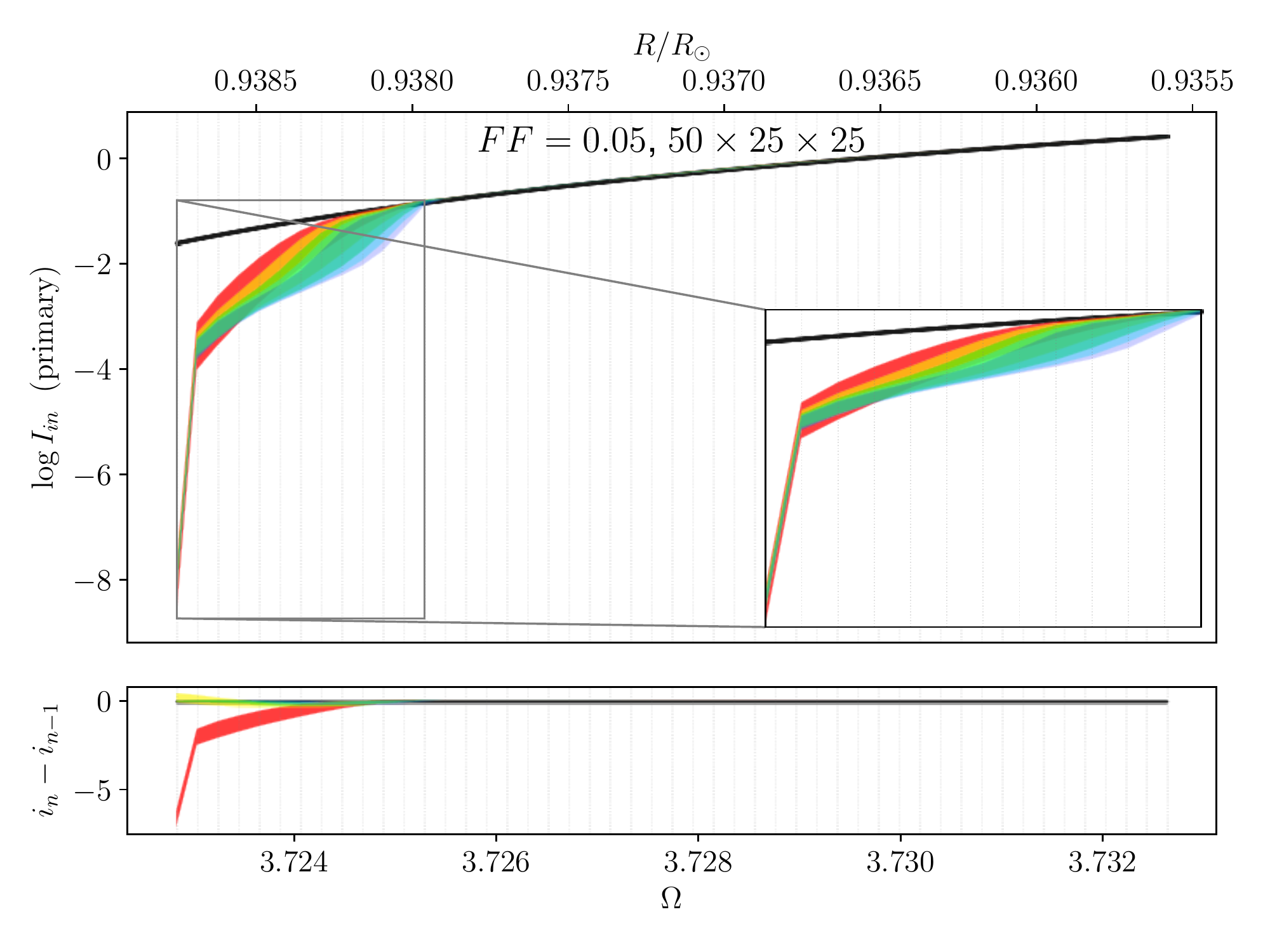}
    \includegraphics[width=0.495\hsize]{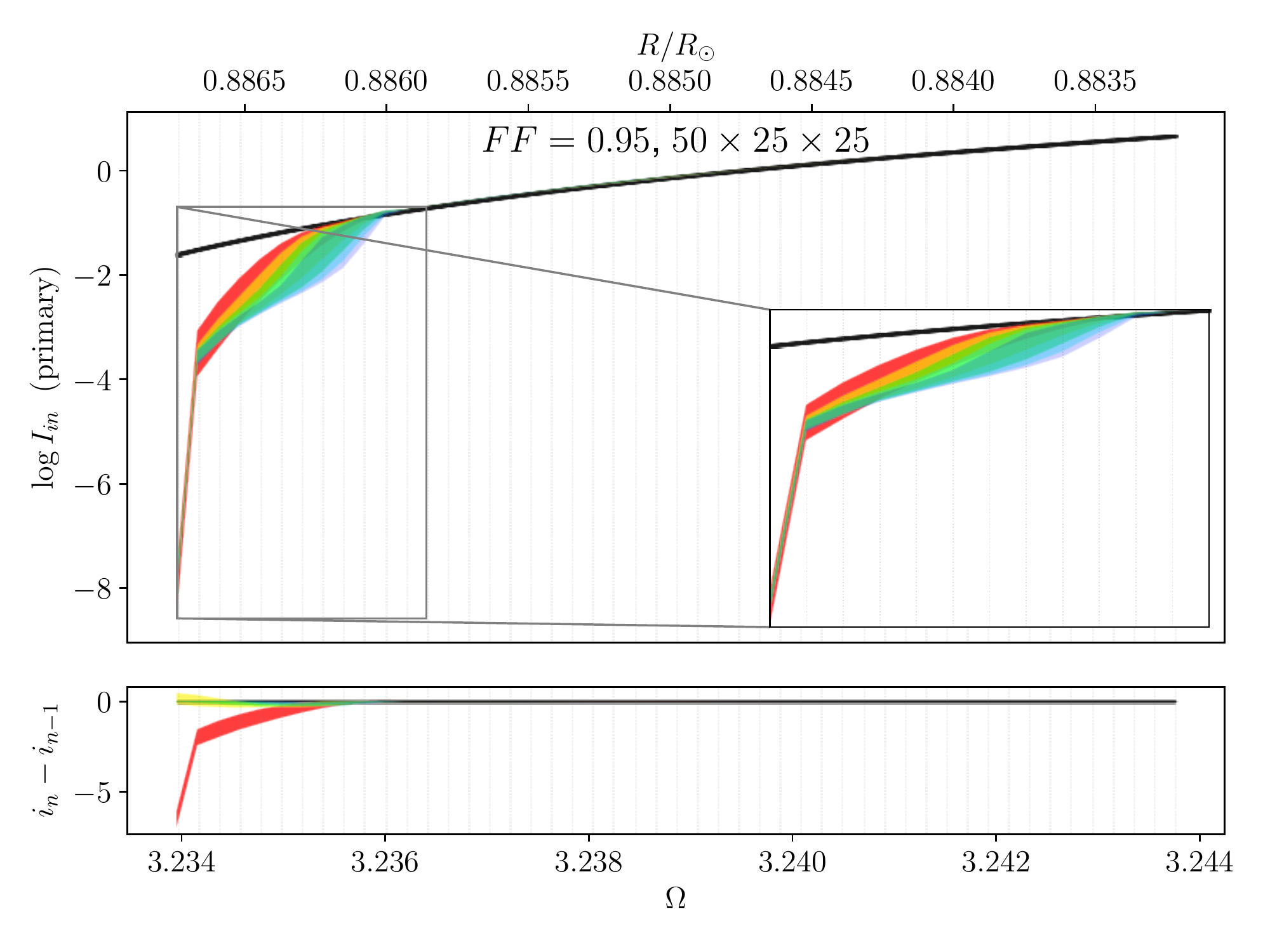}
    
    \includegraphics[width=0.495\hsize]{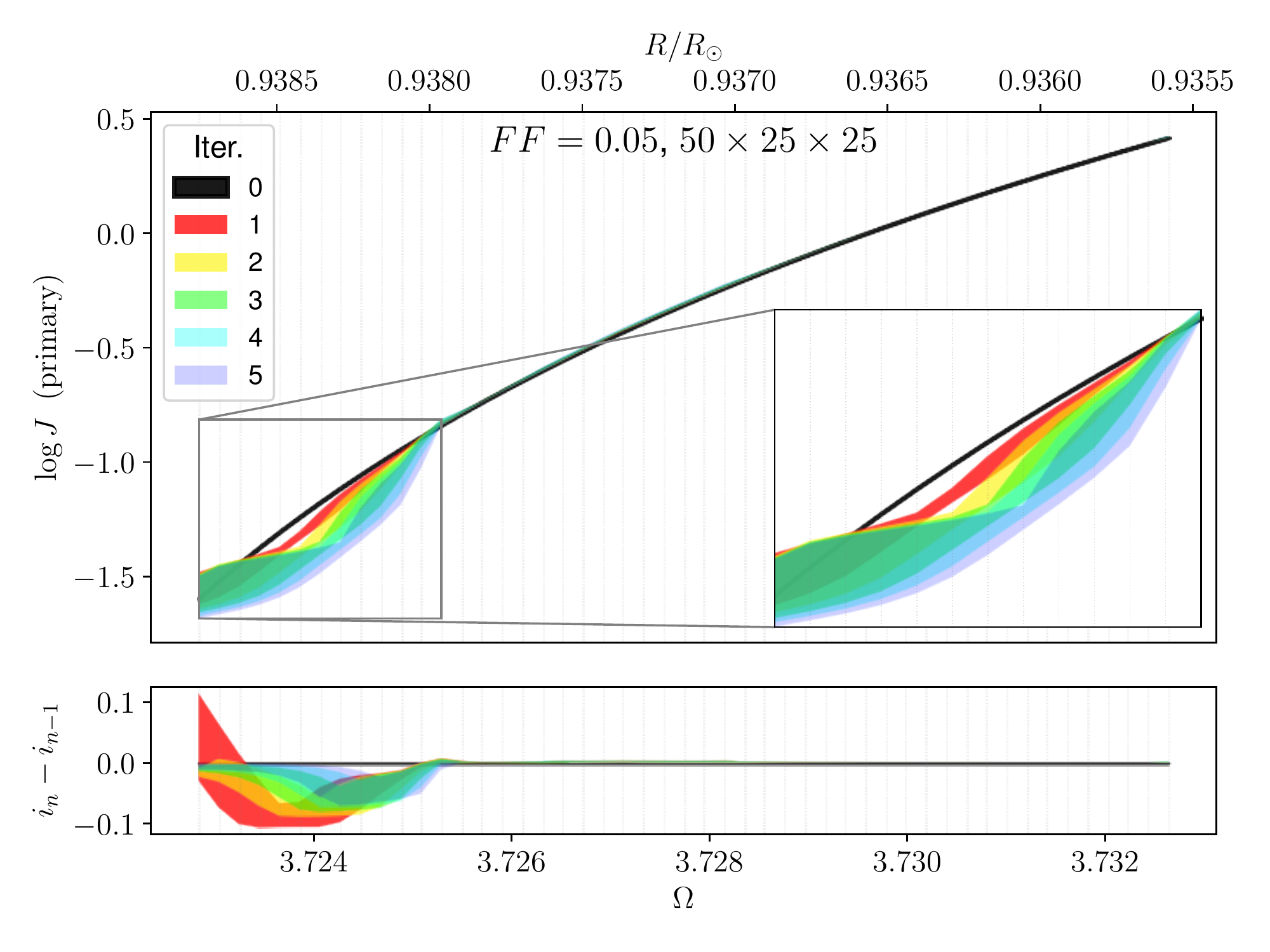}
    \includegraphics[width=0.495\hsize]{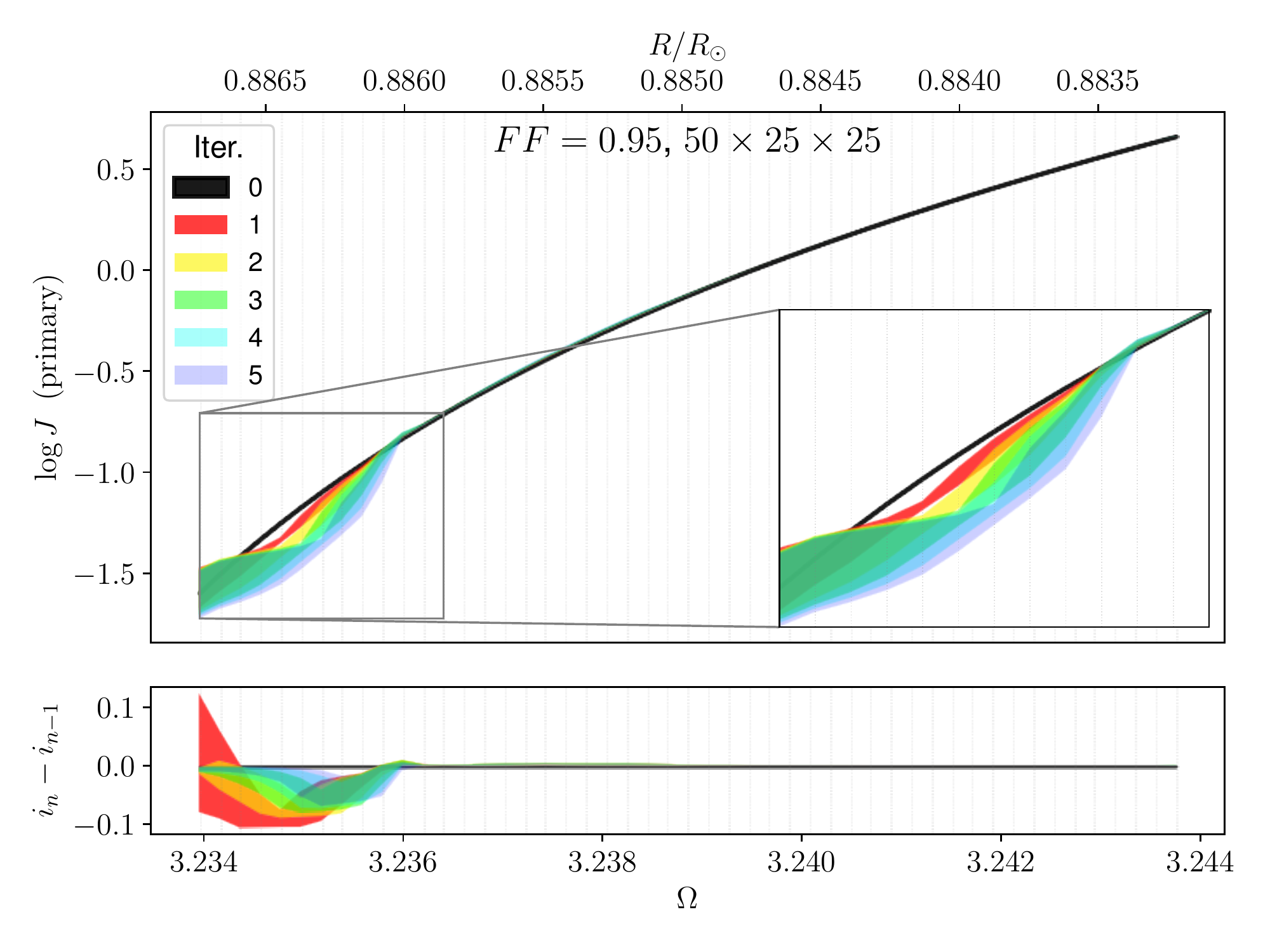}
    
    \caption{Top to bottom: outward, inward and mean intensity as a function of the potential/radius of a contact binary with $(N_{\Omega}, N_{\theta}, N_{\phi}) = (50, 25, 25)$. Left panels: $FF=0.05$, right panels: $FF=0.95$. The bottom panel of each plot shows the differences between successive iterations.}
    \label{fig:dims502525}
\end{figure}

\begin{figure}[h]
    \centering
       \includegraphics[width=0.495\hsize]{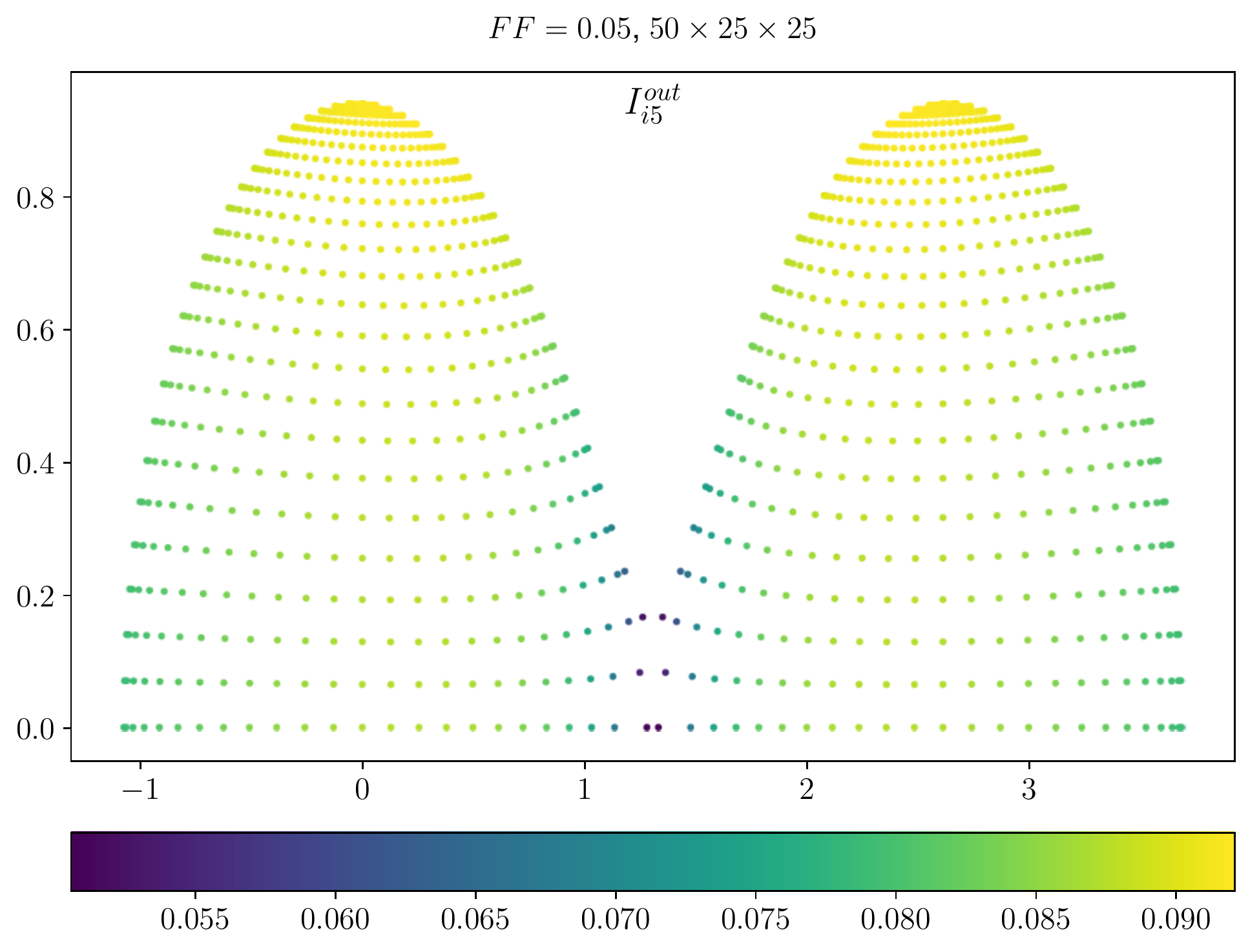}
    \includegraphics[width=0.495\hsize]{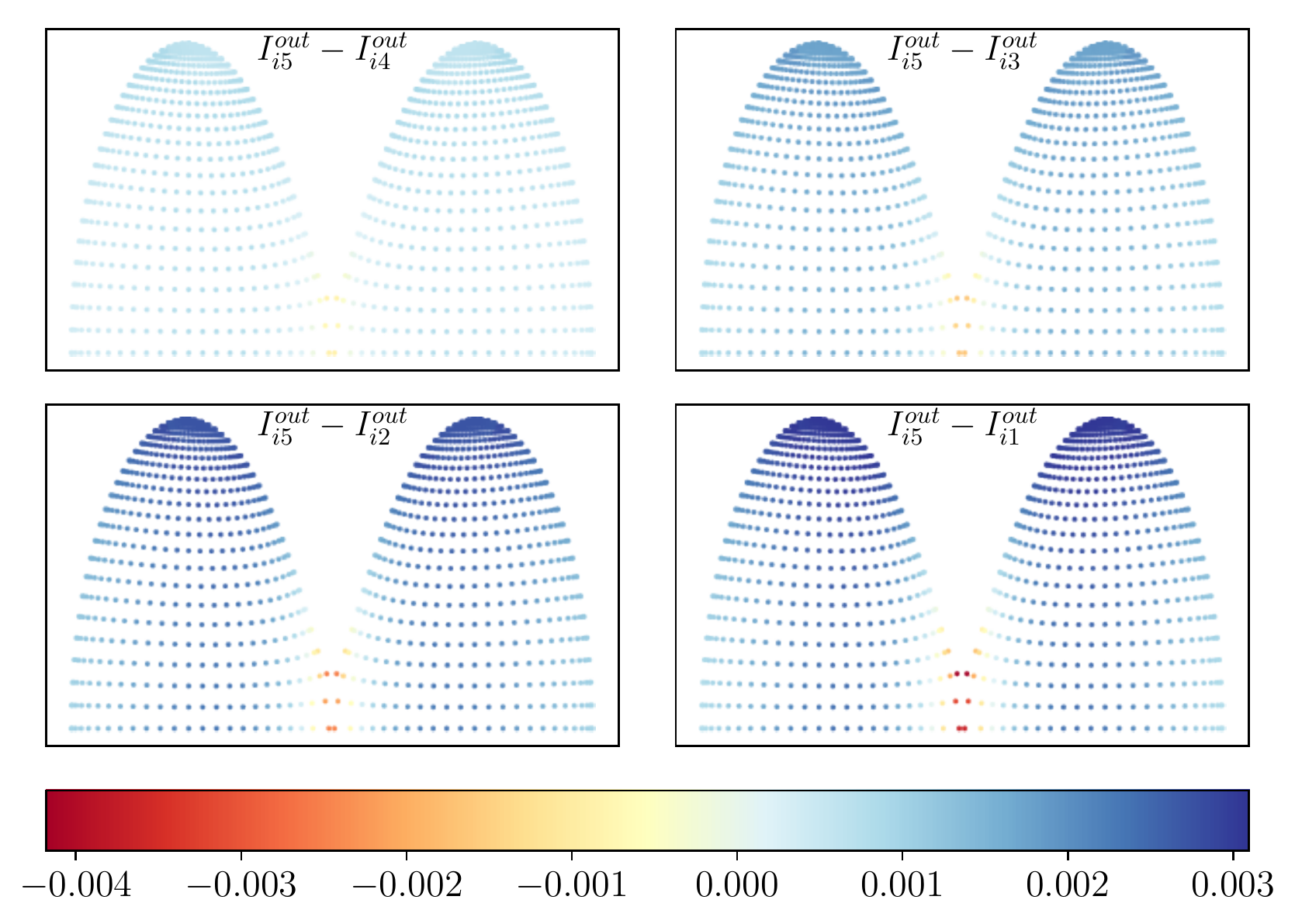}
    
        \includegraphics[width=0.495\hsize]{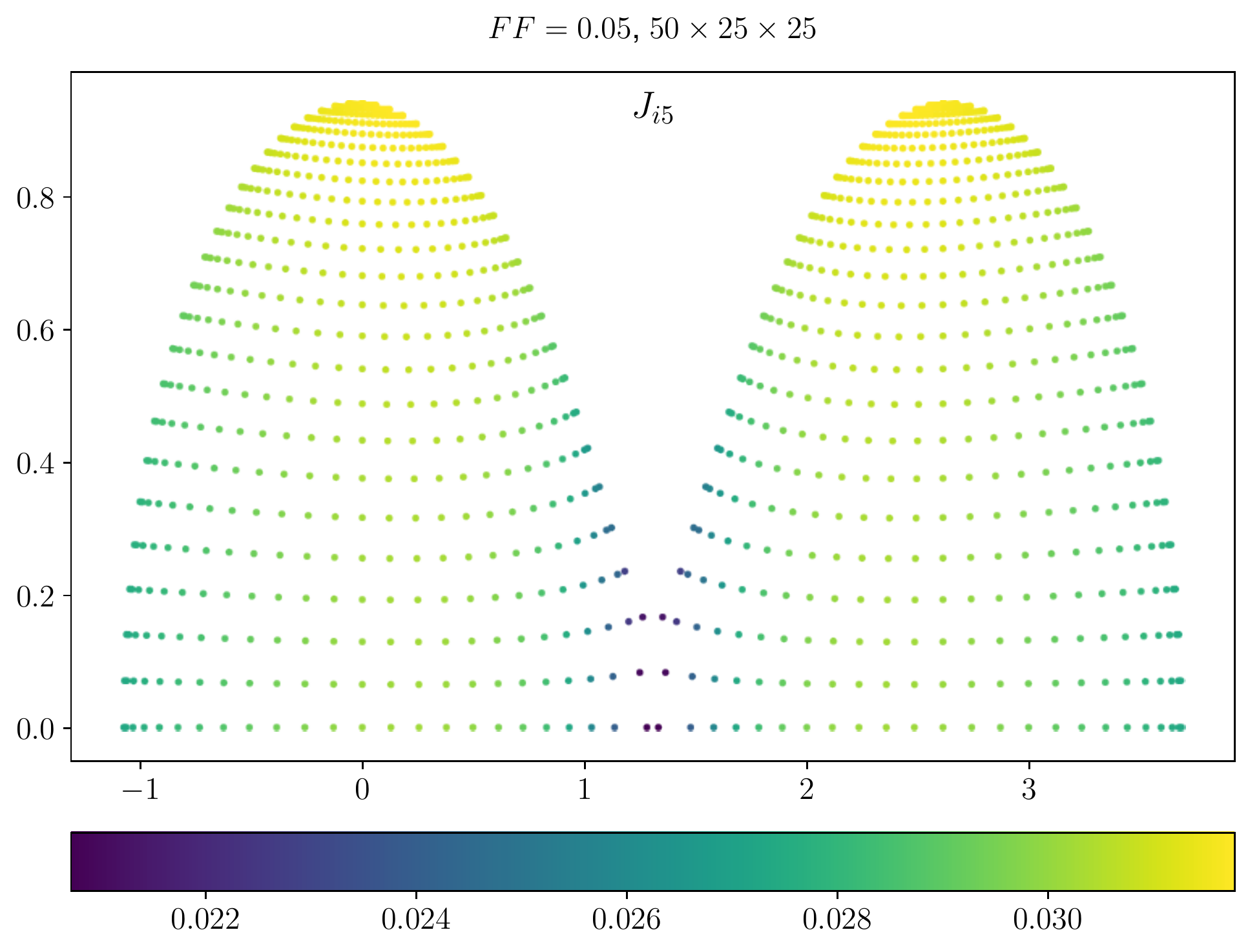}
    \includegraphics[width=0.495\hsize]{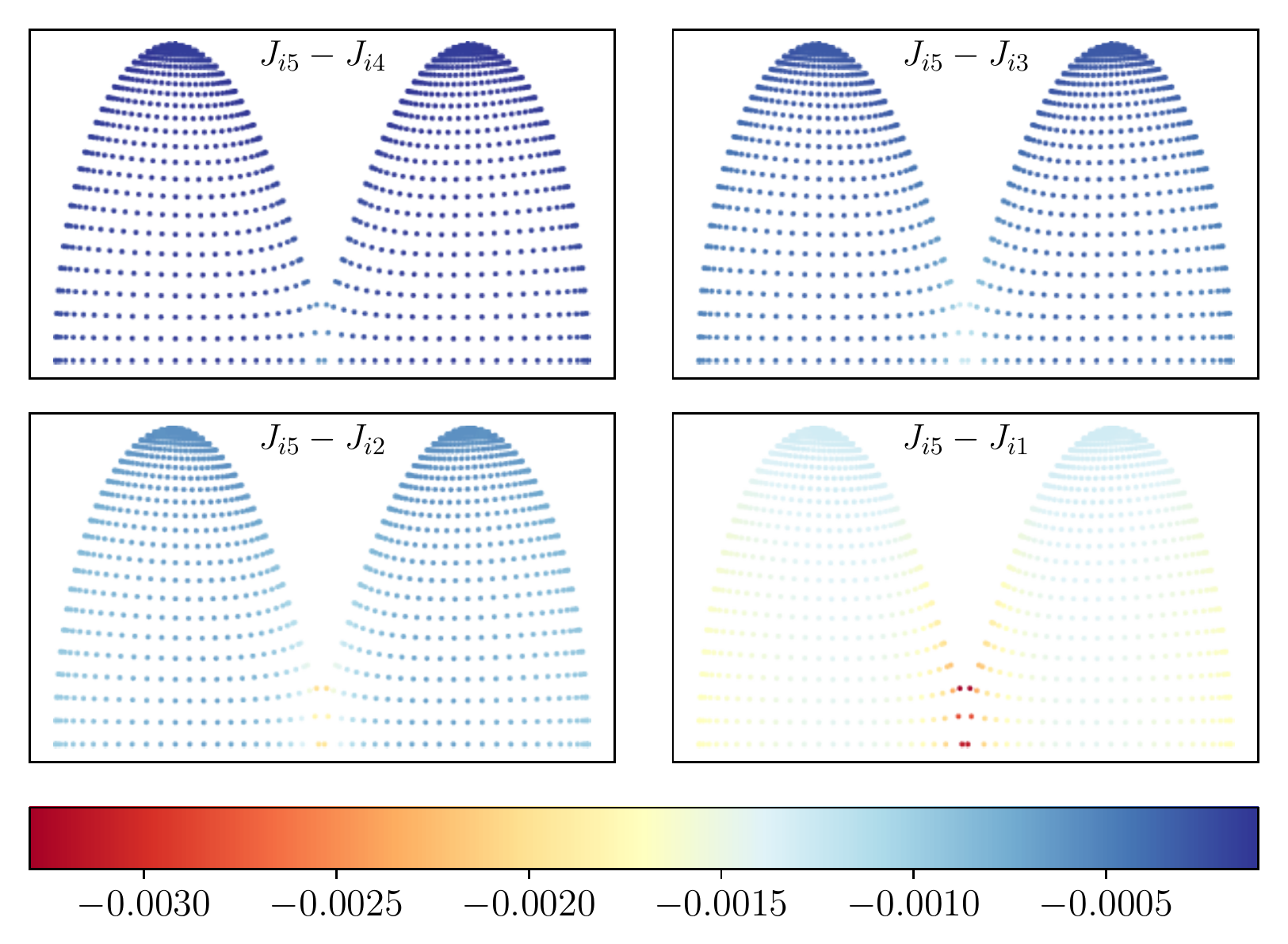}

    \caption{Surface distribution of the outward (top) and mean (bottom) intensity of a contact binary with $(N_{\Omega}, N_{\theta}, N_{\phi}) = (50, 25, 25)$ and $FF=0.05$ after the fifth iteration. Right panels show differences in the surface distribution between the final and each previous iteration.}
    \label{fig:dims502525_s}
\end{figure}

\begin{figure}[h]
    \centering
    \includegraphics[width=0.495\hsize]{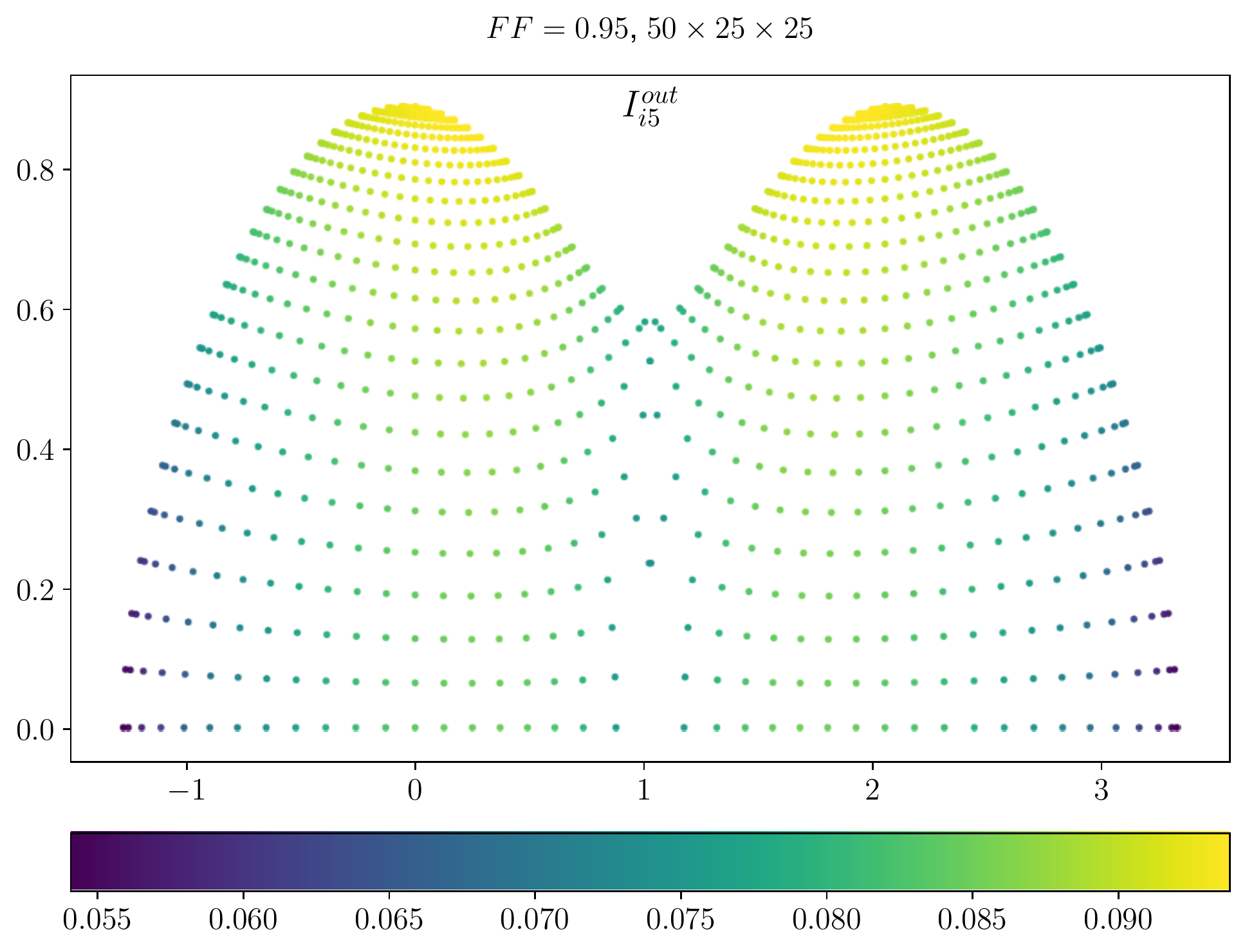}
    \includegraphics[width=0.495\hsize]{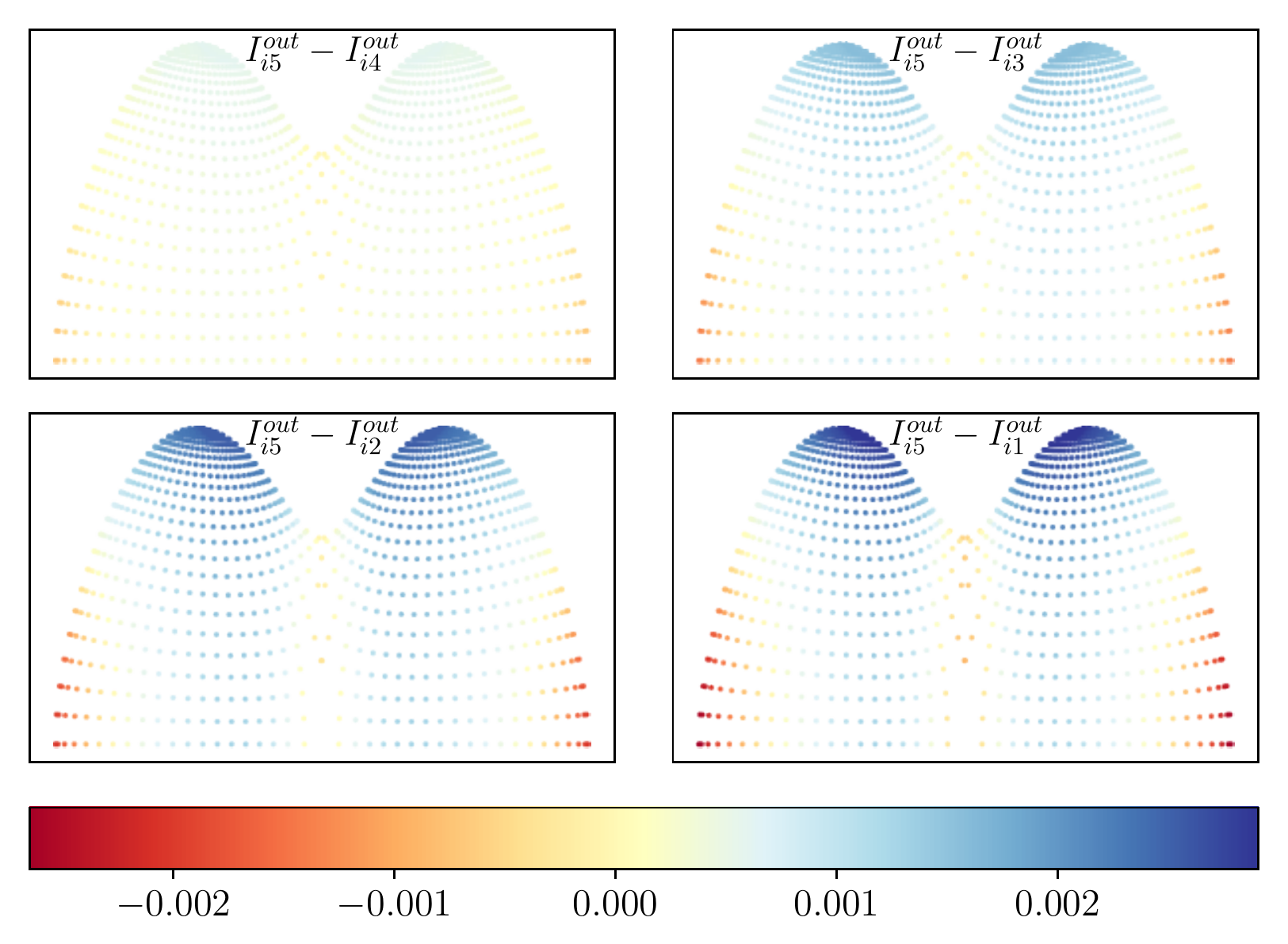}
    
        \includegraphics[width=0.495\hsize]{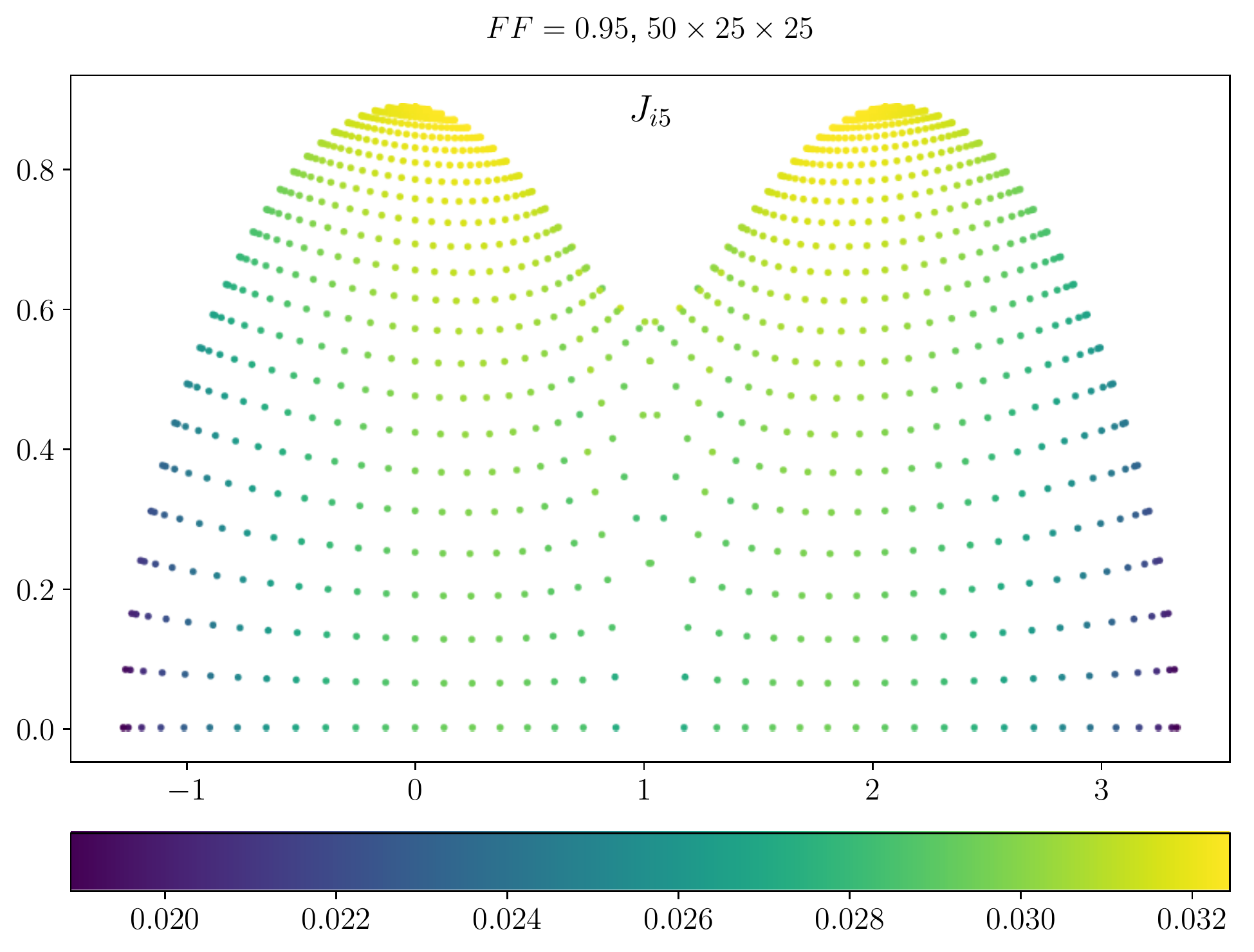}
    \includegraphics[width=0.495\hsize]{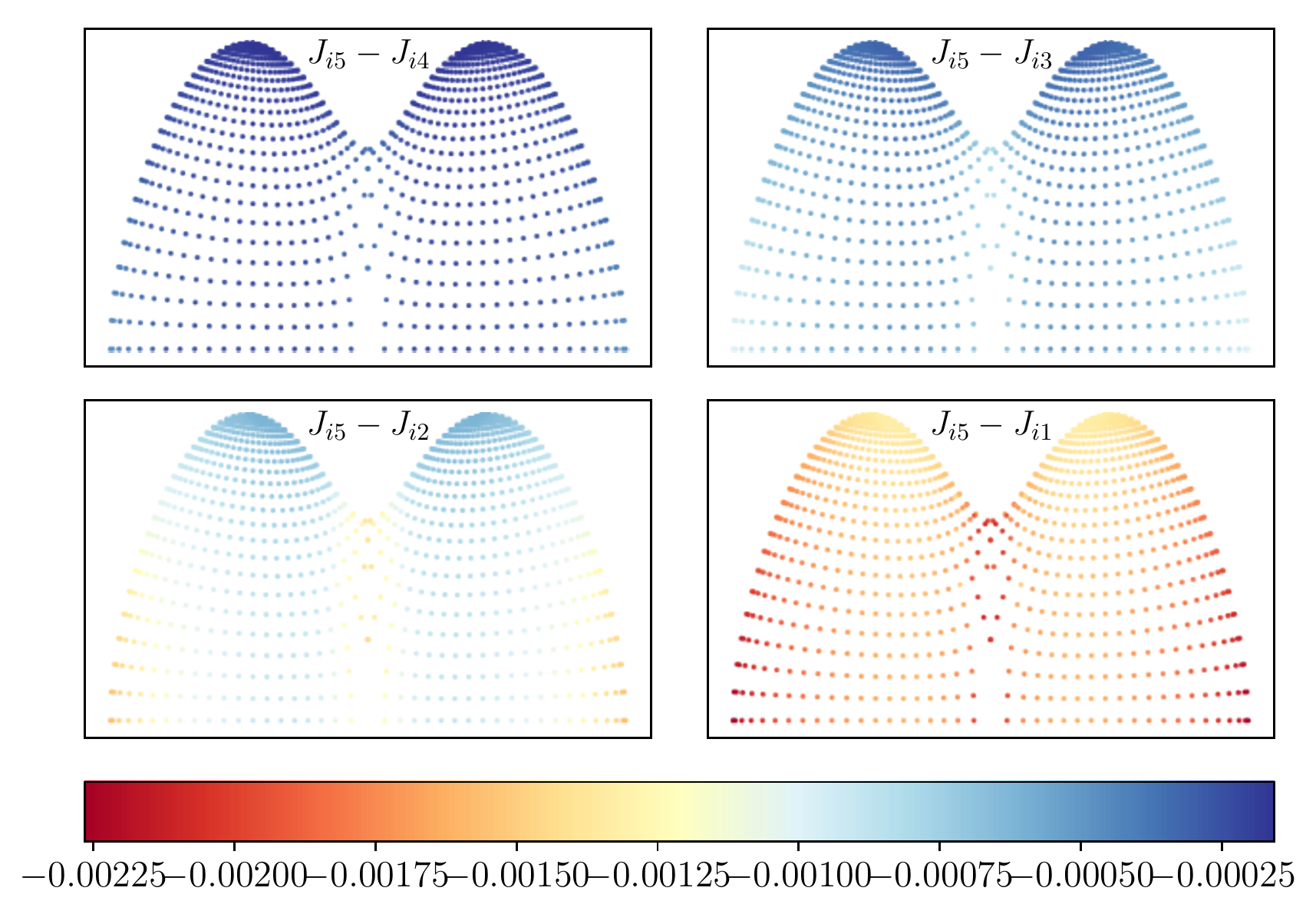}

    \caption{Surface distribution of the outward (top) and mean (bottom) intensity of a contact binary with $(N_{\Omega}, N_{\theta}, N_{\phi}) = (50, 25, 25)$ and $FF=0.95$ after the fifth iteration. Right panels show differences in the surface distribution between the final and each previous iteration.}
    \label{fig:dims502525_ff_s}
\end{figure}

\begin{figure}[h]
    \centering
    \includegraphics[width=0.495\hsize]{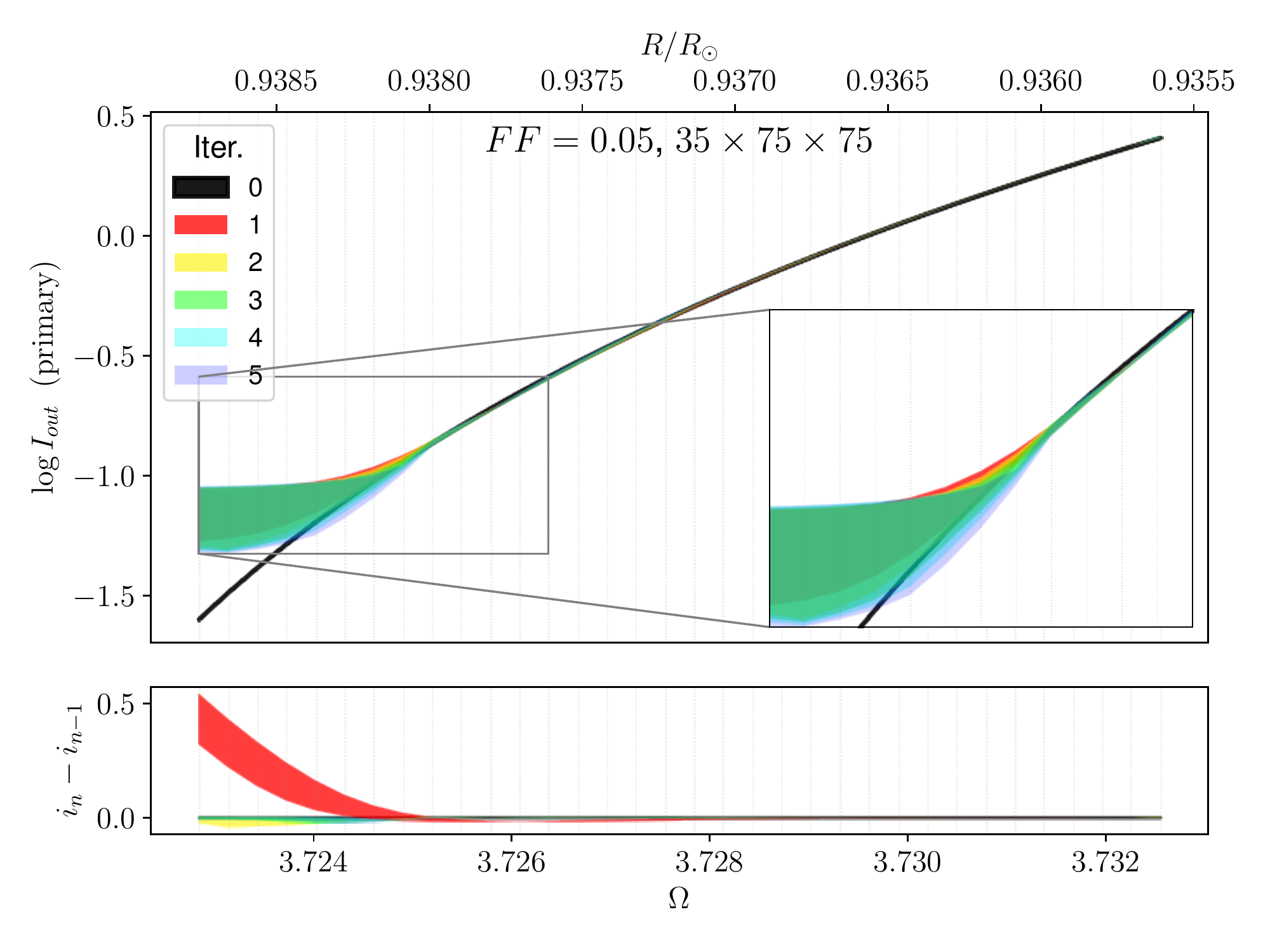}
    \includegraphics[width=0.495\hsize]{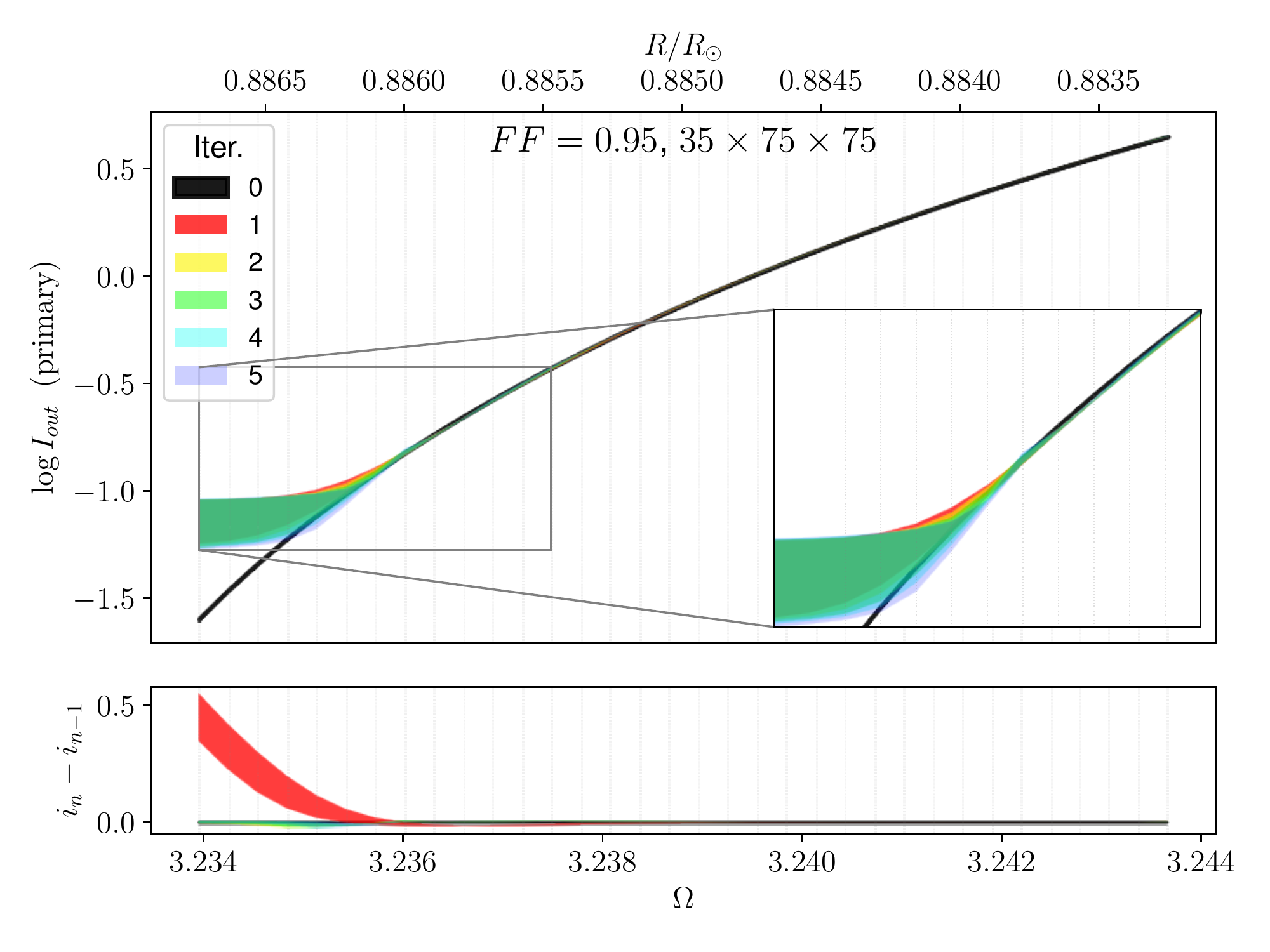}
    
    \includegraphics[width=0.495\hsize]{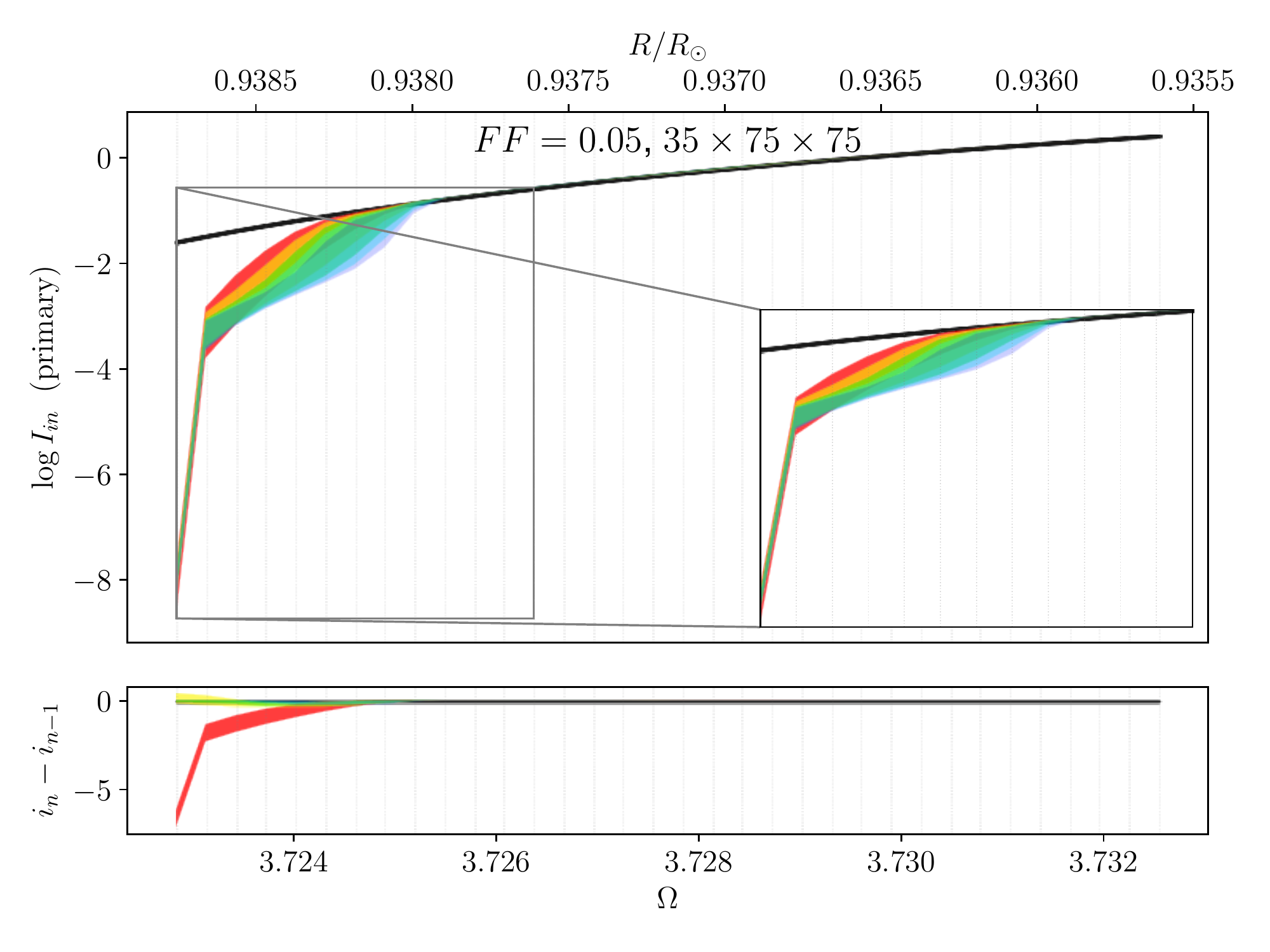}
    \includegraphics[width=0.495\hsize]{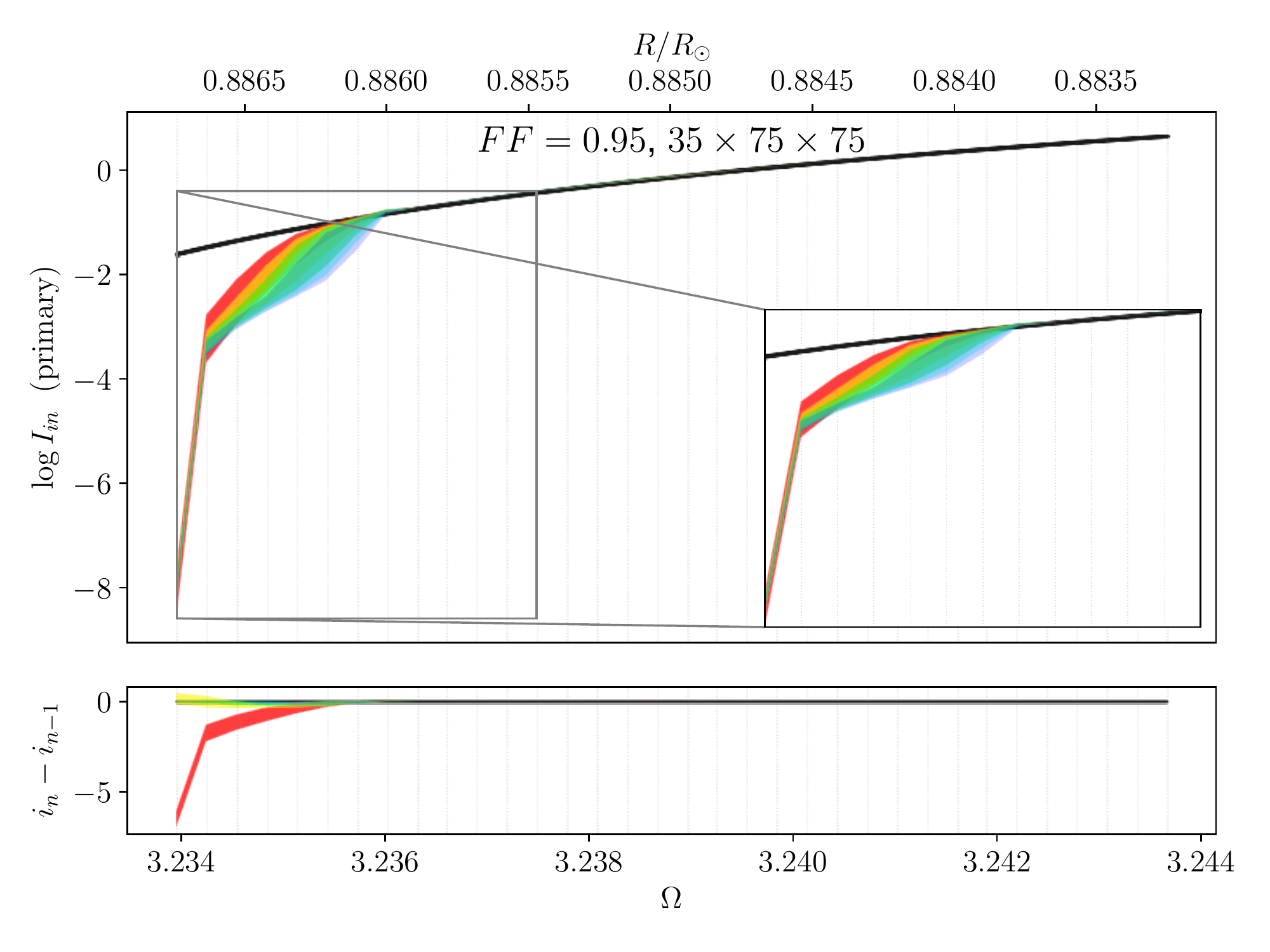}
    
    \includegraphics[width=0.495\hsize]{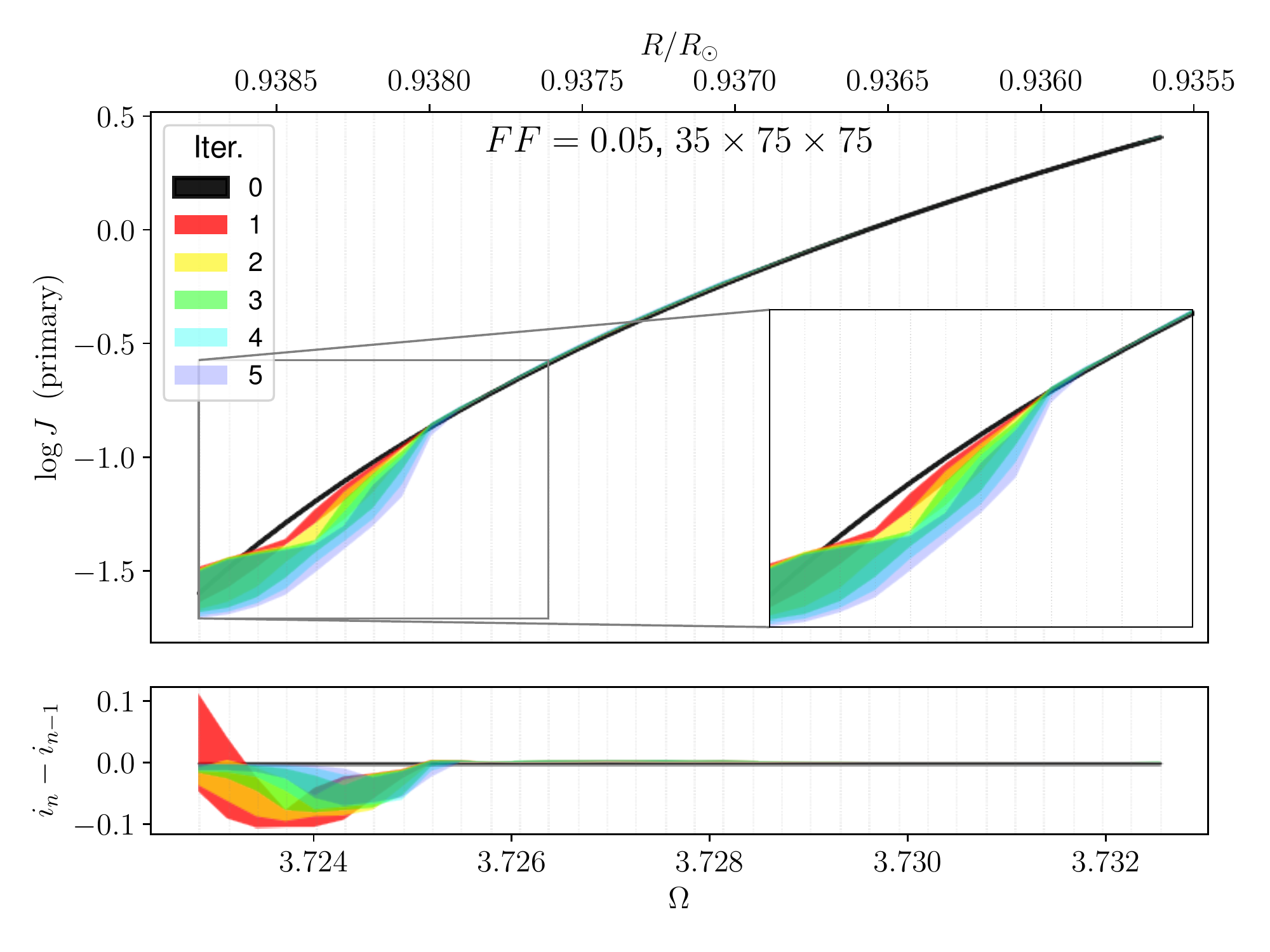}
    \includegraphics[width=0.495\hsize]{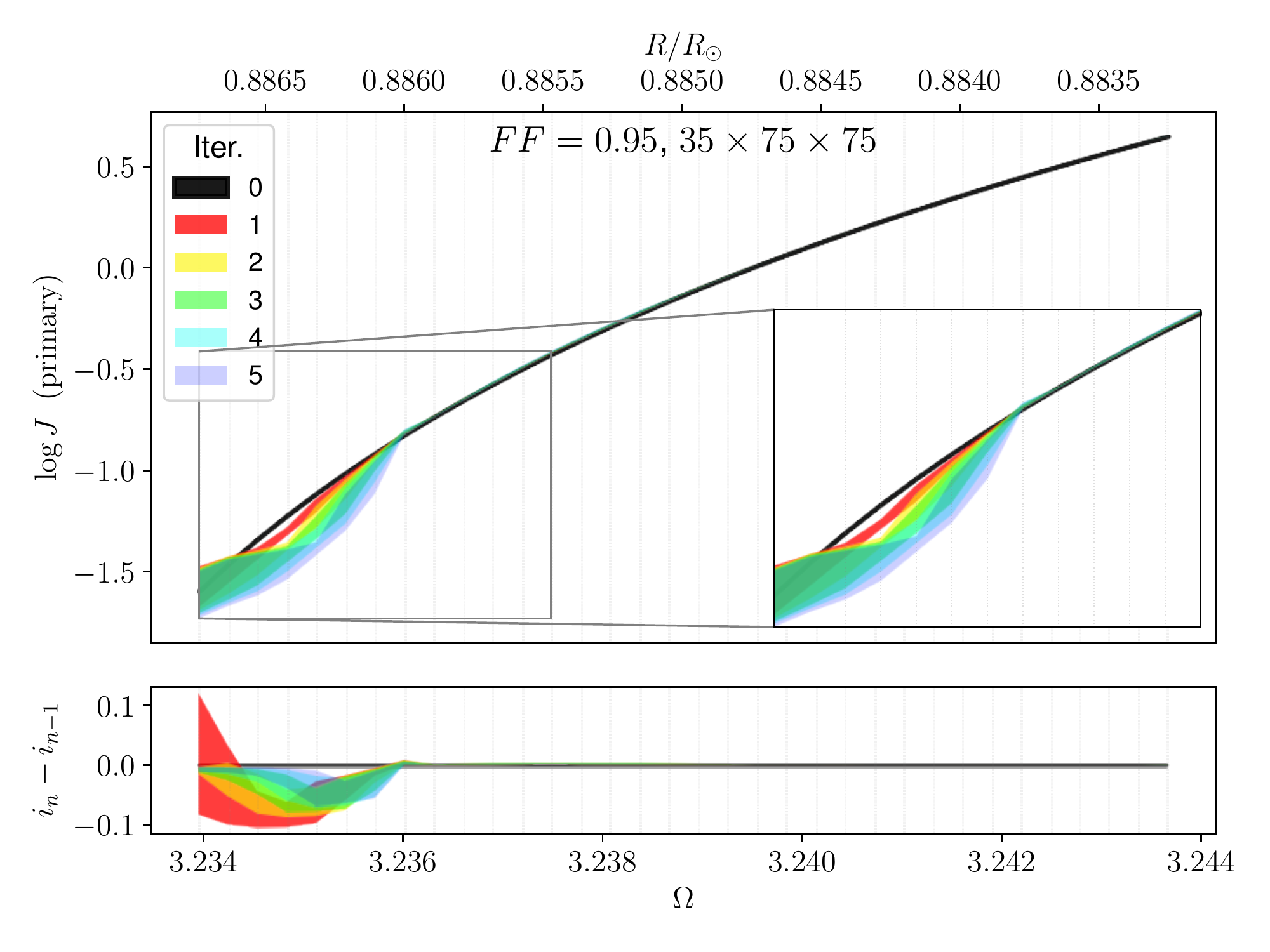}
    
    \caption{Top to bottom: outward, inward and mean intensity as a function of the potential/radius of a contact binary with $(N_{\Omega}, N_{\theta}, N_{\phi}) = (35, 75, 75)$. Left panels: $FF=0.05$, right panels: $FF=0.95$. The bottom panel of each plot shows the differences between successive iterations.}
    \label{fig:dims357575}
\end{figure}

\begin{figure}[h]
    \centering
          \includegraphics[width=0.495\hsize]{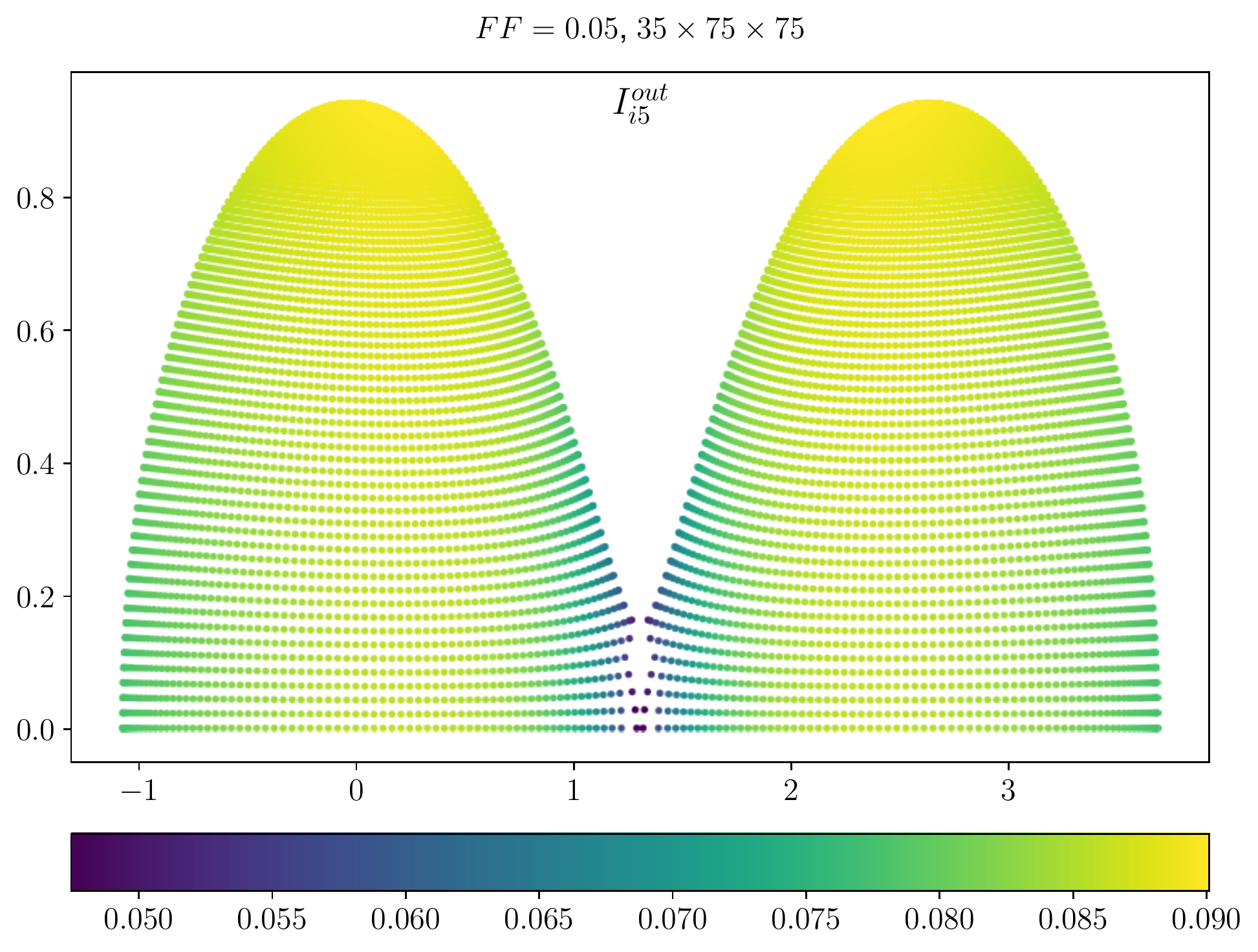}
    \includegraphics[width=0.495\hsize]{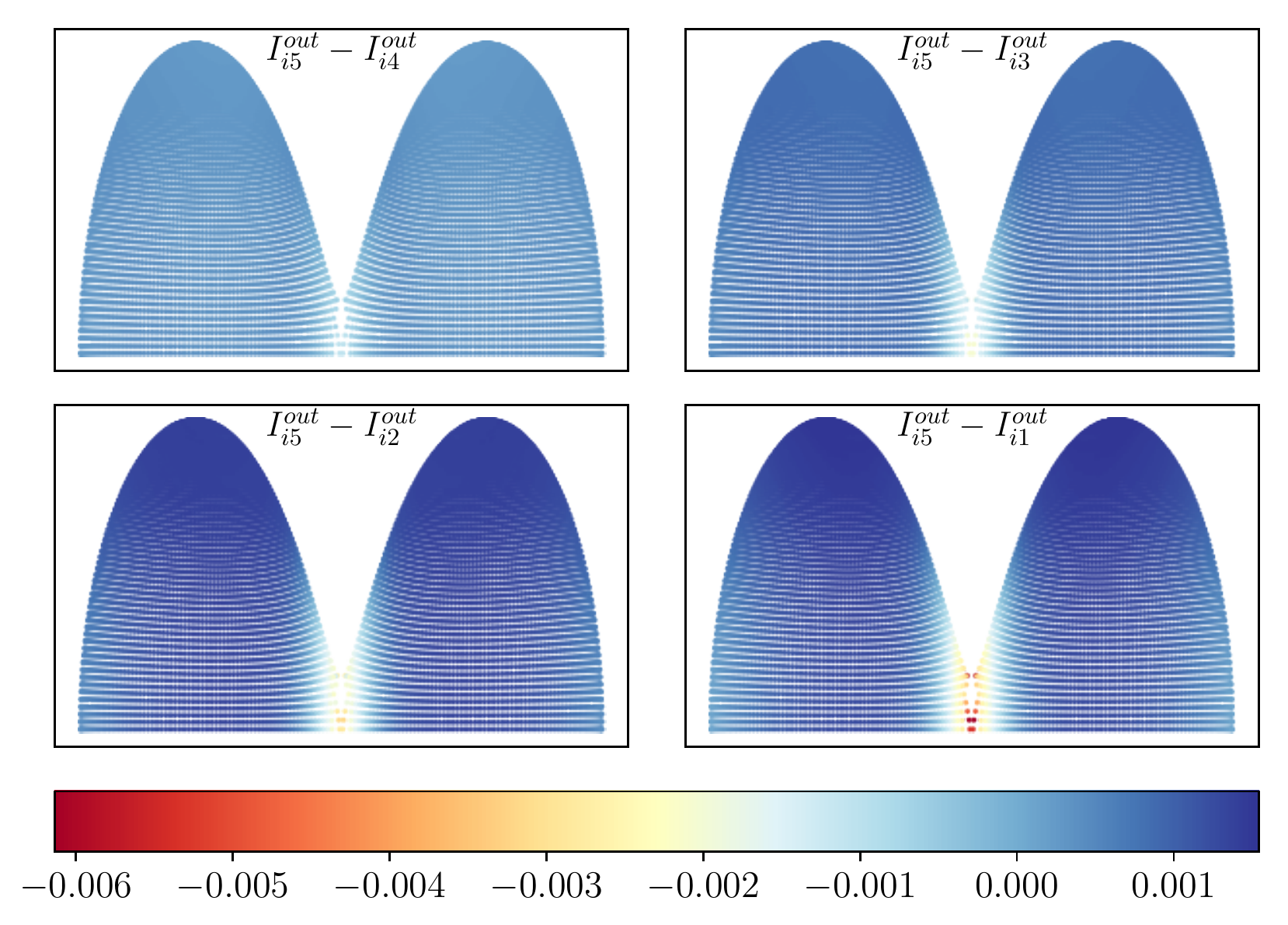}
    
        \includegraphics[width=0.495\hsize]{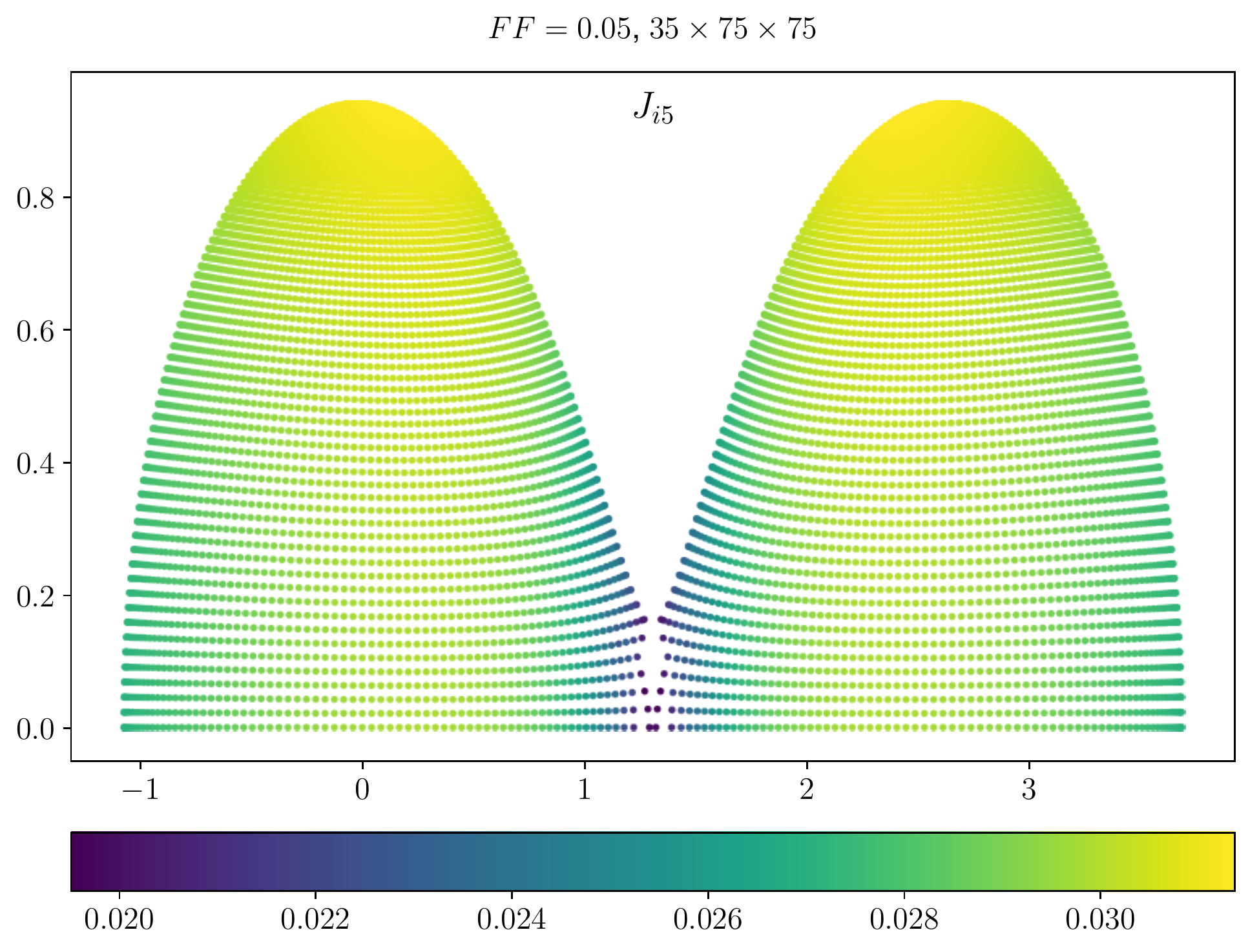}
    \includegraphics[width=0.495\hsize]{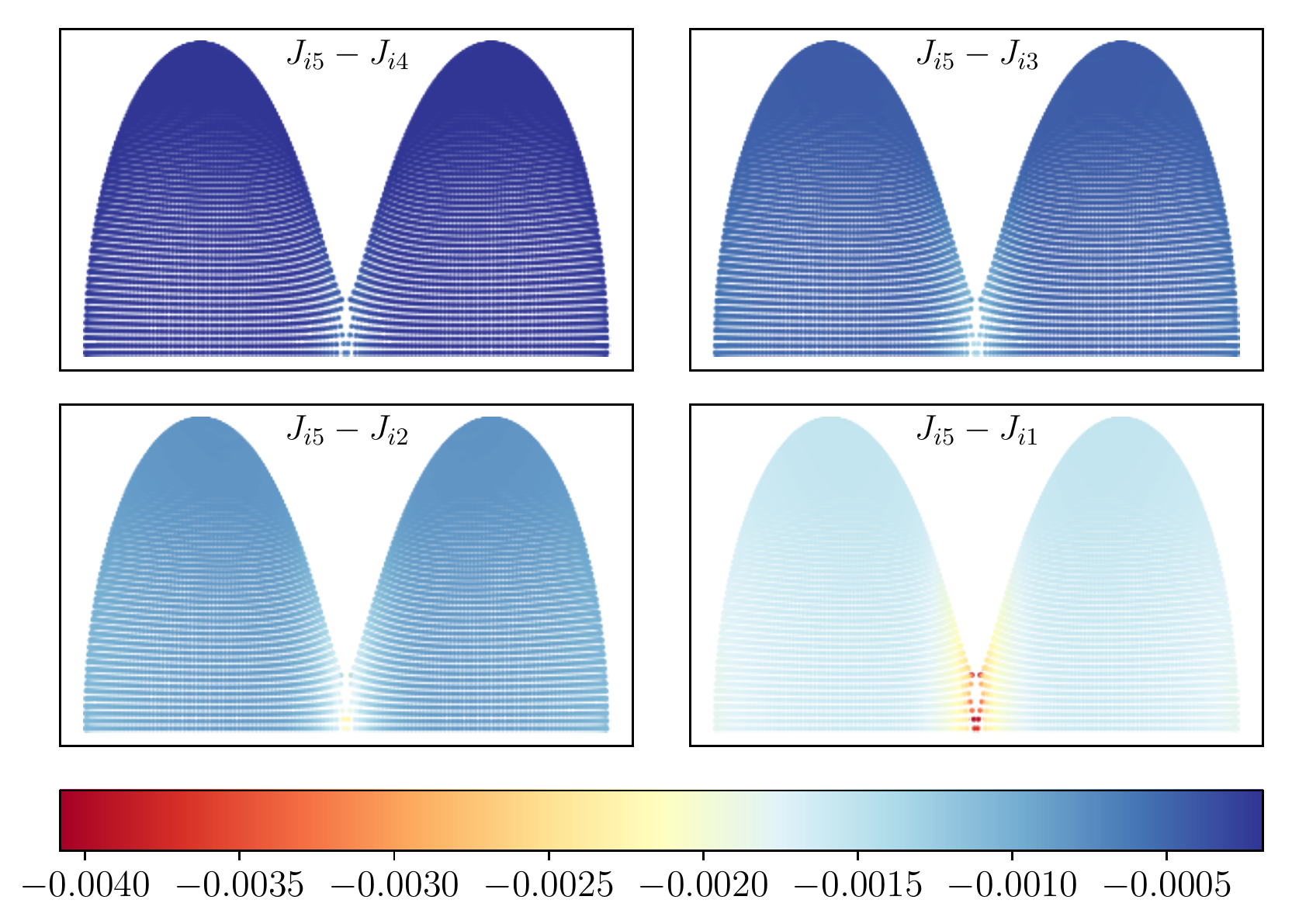}

    \caption{Surface distribution of the outward (top) and mean (bottom) intensity of a contact binary with $(N_{\Omega}, N_{\theta}, N_{\phi}) = (35, 75, 75)$ and $FF=0.05$ after the fifth iteration. Right panels show differences in the surface distribution between the final and each previous iteration.}
    \label{fig:dims357575_s}
\end{figure}

\begin{figure}[h]
    \centering
       \includegraphics[width=0.495\hsize]{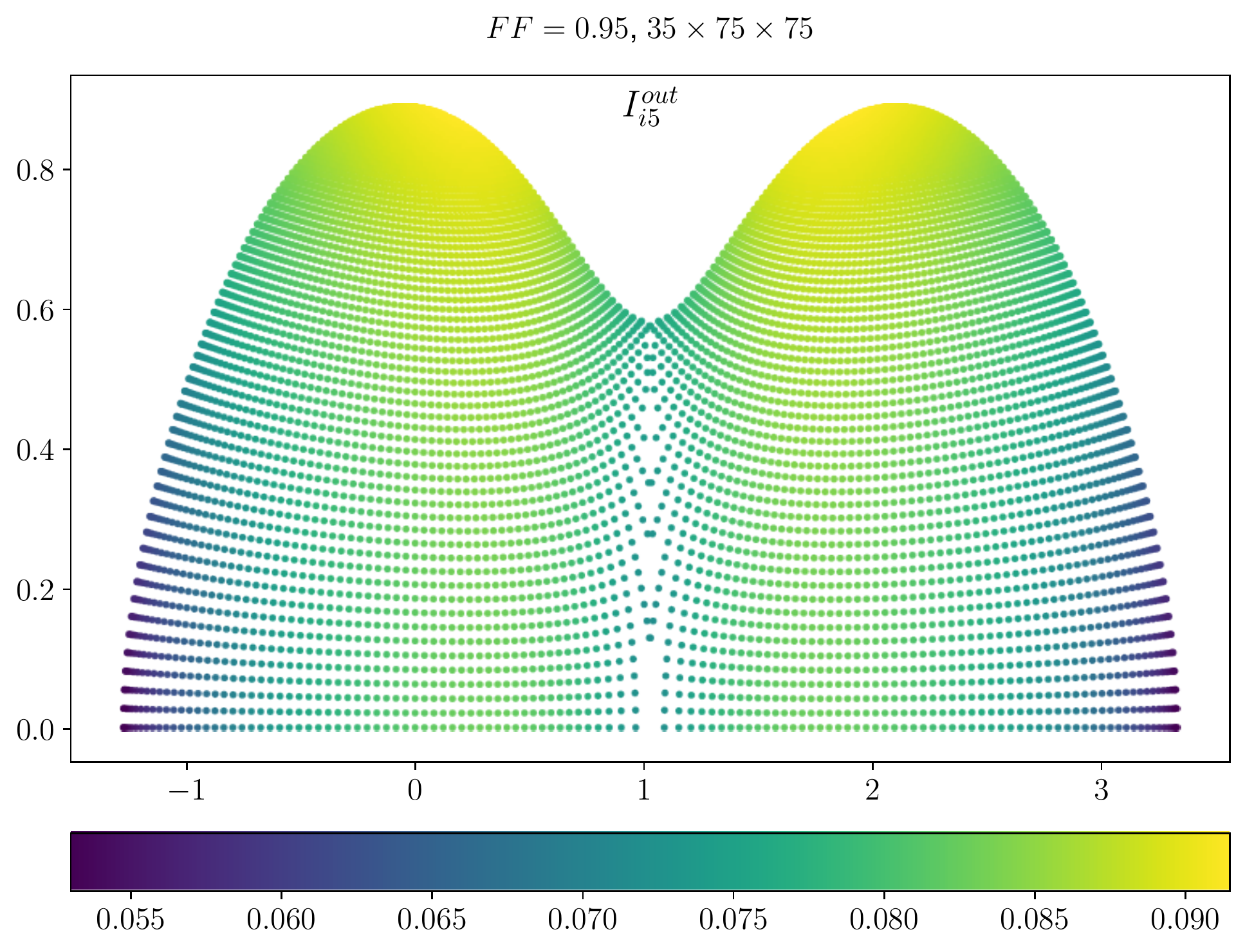}
    \includegraphics[width=0.495\hsize]{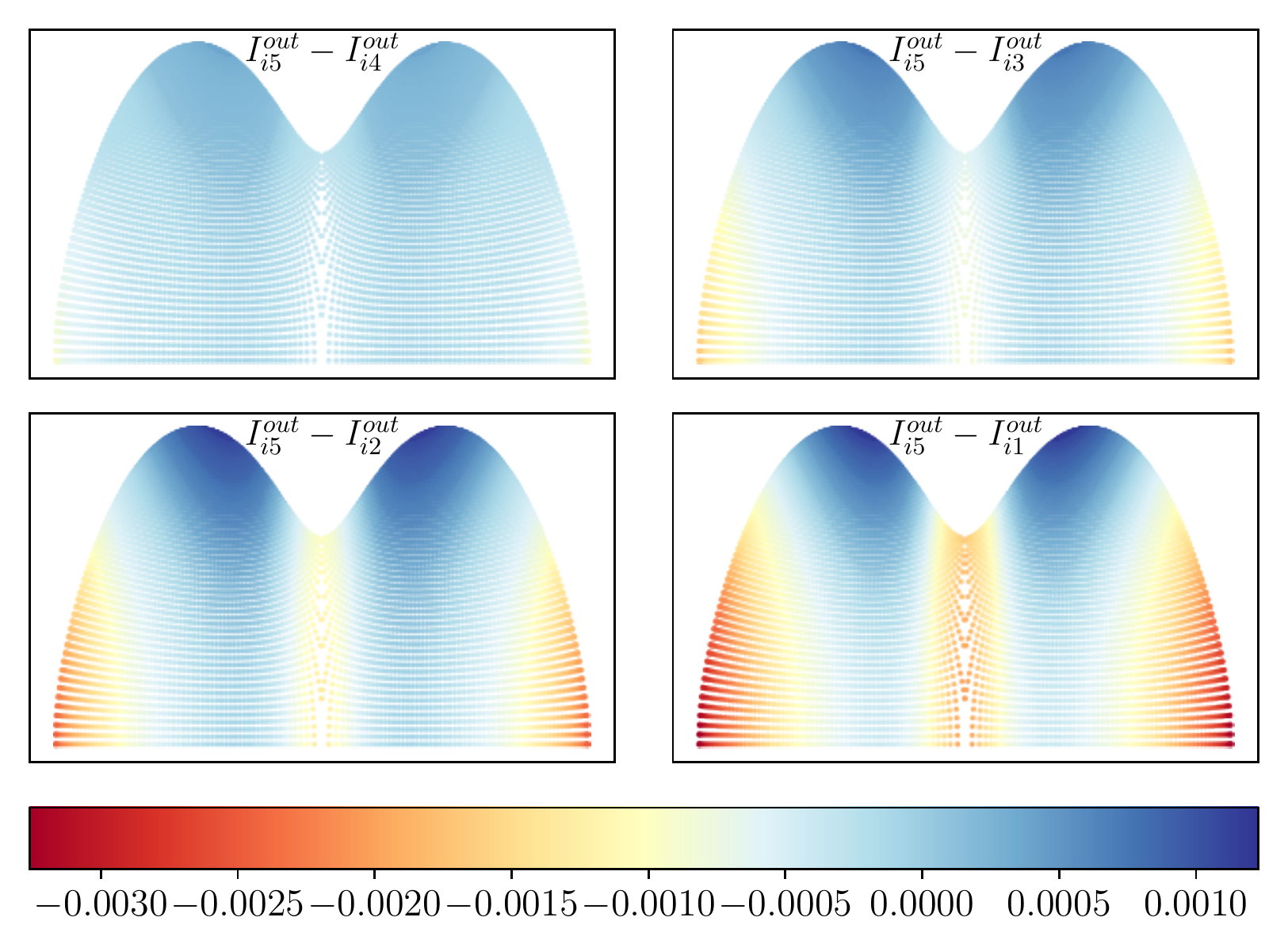}
    
        \includegraphics[width=0.495\hsize]{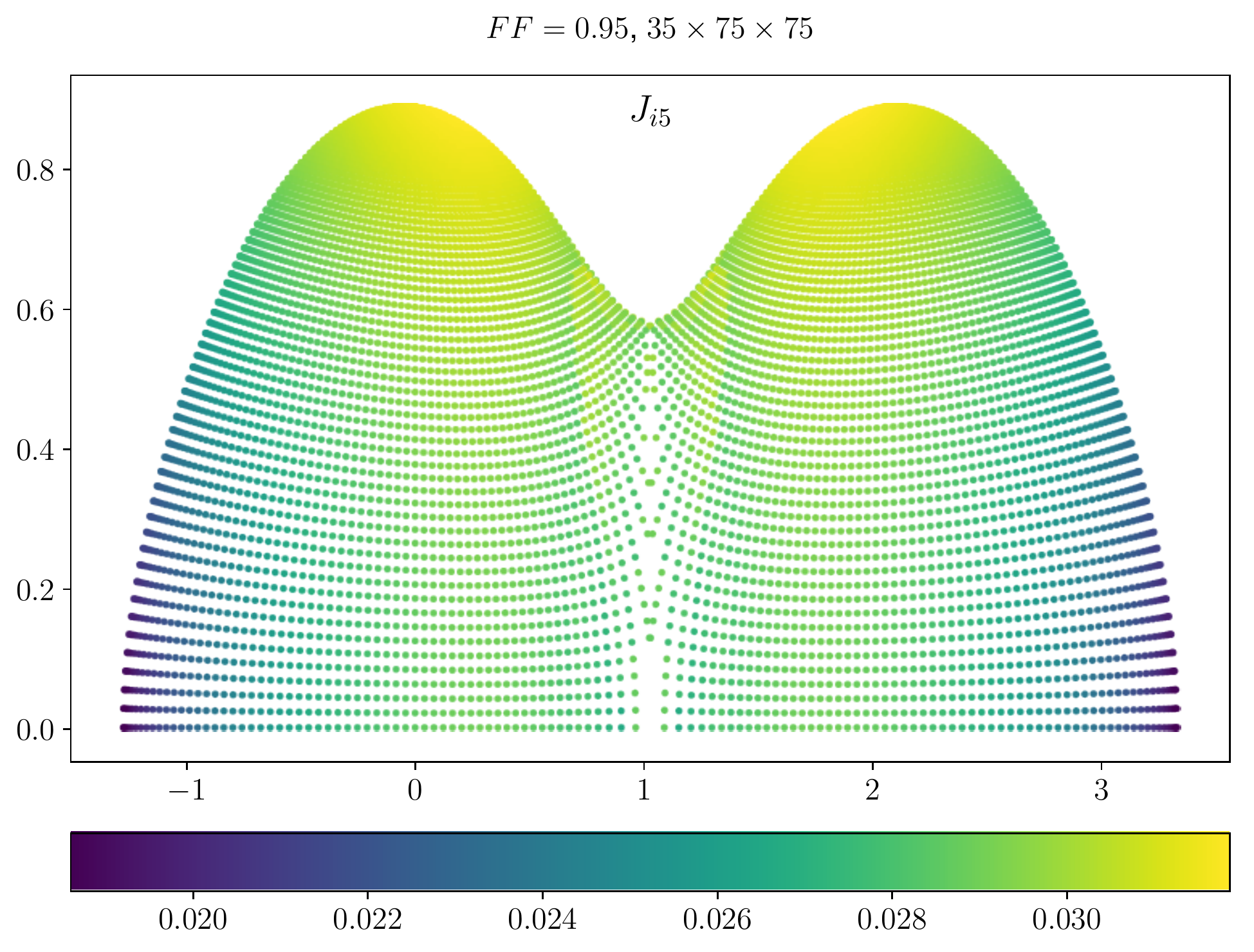}
    \includegraphics[width=0.495\hsize]{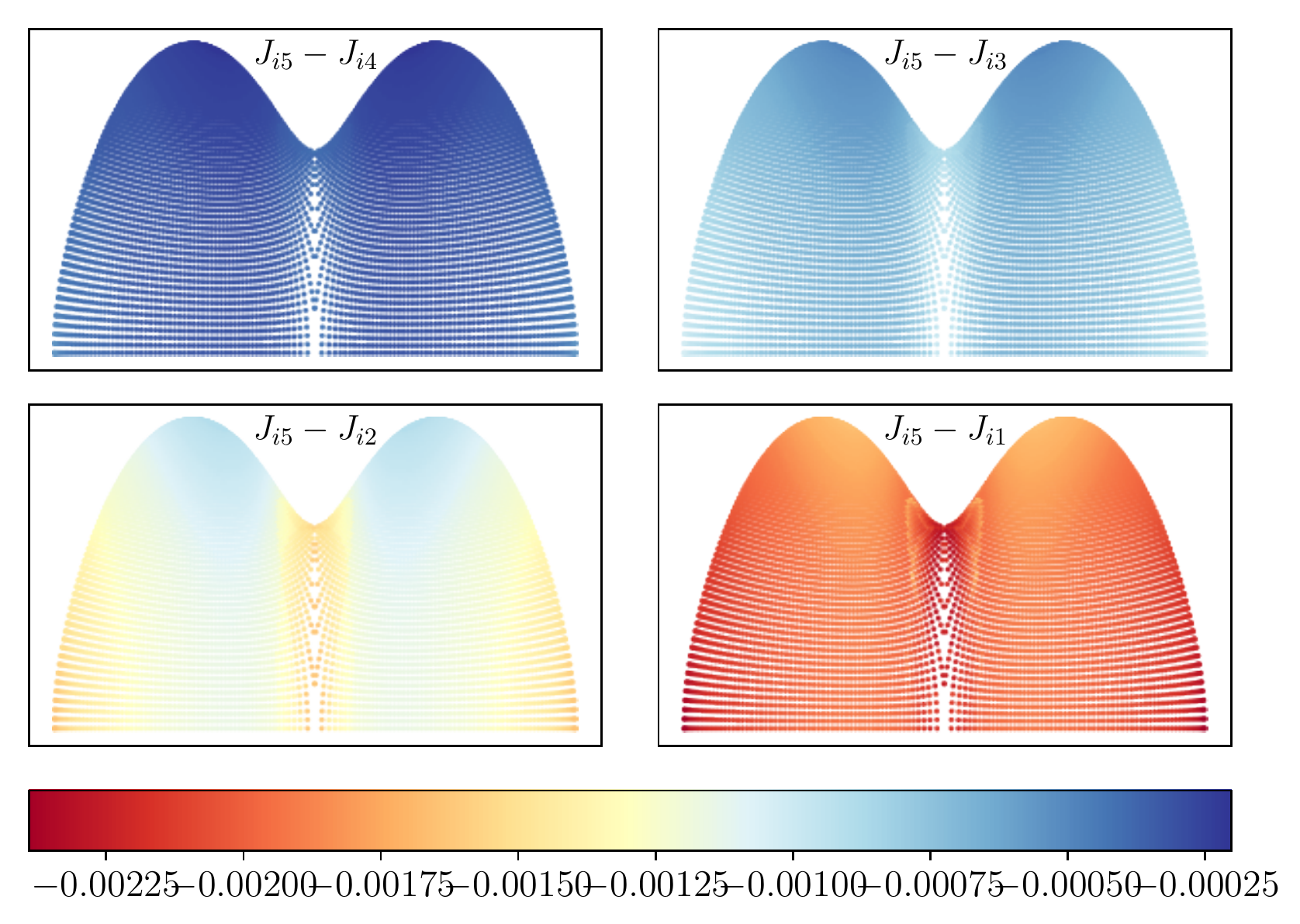}

    \caption{Surface distribution of the outward (top) and mean (bottom) intensity of a contact binary with $(N_{\Omega}, N_{\theta}, N_{\phi}) = (35, 75, 75)$ and $FF=0.95$ after the fifth iteration. Right panels show differences in the surface distribution between the final and each previous iteration.}
    \label{fig:dims357575_ff_s}
\end{figure}

\begin{figure}[h]
    \centering
    \includegraphics[width=0.495\hsize]{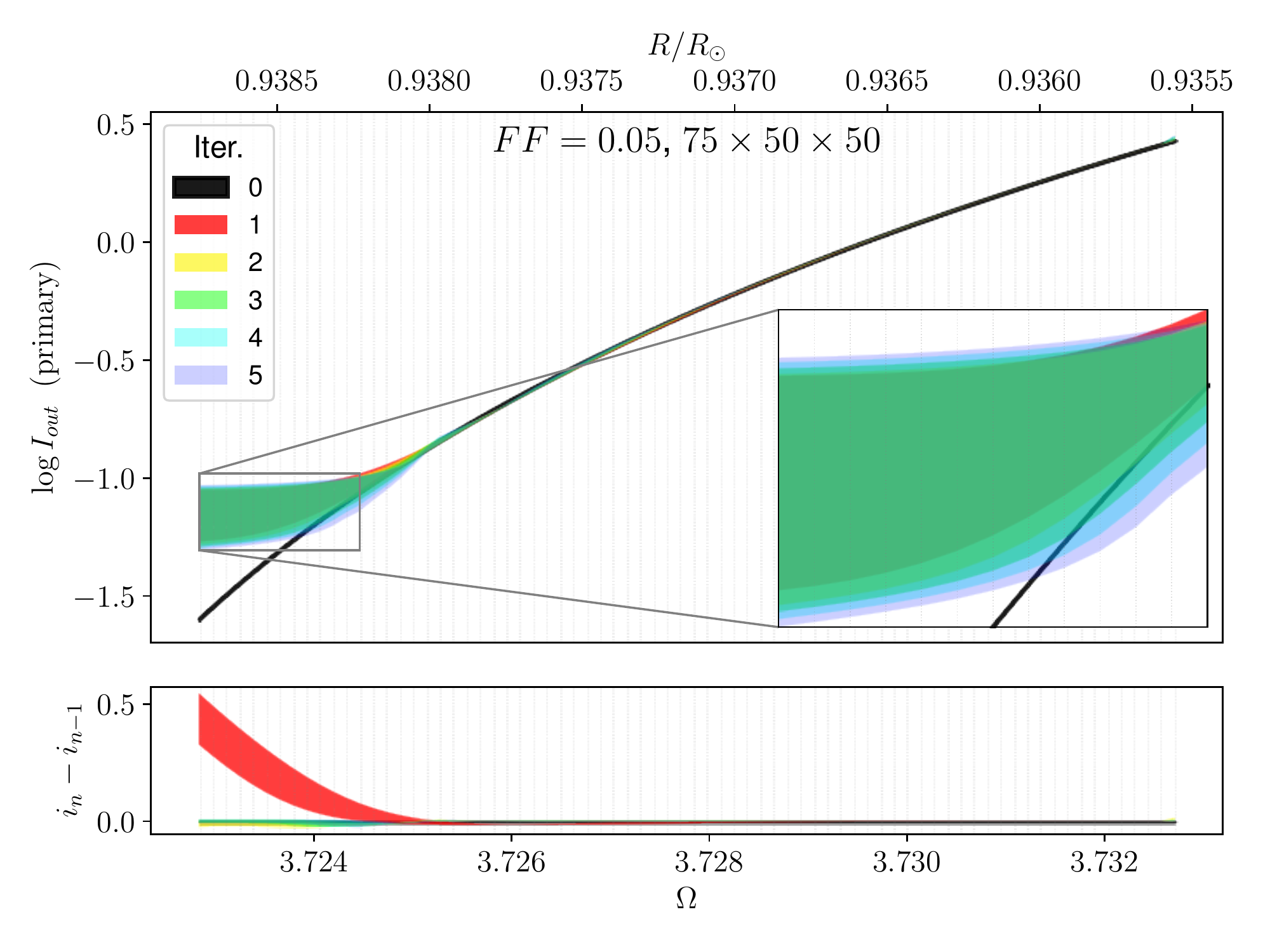}
    \includegraphics[width=0.495\hsize]{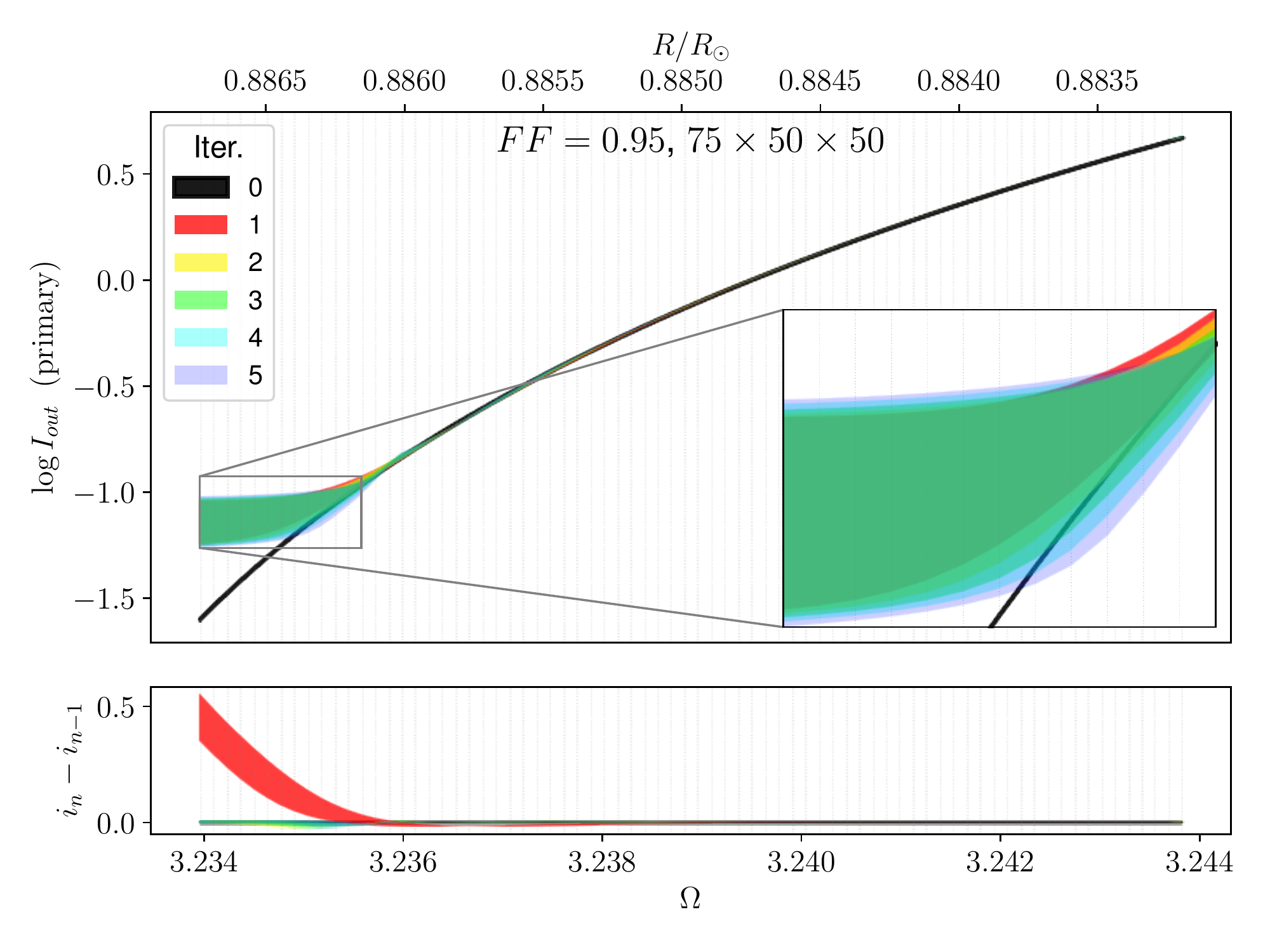}
    
    \includegraphics[width=0.495\hsize]{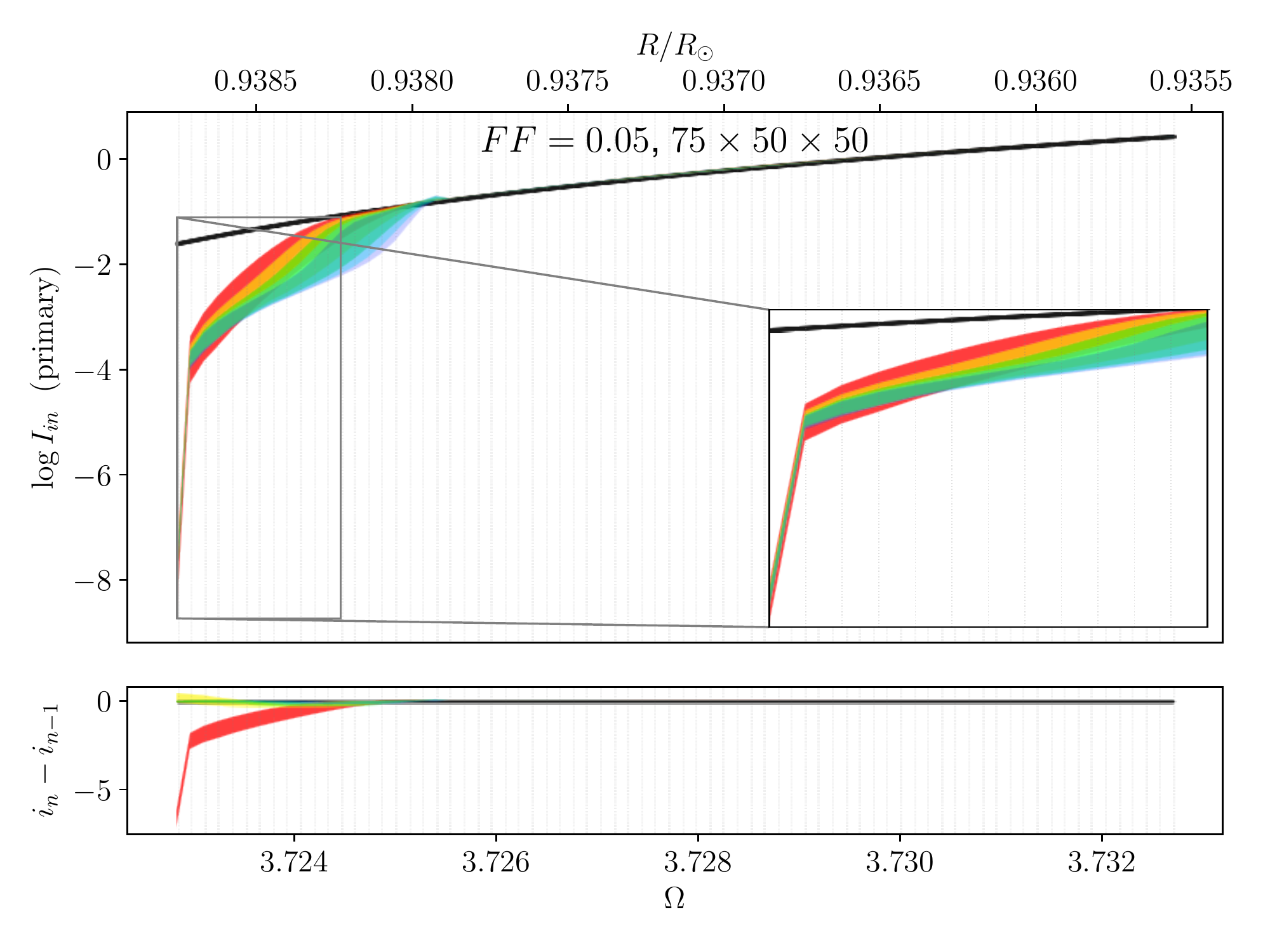}
    \includegraphics[width=0.495\hsize]{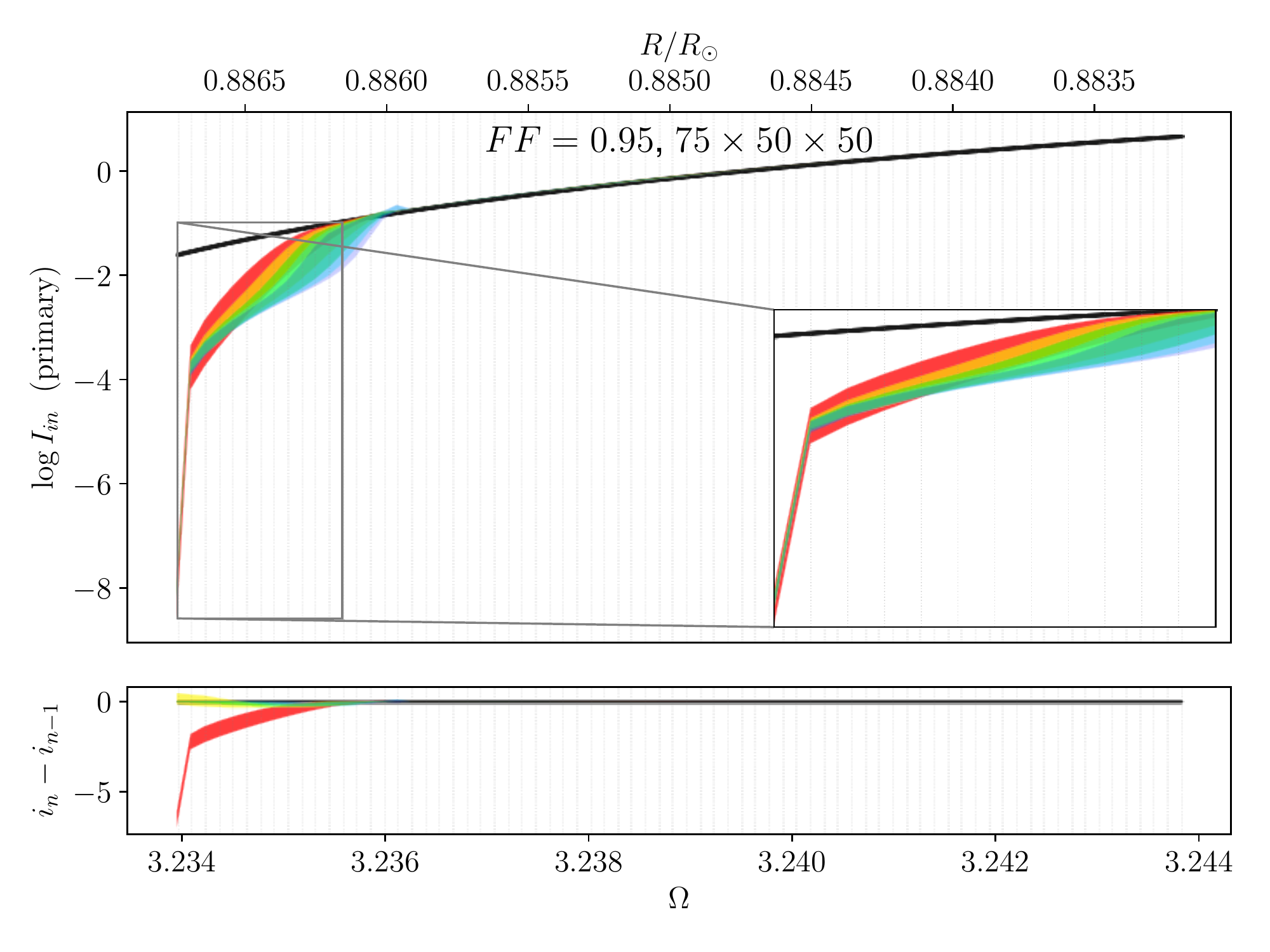}
    
    \includegraphics[width=0.495\hsize]{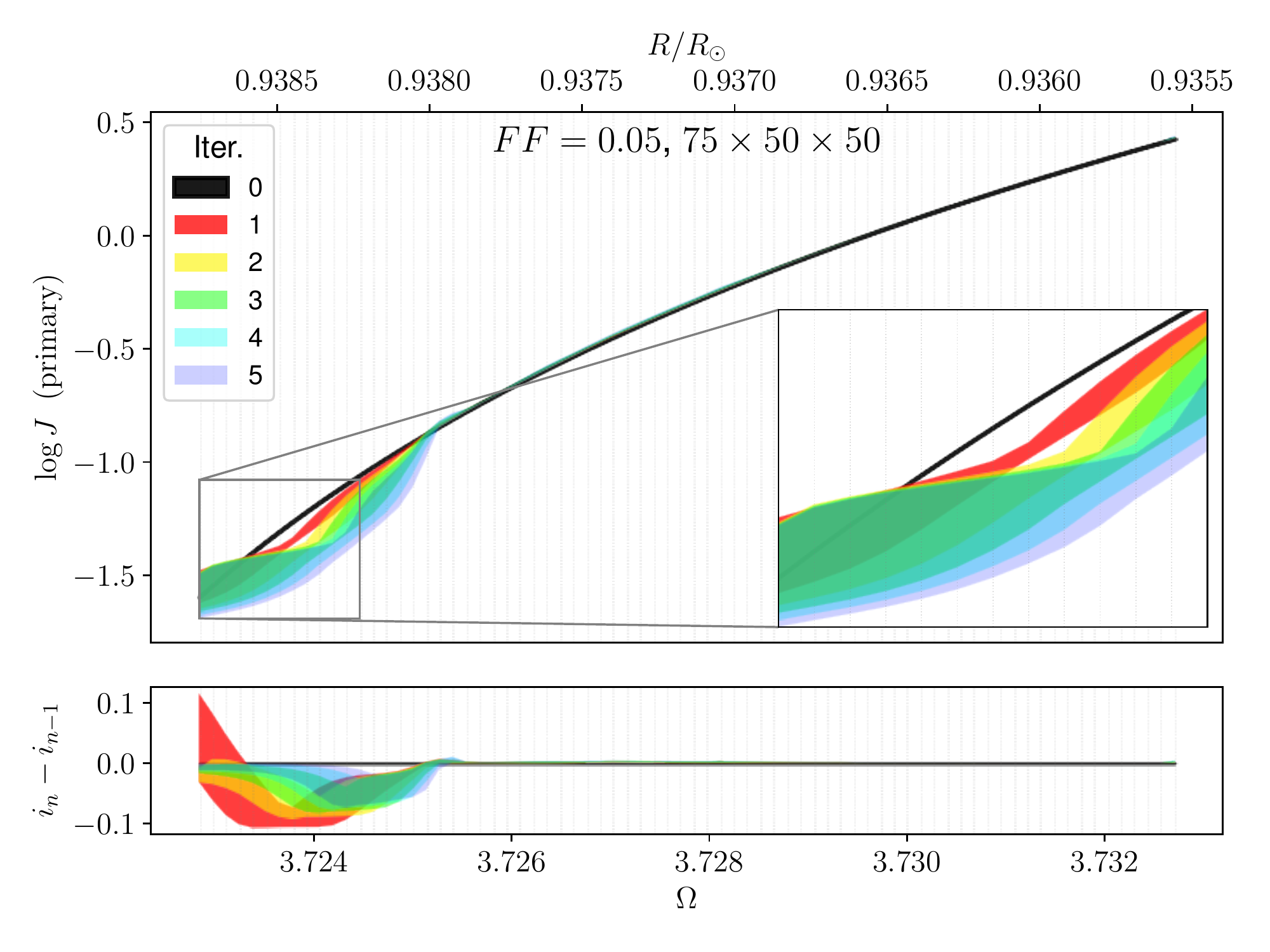}
    \includegraphics[width=0.495\hsize]{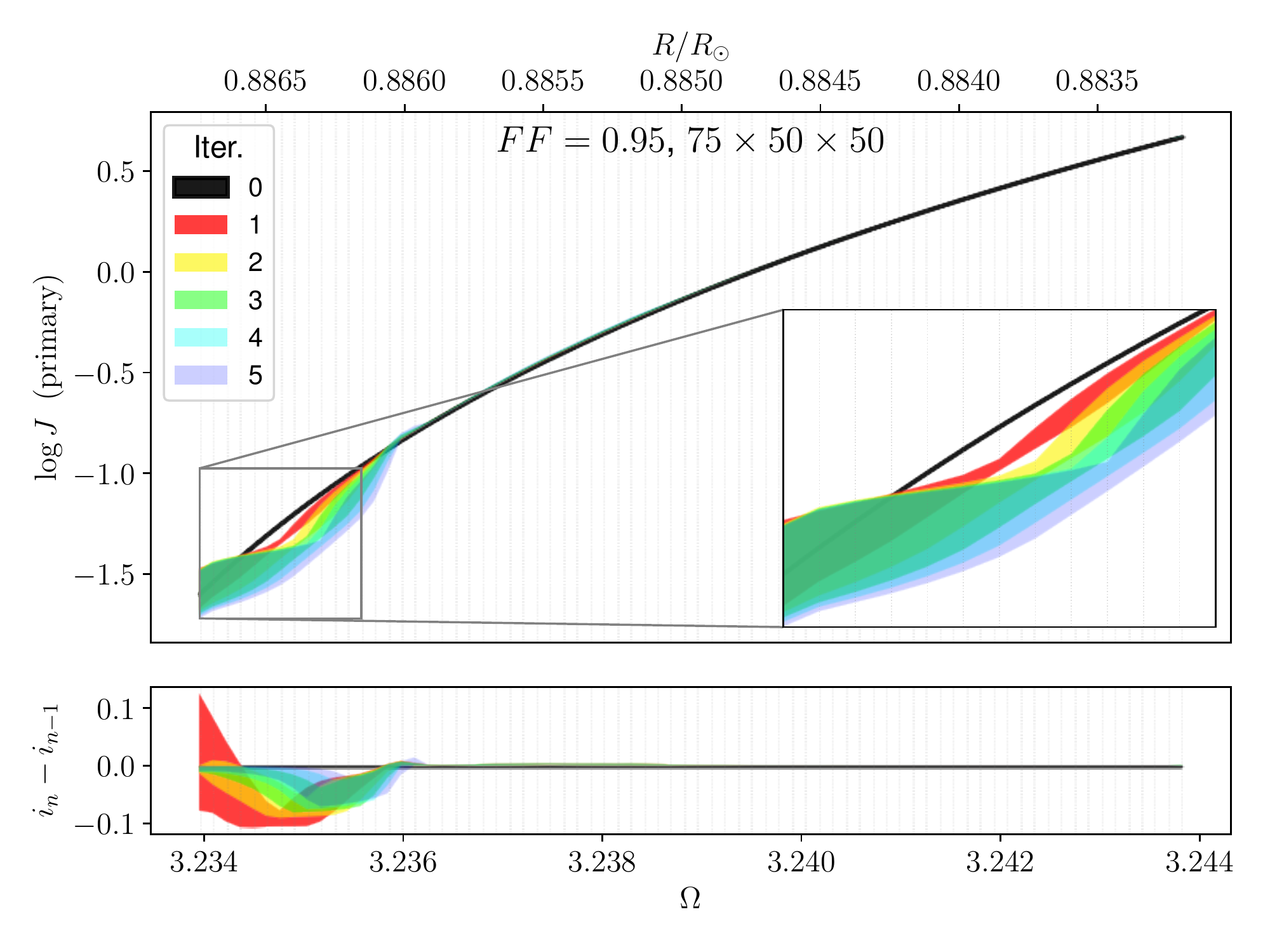}
    
    \caption{Top to bottom: outward, inward and mean intensity as a function of the potential/radius of a contact binary with $(N_{\Omega}, N_{\theta}, N_{\phi}) = (75, 50, 50)$. Left panels: $FF=0.05$, right panels: $FF=0.95$. The bottom panel of each plot shows the differences between successive iterations.}
    \label{fig:dims755050}
\end{figure}

\begin{figure}[h]
    \centering
    \includegraphics[width=0.495\hsize]{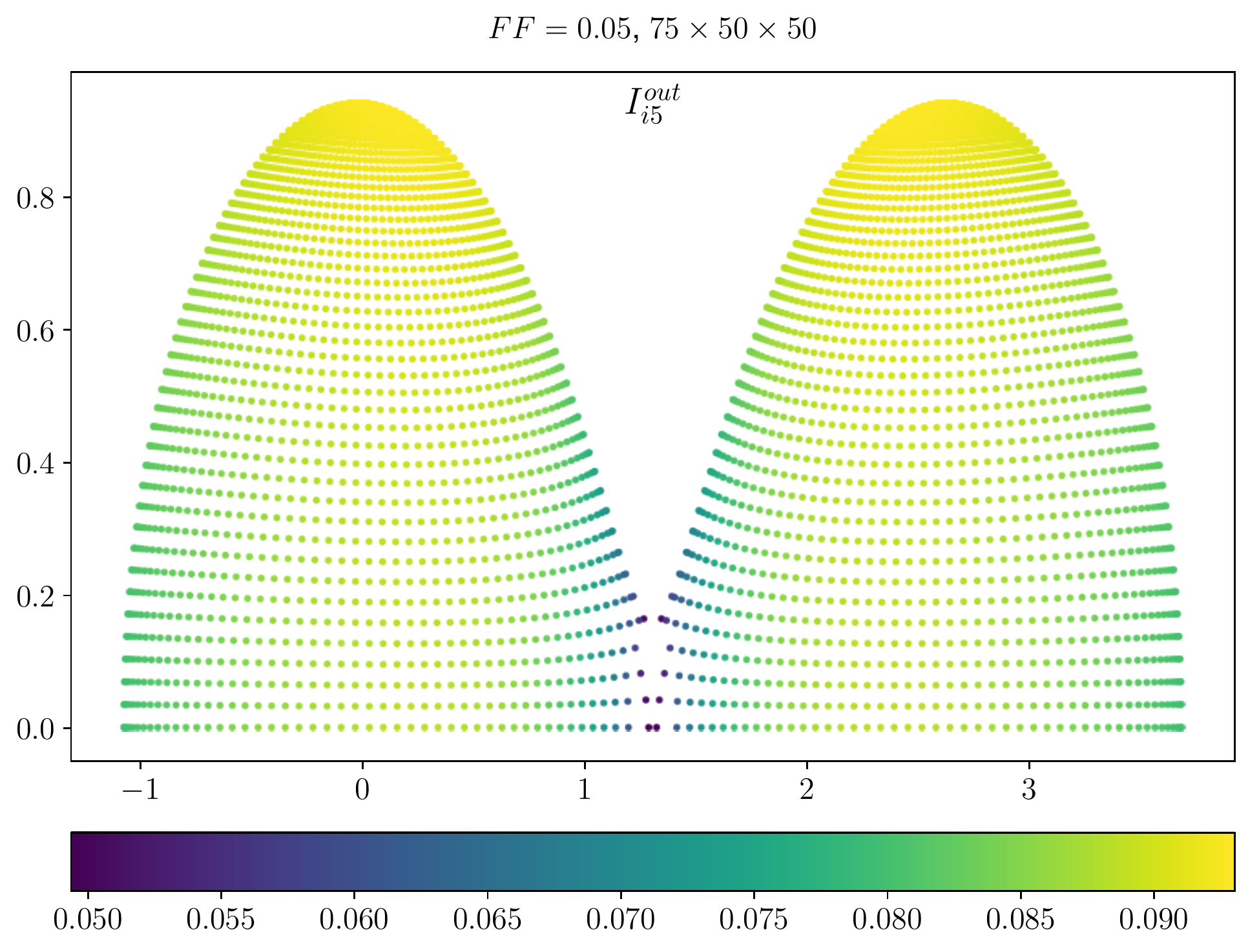}
    \includegraphics[width=0.495\hsize]{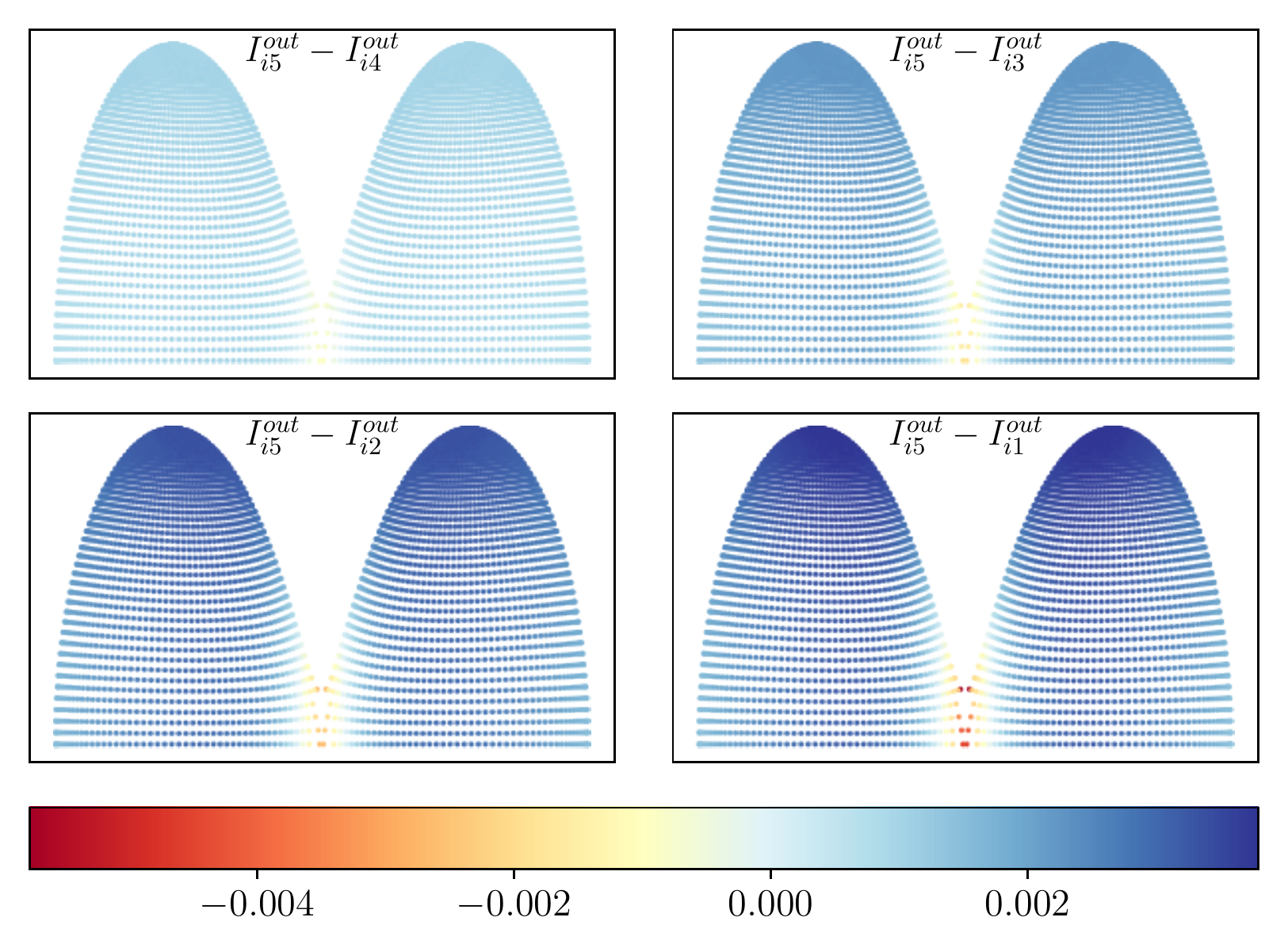}
    
        \includegraphics[width=0.495\hsize]{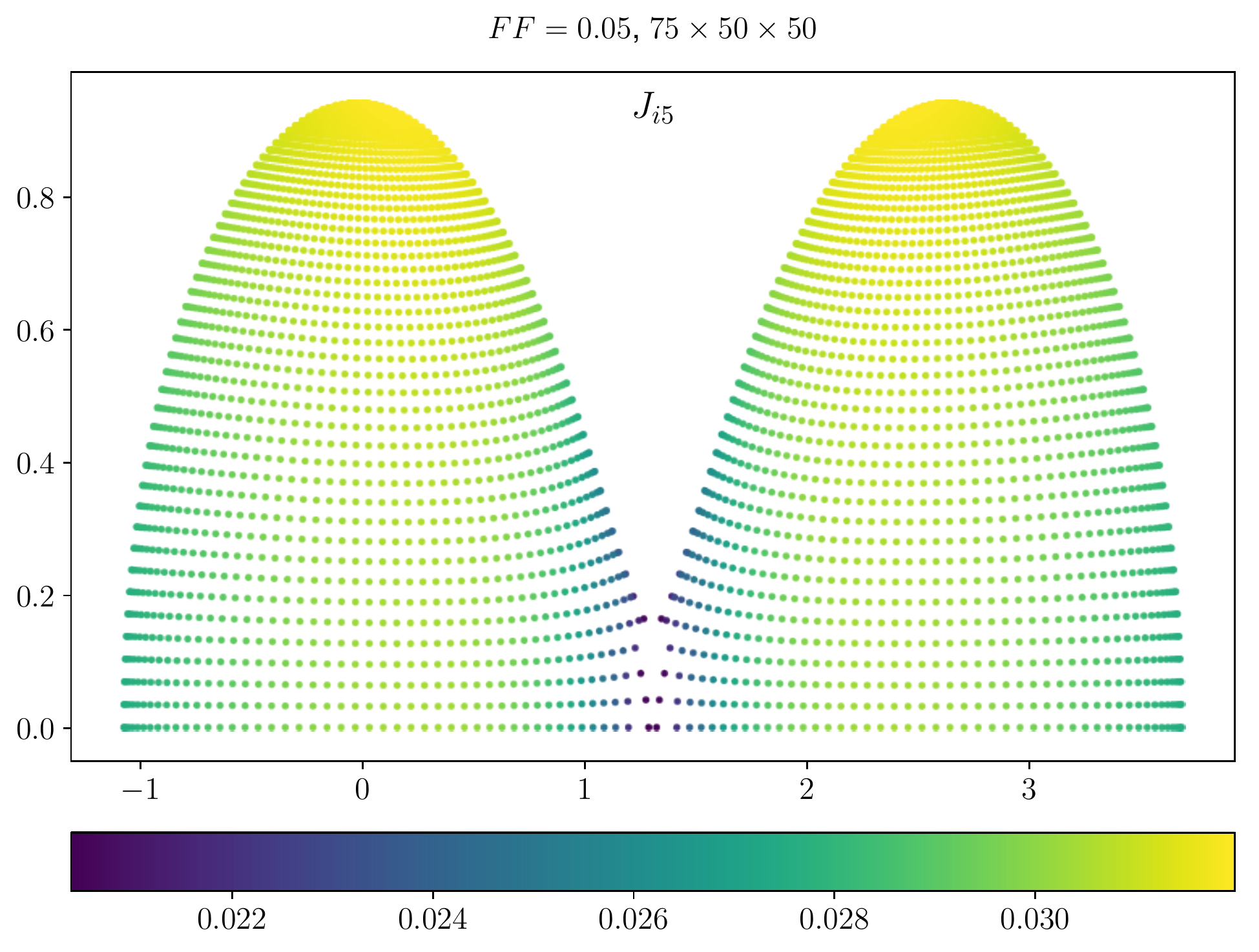}
    \includegraphics[width=0.495\hsize]{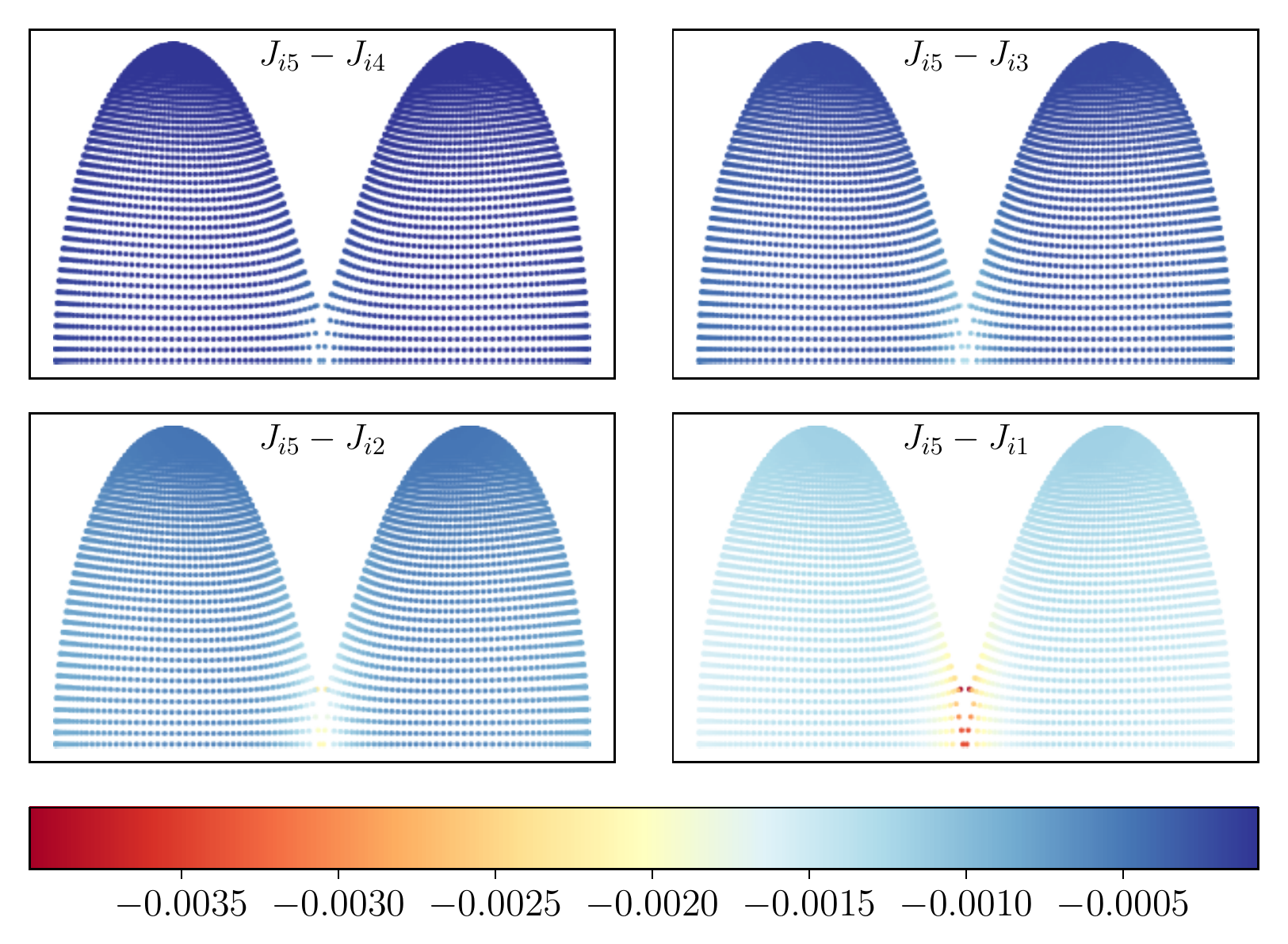}

    \caption{Surface distribution of the outward (top) and mean (bottom) intensity of a contact binary with $(N_{\Omega}, N_{\theta}, N_{\phi}) = (75, 50, 50)$ and $FF=0.05$ after the fifth iteration. Right panels show differences in the surface distribution between the final and each previous iteration.}
    \label{fig:dims755050_s}
\end{figure}

\begin{figure}[h]
    \centering
    \includegraphics[width=0.495\hsize]{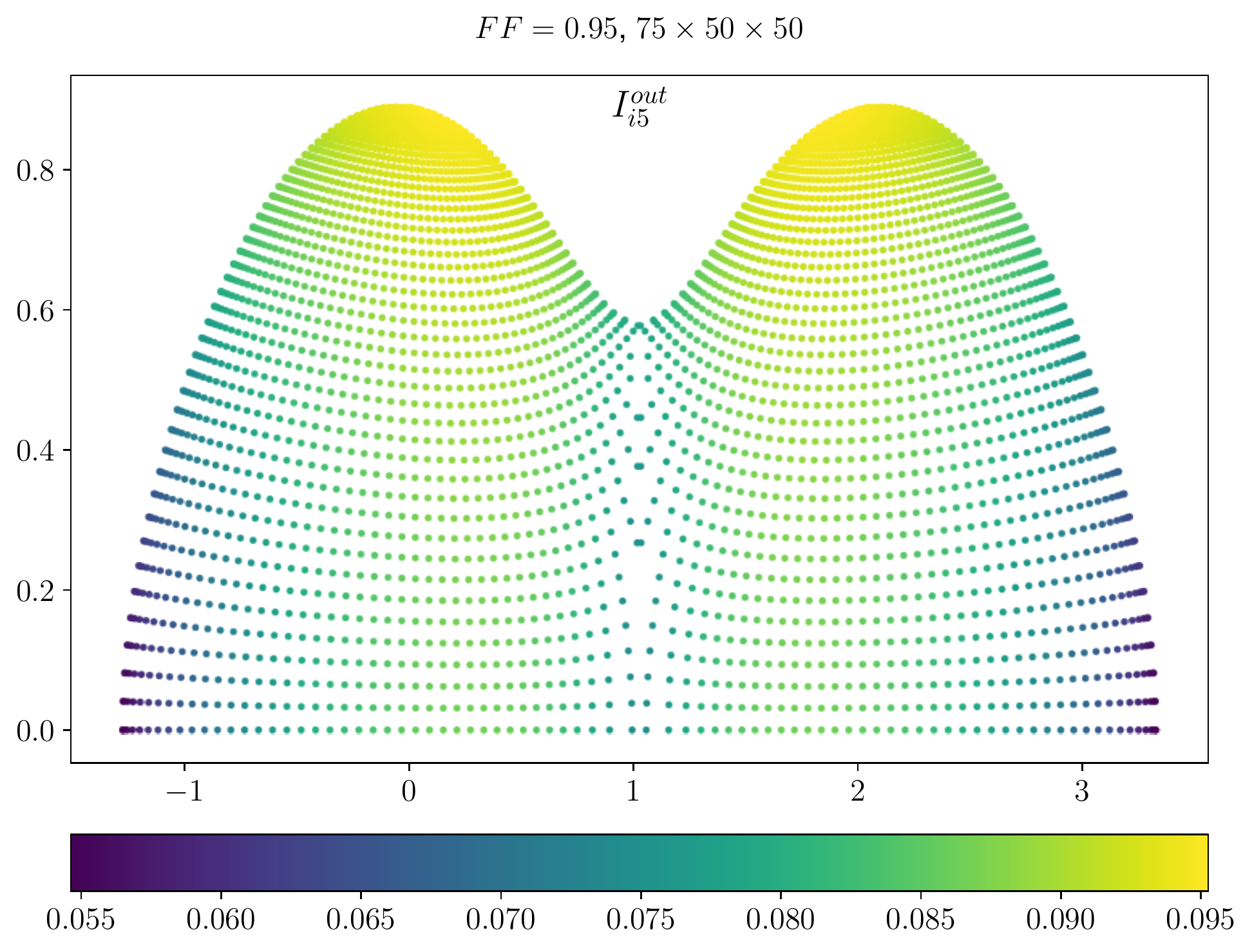}
    \includegraphics[width=0.495\hsize]{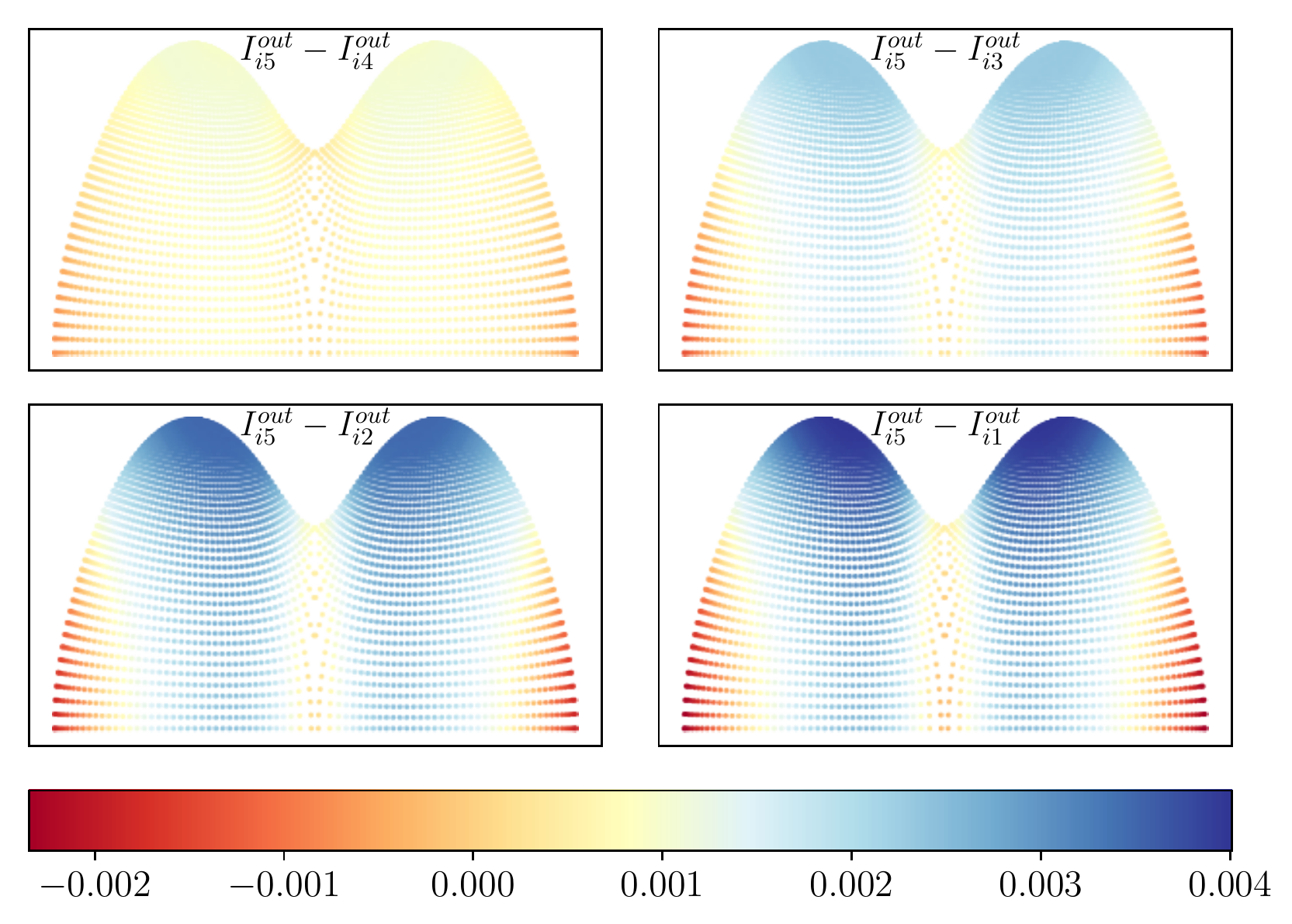}
    
        \includegraphics[width=0.495\hsize]{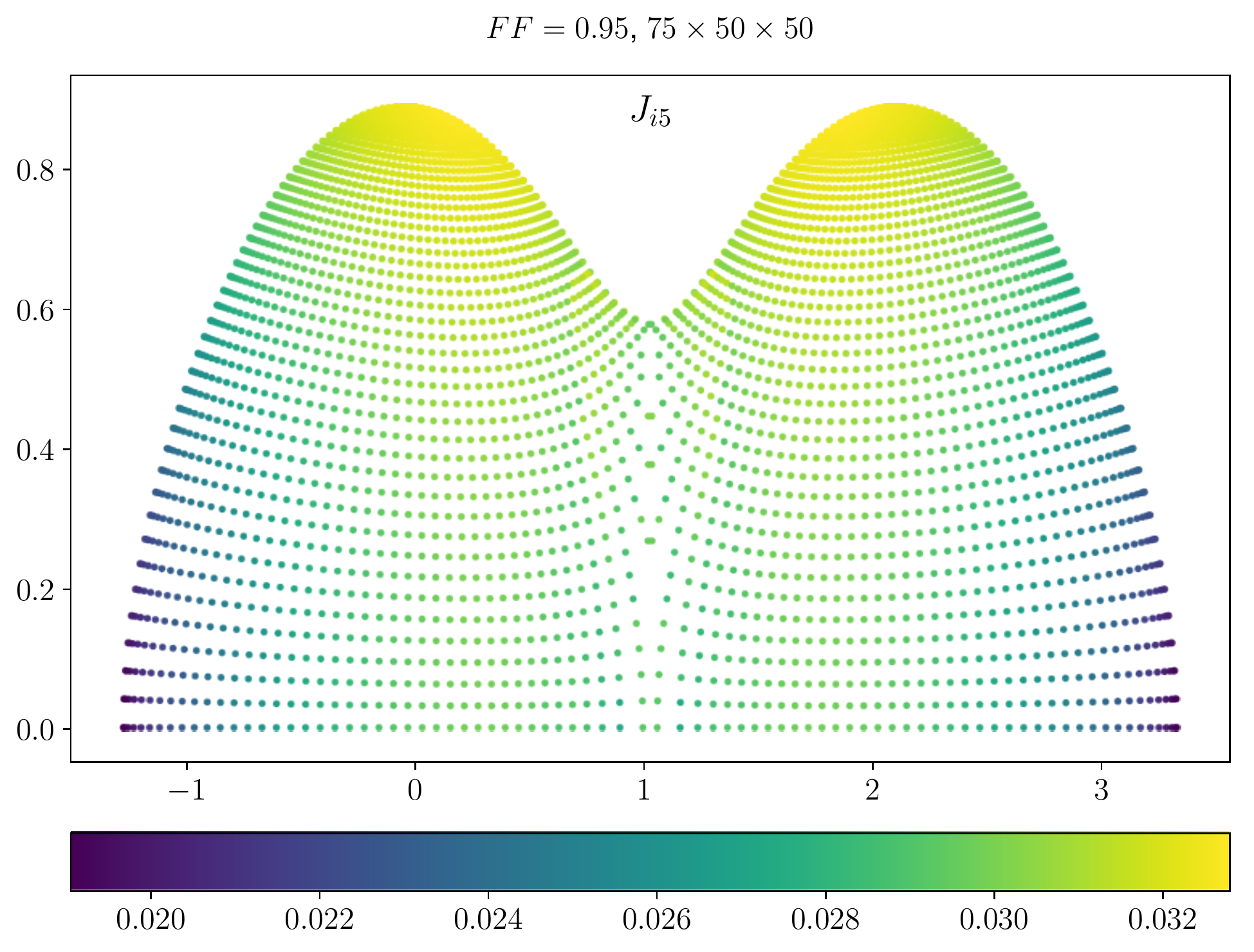}
    \includegraphics[width=0.495\hsize]{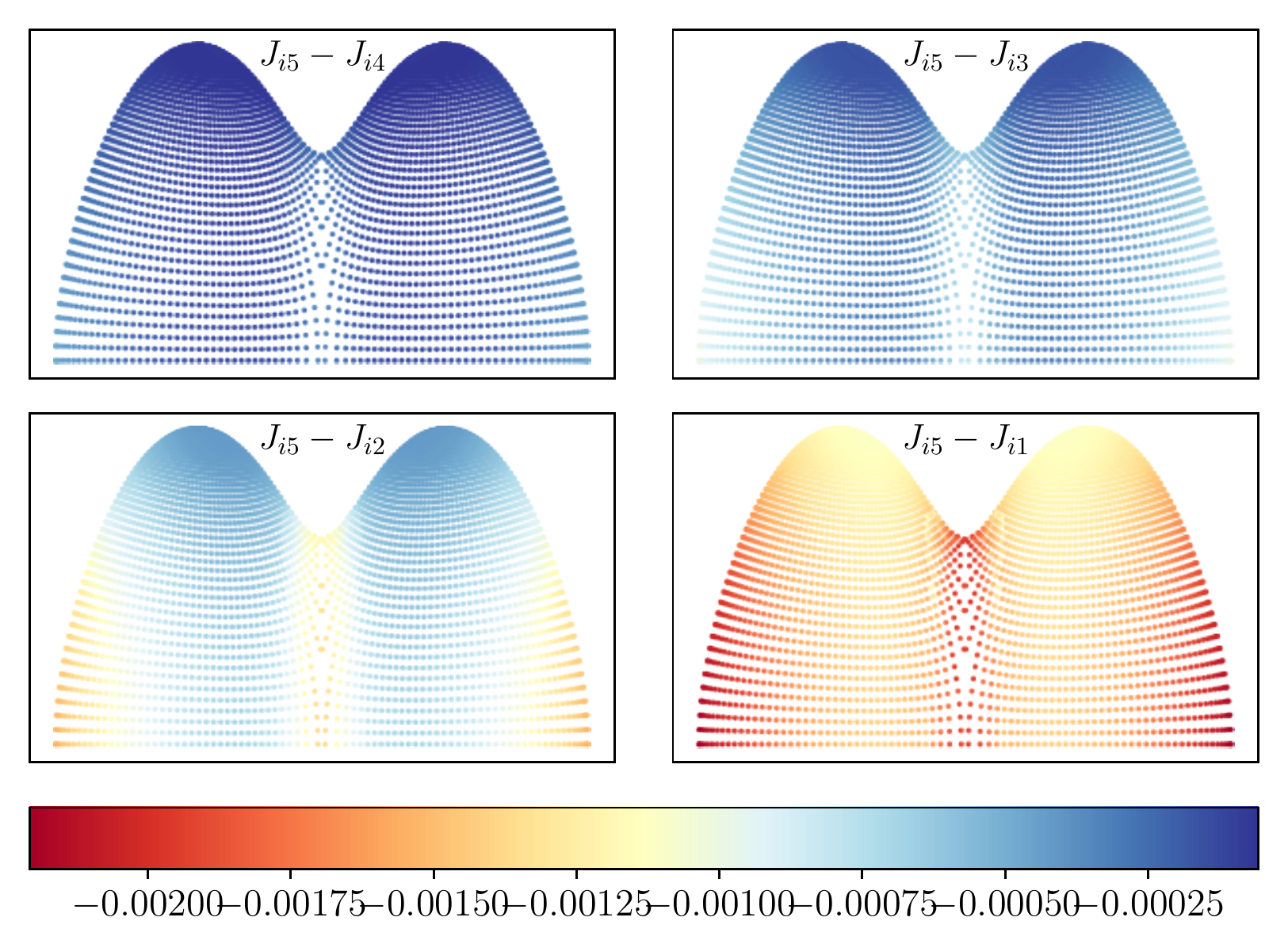}

    \caption{Surface distribution of the outward (top) and mean (bottom) intensity of a contact binary with $(N_{\Omega}, N_{\theta}, N_{\phi}) = (75, 50, 50)$ and $FF=0.95$ after the fifth iteration. Right panels show differences in the surface distribution between the final and each previous iteration.}
    \label{fig:dims755050_ff_s}
\end{figure}

The resulting distributions after five iterations (Figures~\ref{fig:dims502525}-\ref{fig:dims755050_ff_s}) are similar for all grid configurations and converge within the threshold of 0.01\% after five iterations (Figure~\ref{fig:iters_dims}), therefore it is safe to conclude that the interpolation works well enough for the chosen default grid resolution of  $(N_{\Omega}, N_{\theta}, N_{\phi}) = (50, 50, 50)$. 

\subsection{Mass ratio}

\begin{figure}[h]
    \centering
    \includegraphics[width=\hsize]{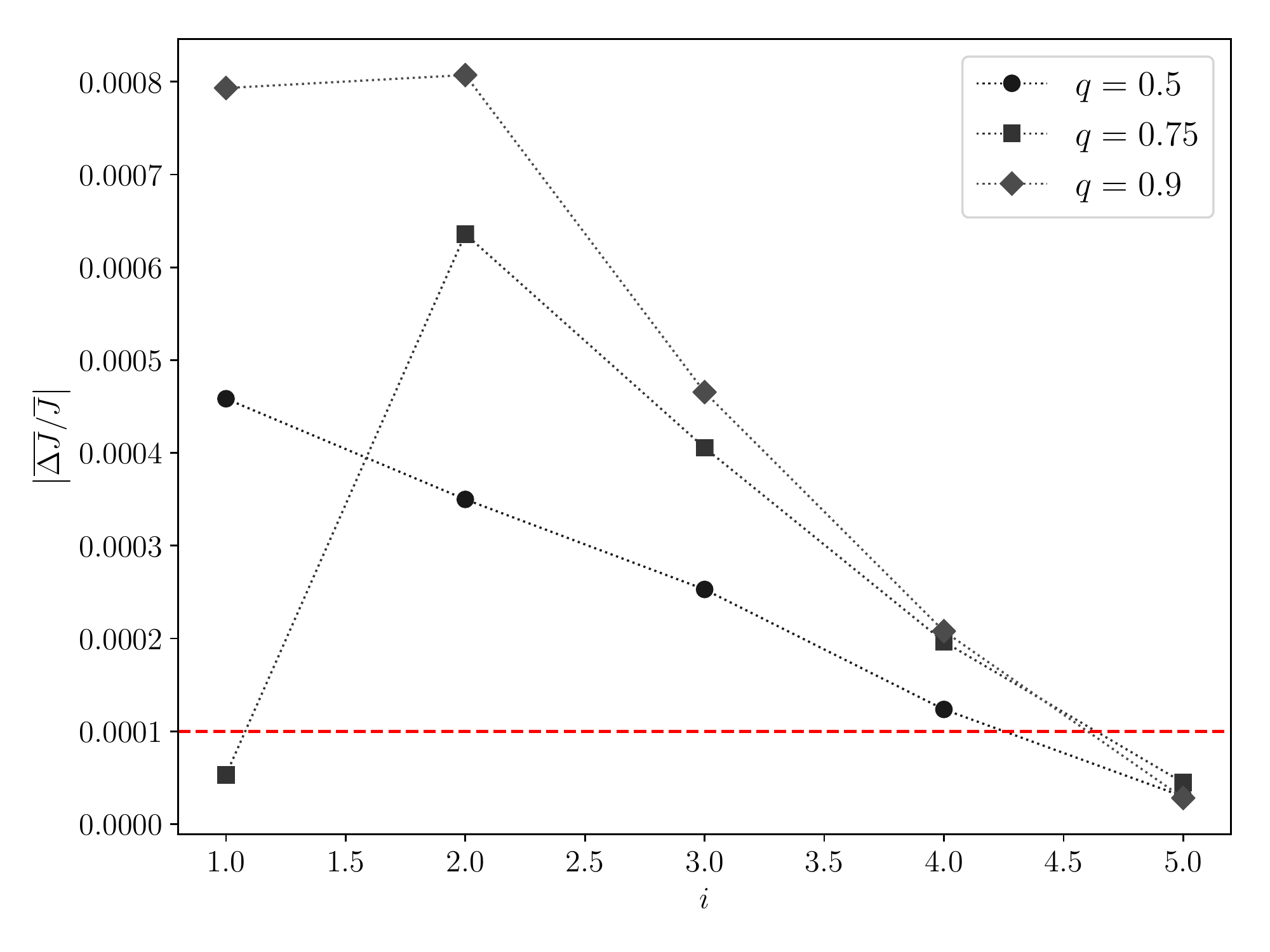}
    \caption{Convergence of the mean intensity in test systems with variable mass ratio.}
    \label{fig:iters_q}
\end{figure}

\begin{figure}[h]
    \centering
    \includegraphics[width=0.495\hsize]{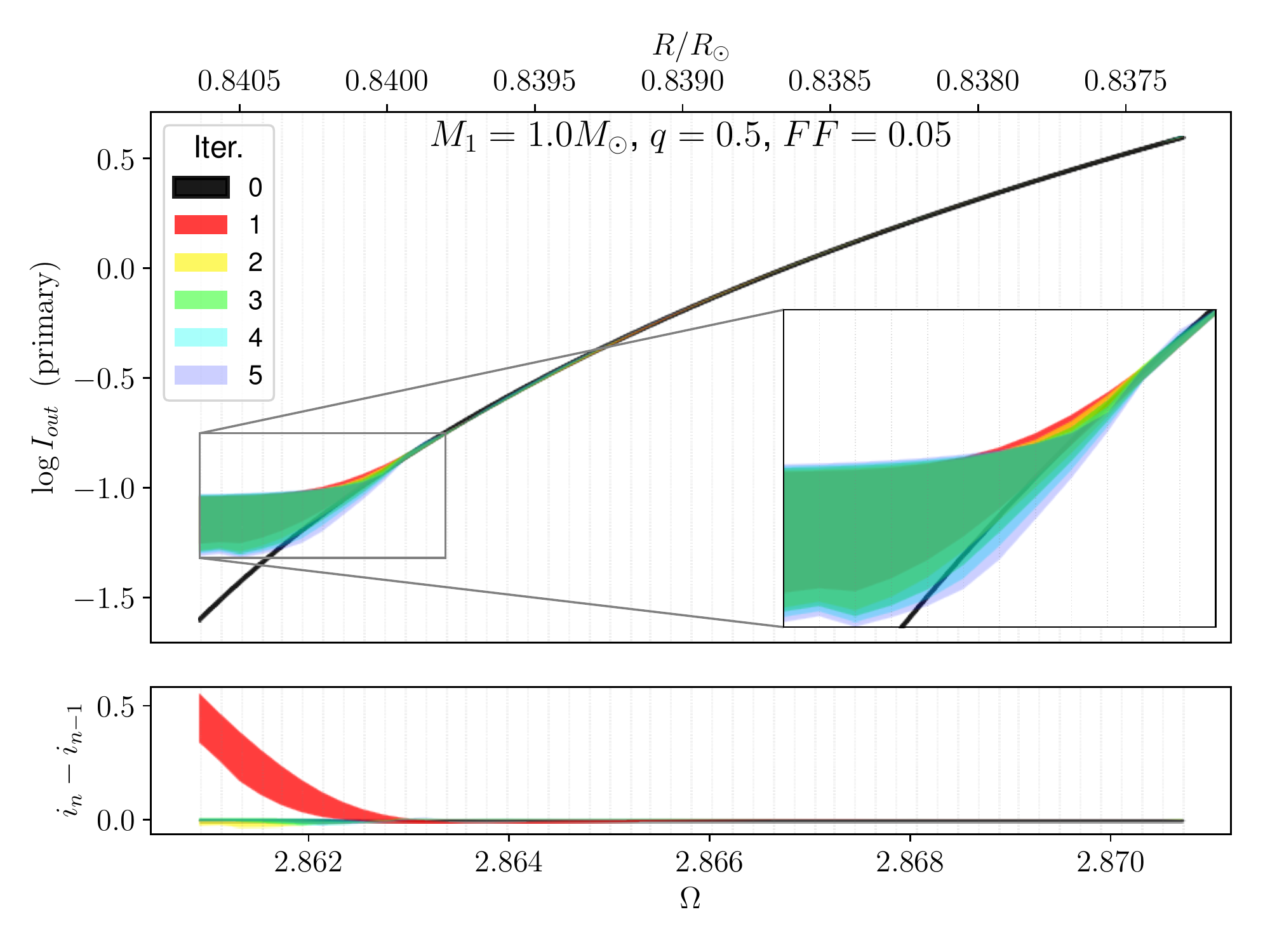}
    \includegraphics[width=0.495\hsize]{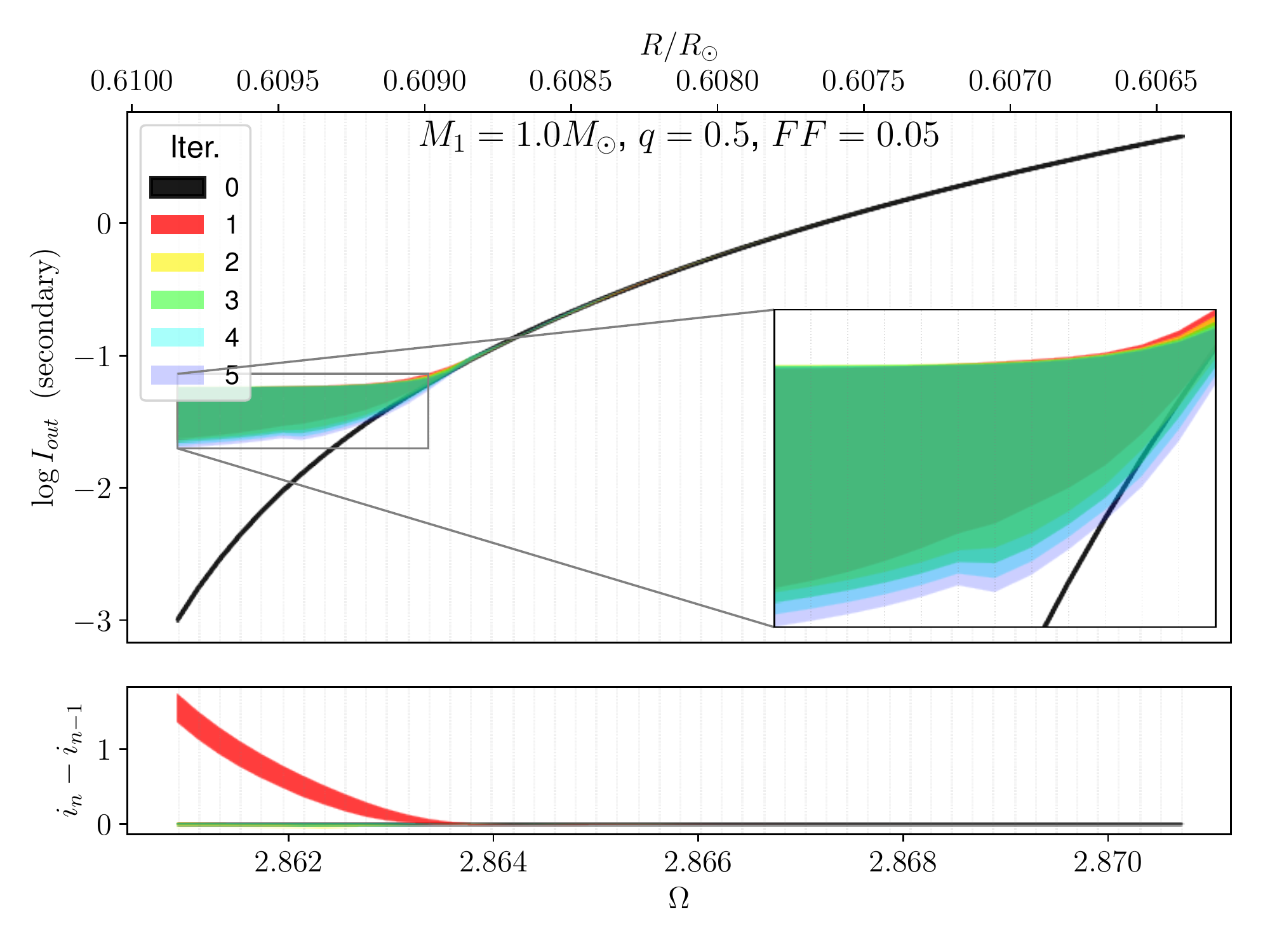}
    
    \includegraphics[width=0.495\hsize]{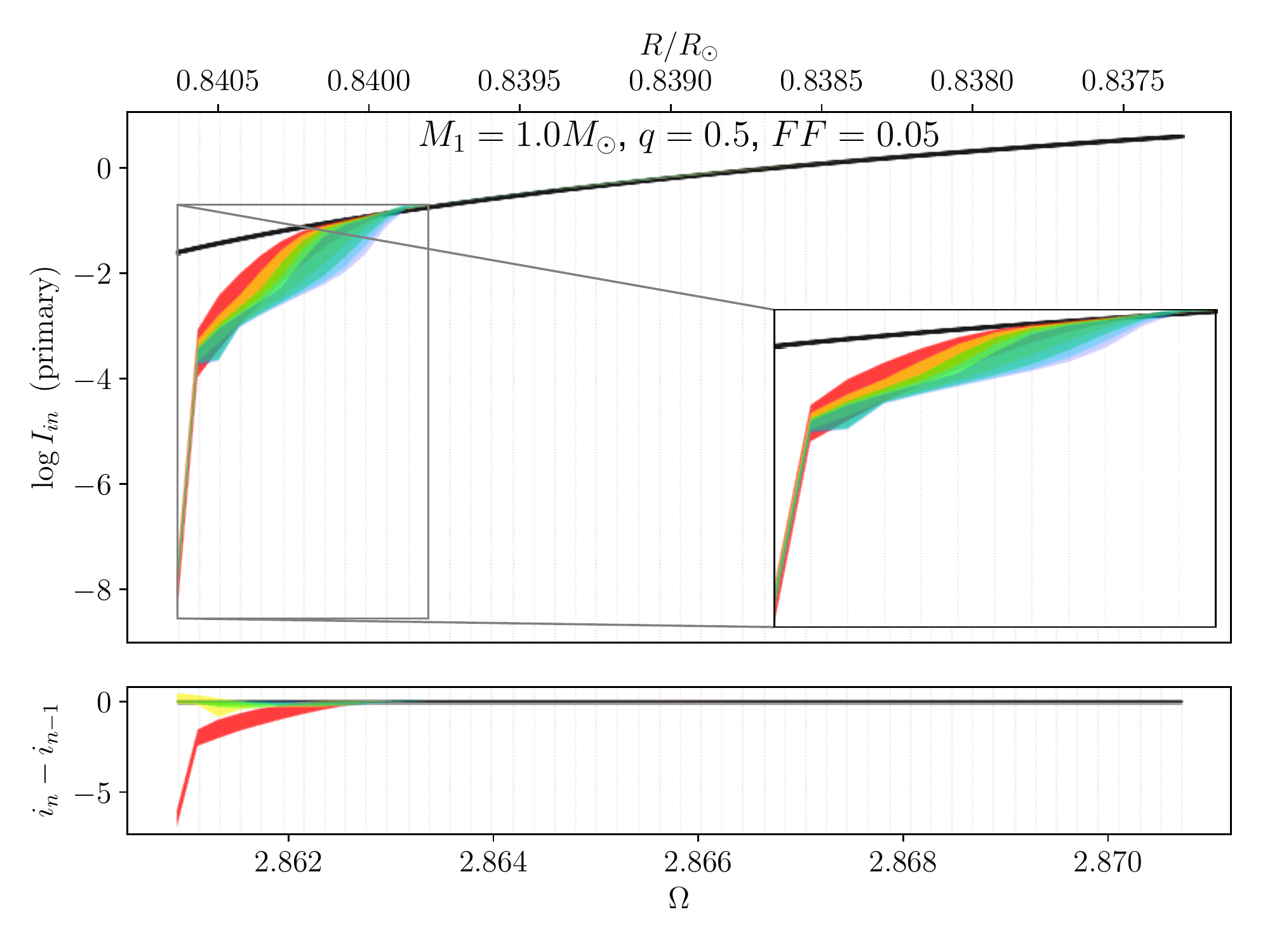}
    \includegraphics[width=0.495\hsize]{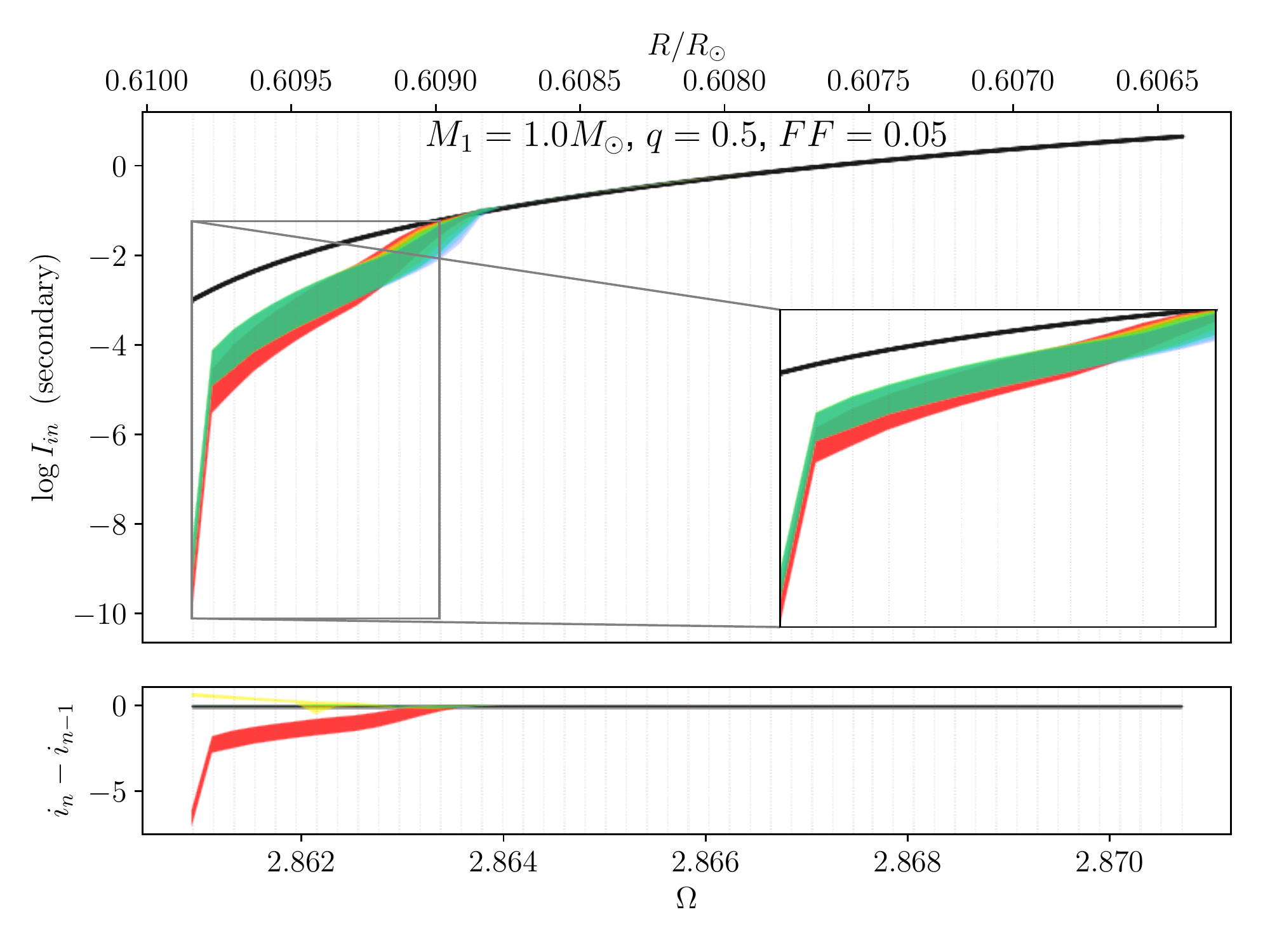}
    
    \includegraphics[width=0.495\hsize]{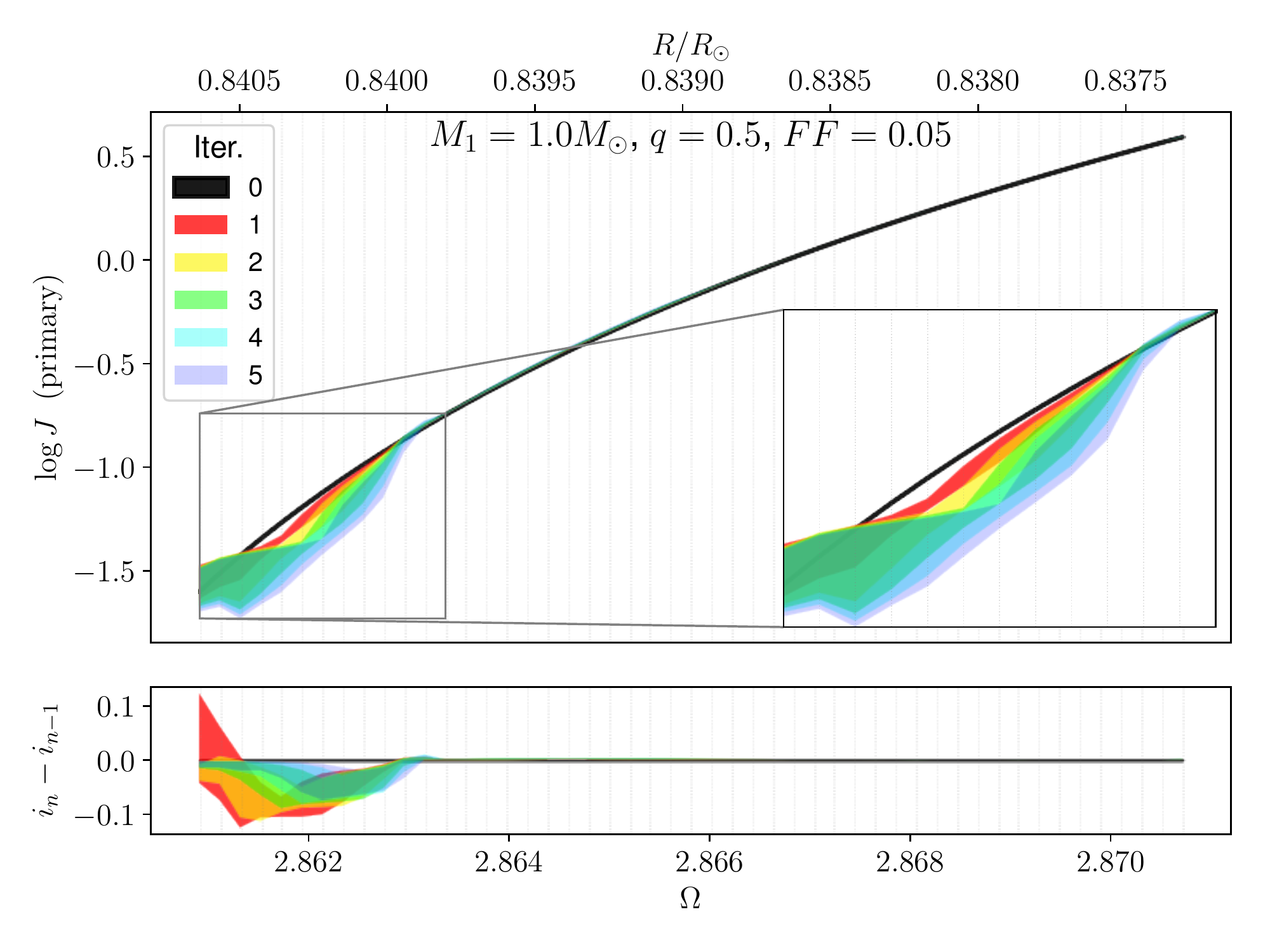}
    \includegraphics[width=0.495\hsize]{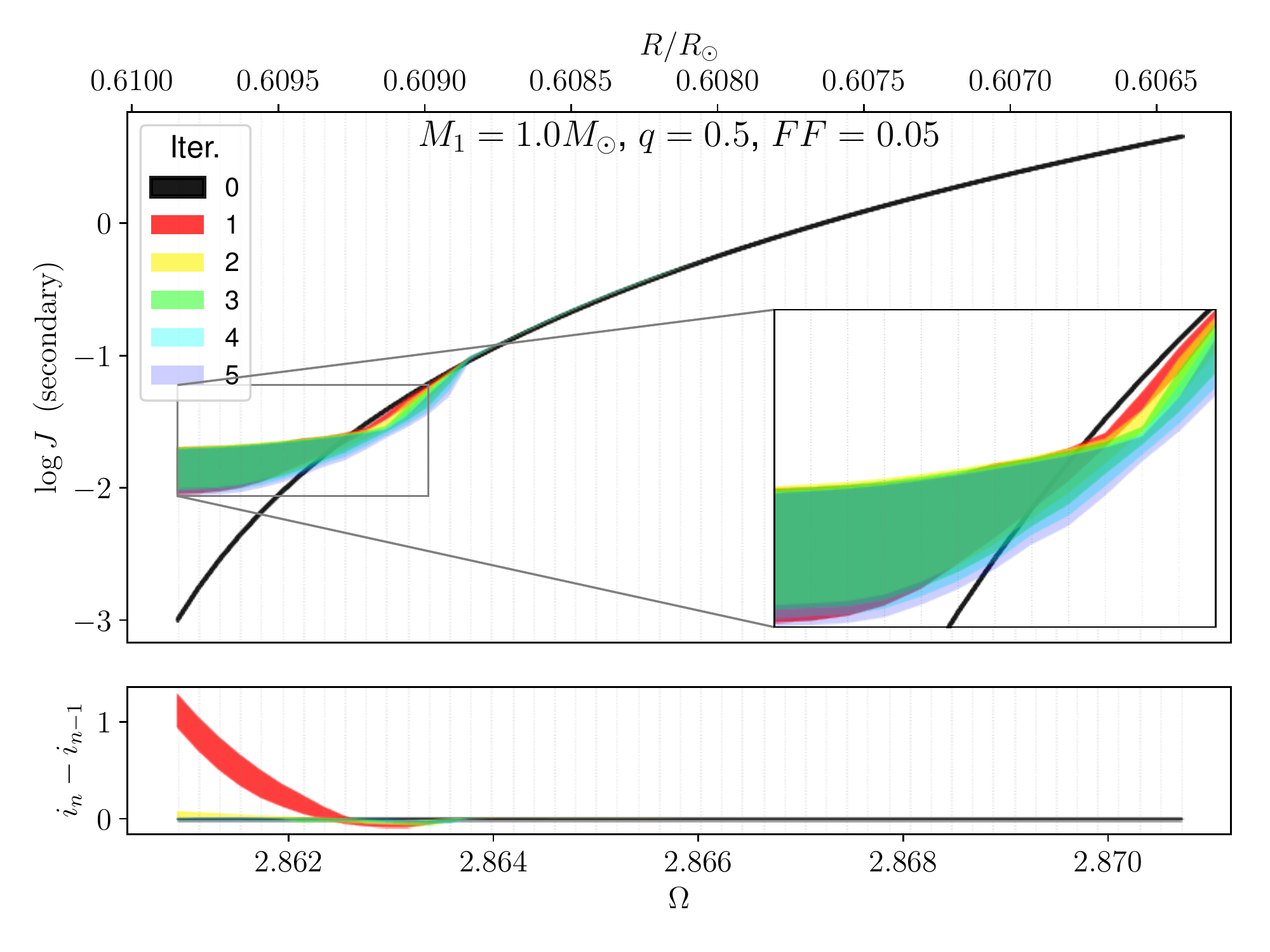}
    
    \caption{Top to bottom: outward, inward and mean intensity as a function of the potential/radius of a contact binary with $q=0.5$. Left panels: primary, right panels: secondary component. The bottom panel of each plot shows the differences between successive iterations.}
    \label{fig:q05}
\end{figure}

\begin{figure}[h]
    \centering
            \includegraphics[width=0.495\hsize]{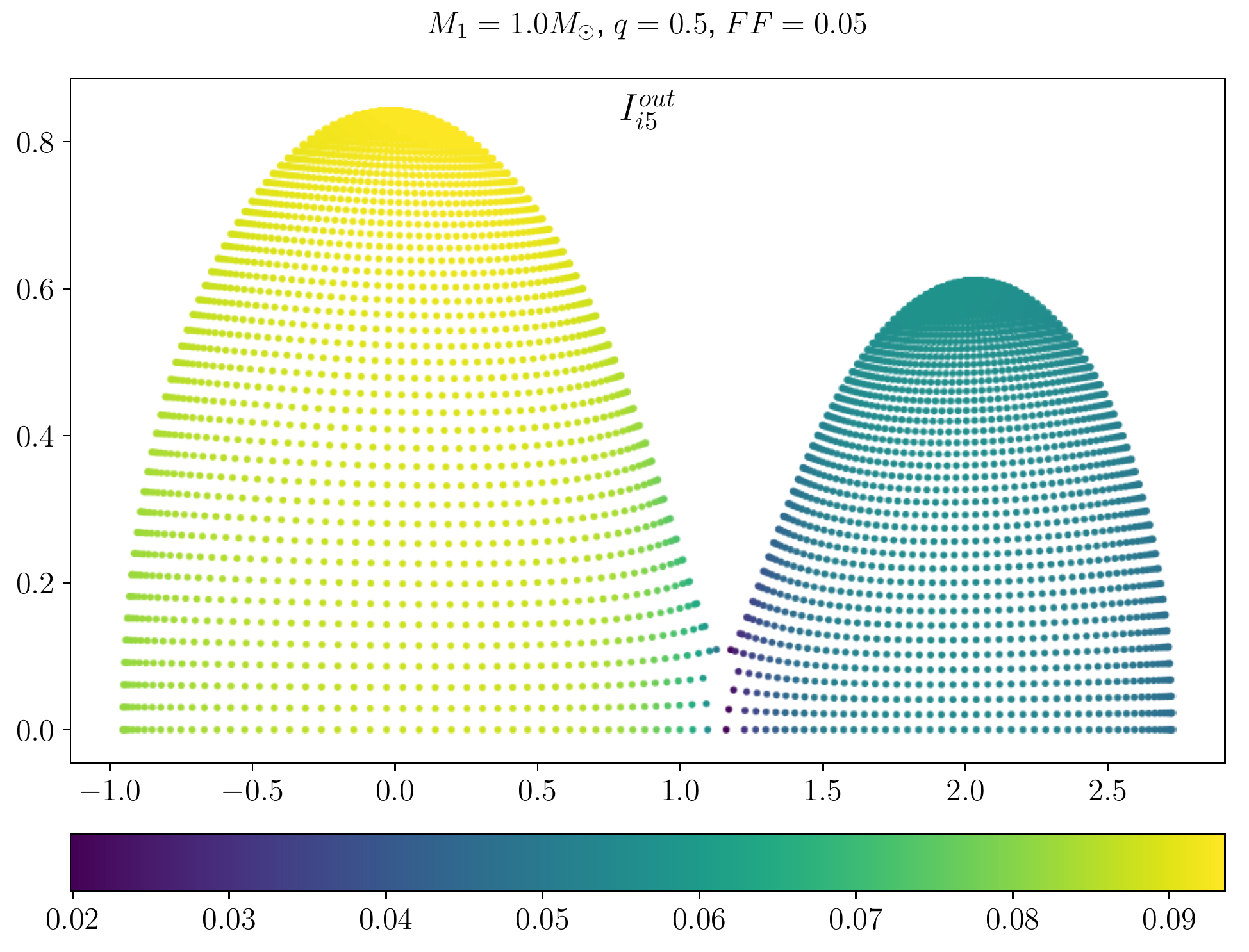}
    \includegraphics[width=0.495\hsize]{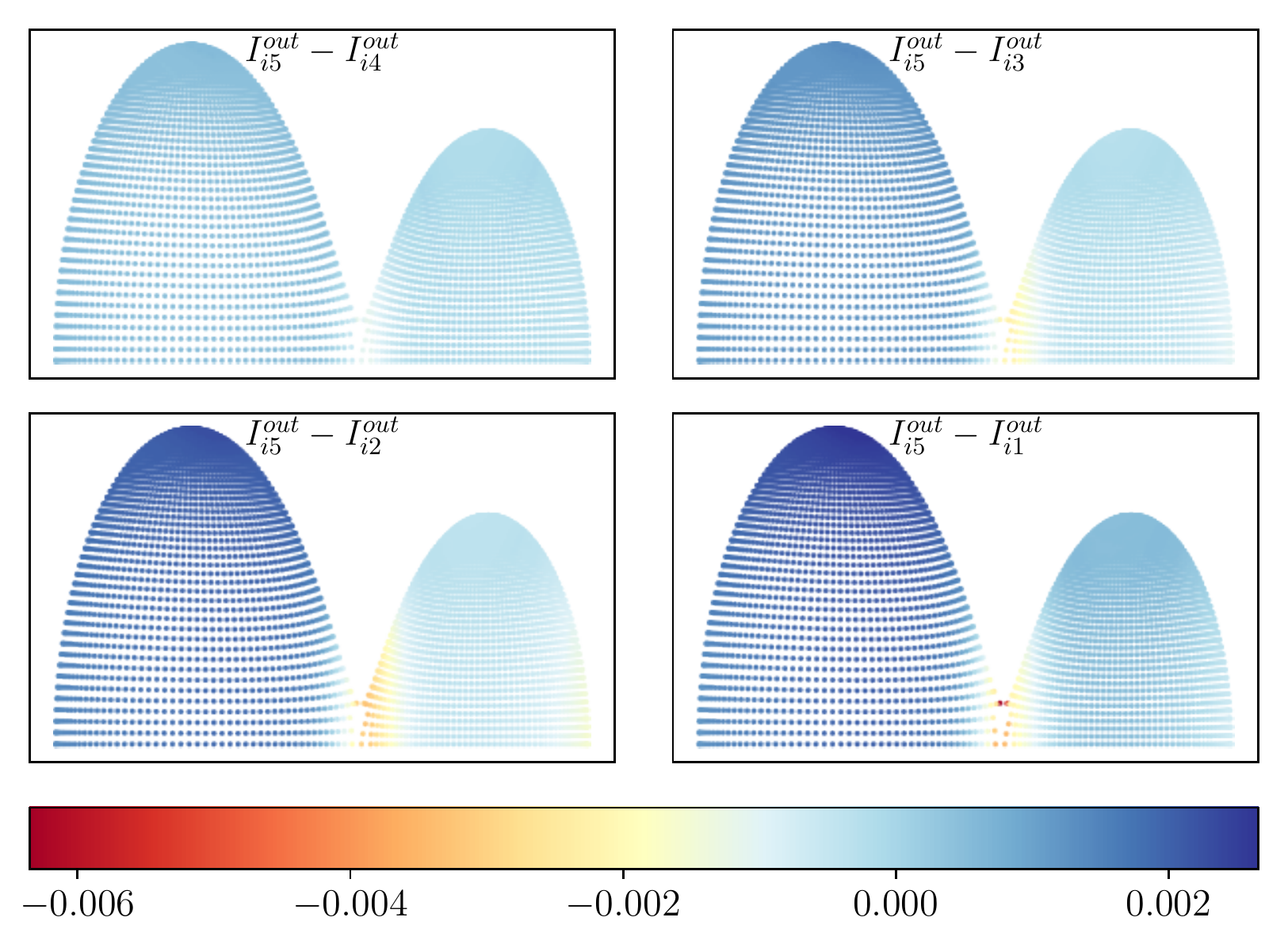}
    
        \includegraphics[width=0.495\hsize]{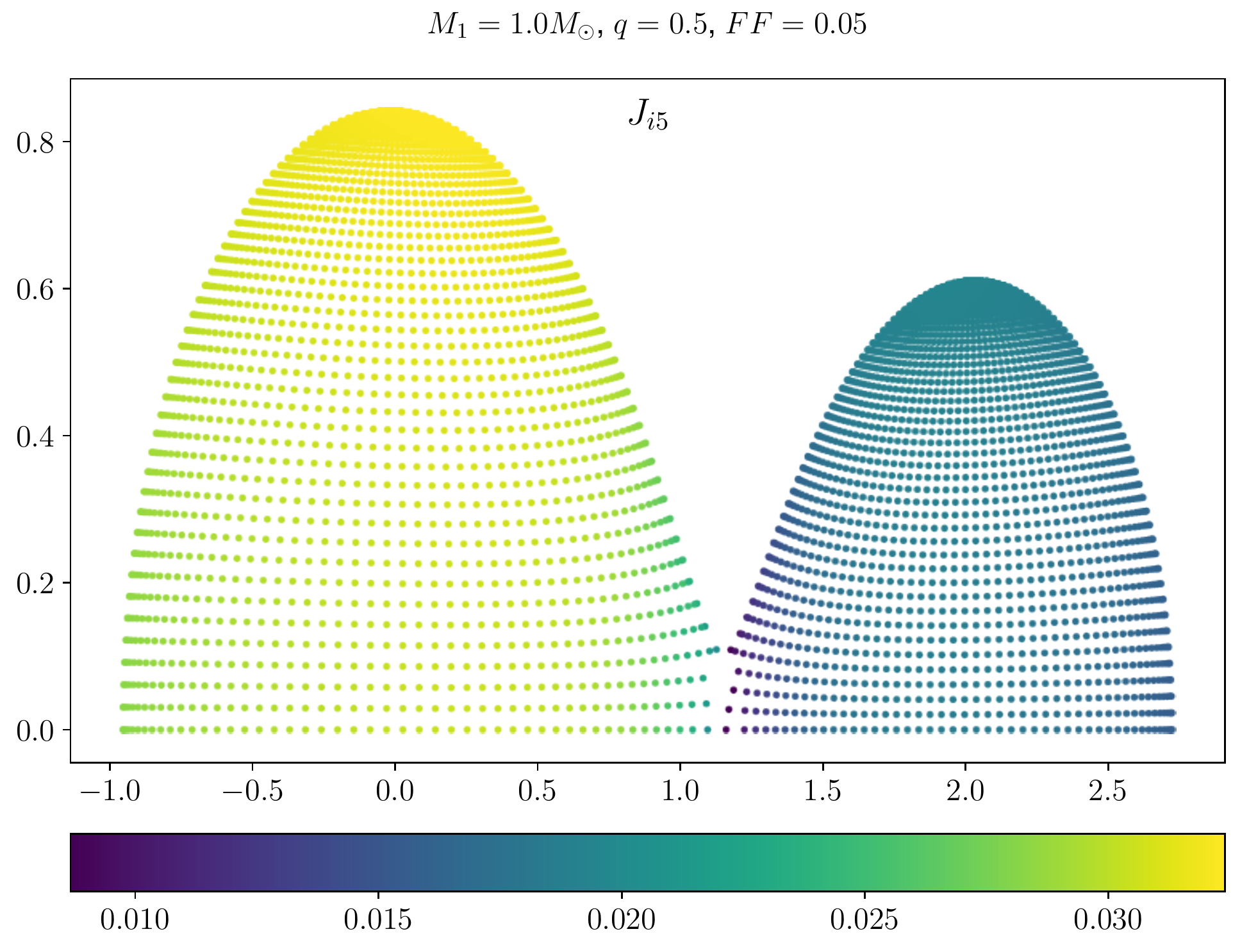}
    \includegraphics[width=0.495\hsize]{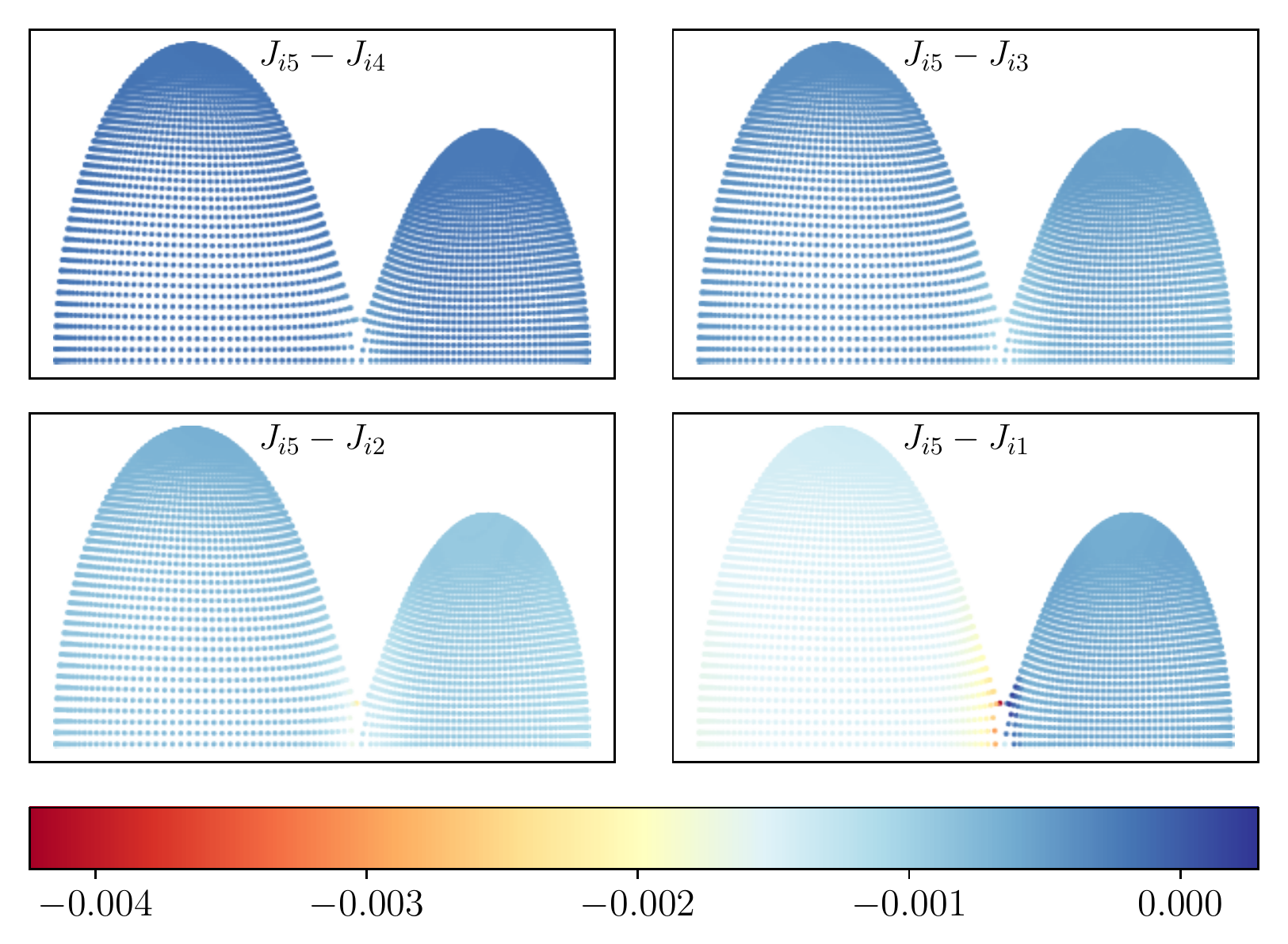}

    \caption{Surface distribution of the outward (top) and mean (bottom) intensity of a contact binary with $q=0.5$ after the fifth iteration. Right panels show the differences in the surface distribution between the final and each previous iteration.}
    \label{fig:q05_s}
\end{figure}

\begin{figure}[h]
    \centering
    \includegraphics[width=0.495\hsize]{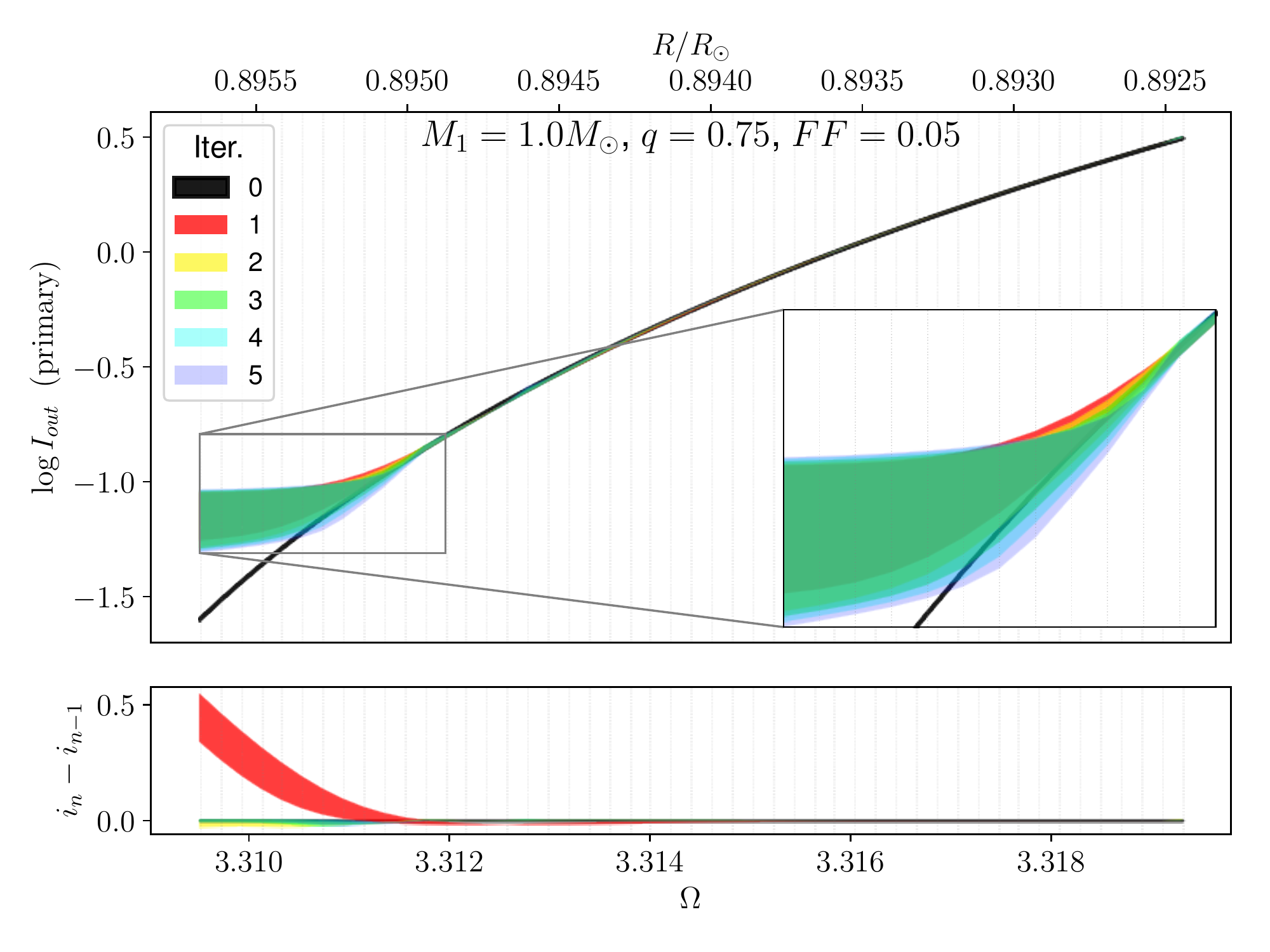}
    \includegraphics[width=0.495\hsize]{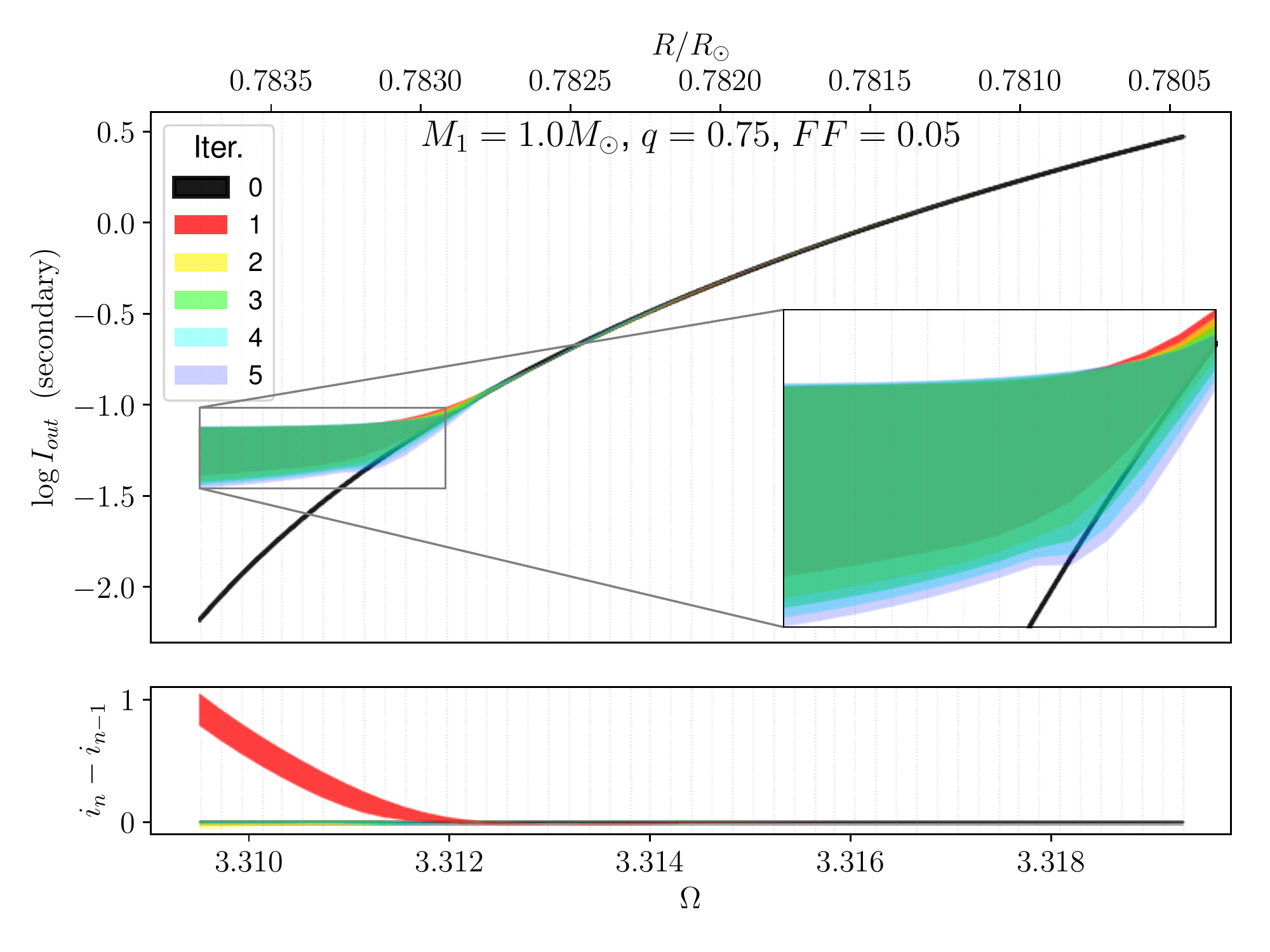}
    
    \includegraphics[width=0.495\hsize]{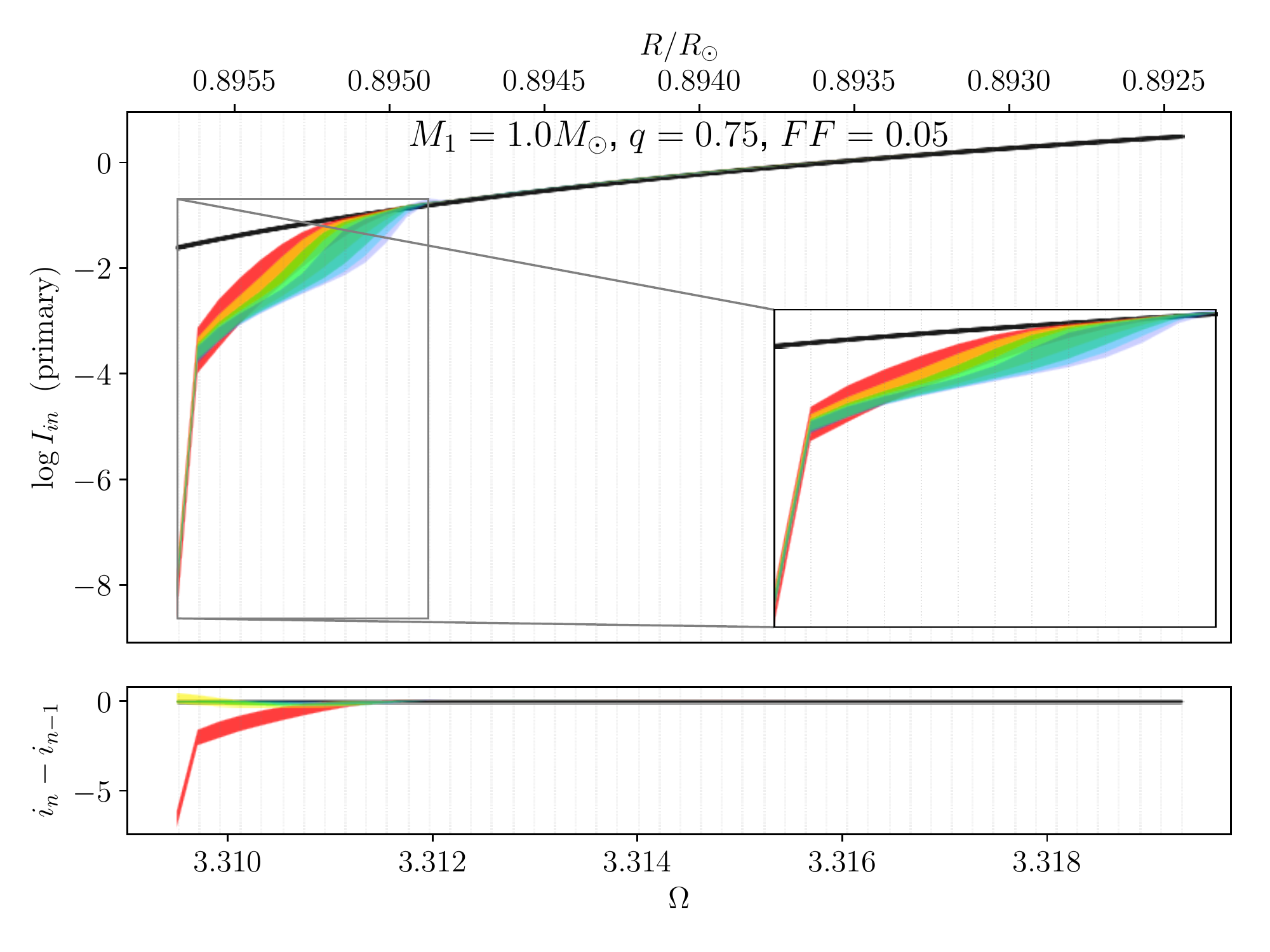}
    \includegraphics[width=0.495\hsize]{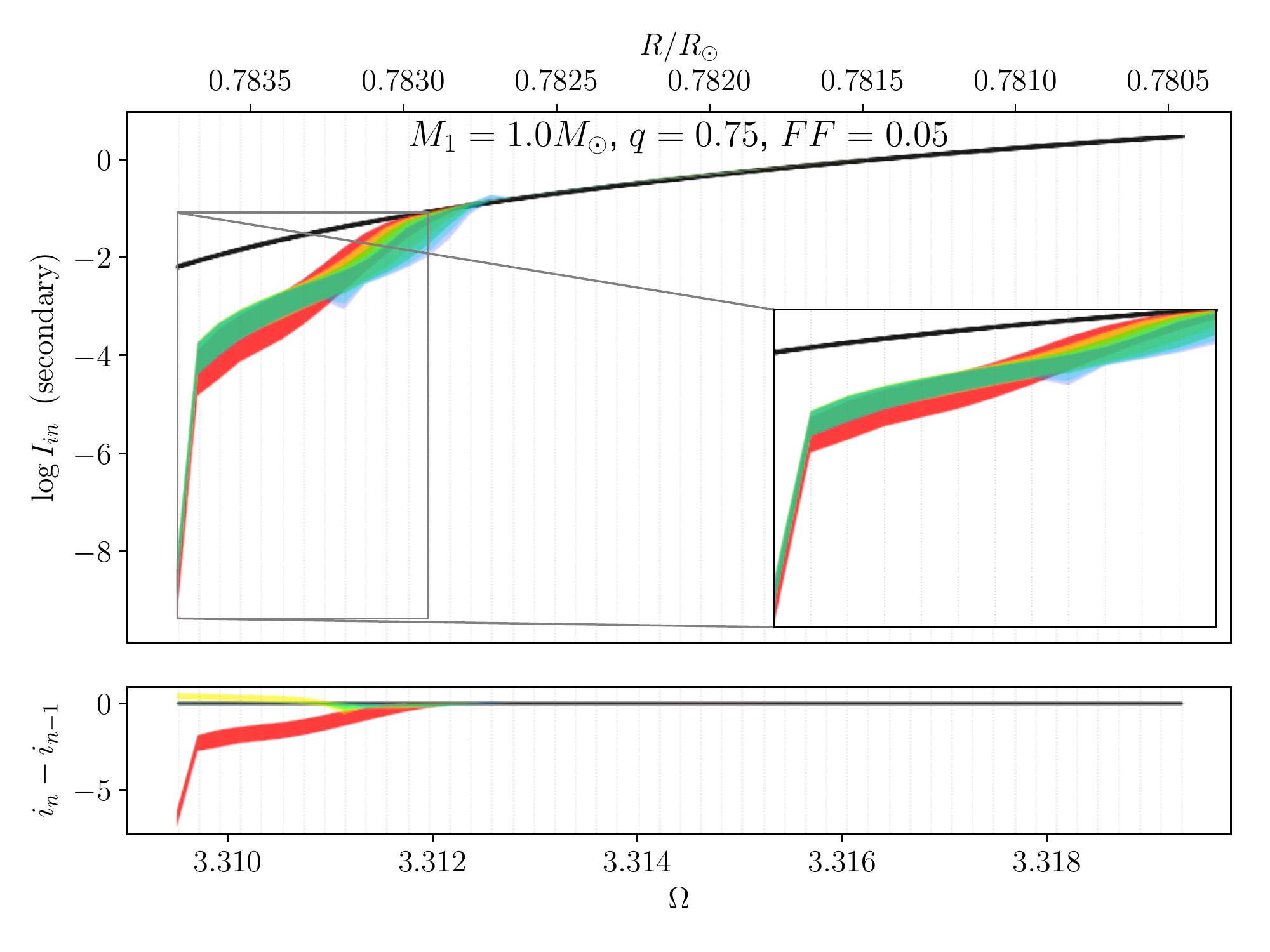}
    
    \includegraphics[width=0.495\hsize]{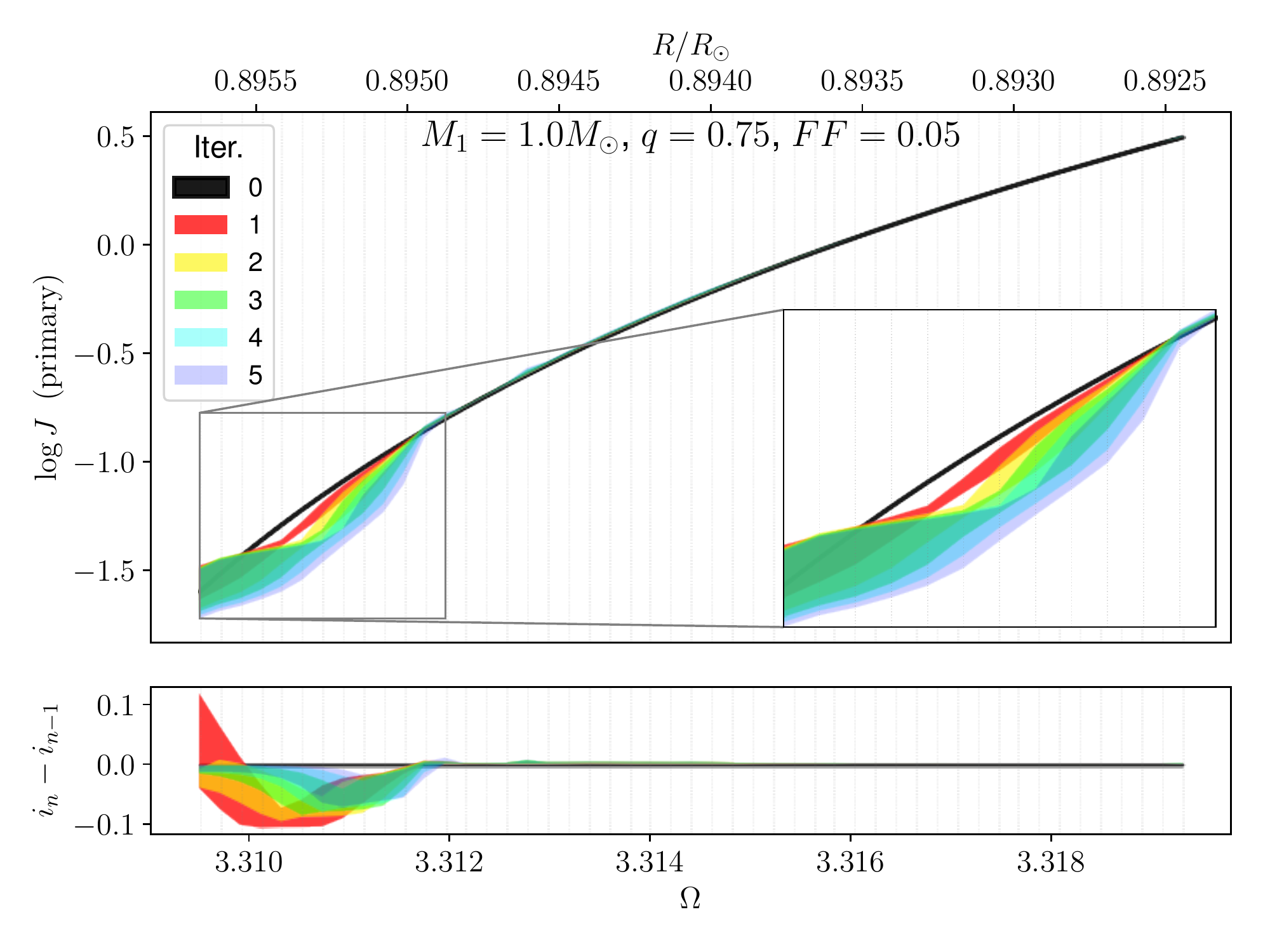}
    \includegraphics[width=0.495\hsize]{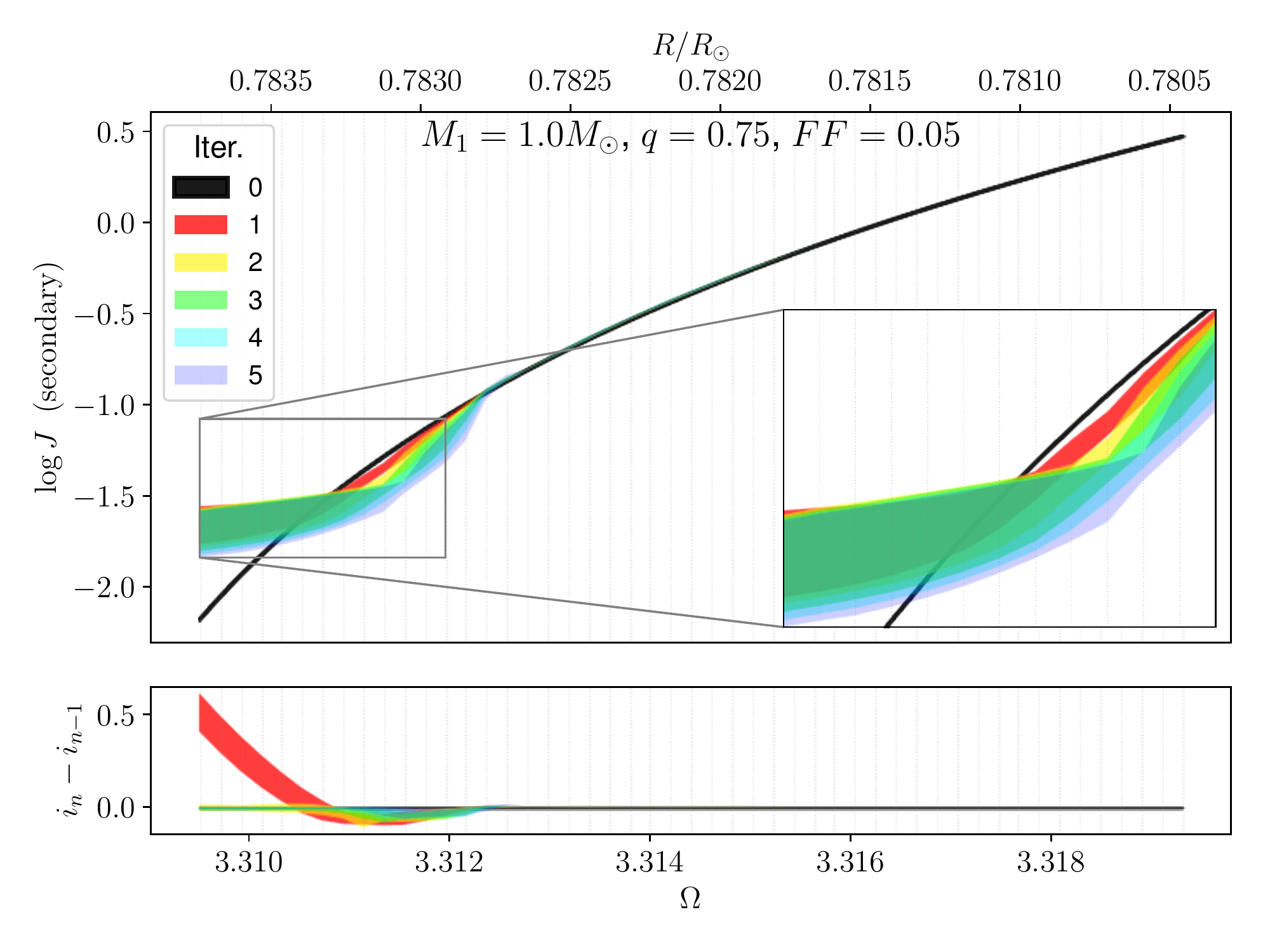}
    
    \caption{Top to bottom: outward, inward and mean intensity as a function of the potential/radius of a contact binary with $q=0.75$. Left panels: primary, right panels: secondary component. The bottom panel of each plot shows the differences between successive iterations.}
    \label{fig:q075}
\end{figure}

\begin{figure}[h]
    \centering
            \includegraphics[width=0.495\hsize]{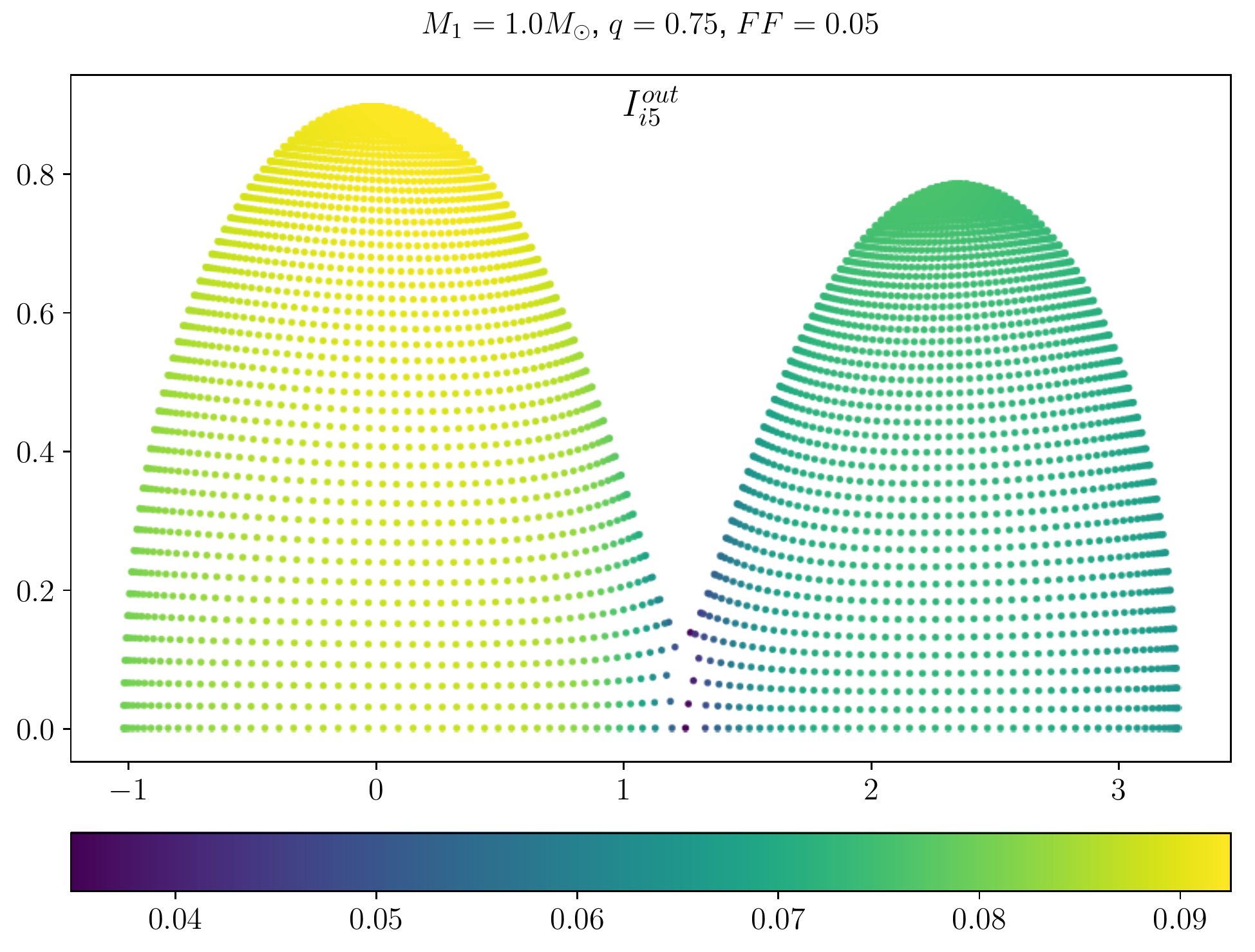}
    \includegraphics[width=0.495\hsize]{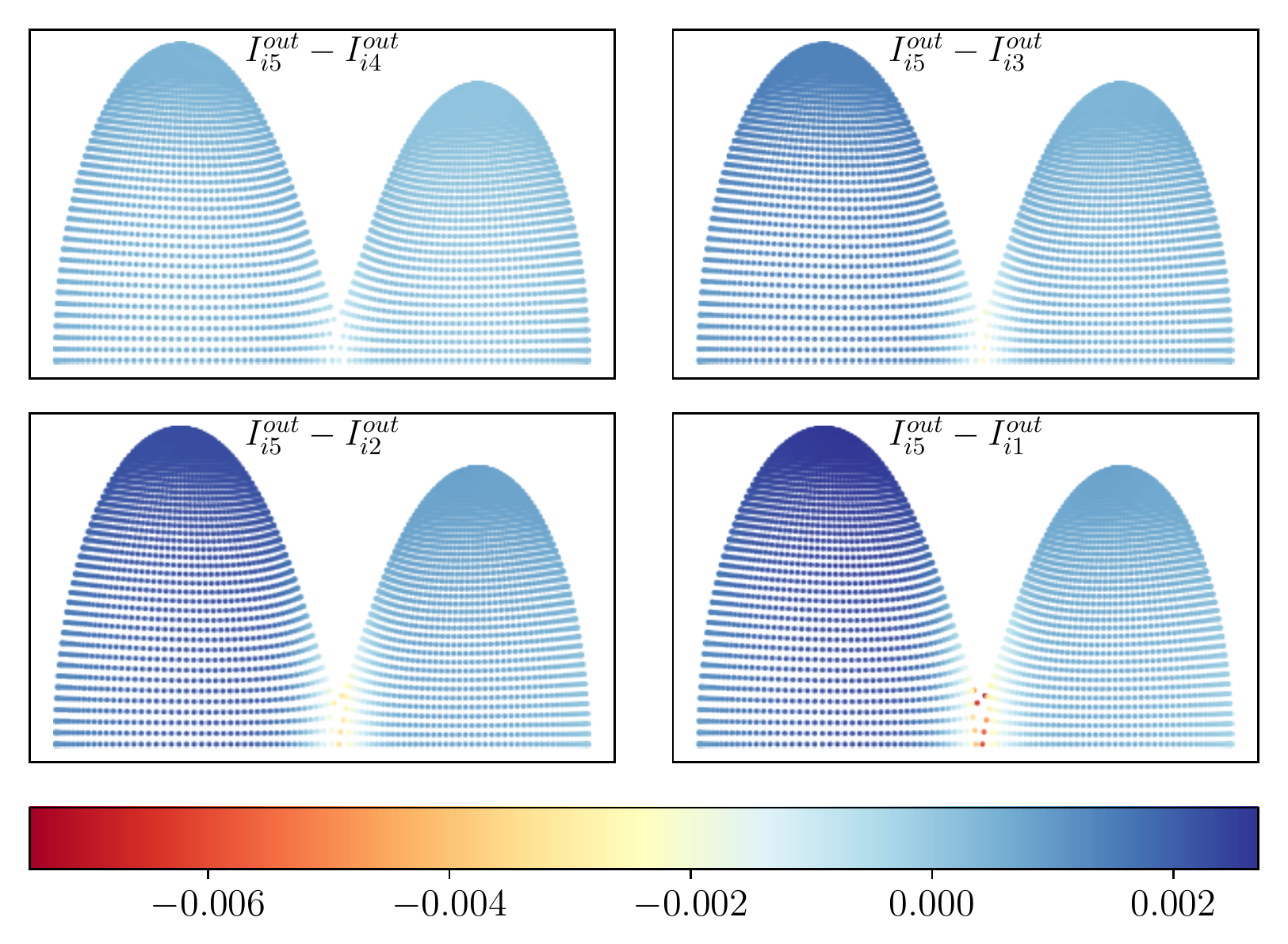}
    
        \includegraphics[width=0.495\hsize]{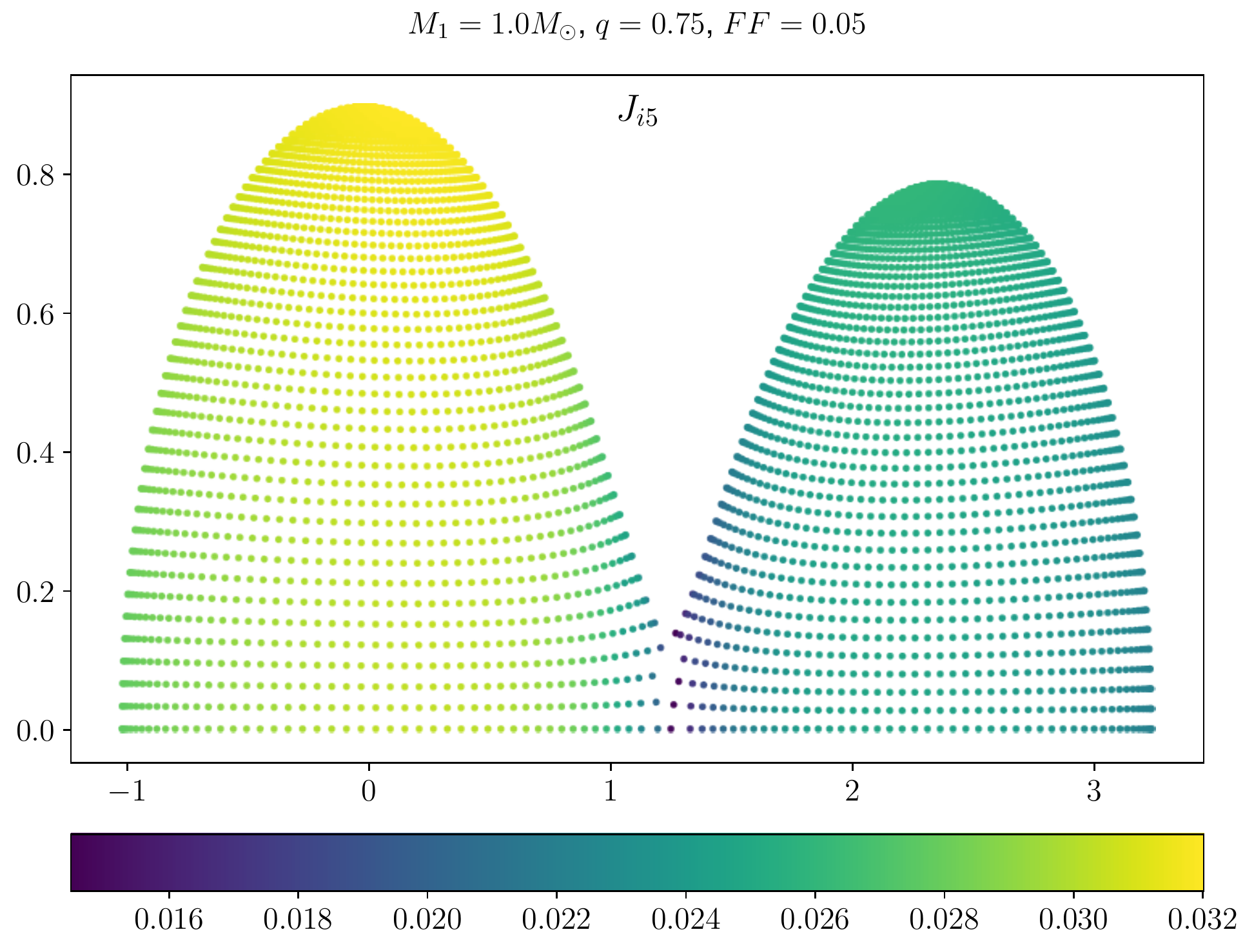}
    \includegraphics[width=0.495\hsize]{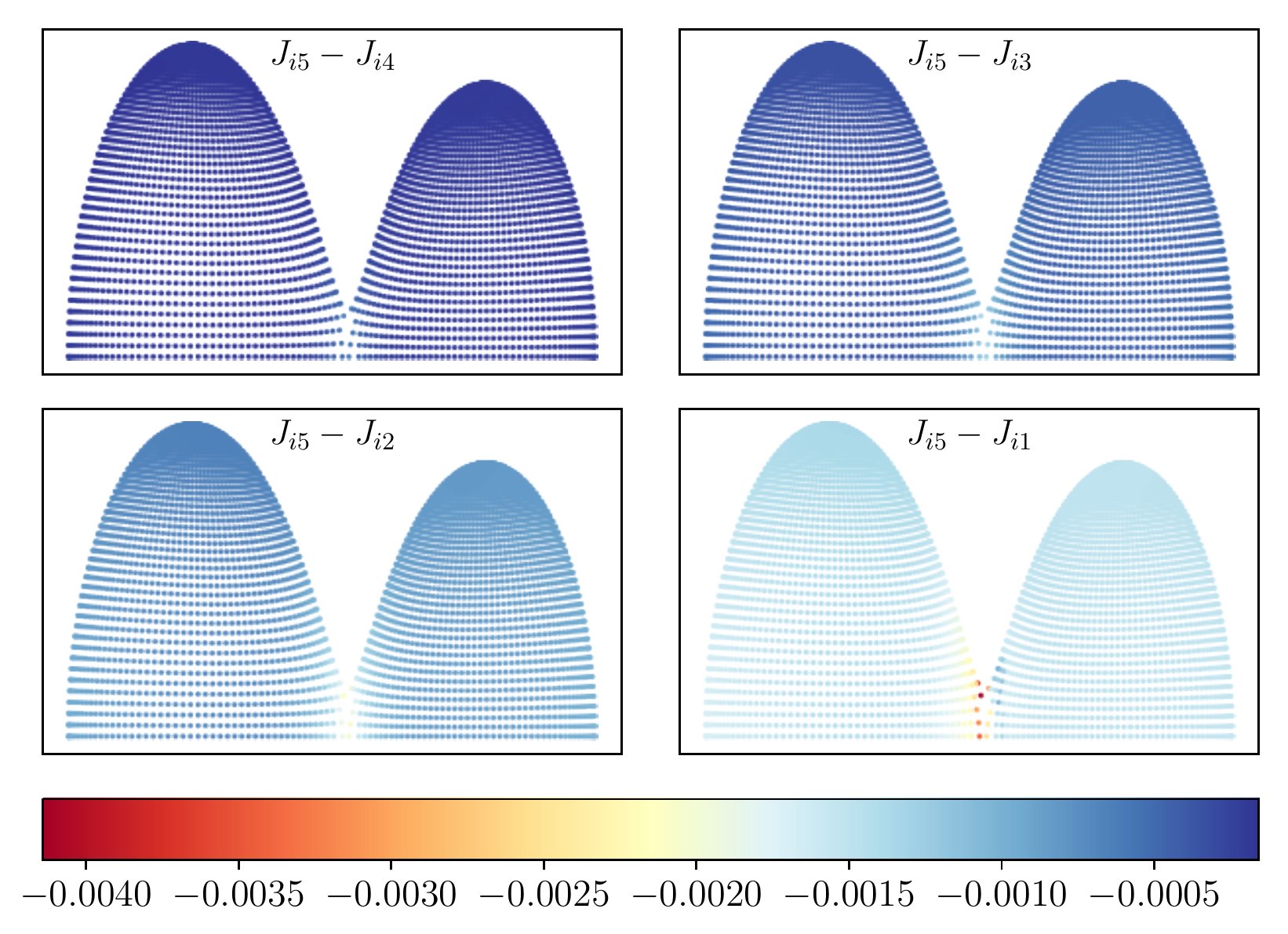}

    \caption{Surface distribution of the outward (top) and mean (bottom) intensity of a contact binary with $q=0.75$ after the fifth iteration. Right panels show the differences in the surface distribution between the final and each previous iteration.}
    \label{fig:q075_s}
\end{figure}

\begin{figure}[h]
    \centering
    \includegraphics[width=0.495\hsize]{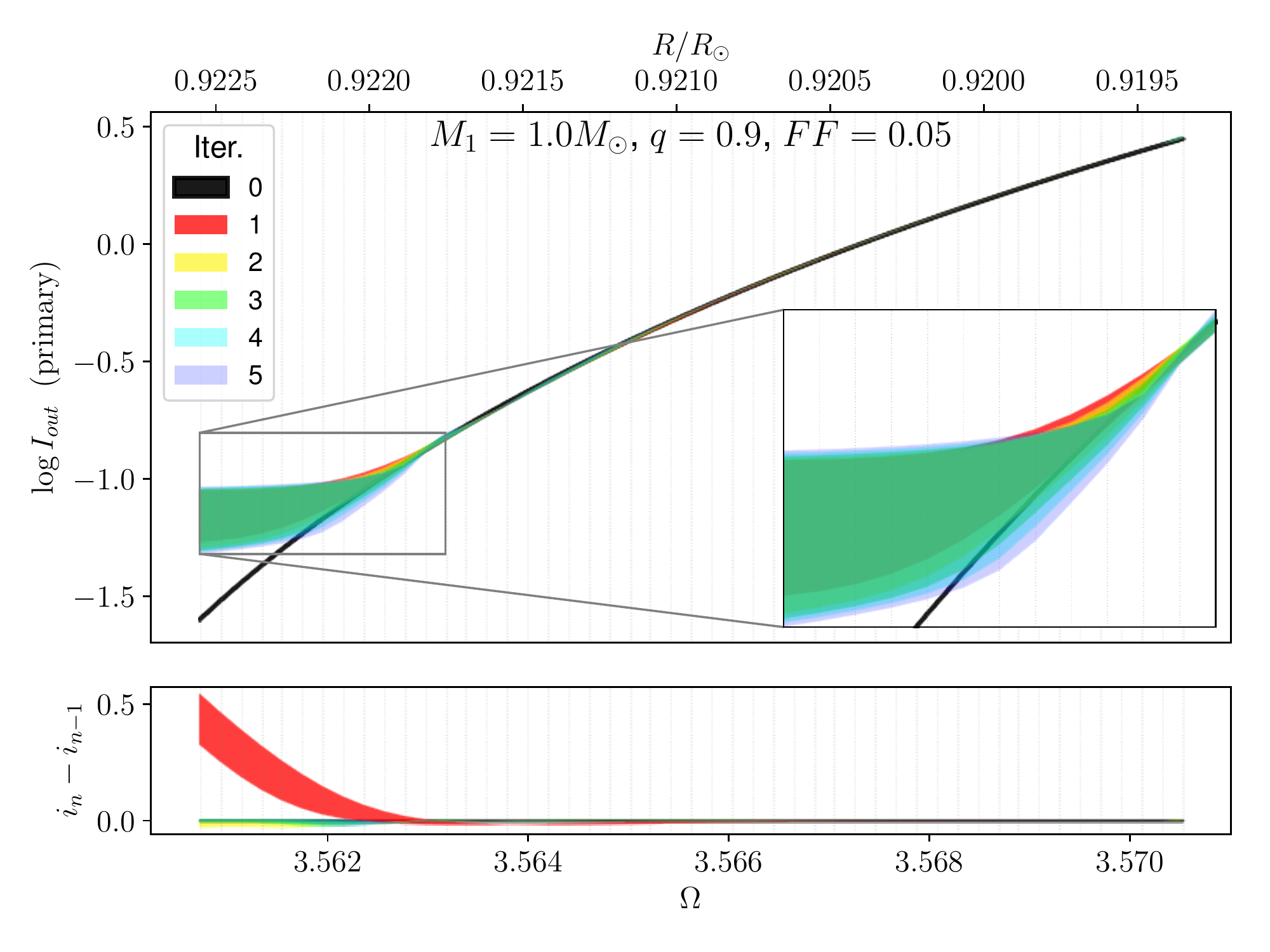}
    \includegraphics[width=0.495\hsize]{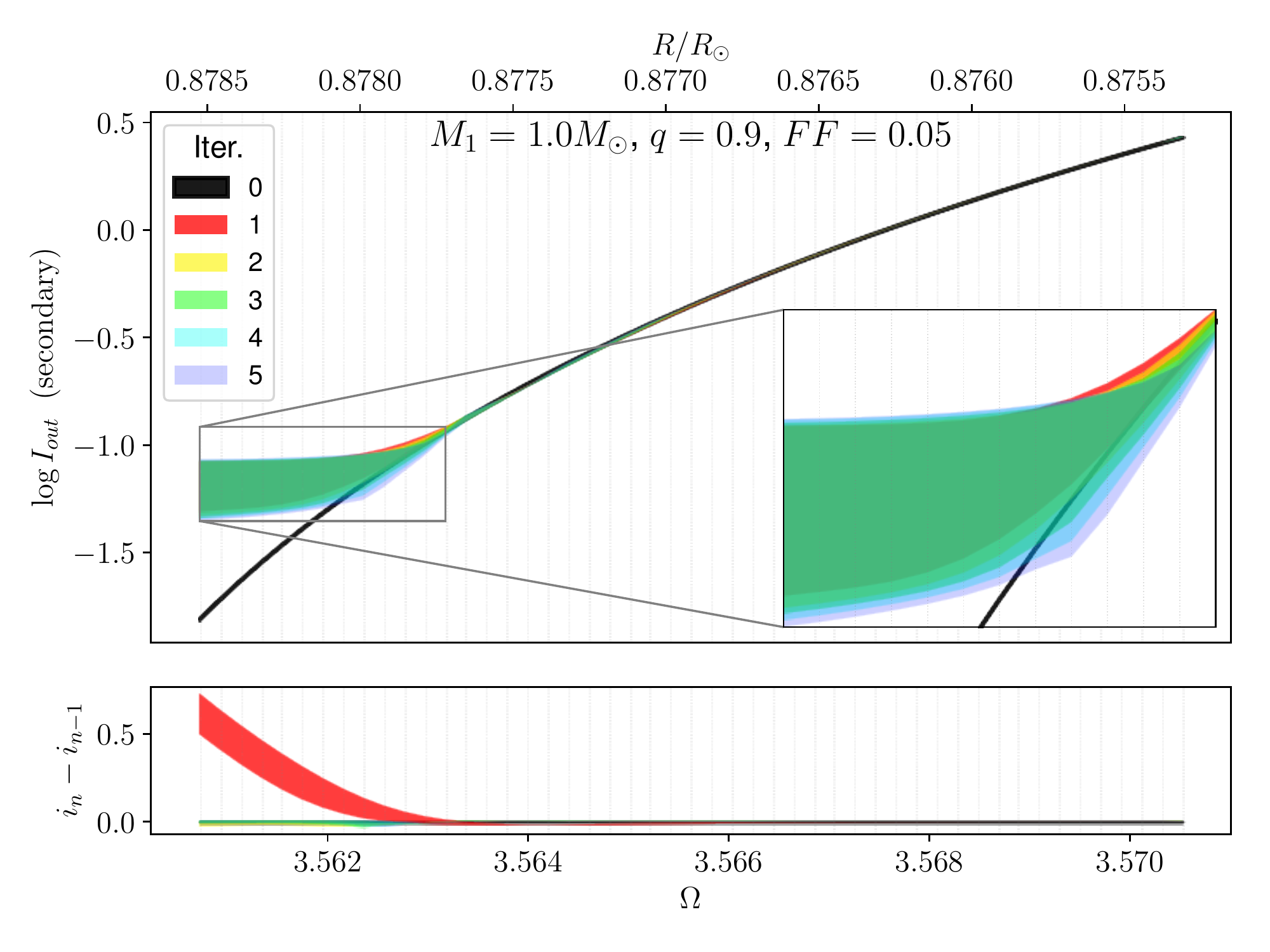}
    
    \includegraphics[width=0.495\hsize]{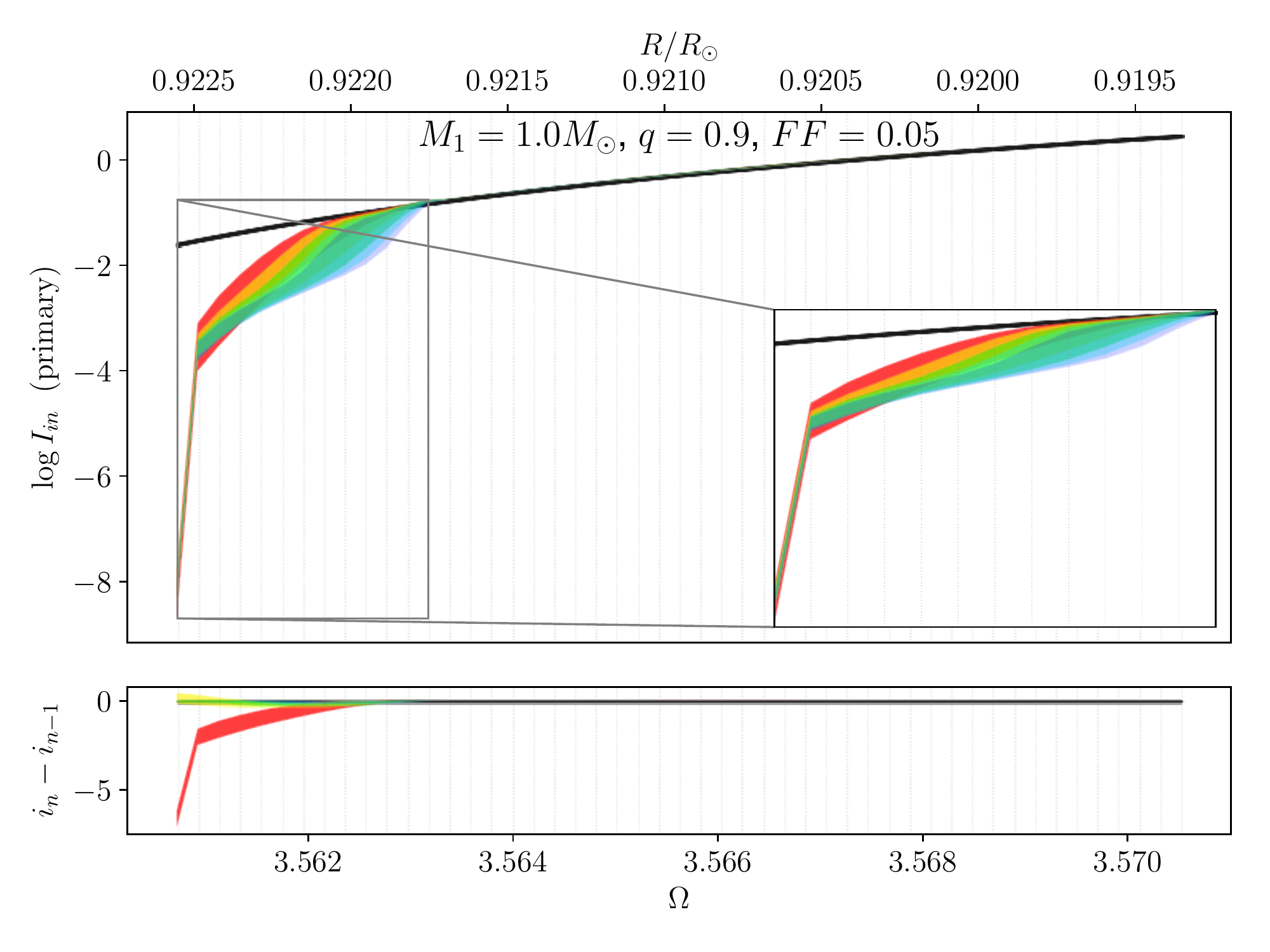}
    \includegraphics[width=0.495\hsize]{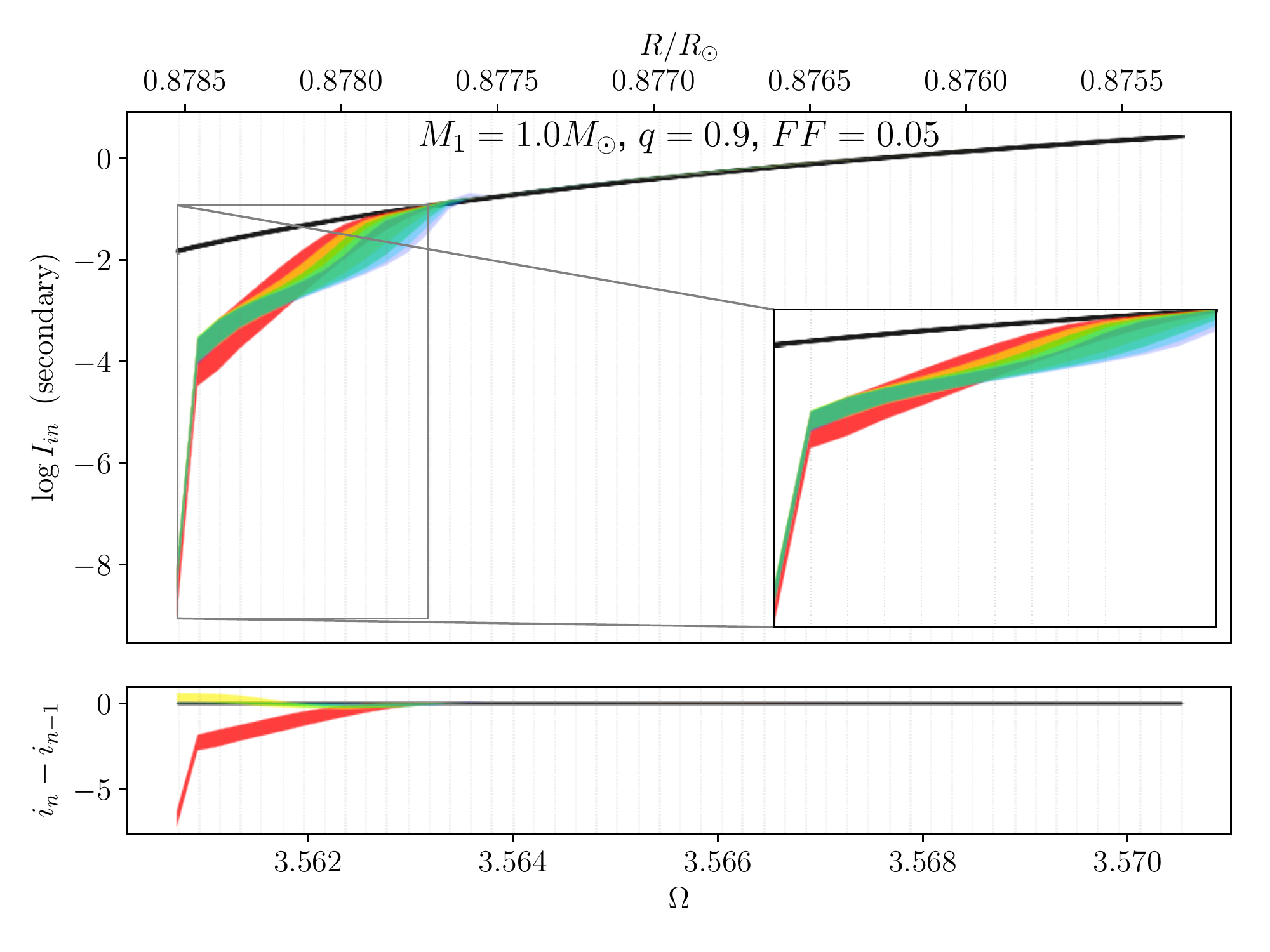}
    
    \includegraphics[width=0.495\hsize]{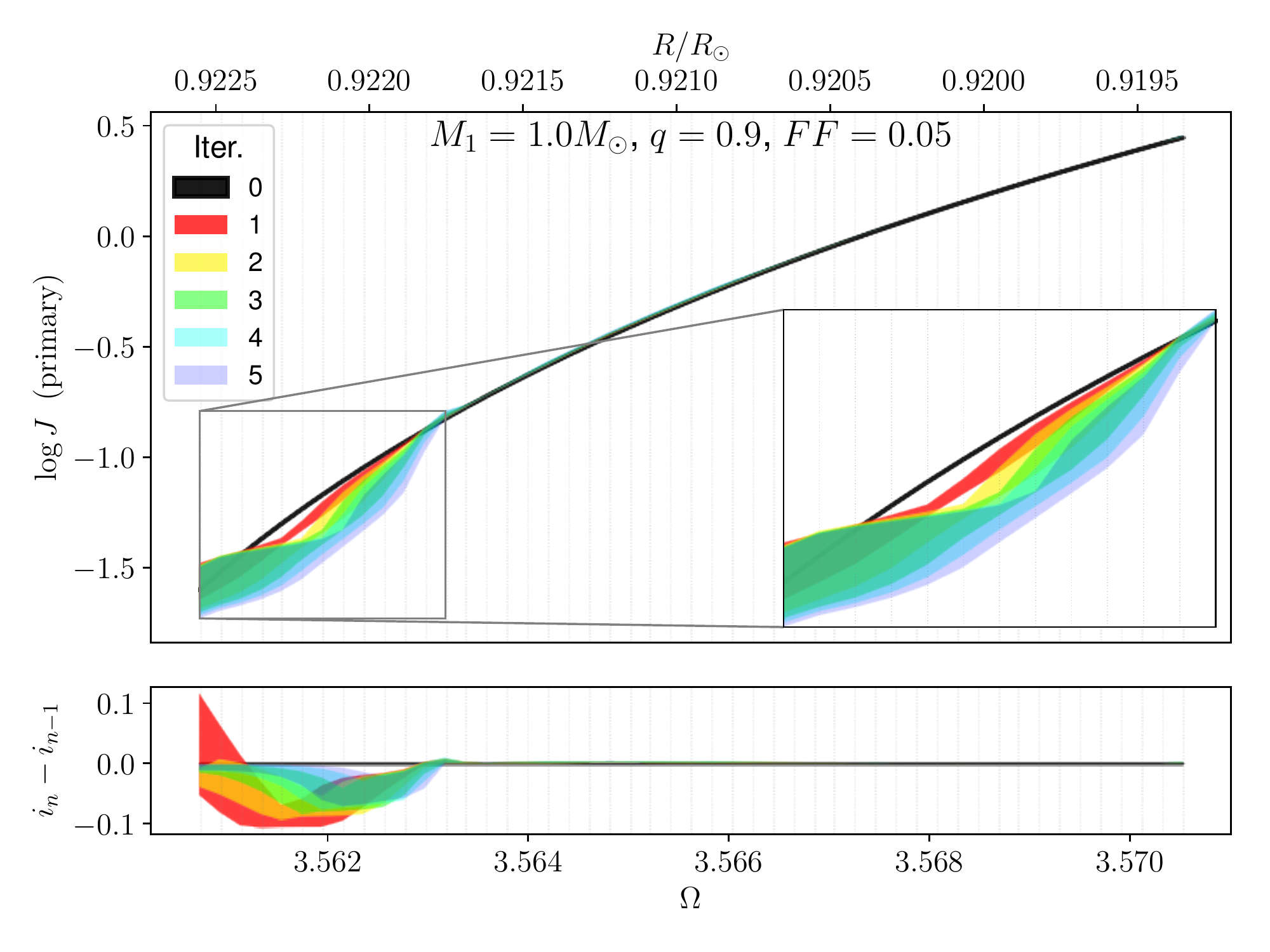}
    \includegraphics[width=0.495\hsize]{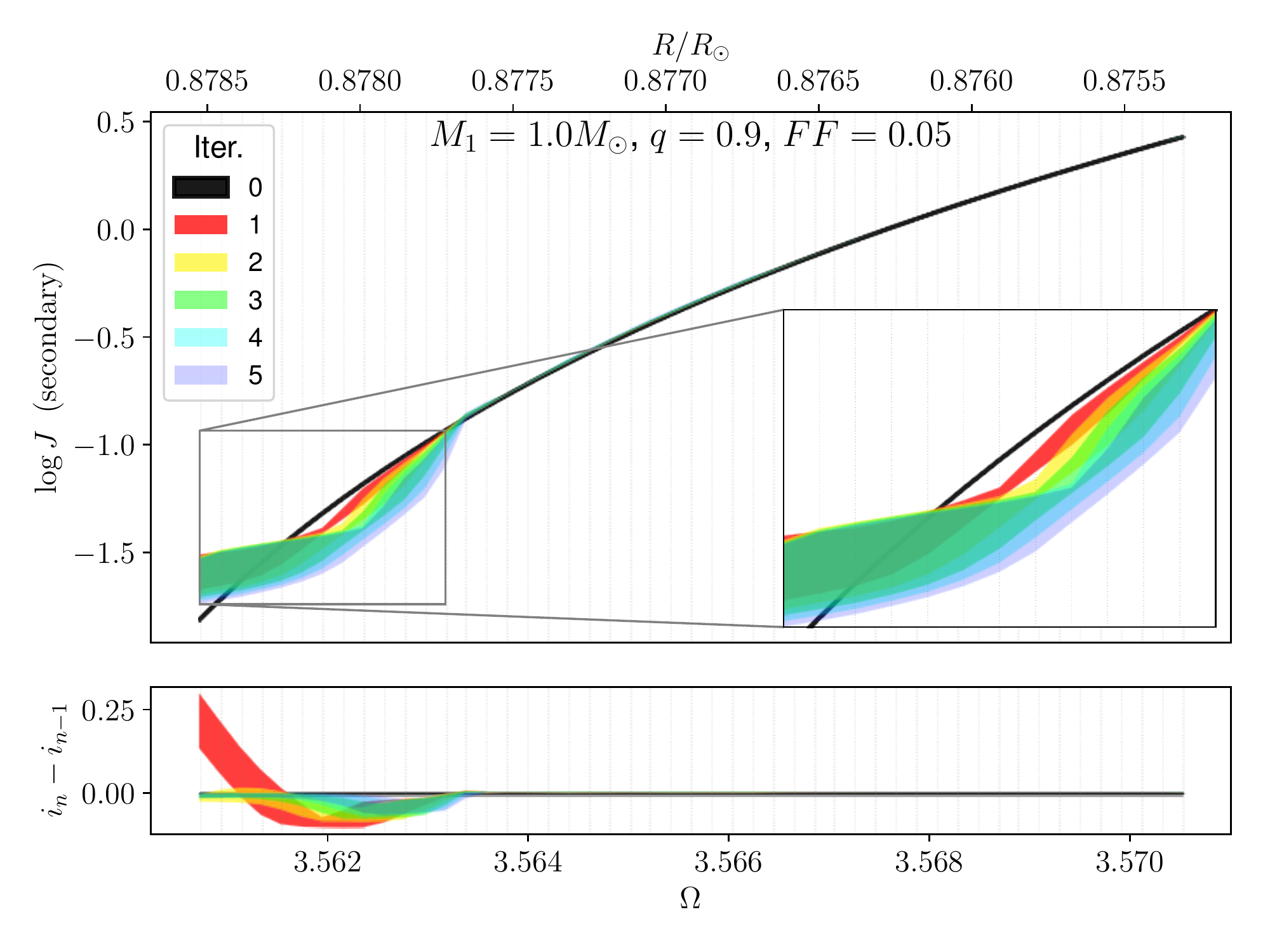}
    
    \caption{Top to bottom: outward, inward and mean intensity as a function of the potential/radius of a contact binary with $q=0.9$. Left panels: primary, right panels: secondary component. The bottom panel of each plot shows the differences between successive iterations.}
    \label{fig:q09}
\end{figure}

\begin{figure}[h]
    \centering
        \includegraphics[width=0.495\hsize]{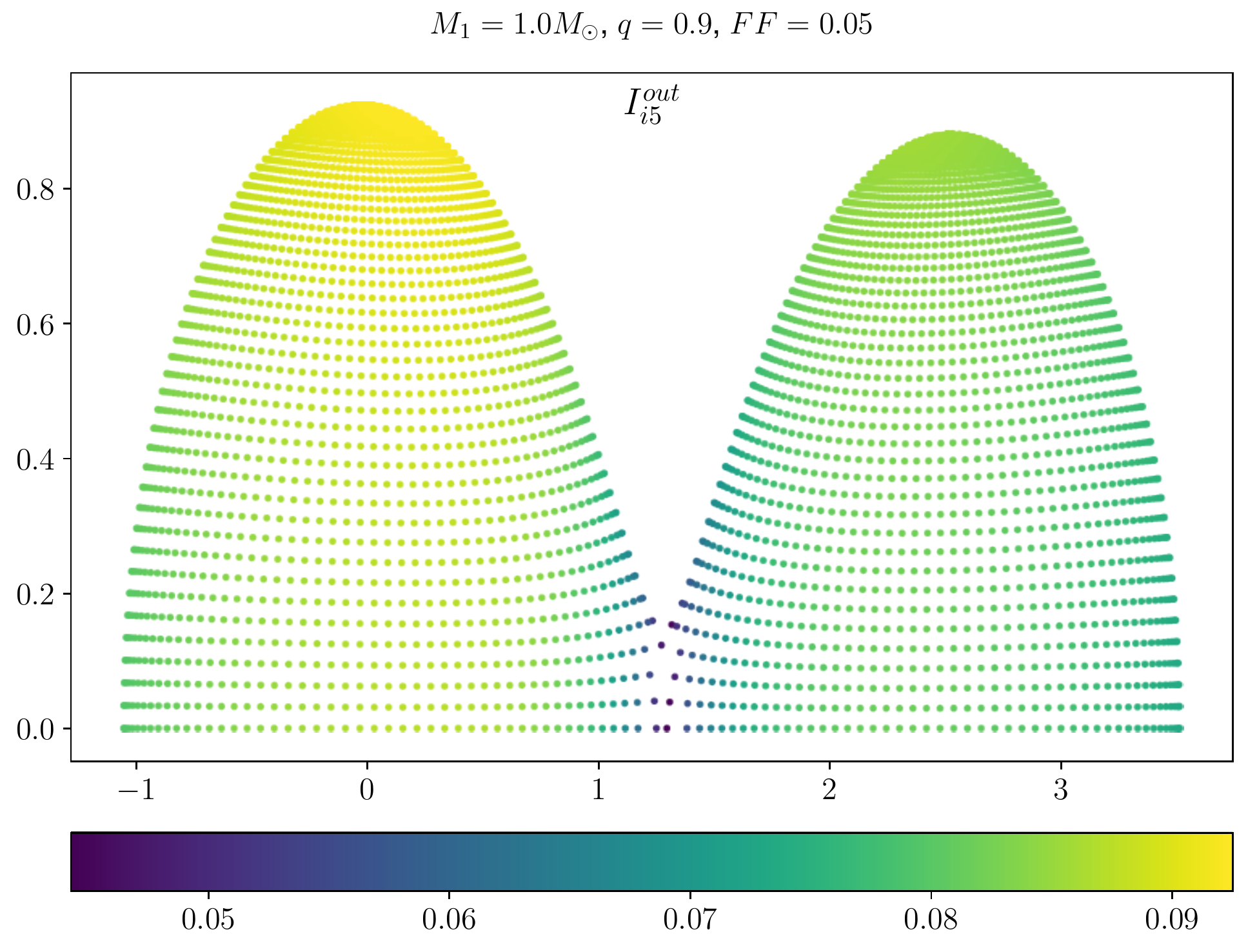}
    \includegraphics[width=0.495\hsize]{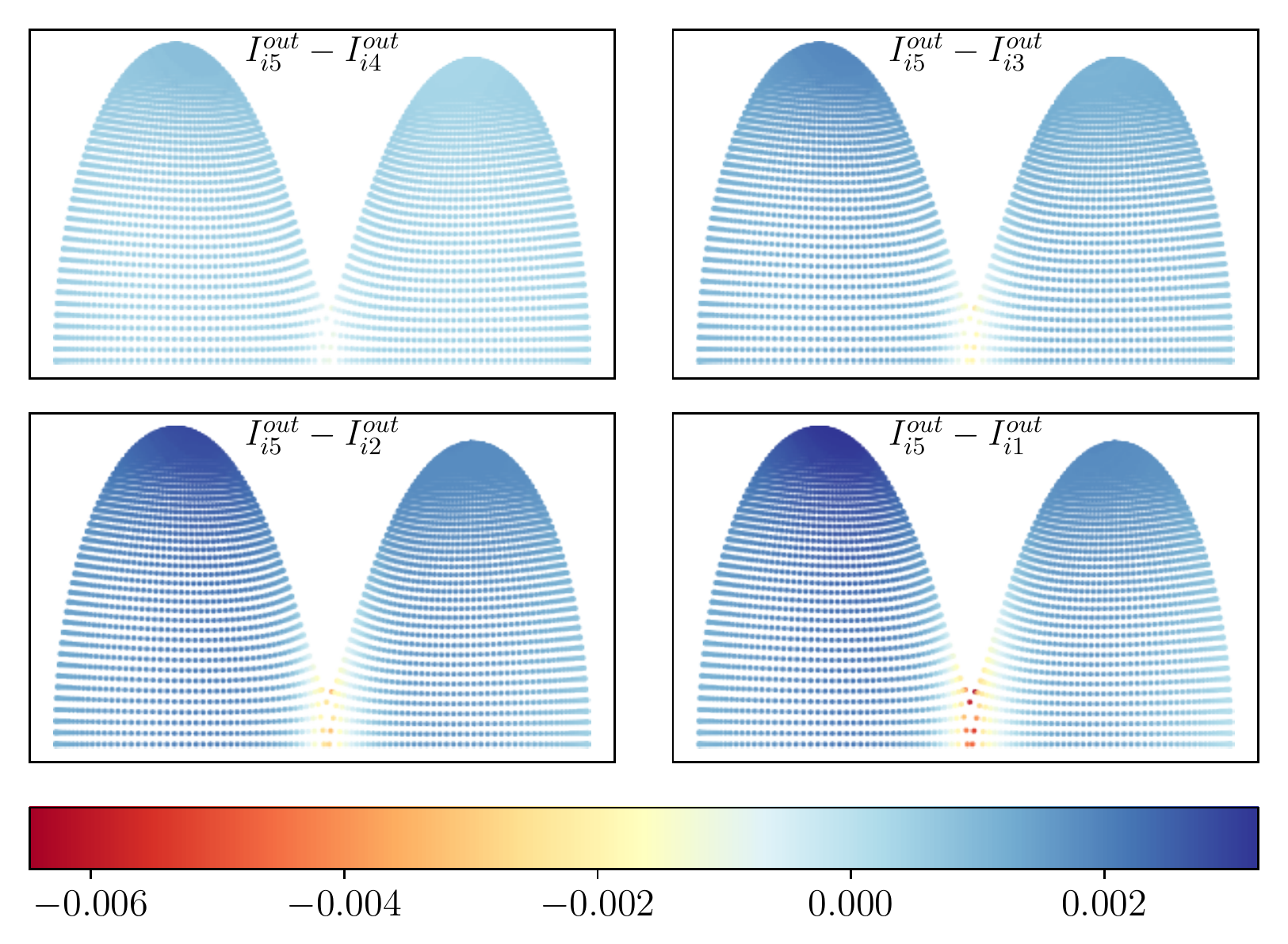}
    
        \includegraphics[width=0.495\hsize]{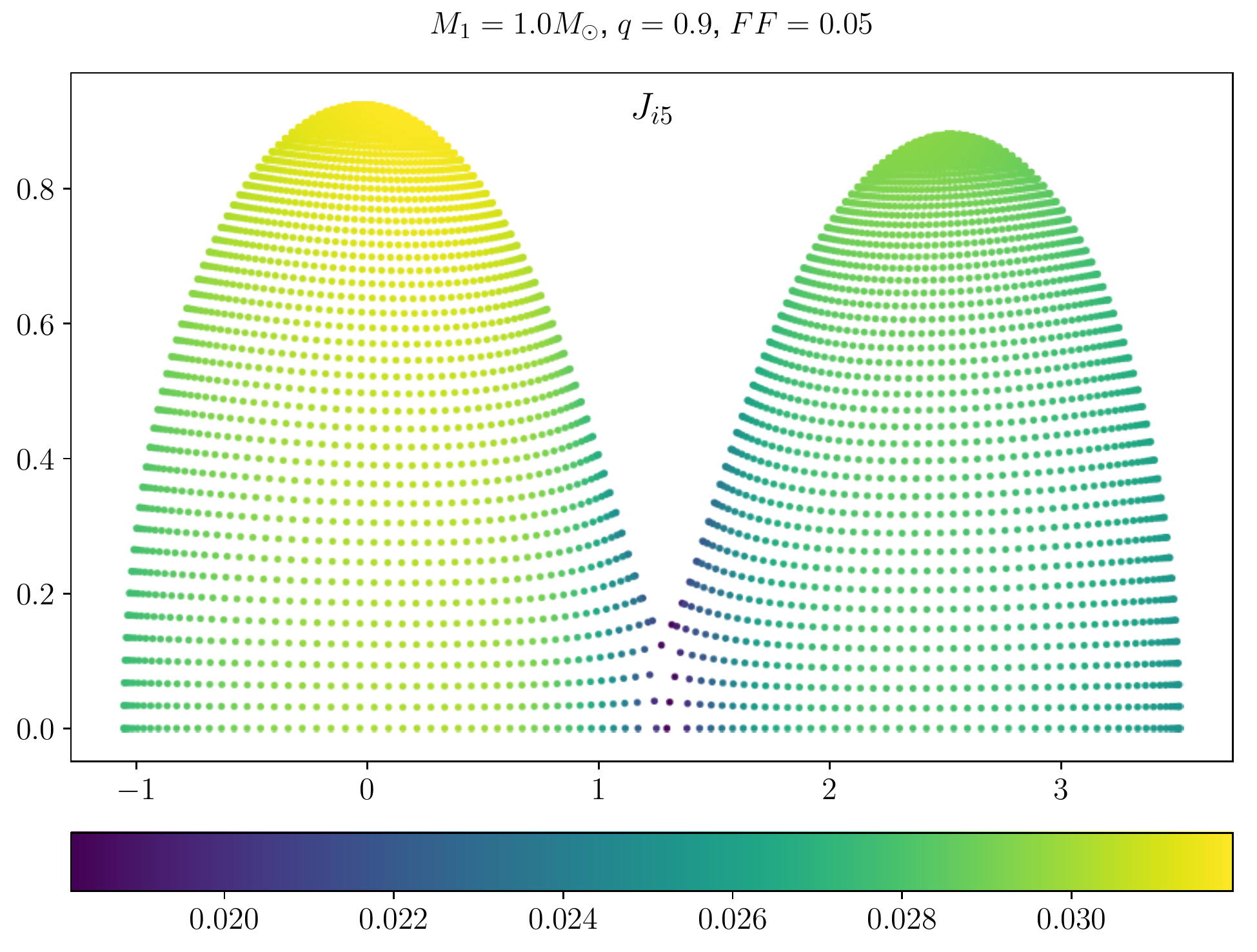}
    \includegraphics[width=0.495\hsize]{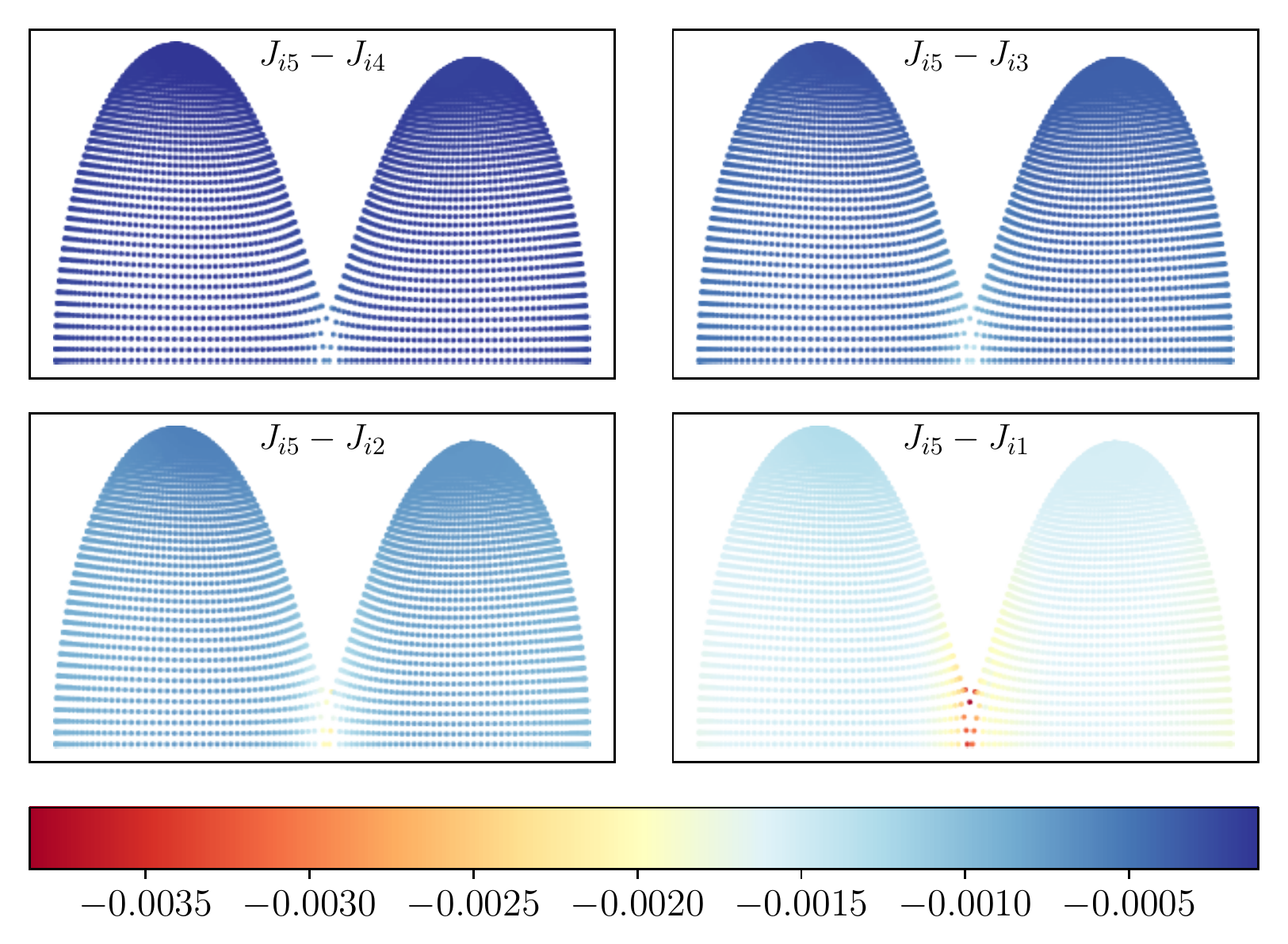}

    \caption{Surface distribution of the outward (top) and mean (bottom) intensity of a contact binary with $q=0.9$ after the fifth iteration. Right panels show the differences in the surface distribution between the final and each previous iteration.}
    \label{fig:q09_s}
\end{figure}

The radiative transfer simulation results for different mass ratios are depicted in Figures~\ref{fig:q05}-\ref{fig:q09_s}. The resulting distributions are predominantly smooth and all converge within the threshold of 0.01\% after five iterations (Figure~\ref{fig:iters_q}), however there is a notable "break" in the potential-intensity distribution of $q=0.5$ (Figure~\ref{fig:q05}).
This is a consequence of insufficient coverage in the neck area for that particular equipotential, due to the diverging radii. Therefore, different mass ratios generally require different grid sampling, to assure that the limits of all spherical angles are covered and there are no gaps in an equipotential caused by the diverging radius points. The surface distributions (Figures~\ref{fig:q05_s},~\ref{fig:q075_s} and~\ref{fig:q09_s}) show that the secondary converges faster than the primary in most cases, because of the grid sampling being better suited to the secondary star. This points to the fact that, in cases of extreme mass ratios, different grid resolutions for the two components might be necessary to ensure an optimal rate of convergence.

\subsection{Fillout factor}

\begin{figure}[h]
    \centering
    \includegraphics[width=\hsize]{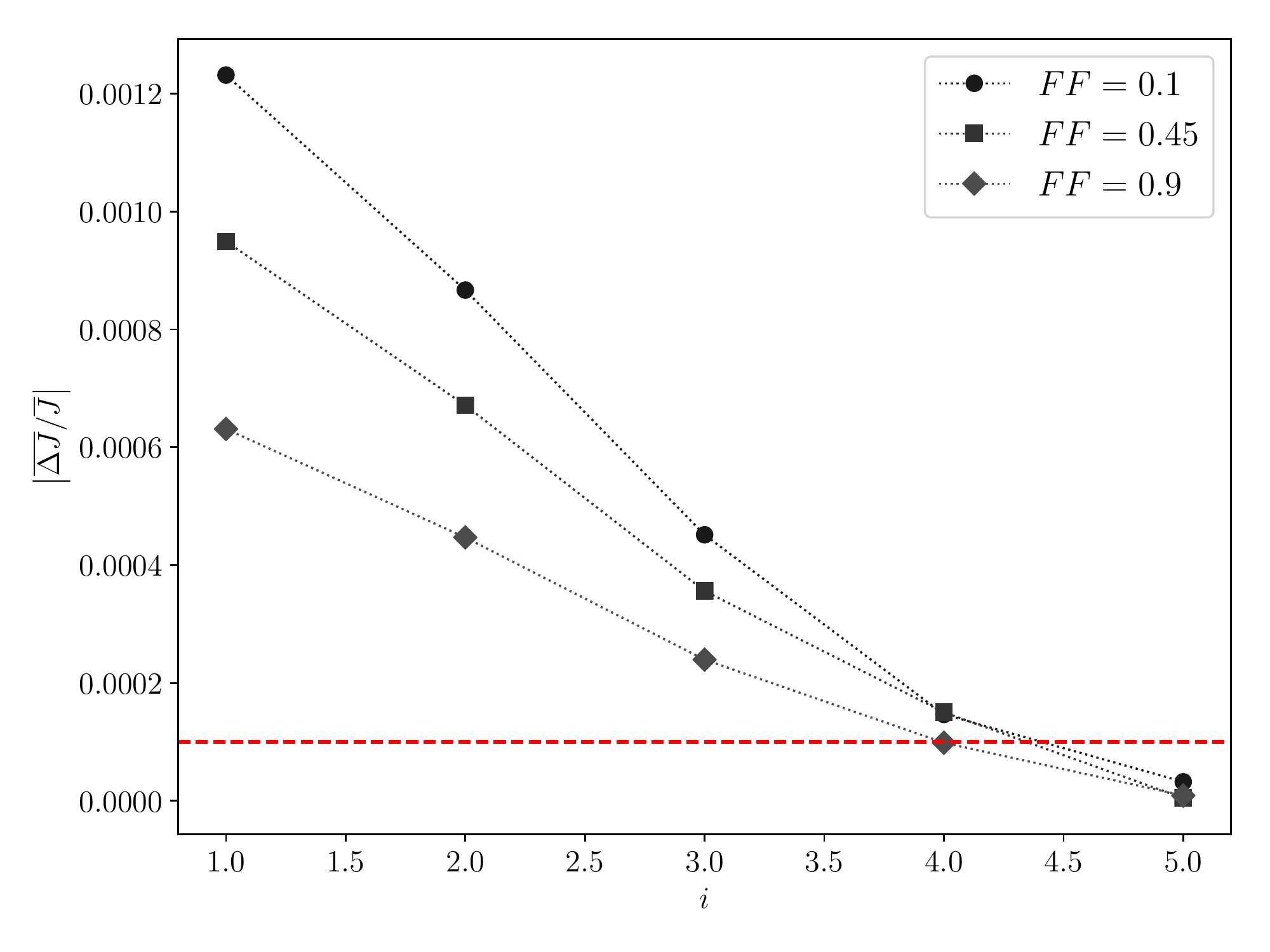}
    \caption{Convergence of the mean intensity in test systems with variable fillout factor.}
    \label{fig:iters_ff}
\end{figure}

\begin{figure}[h]
    \centering
    \includegraphics[width=0.495\hsize]{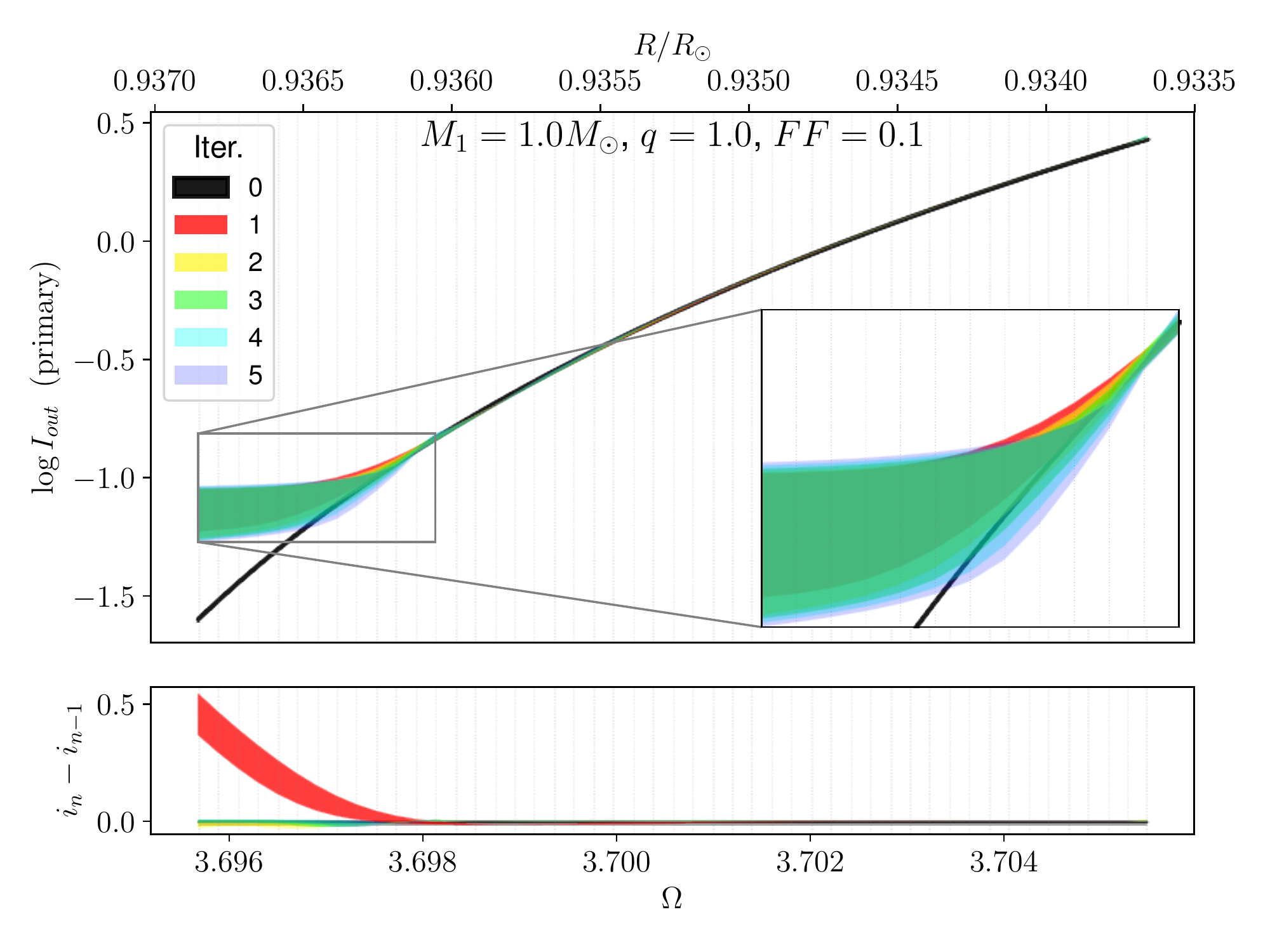}
    \includegraphics[width=0.495\hsize]{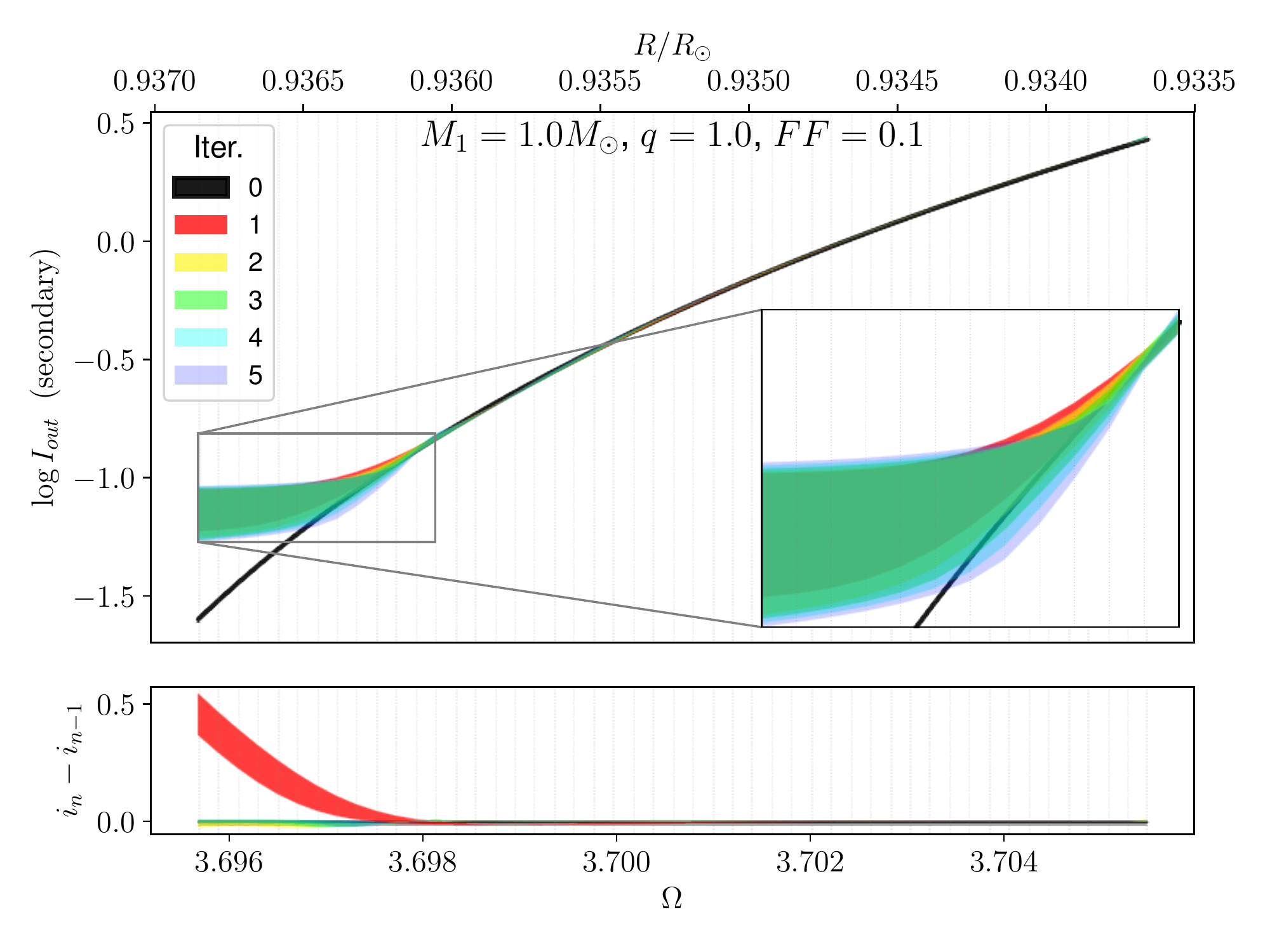}
    
    \includegraphics[width=0.495\hsize]{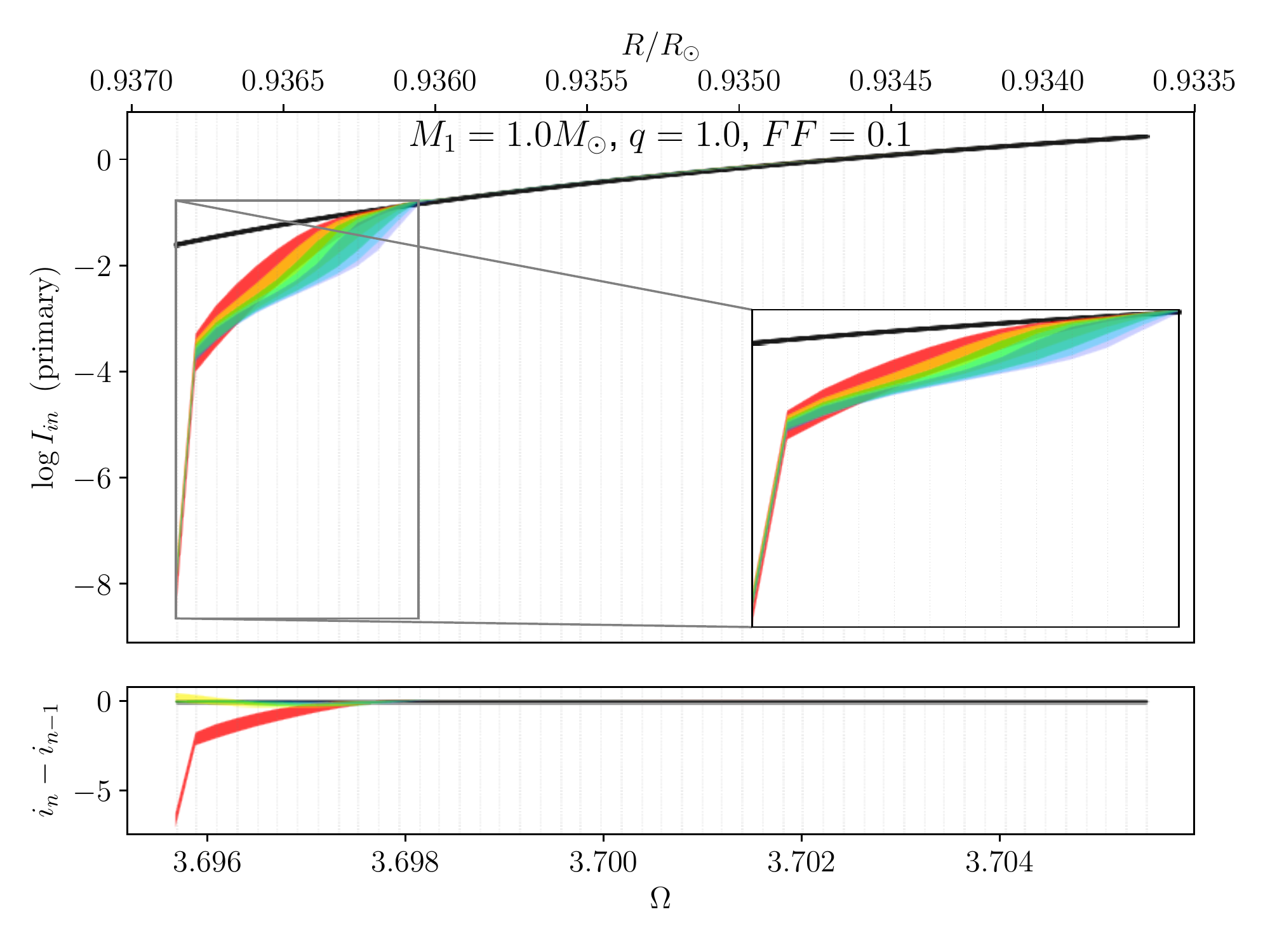}
    \includegraphics[width=0.495\hsize]{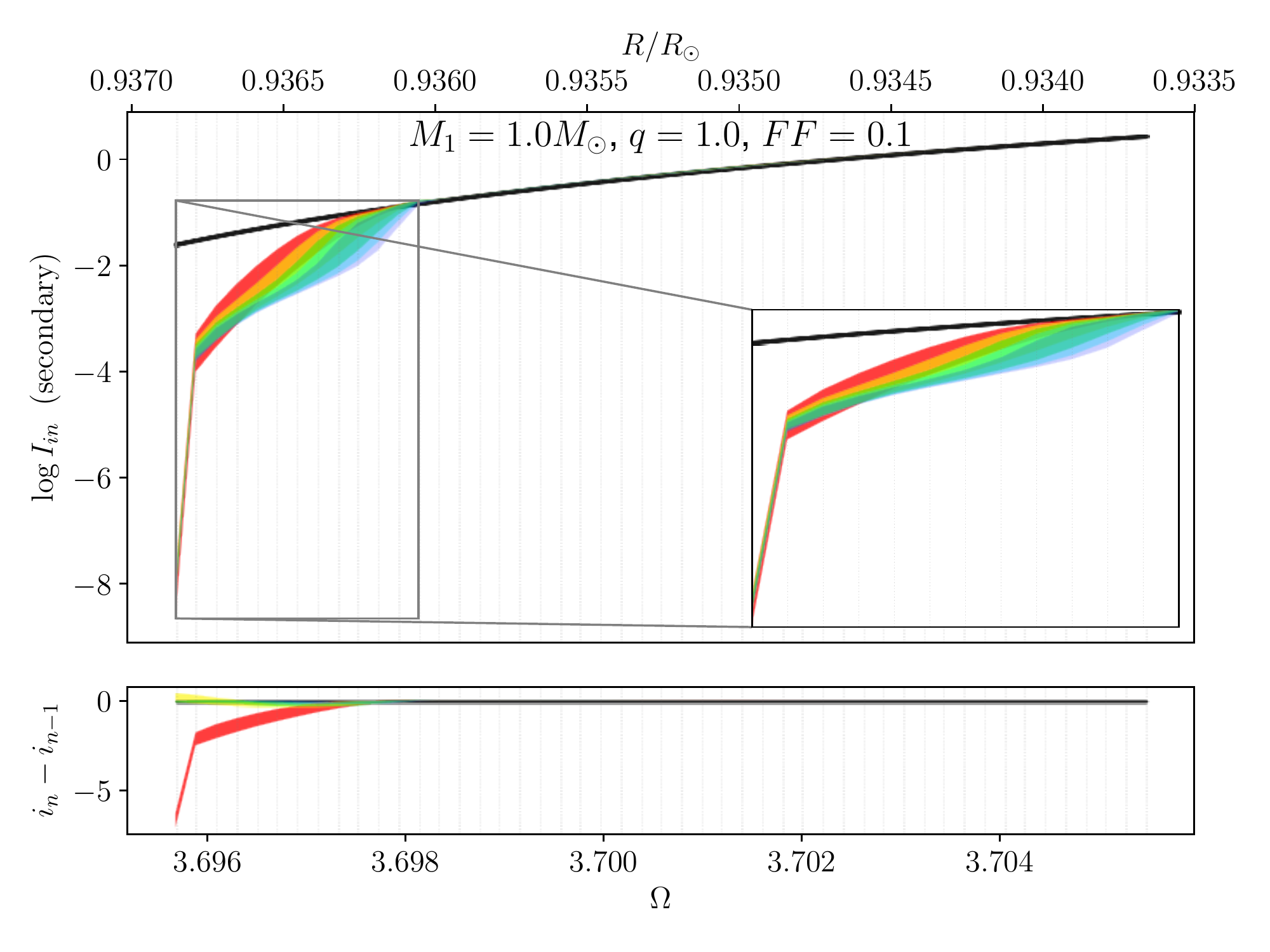}
    
    \includegraphics[width=0.495\hsize]{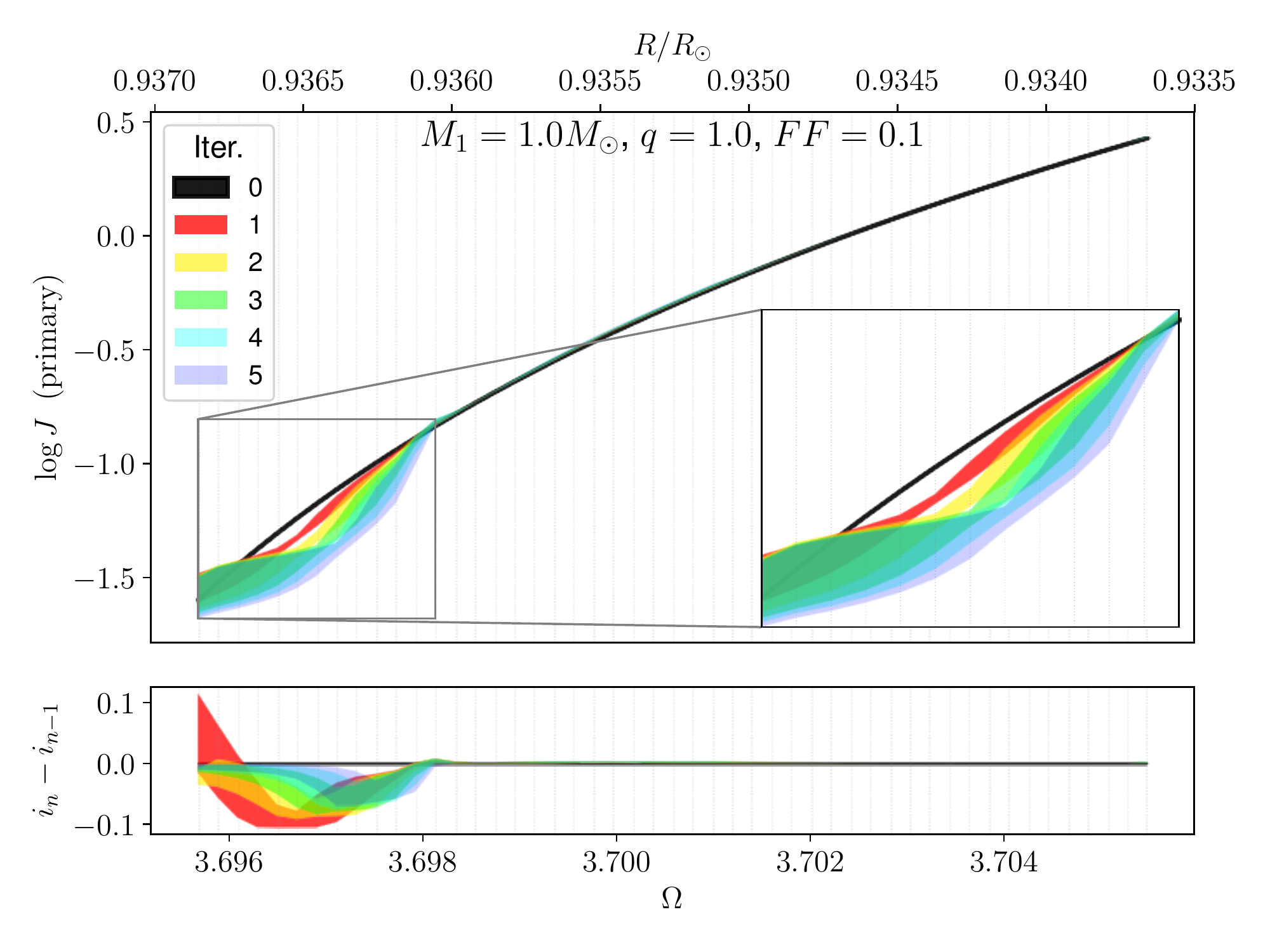}
    \includegraphics[width=0.495\hsize]{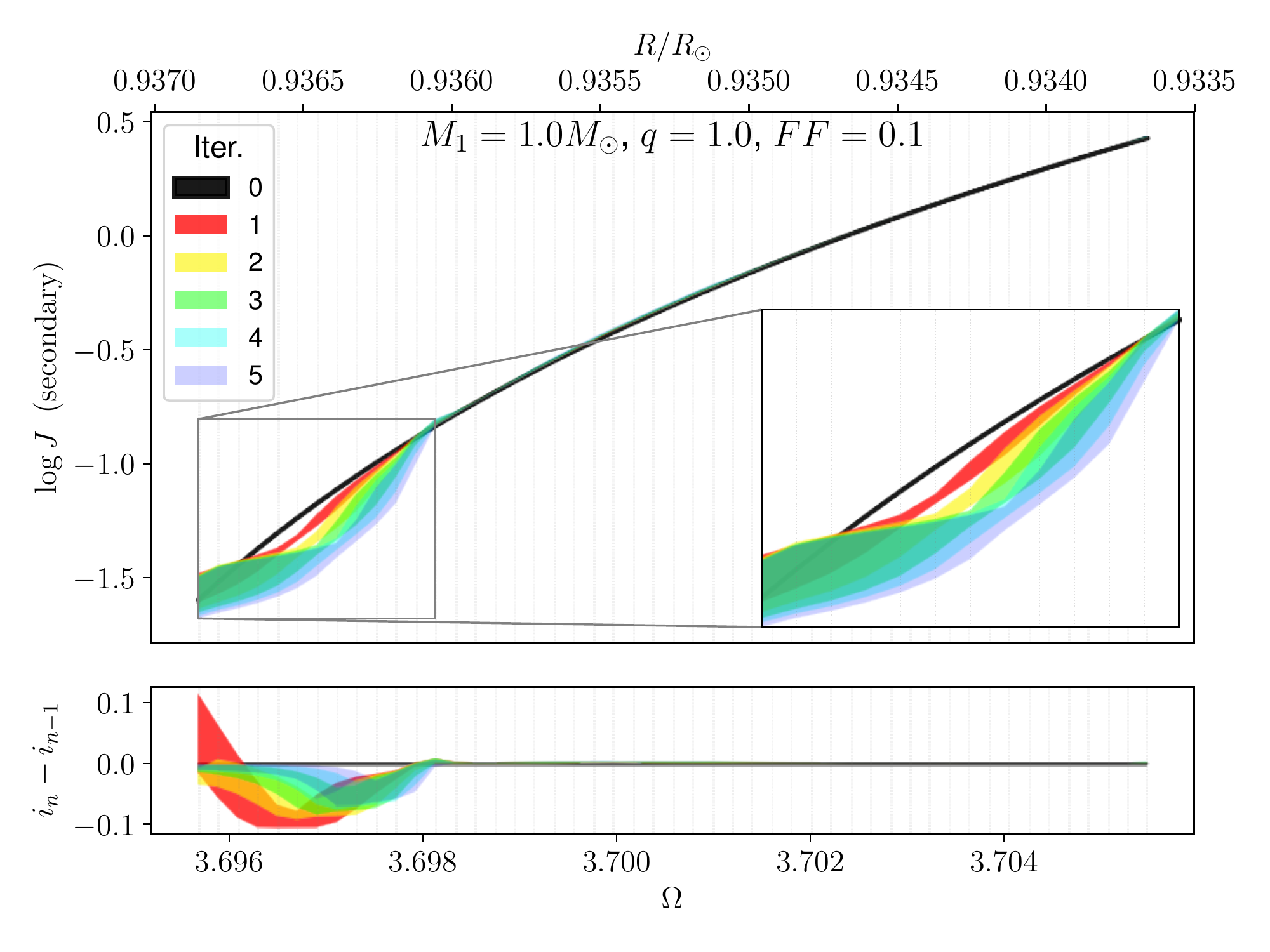}
    
    \caption{Top to bottom: outward, inward and mean intensity as a function of the potential/radius of a contact binary with $FF=0.1$. Left panels: primary, right panels: secondary component. The bottom panel of each plot shows the differences between successive iterations.}
    \label{fig:ff01}
\end{figure}

\begin{figure}[h]
    \centering
       \includegraphics[width=0.495\hsize]{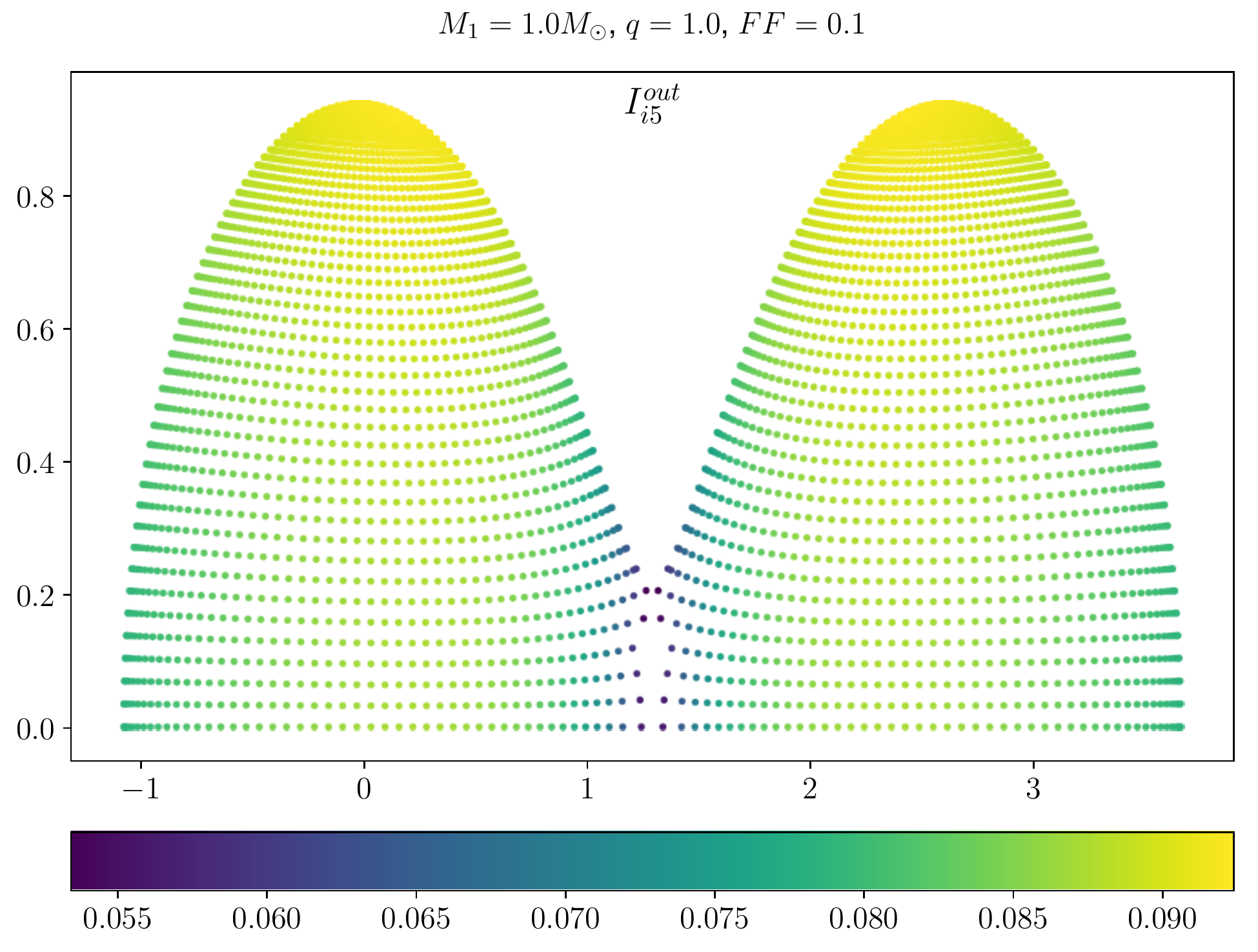}
    \includegraphics[width=0.495\hsize]{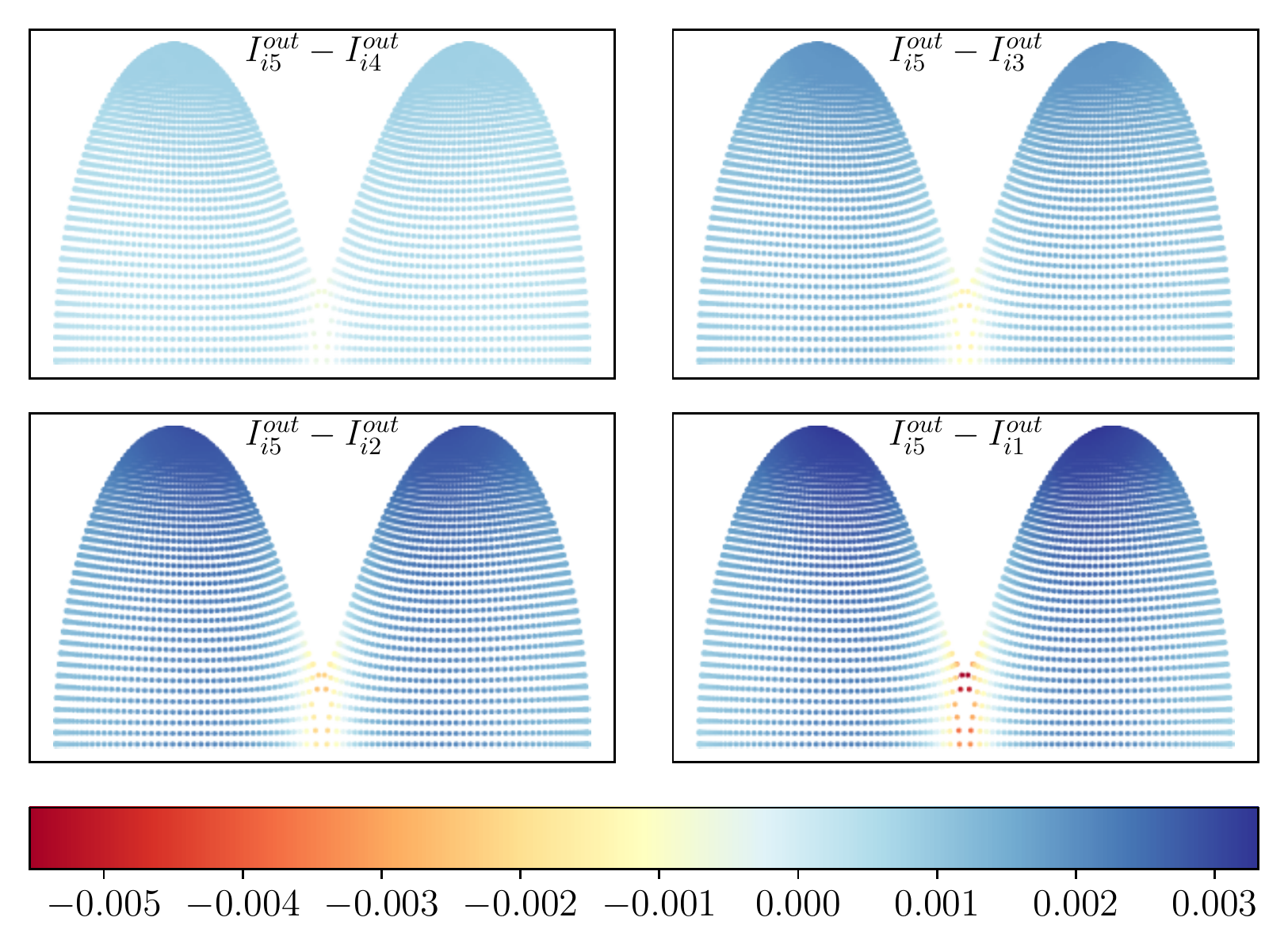}
    
        \includegraphics[width=0.495\hsize]{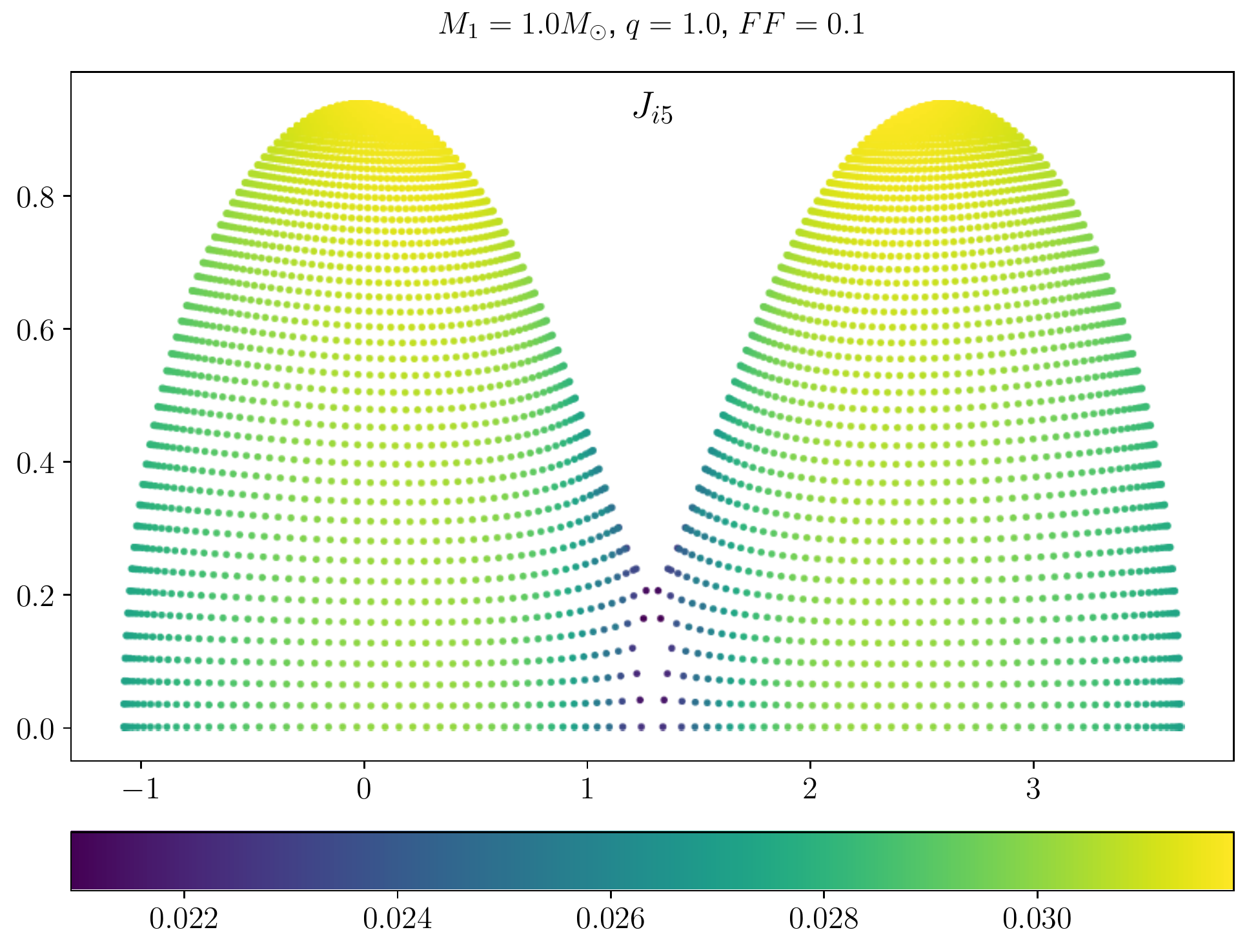}
    \includegraphics[width=0.495\hsize]{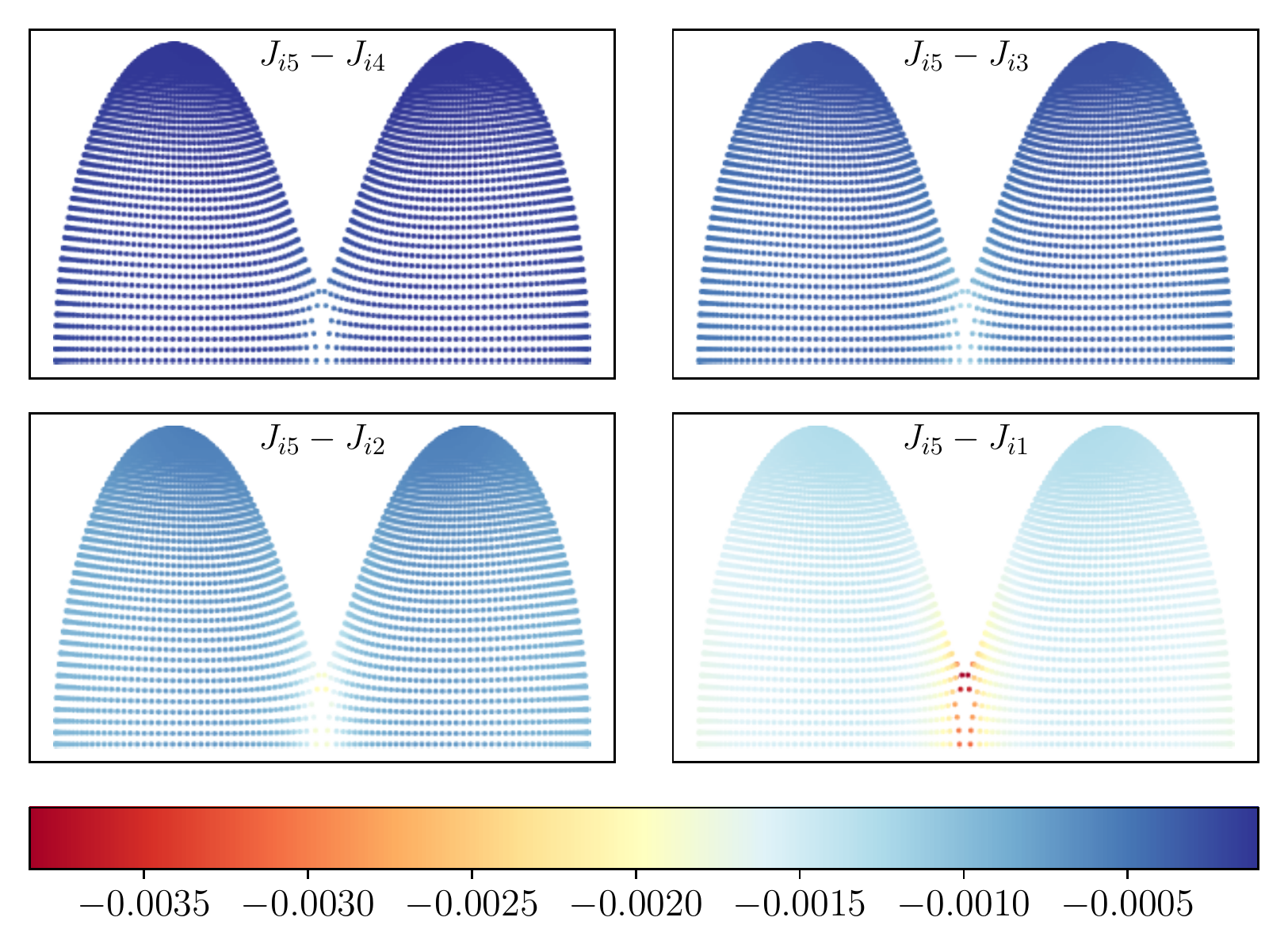}

    \caption{Surface distribution of the outward (top) and mean (bottom) intensity of a contact binary with $FF=0.1$ after the fifth iteration. Right panels show the differences in the surface distribution between the final and each previous iteration.}
    \label{fig:ff01_s}
\end{figure}

\begin{figure}[h]
    \centering
    \includegraphics[width=0.495\hsize]{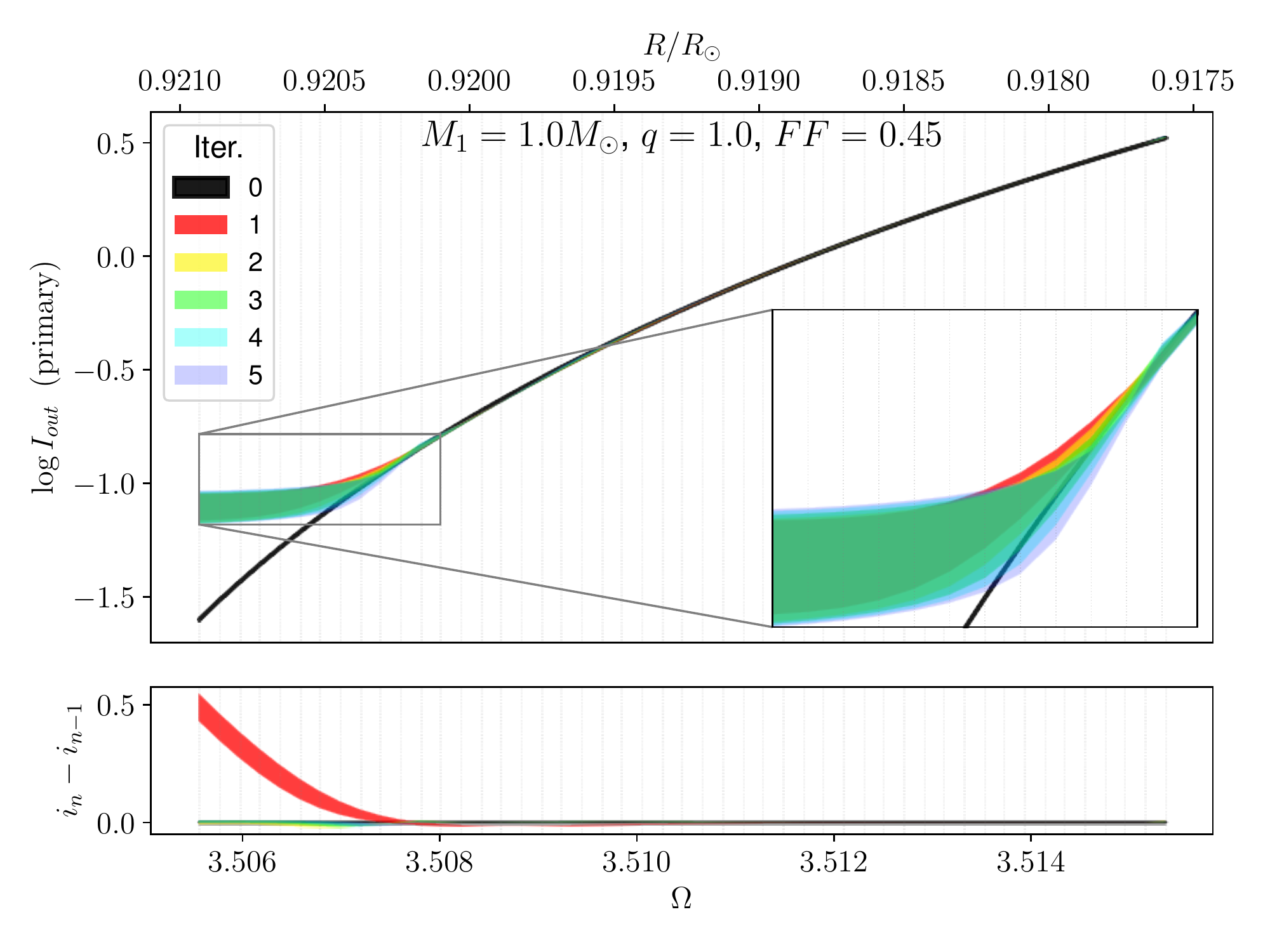}
    \includegraphics[width=0.495\hsize]{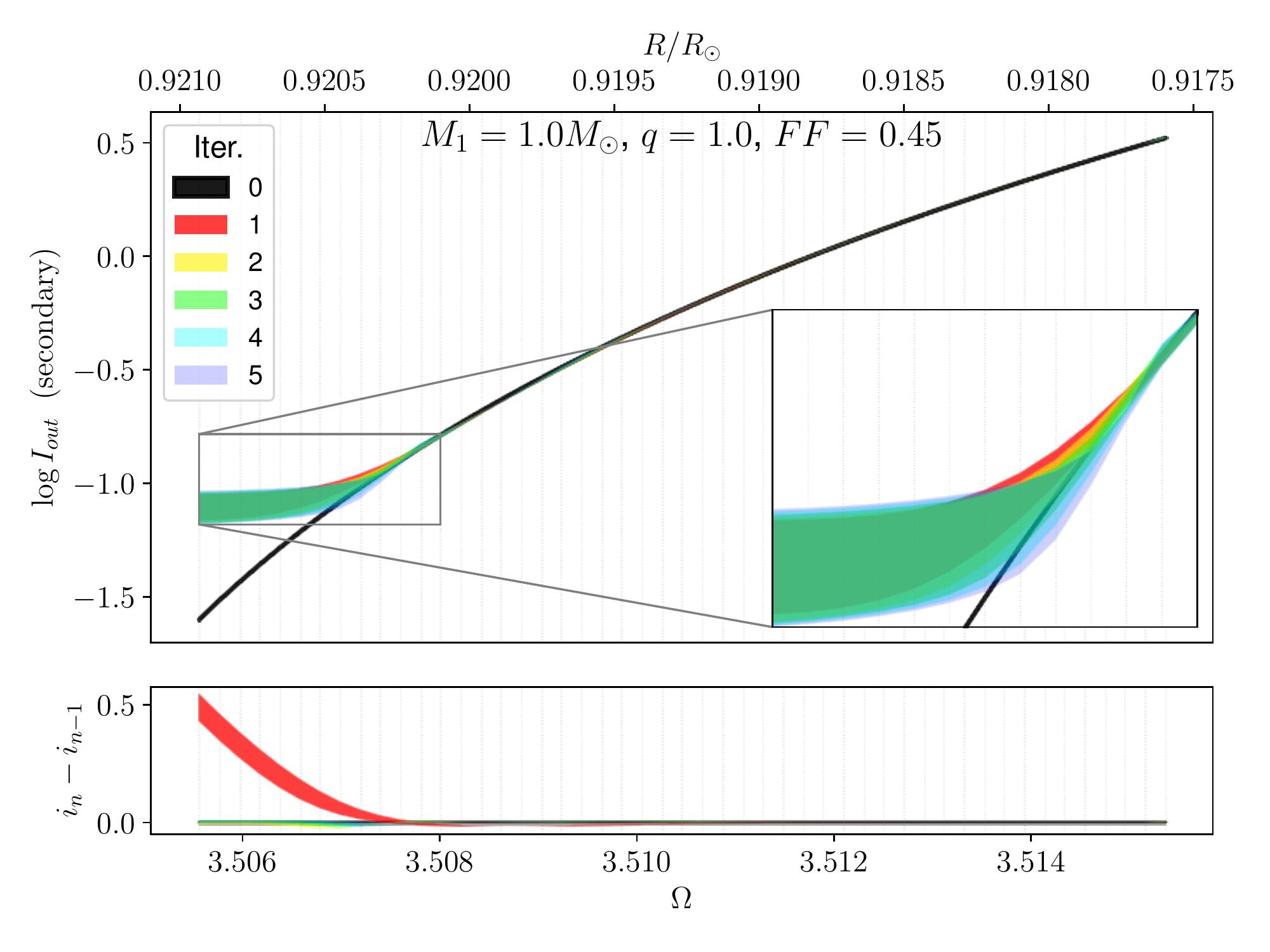}
    
    \includegraphics[width=0.495\hsize]{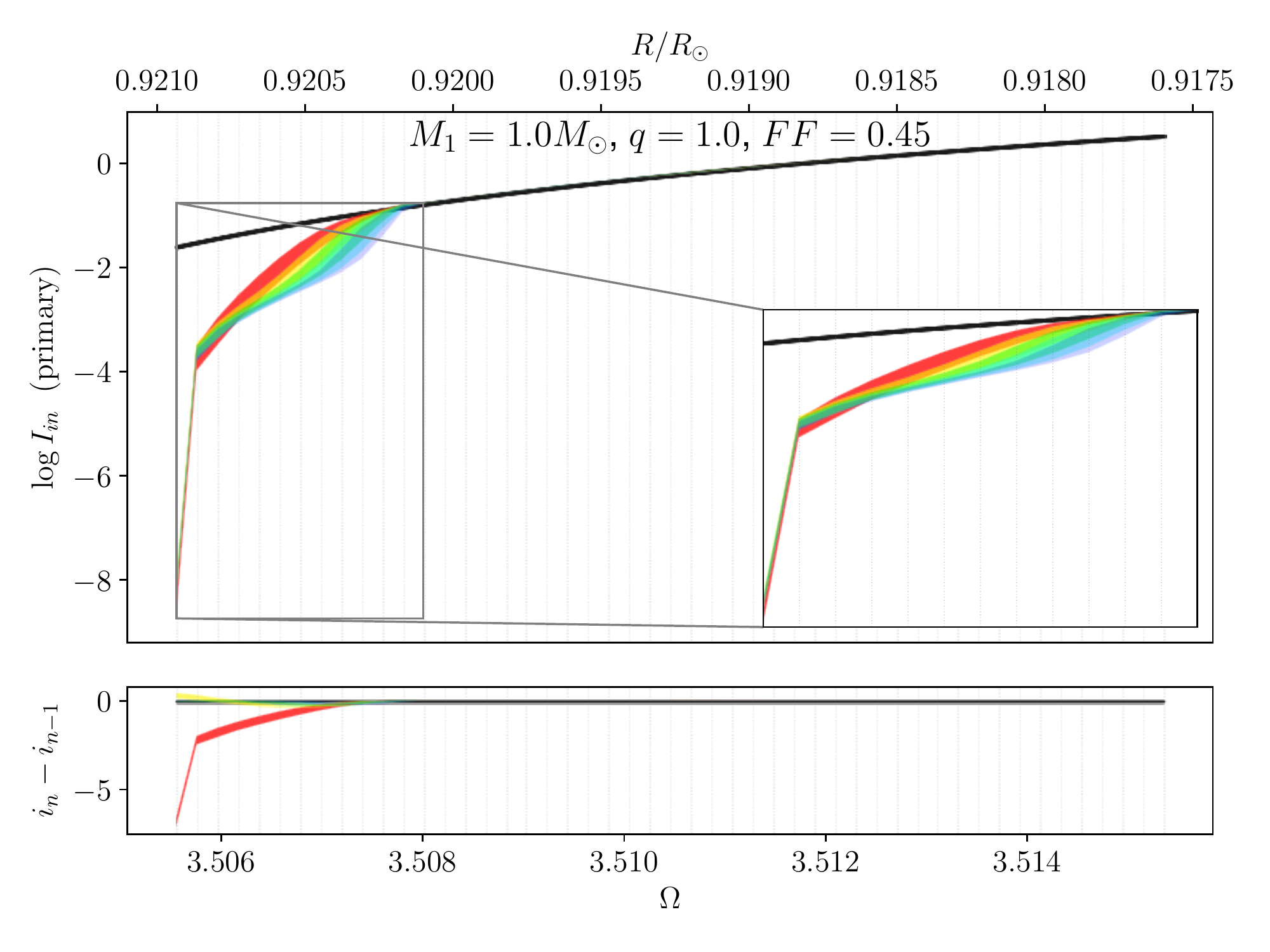}
    \includegraphics[width=0.495\hsize]{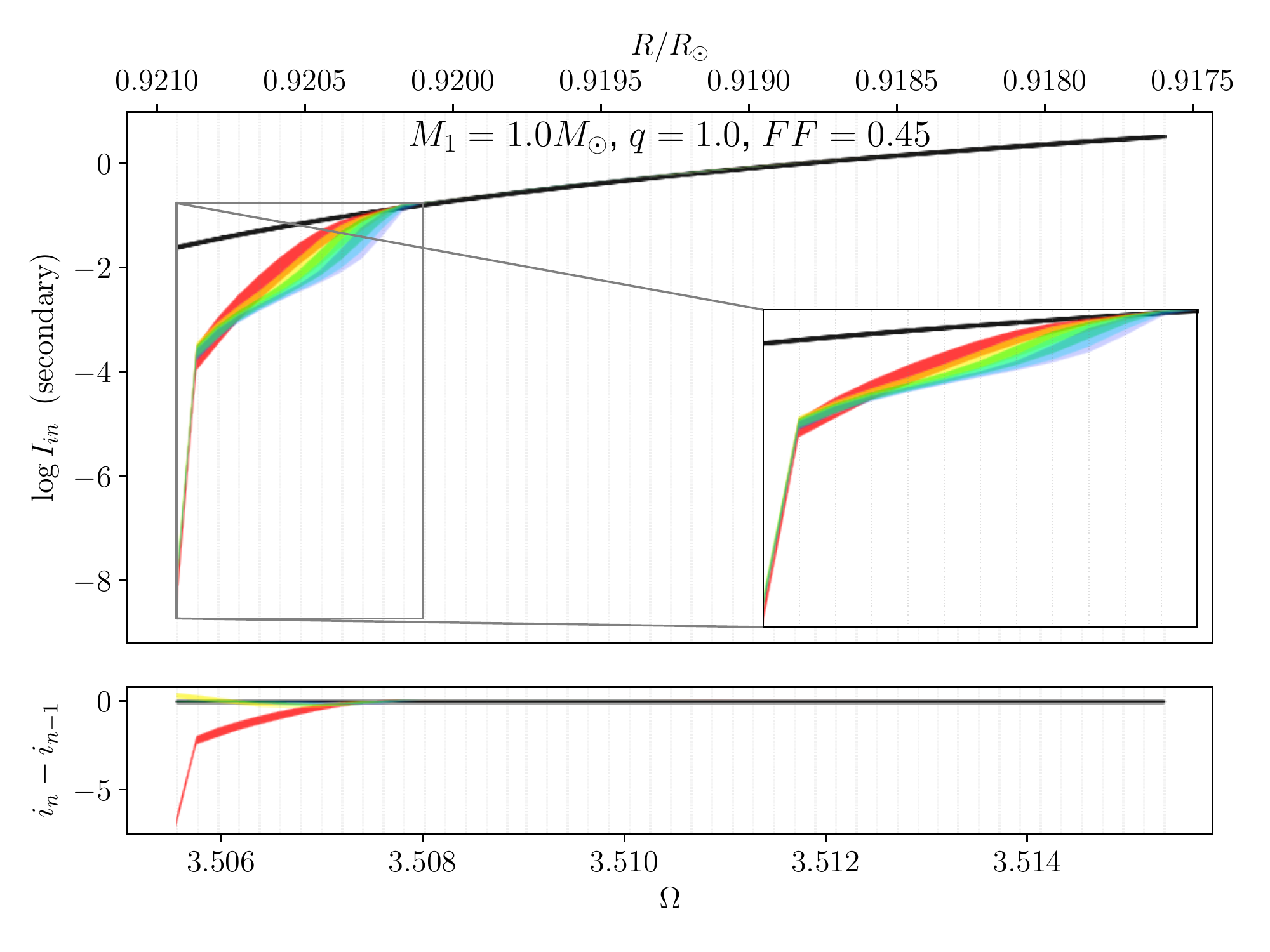}
    
    \includegraphics[width=0.495\hsize]{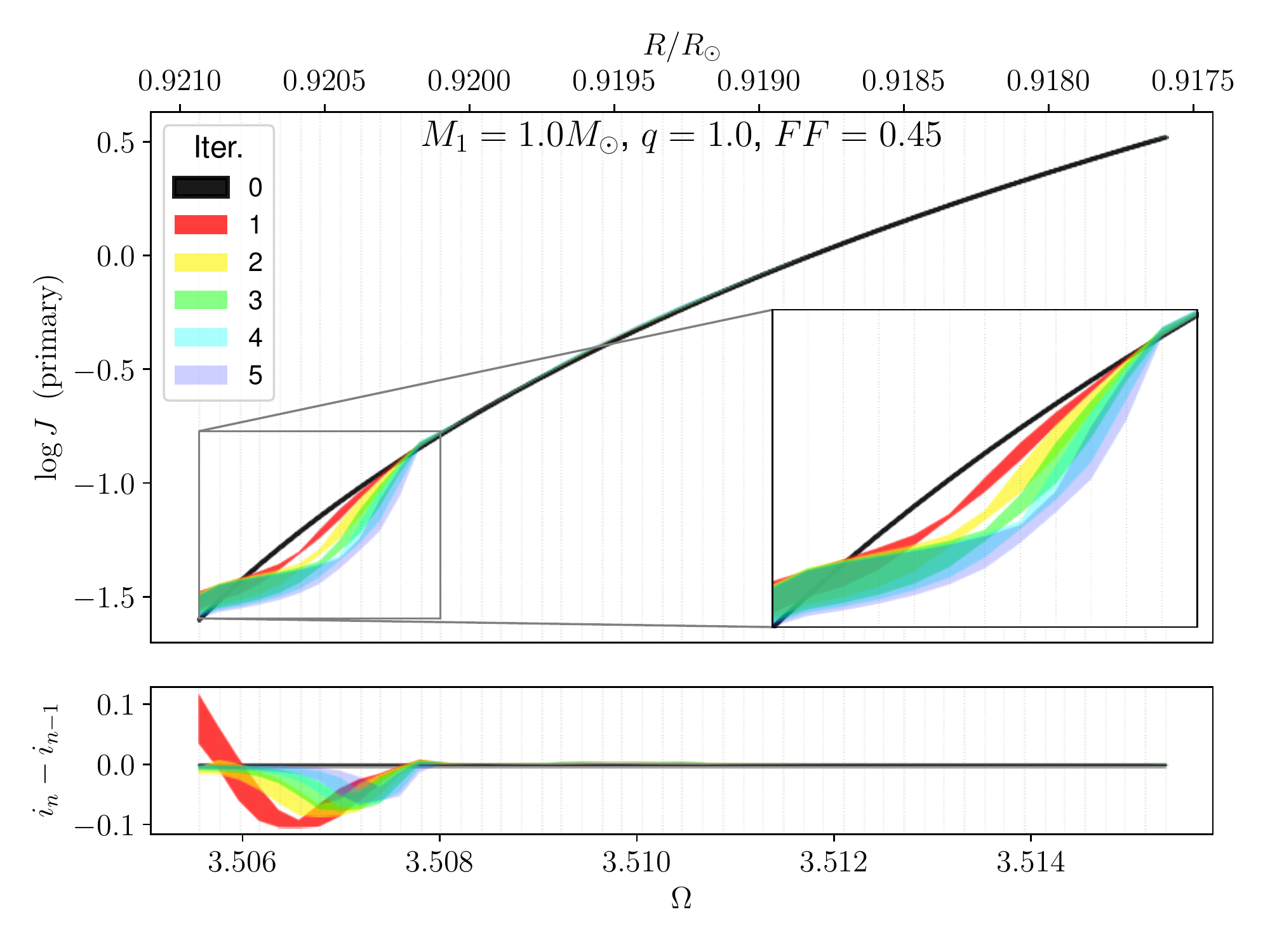}
    \includegraphics[width=0.495\hsize]{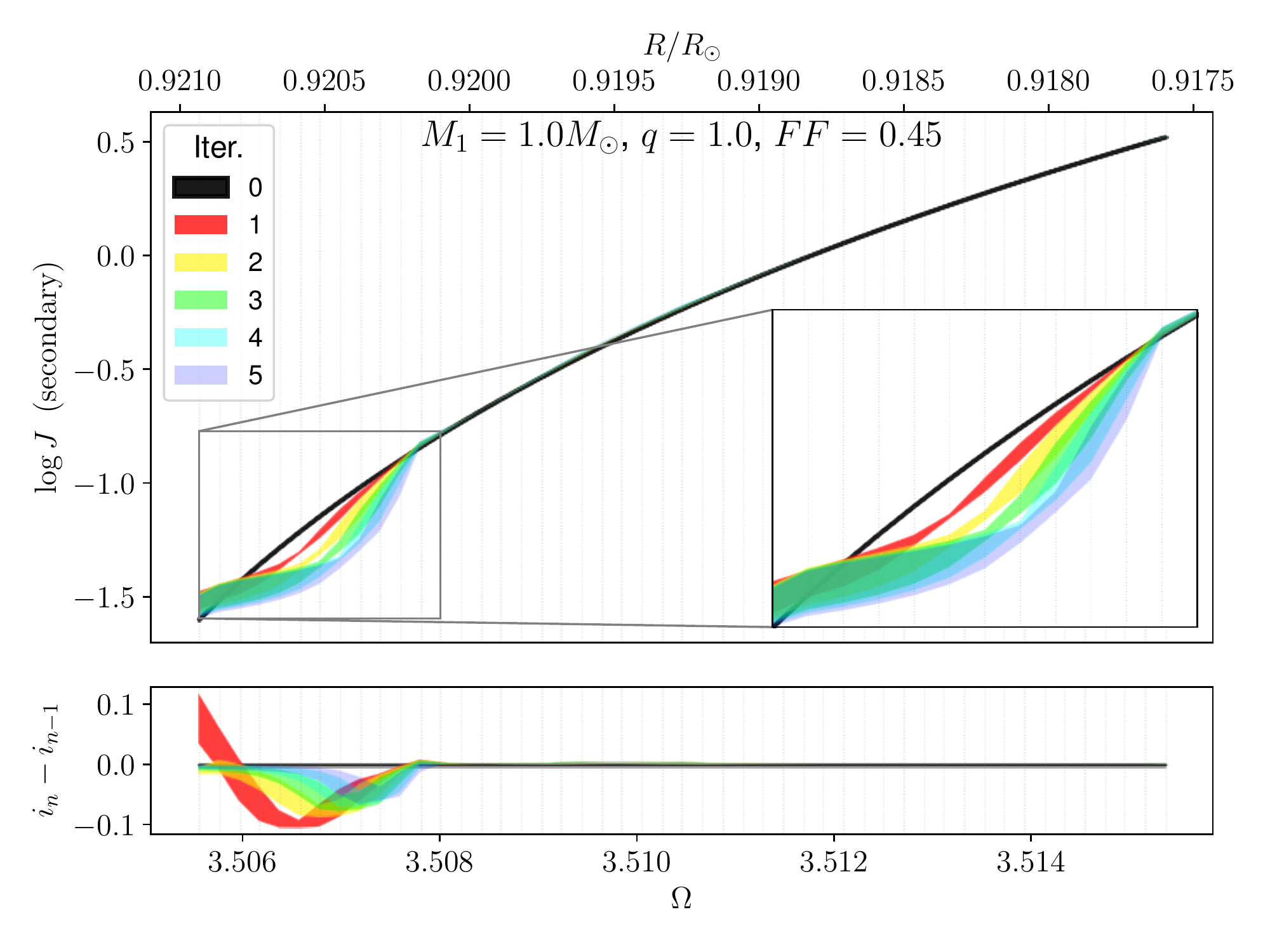}
    
    \caption{Top to bottom: outward, inward and mean intensity as a function of the potential/radius of a contact binary with $FF=0.45$. Left panels: primary, right panels: secondary component. The bottom panel of each plot shows the differences between successive iterations.}
    \label{fig:ff045}
\end{figure}

\begin{figure}[h]
    \centering
        \includegraphics[width=0.495\hsize]{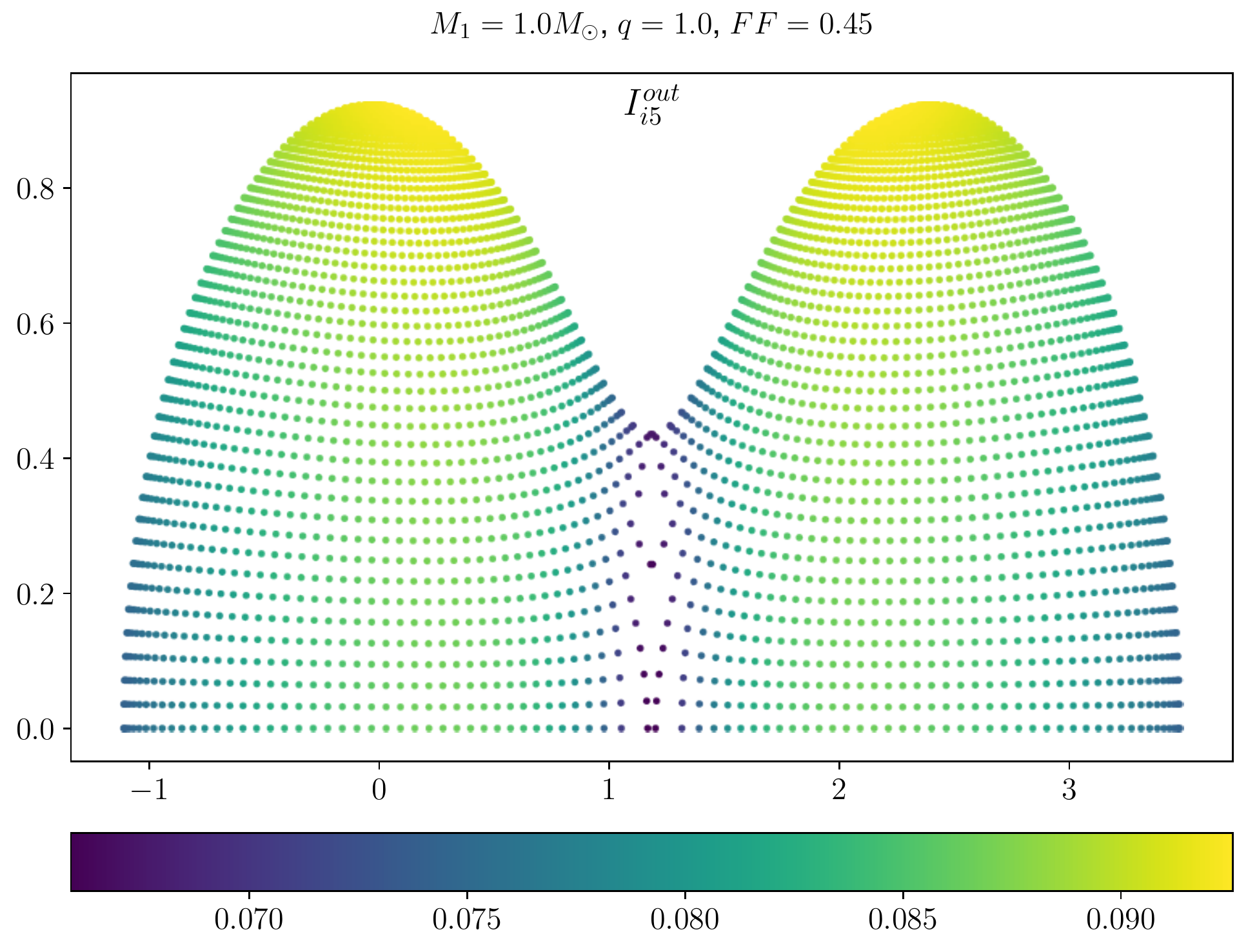}
    \includegraphics[width=0.495\hsize]{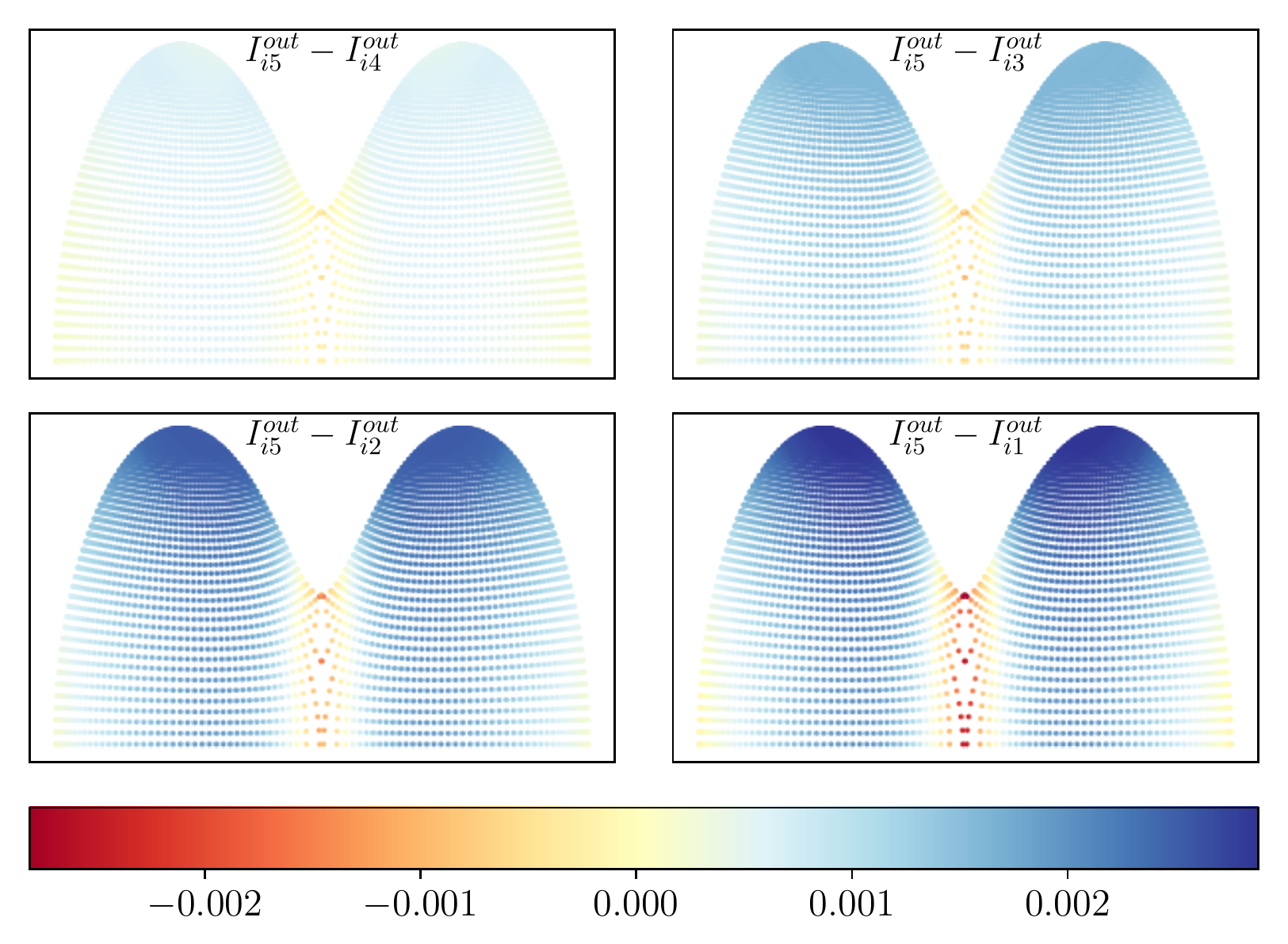}
    
        \includegraphics[width=0.495\hsize]{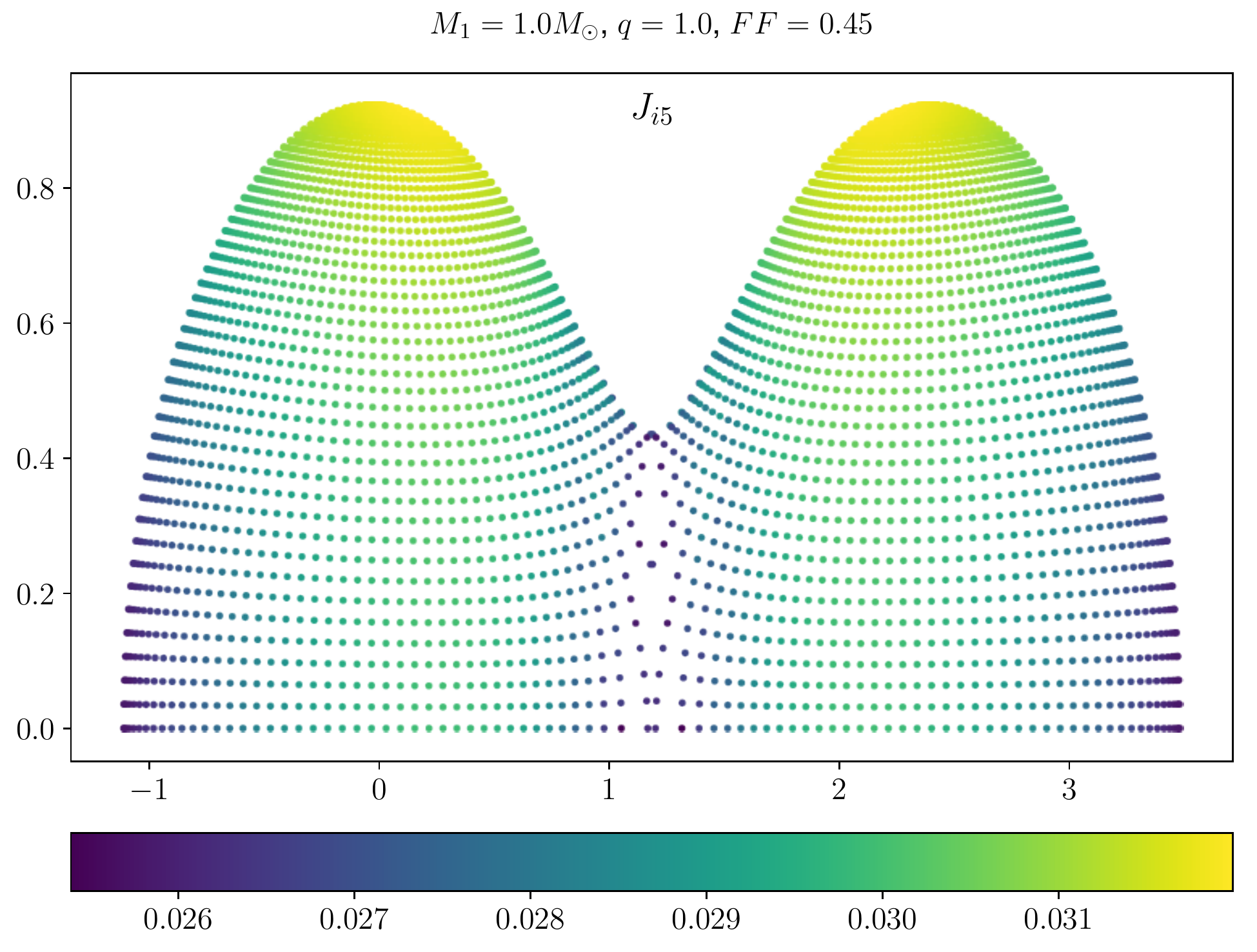}
    \includegraphics[width=0.495\hsize]{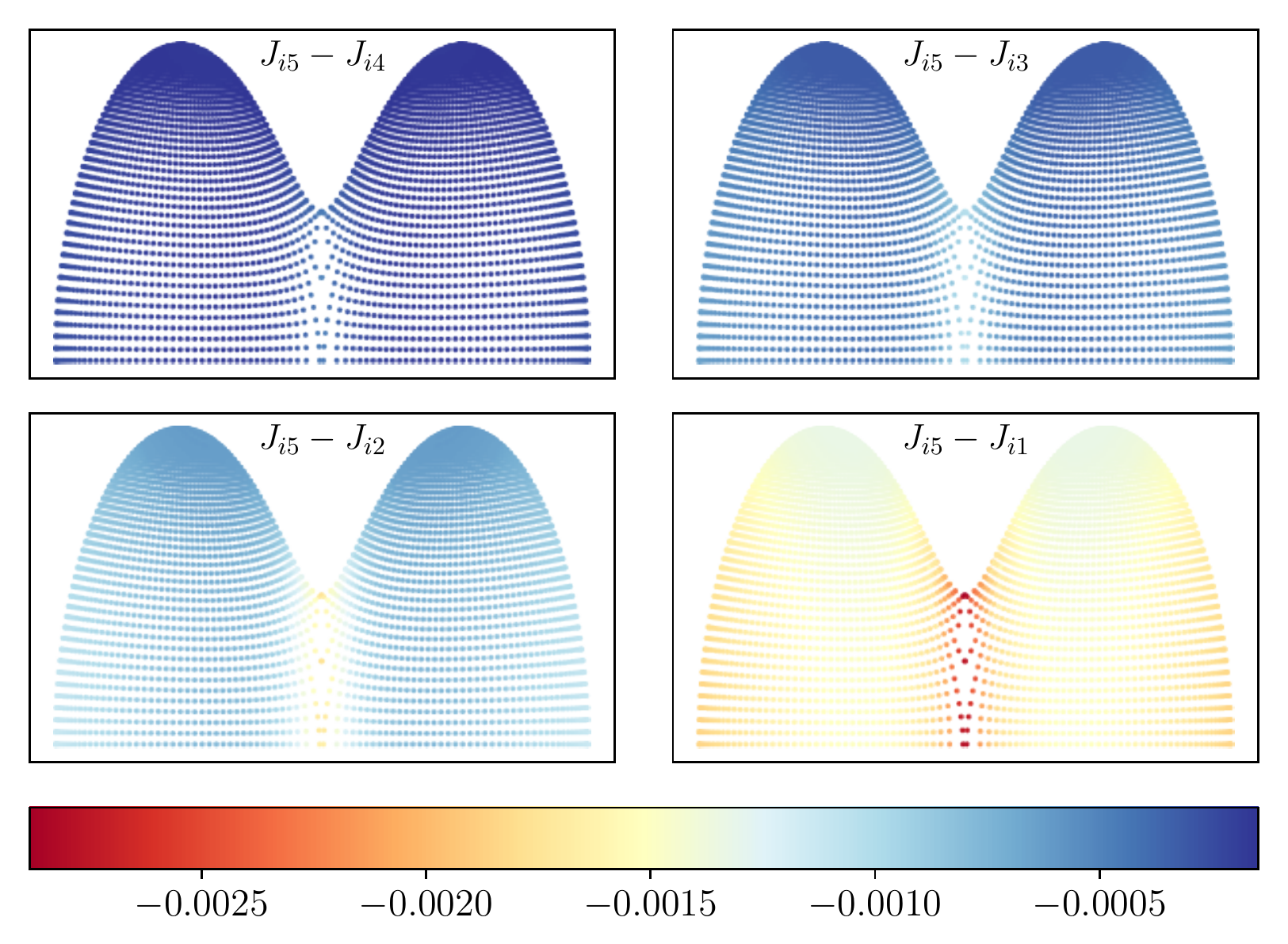}

    \caption{Surface distribution of the outward (top) and mean (bottom) intensity of a contact binary with $FF=0.45$ after the fifth iteration. Right panels show the differences in the surface distribution between the final and each previous iteration.}
    \label{fig:ff045_s}
\end{figure}

\begin{figure}[h]
    \centering
    \includegraphics[width=0.495\hsize]{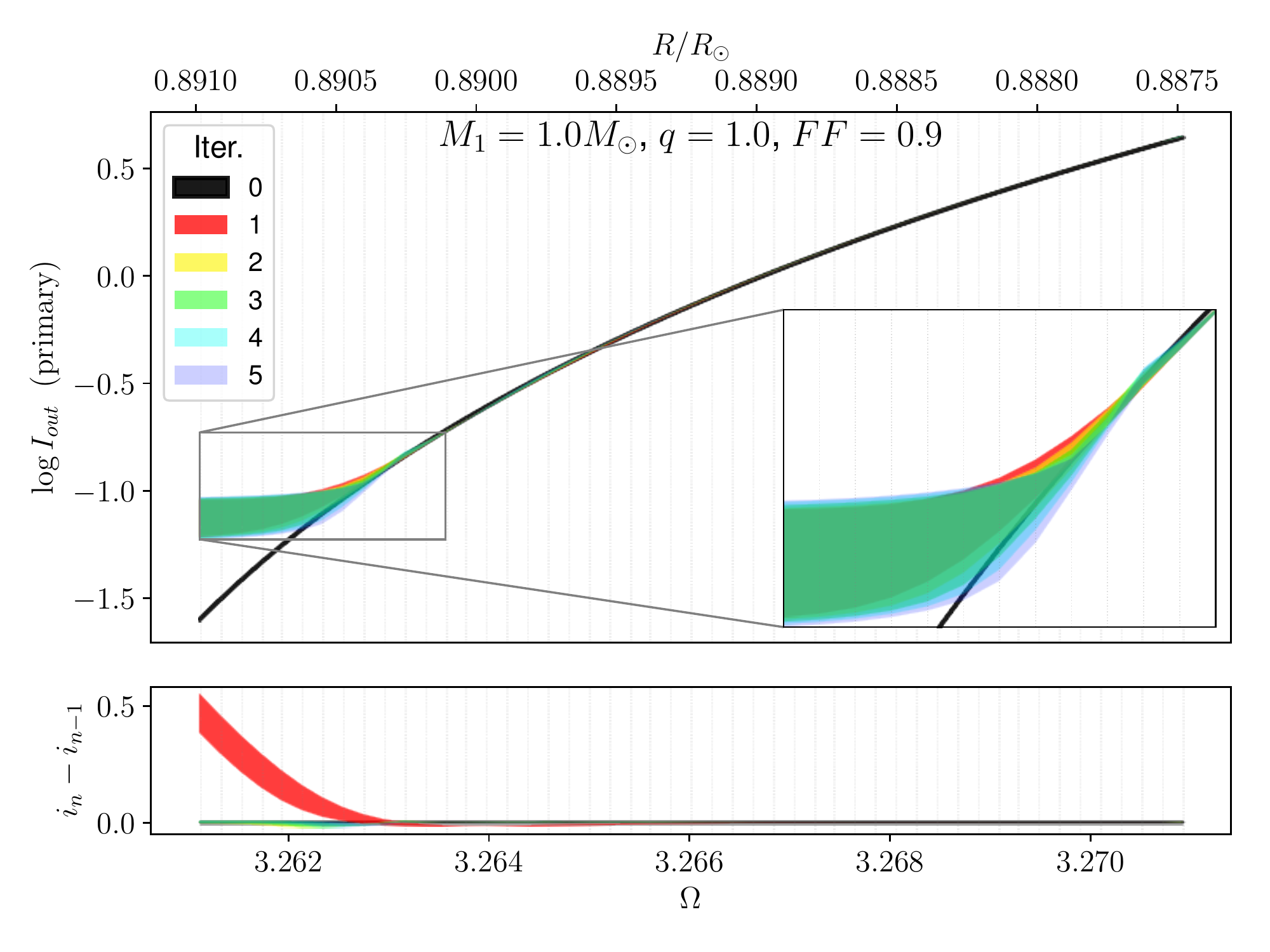}
    \includegraphics[width=0.495\hsize]{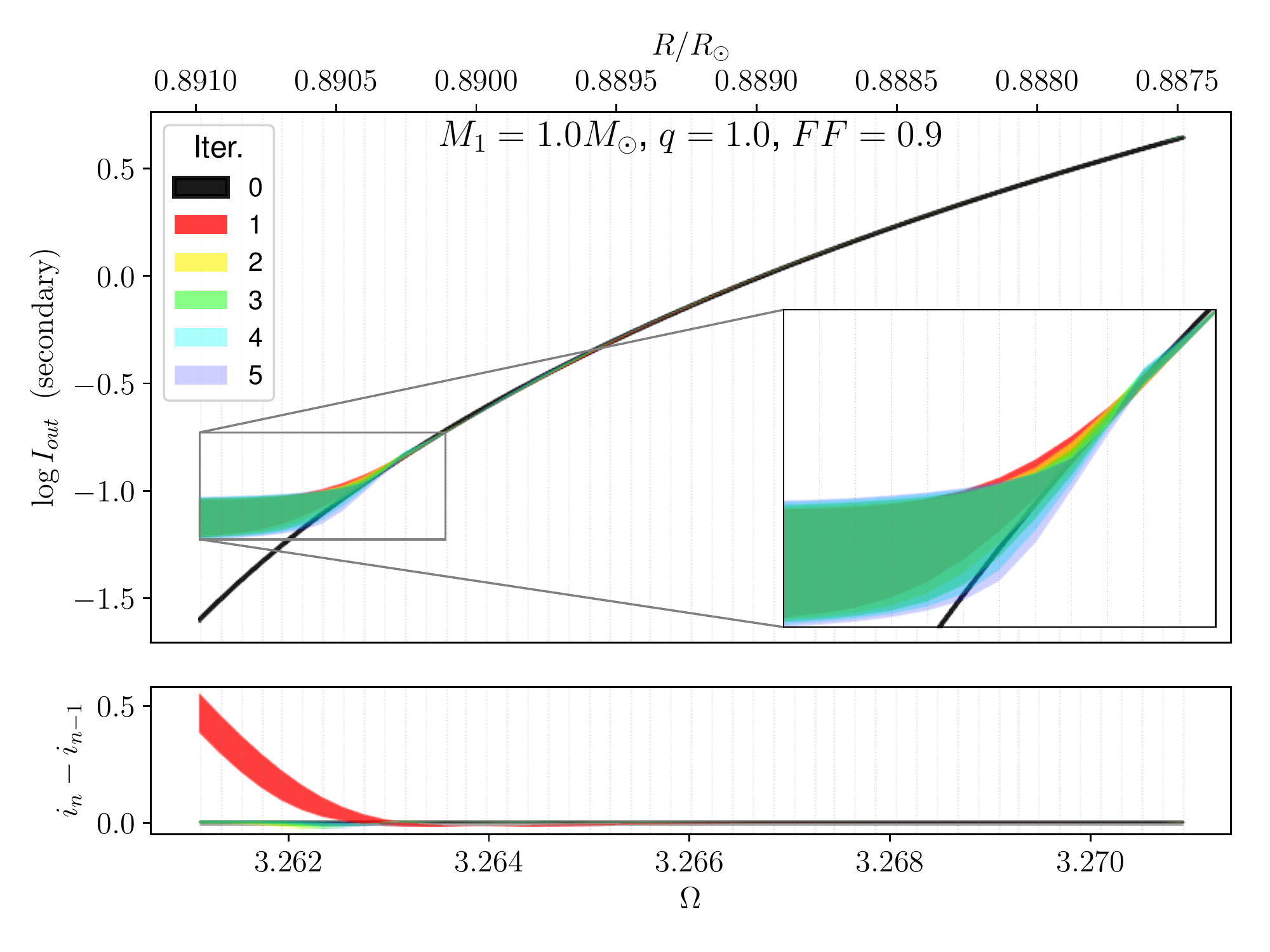}
    
    \includegraphics[width=0.495\hsize]{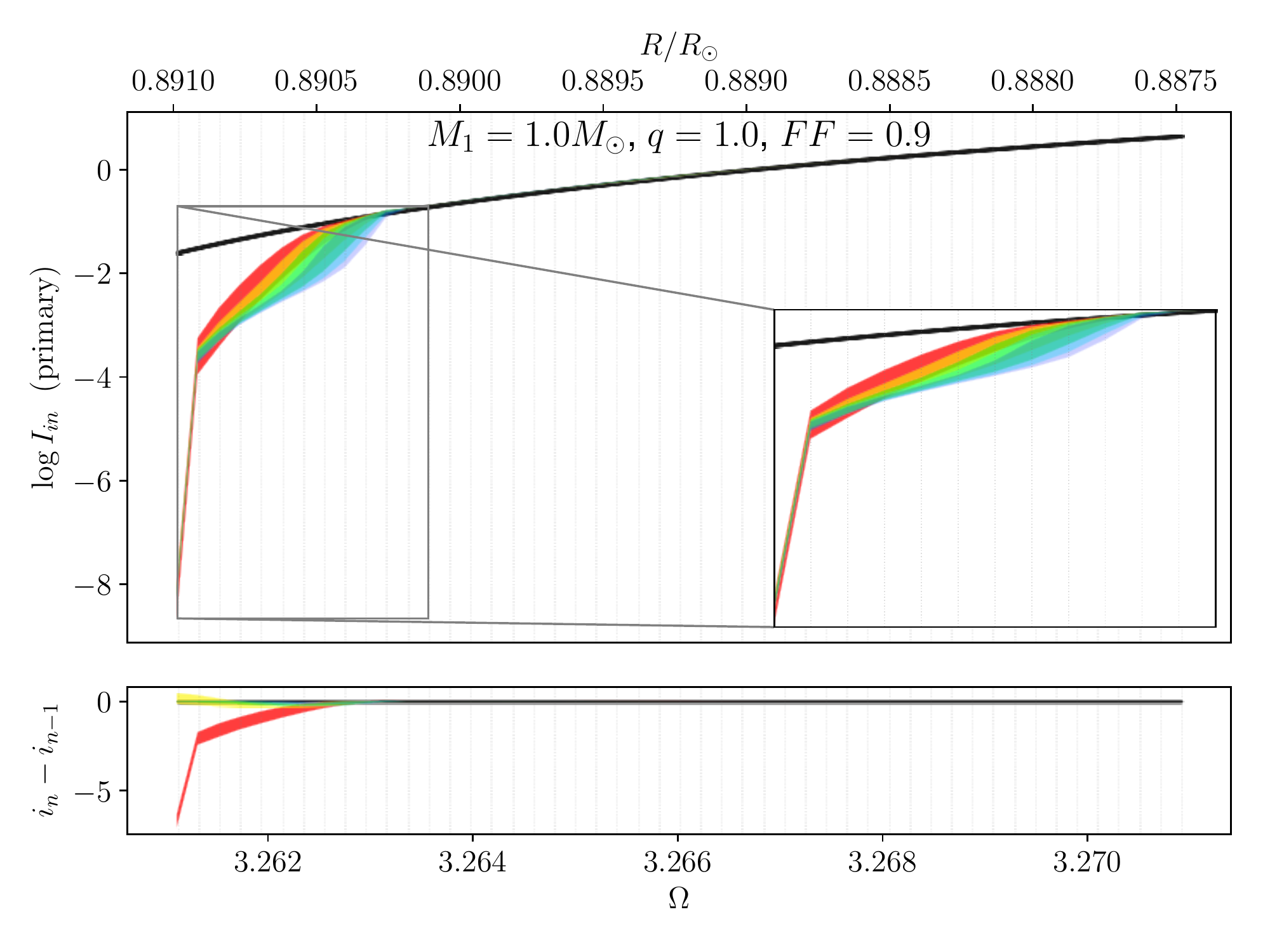}
    \includegraphics[width=0.495\hsize]{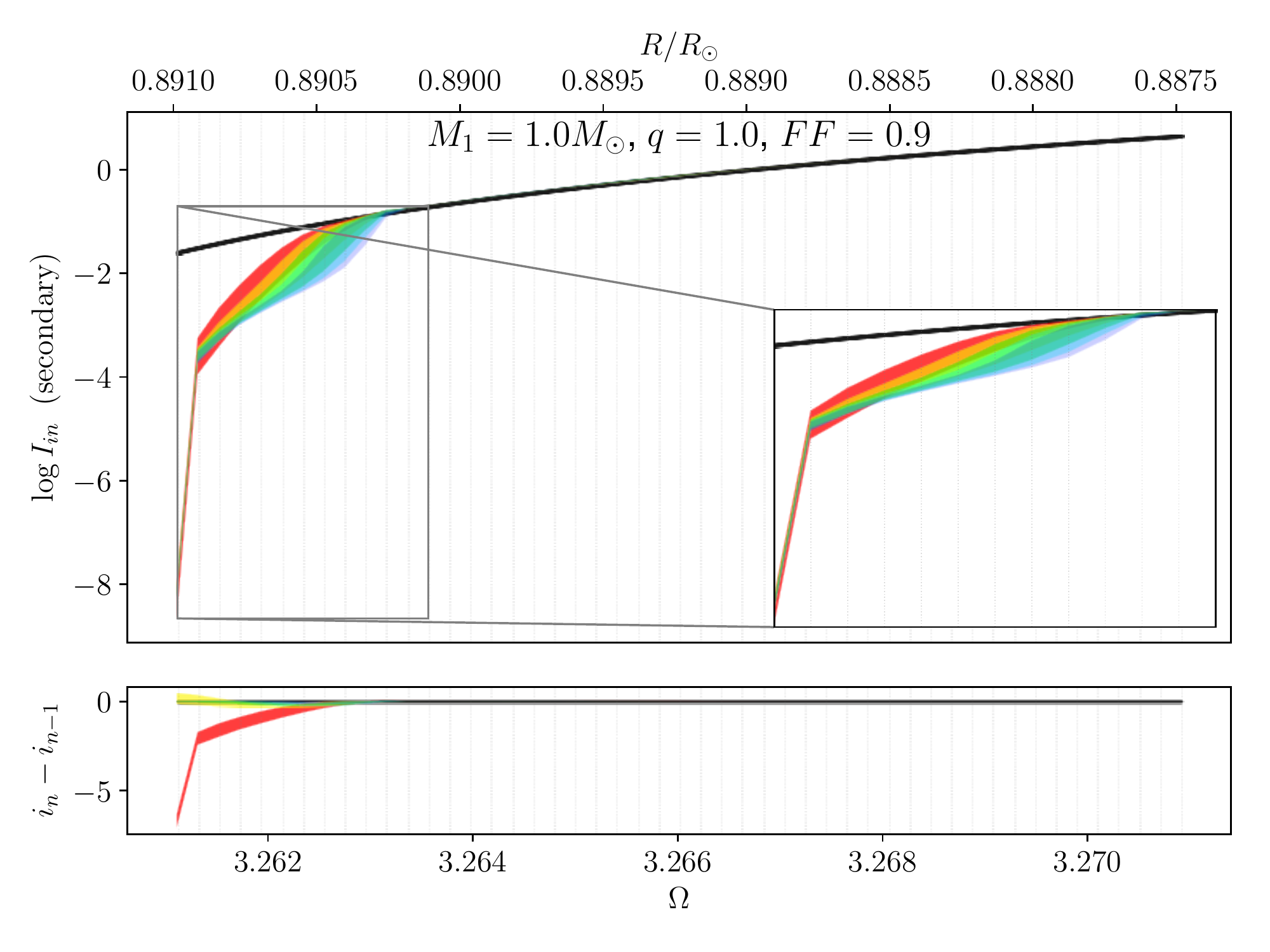}
    
    \includegraphics[width=0.495\hsize]{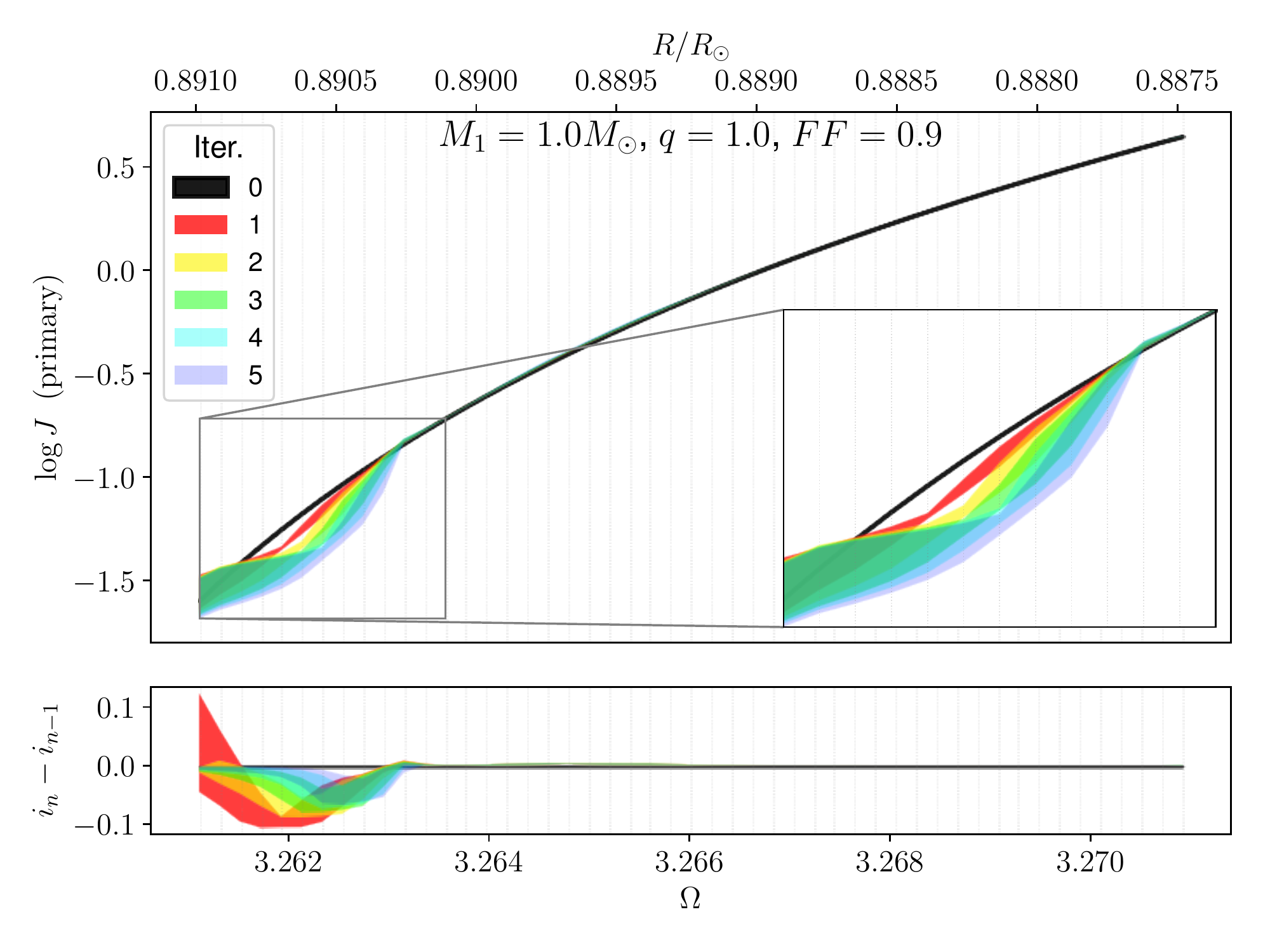}
    \includegraphics[width=0.495\hsize]{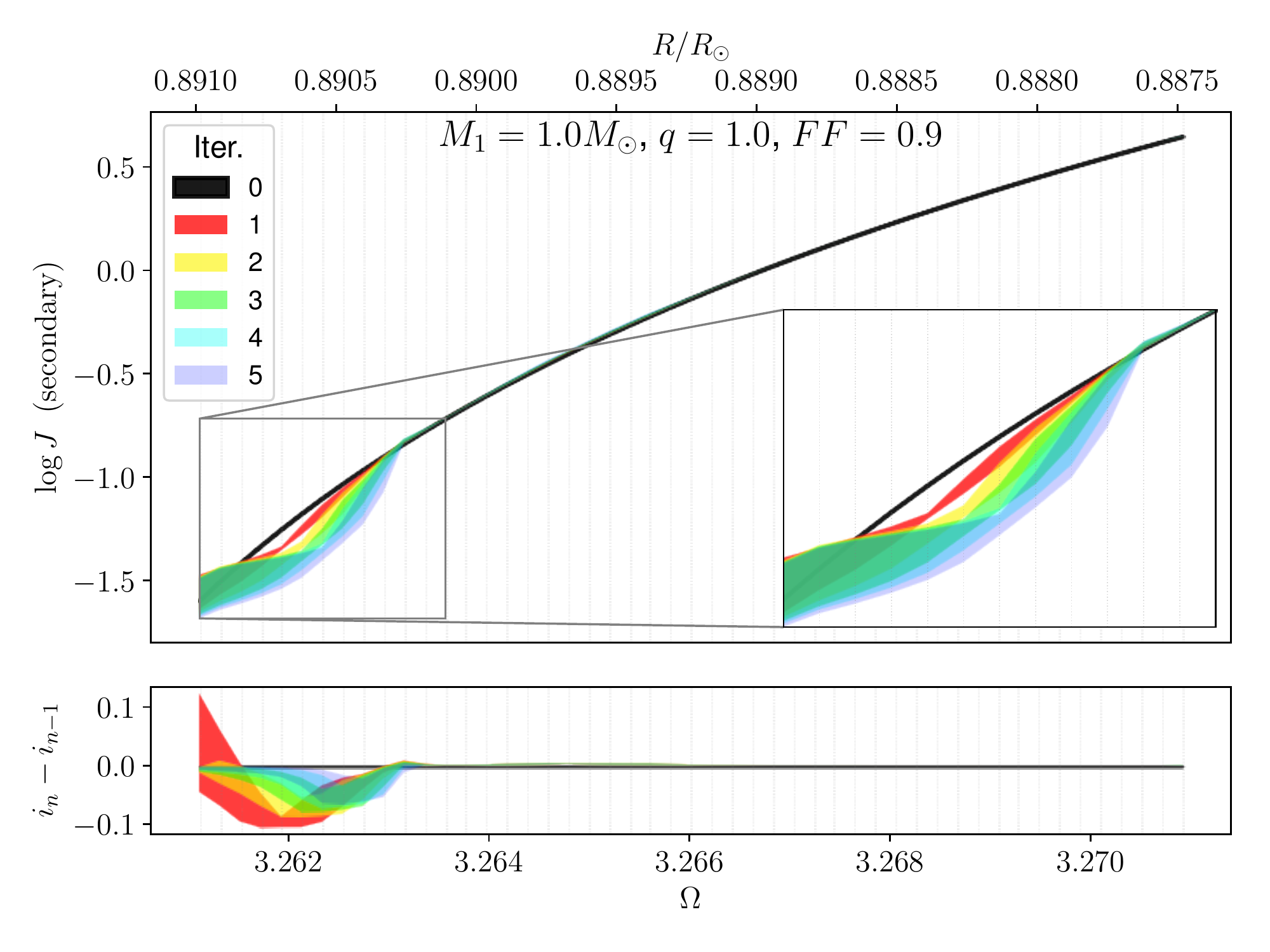}
    
    \caption{Top to bottom: outward, inward and mean intensity as a function of the potential/radius of a contact binary with $ff=0.9$. Left panels: primary, right panels: secondary component. The bottom panel of each plot shows the differences between successive iterations.}
    \label{fig:ff09}
\end{figure}

\begin{figure}[h]
    \centering
    \includegraphics[width=0.495\hsize]{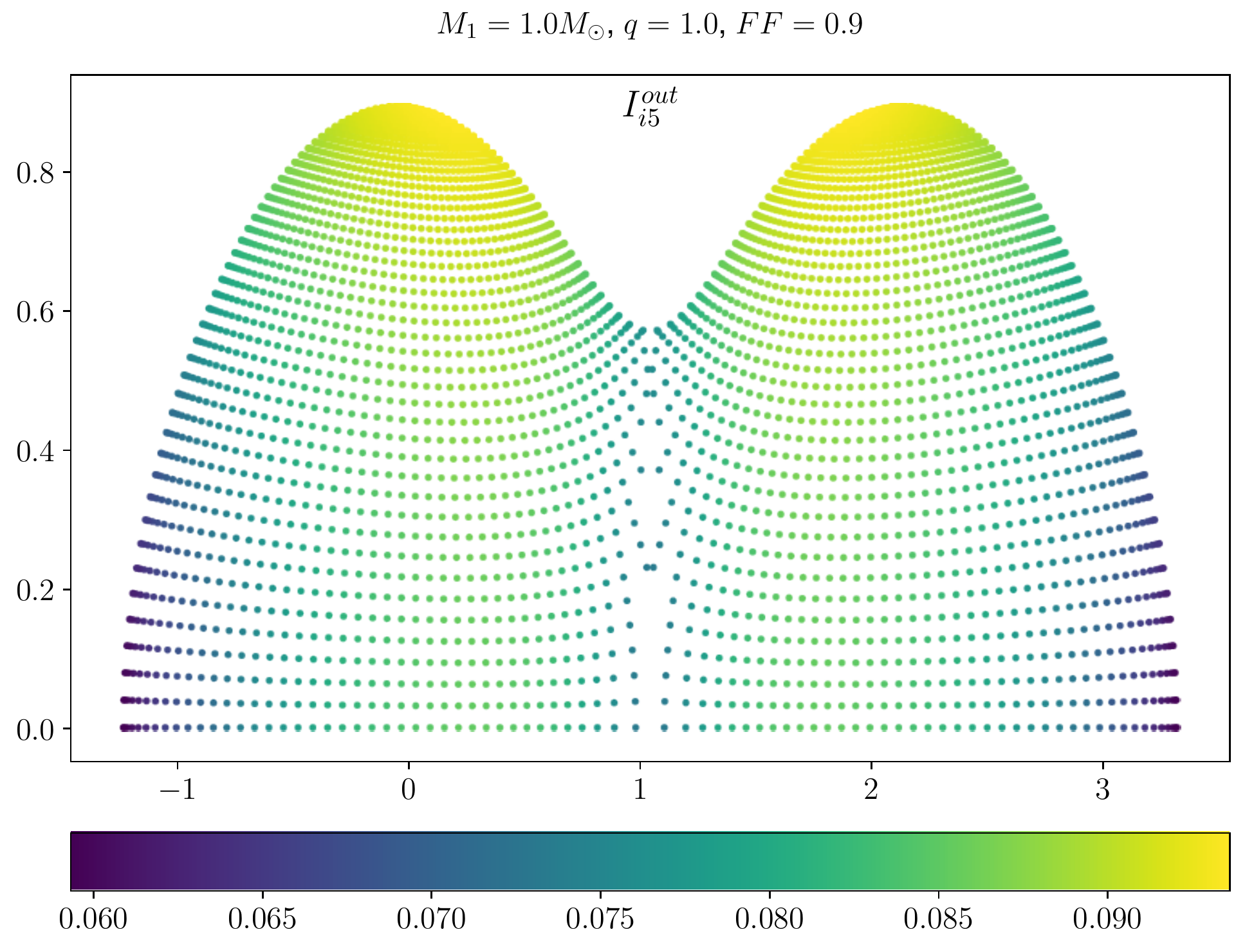}
    \includegraphics[width=0.495\hsize]{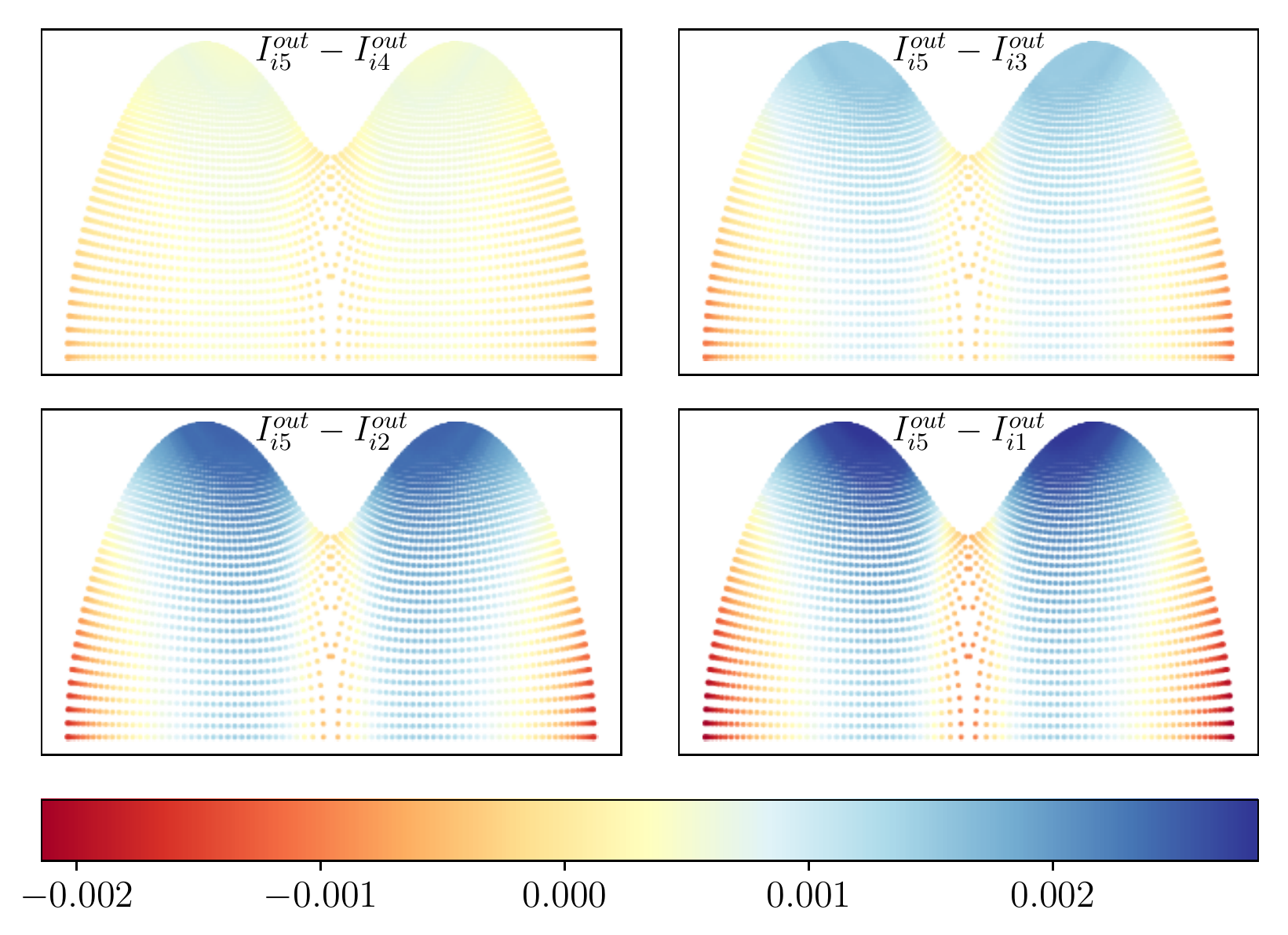}
    
        \includegraphics[width=0.495\hsize]{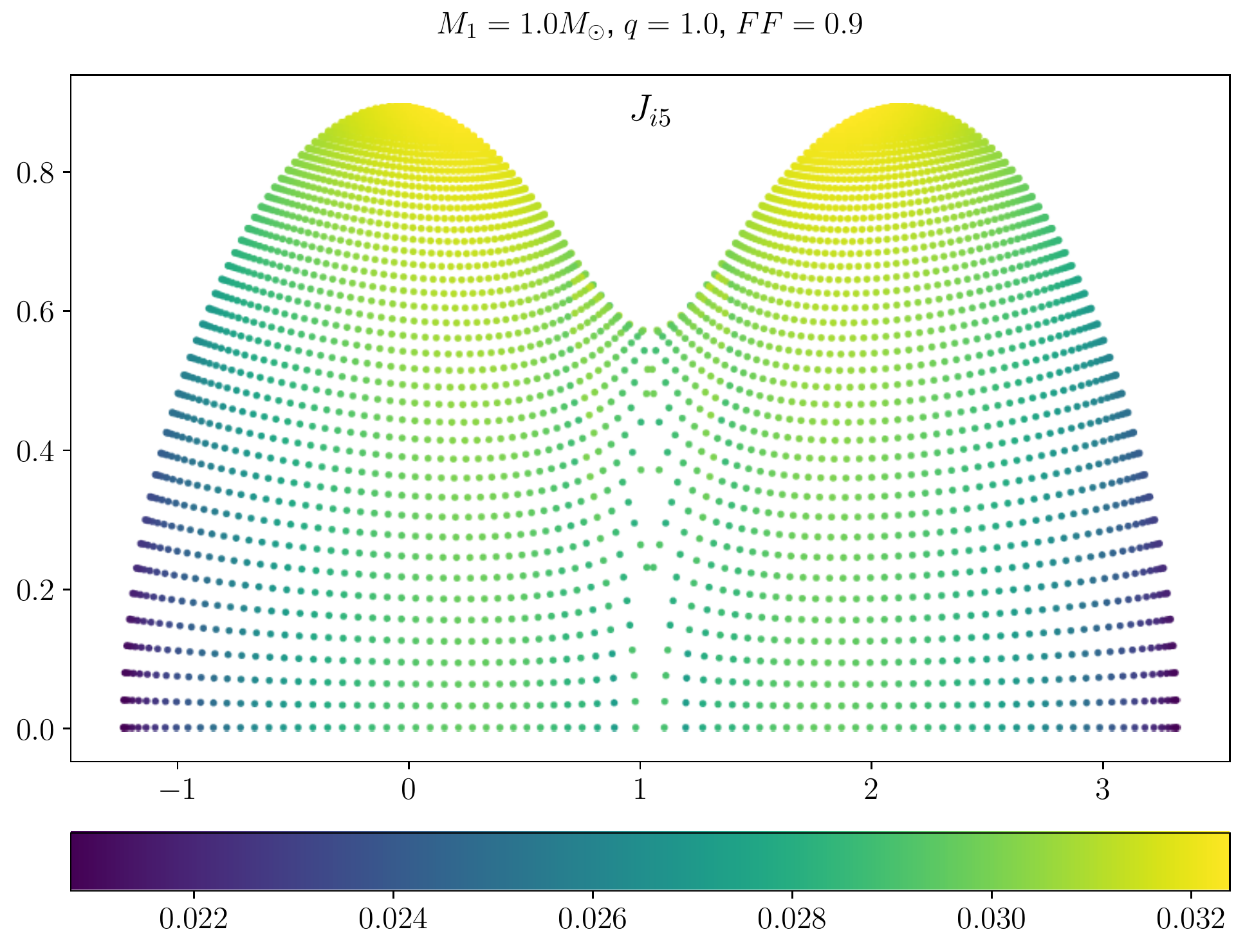}
    \includegraphics[width=0.495\hsize]{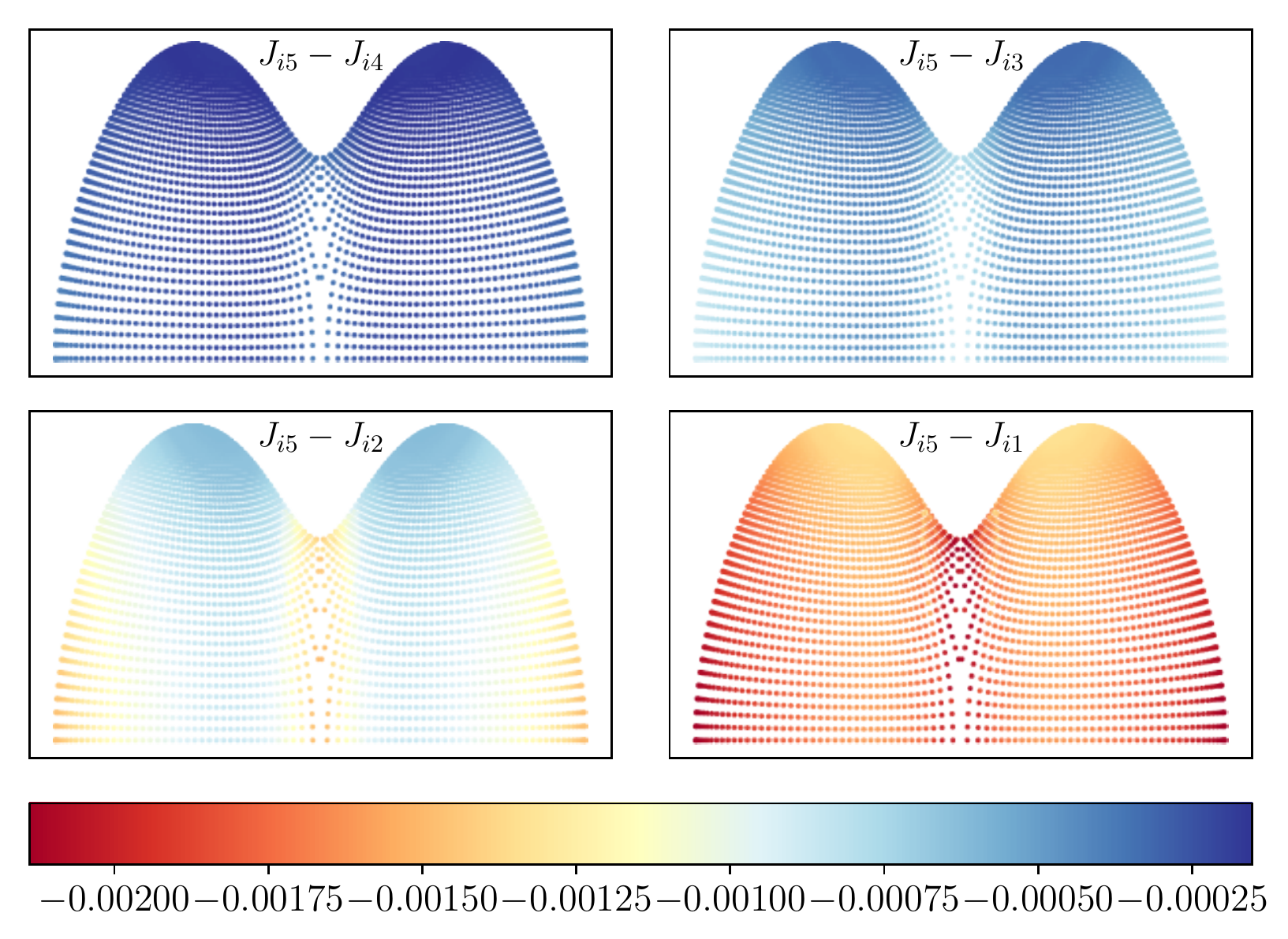}

    \caption{Surface distribution of the outward (top) and mean (bottom) intensity of a contact binary with $FF=0.9$ after the fifth iteration. Right panels show the differences in the surface distribution between the final and each previous iteration.}
    \label{fig:ff09_s}
\end{figure}

Simulation results for different fillout factors are given in Figures~\ref{fig:ff01}-\ref{fig:ff09_s}. The only notable effect of a higher fillout factor on the radiative transfer simulations is the large difference between the initial values and first iteration. However, the differences between successive iterations after the first one tend to equalize and they all converge within the threshold of 0.01\% after five iterations (Figure~\ref{fig:iters_ff}).

\subsection{Mass}

\begin{figure}[h]
    \centering
    \includegraphics[width=0.495\hsize]{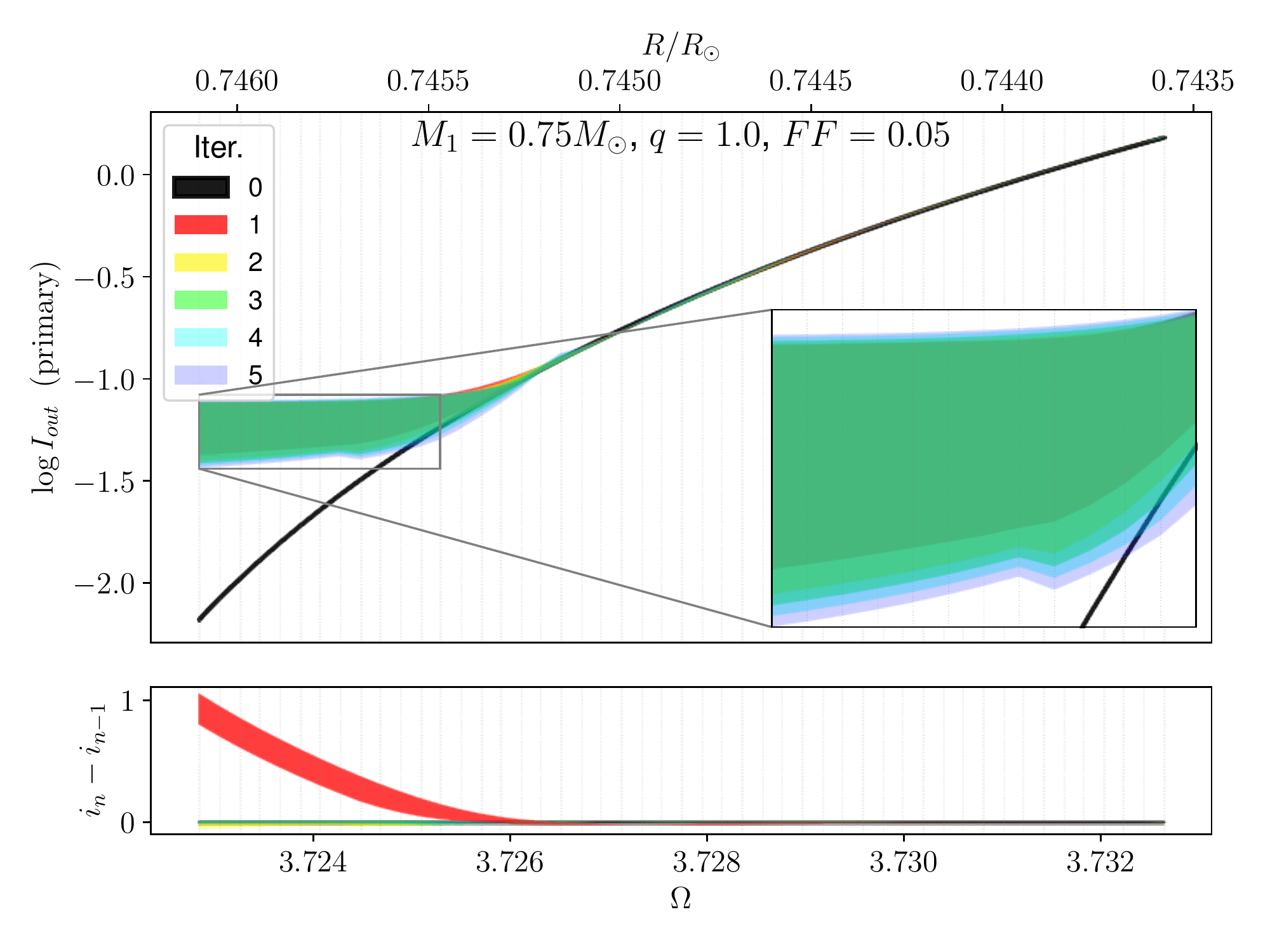}
    \includegraphics[width=0.495\hsize]{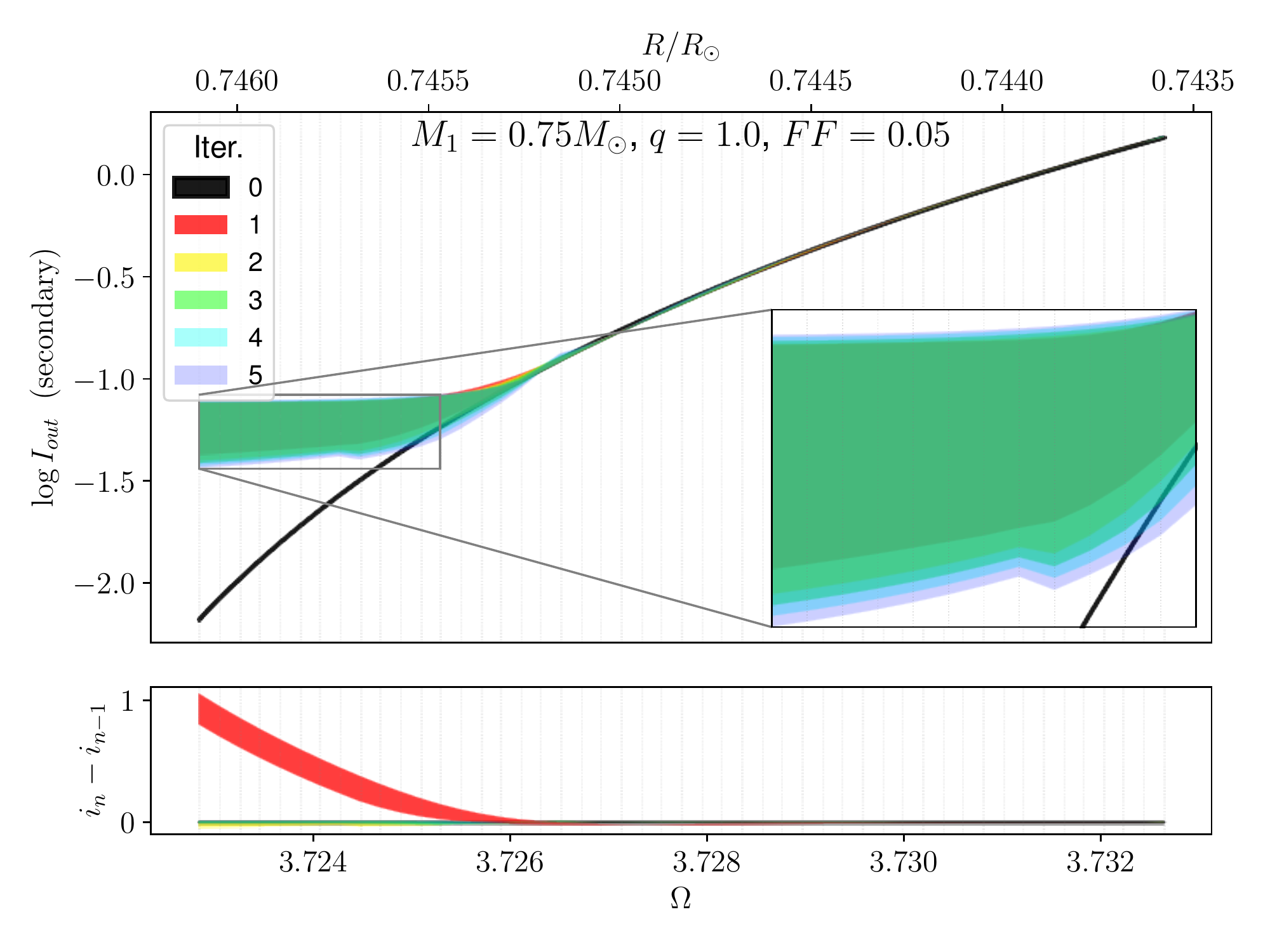}
    
    \includegraphics[width=0.495\hsize]{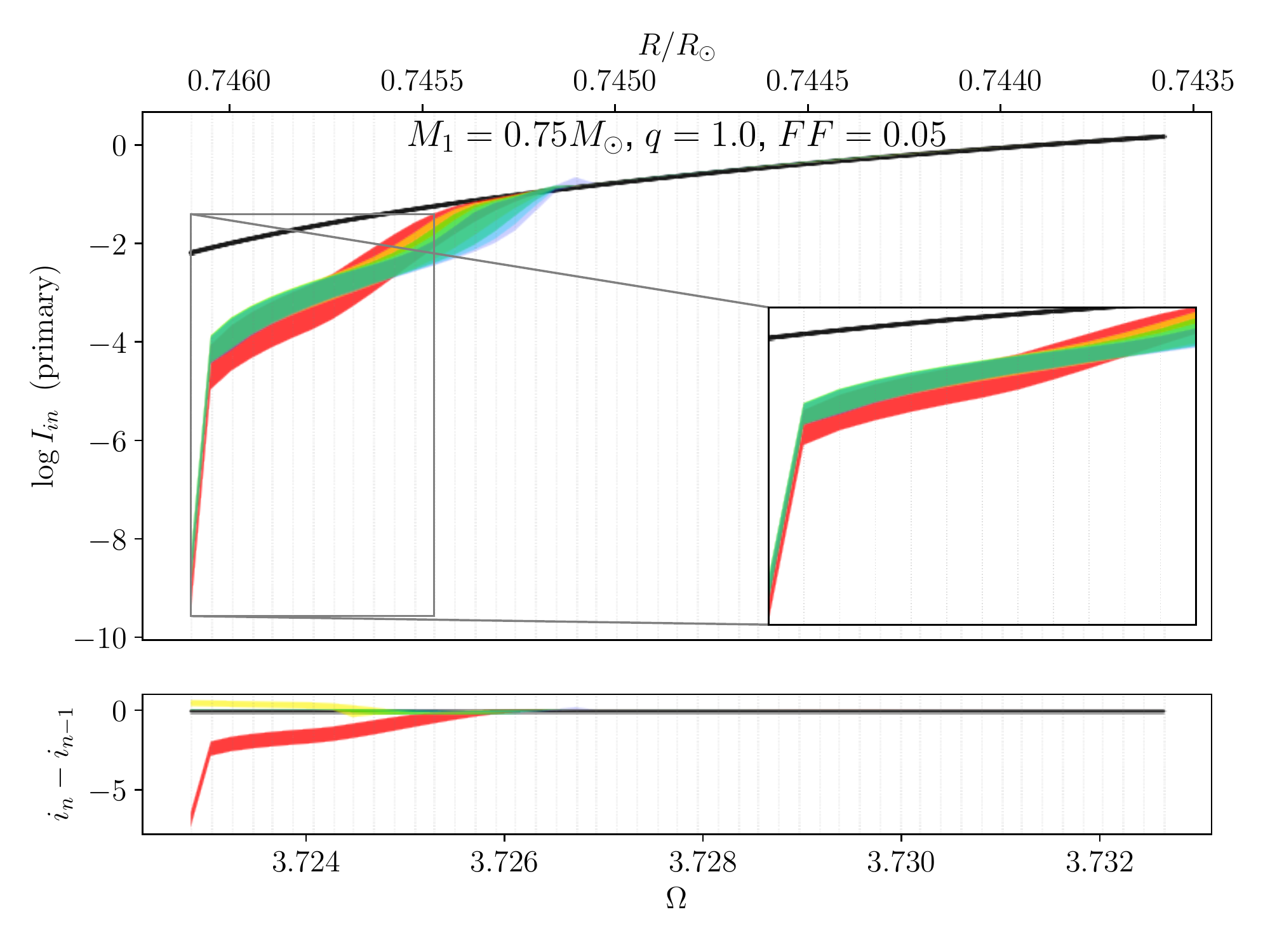}
    \includegraphics[width=0.495\hsize]{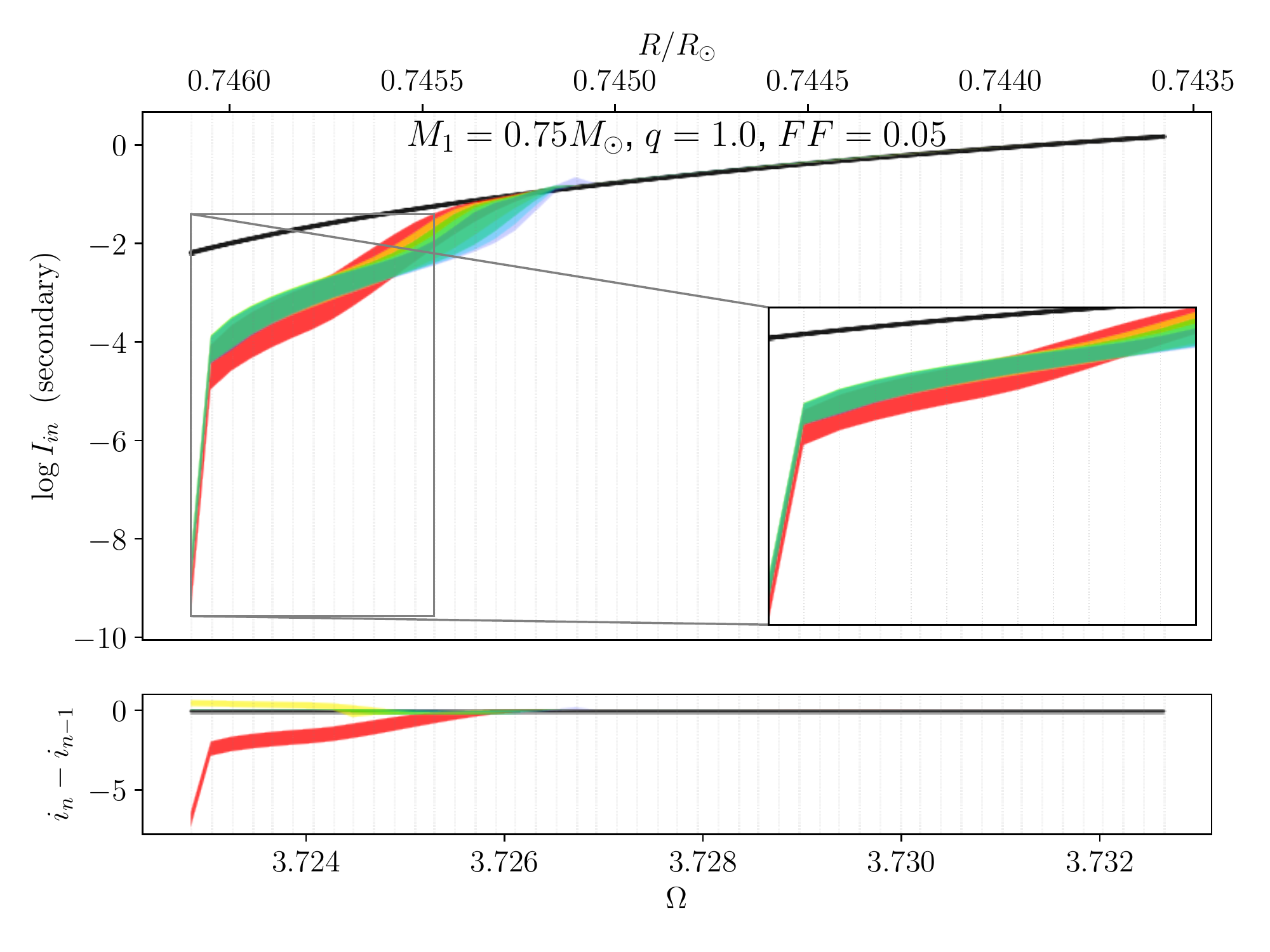}
    
    \includegraphics[width=0.495\hsize]{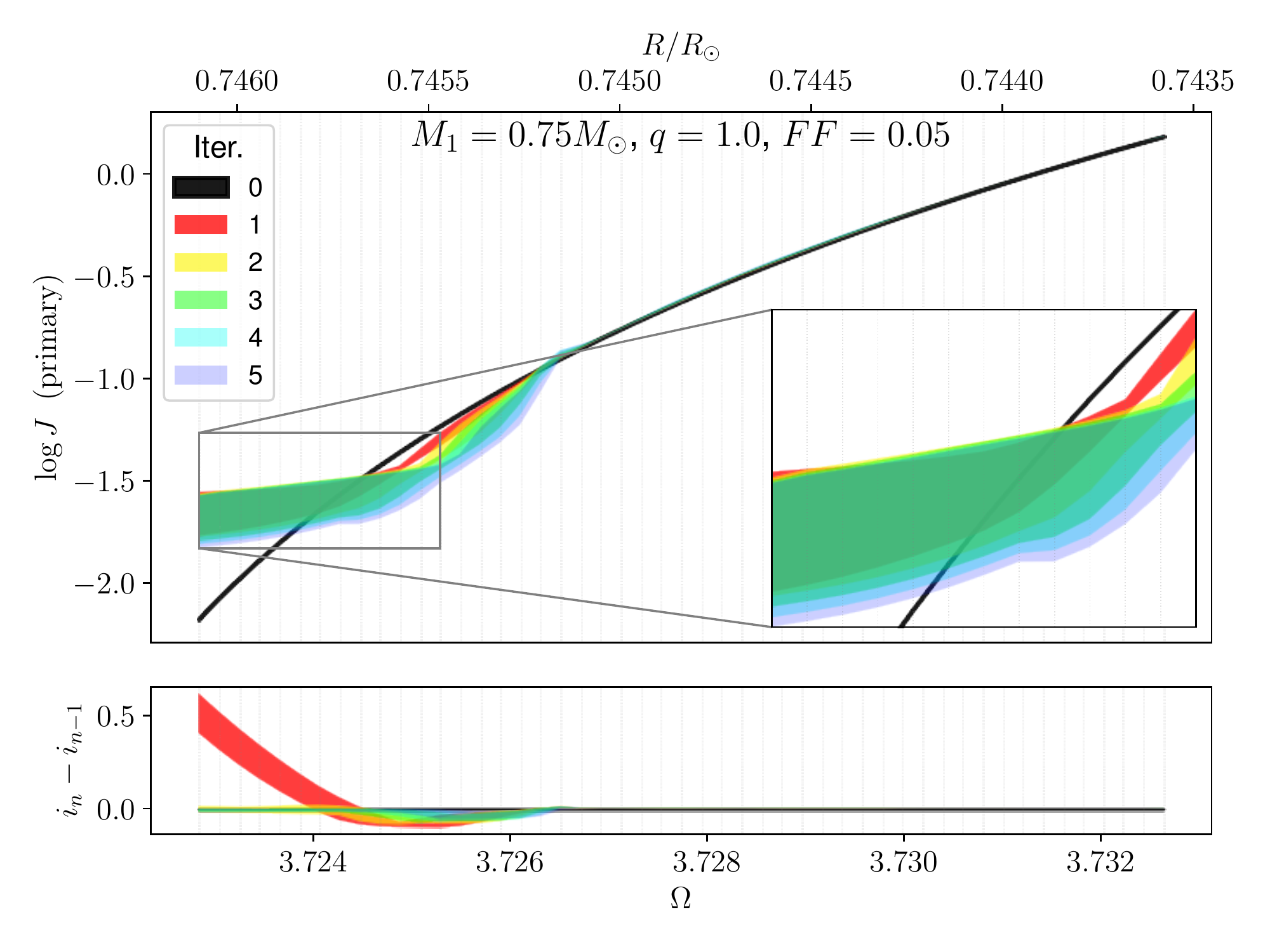}
    \includegraphics[width=0.495\hsize]{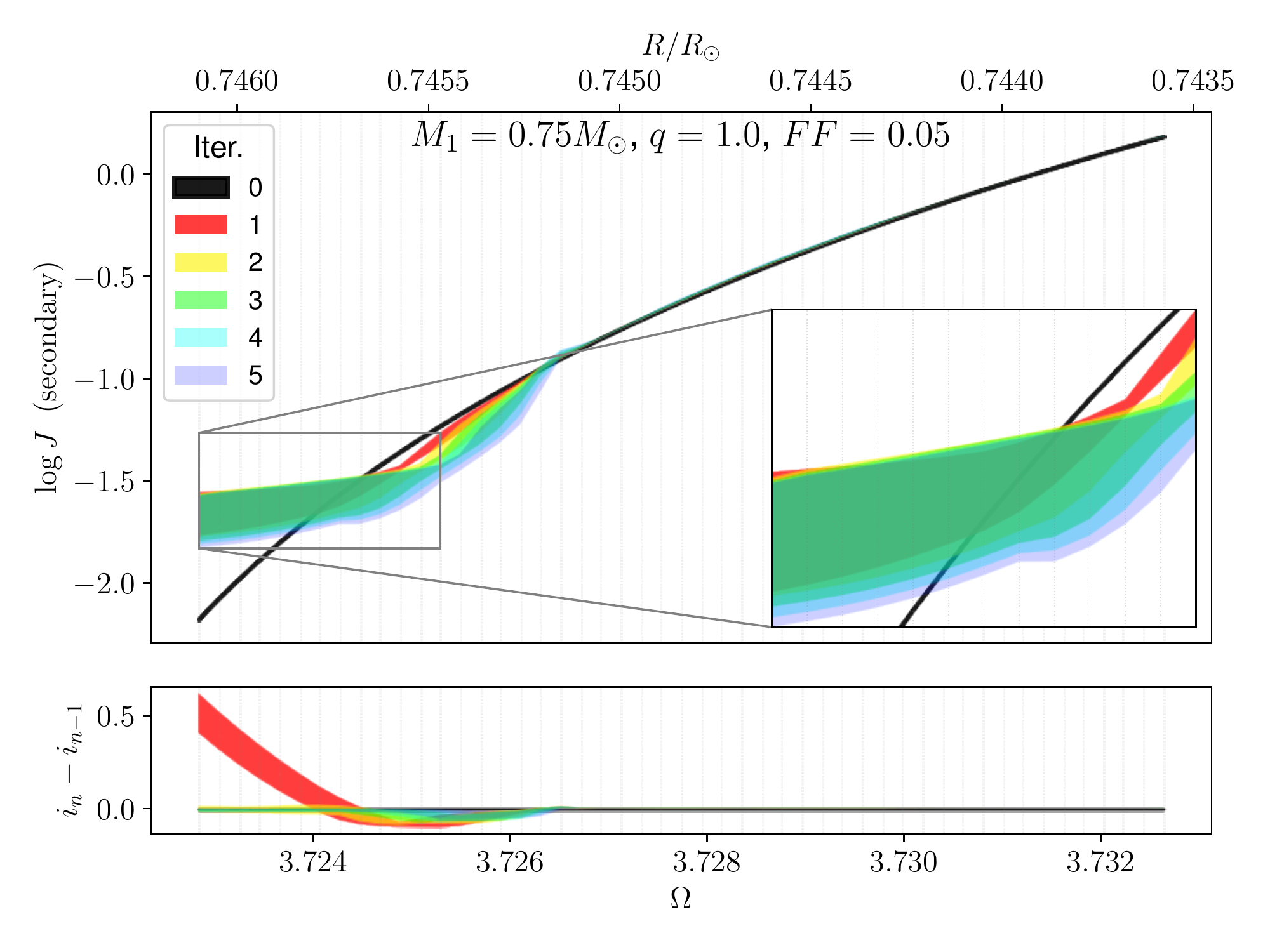}
    
    \caption{Top to bottom: outward, inward and mean intensity as a function of the potential/radius of a contact binary with $M_1=0.75 M_{\odot}$. Left panels: primary, right panels: secondary component. The bottom panel of each plot shows the differences between successive iterations.}
    \label{fig:m075}
\end{figure}

\begin{figure}[h]
    \centering
    \includegraphics[width=0.495\hsize]{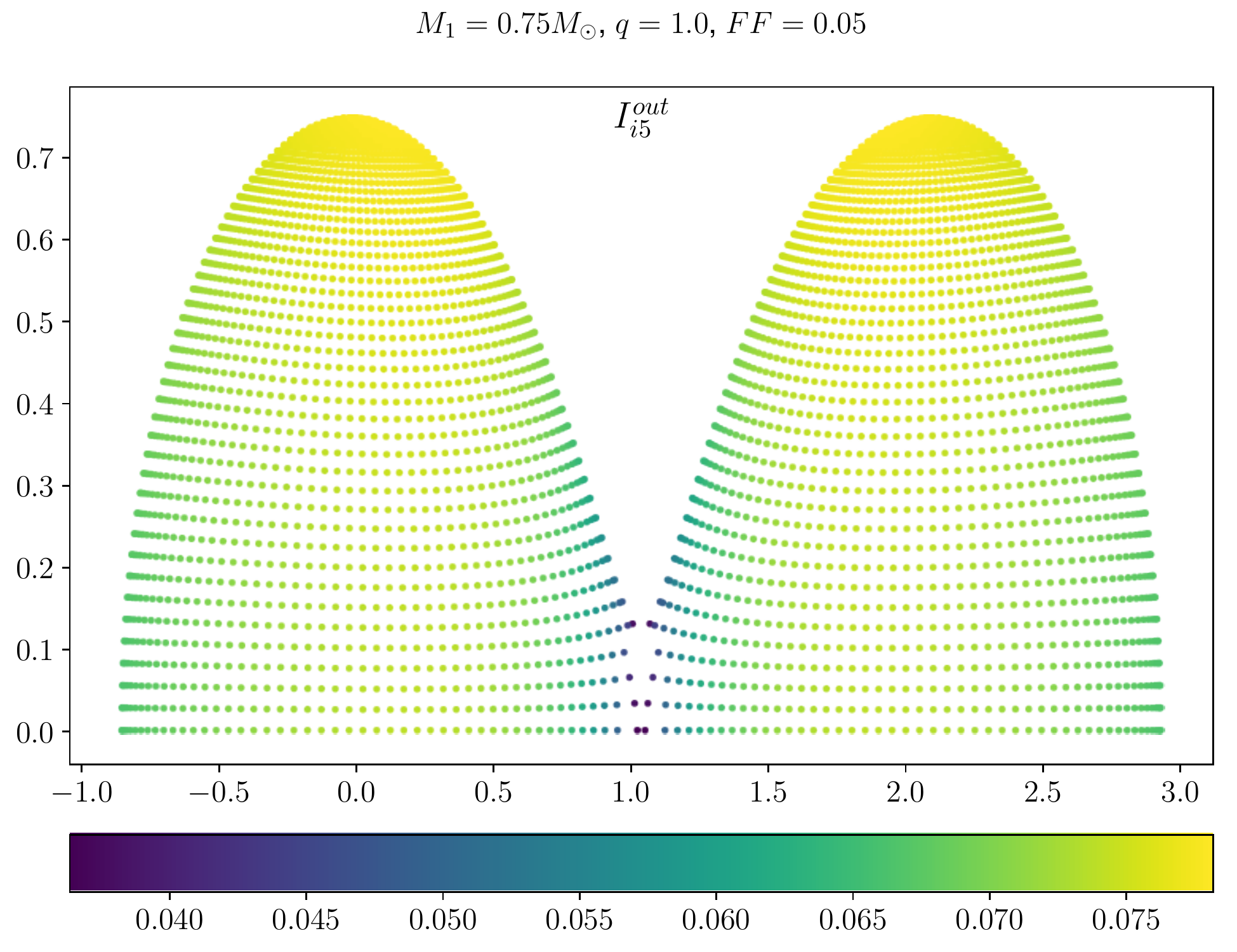}
    \includegraphics[width=0.495\hsize]{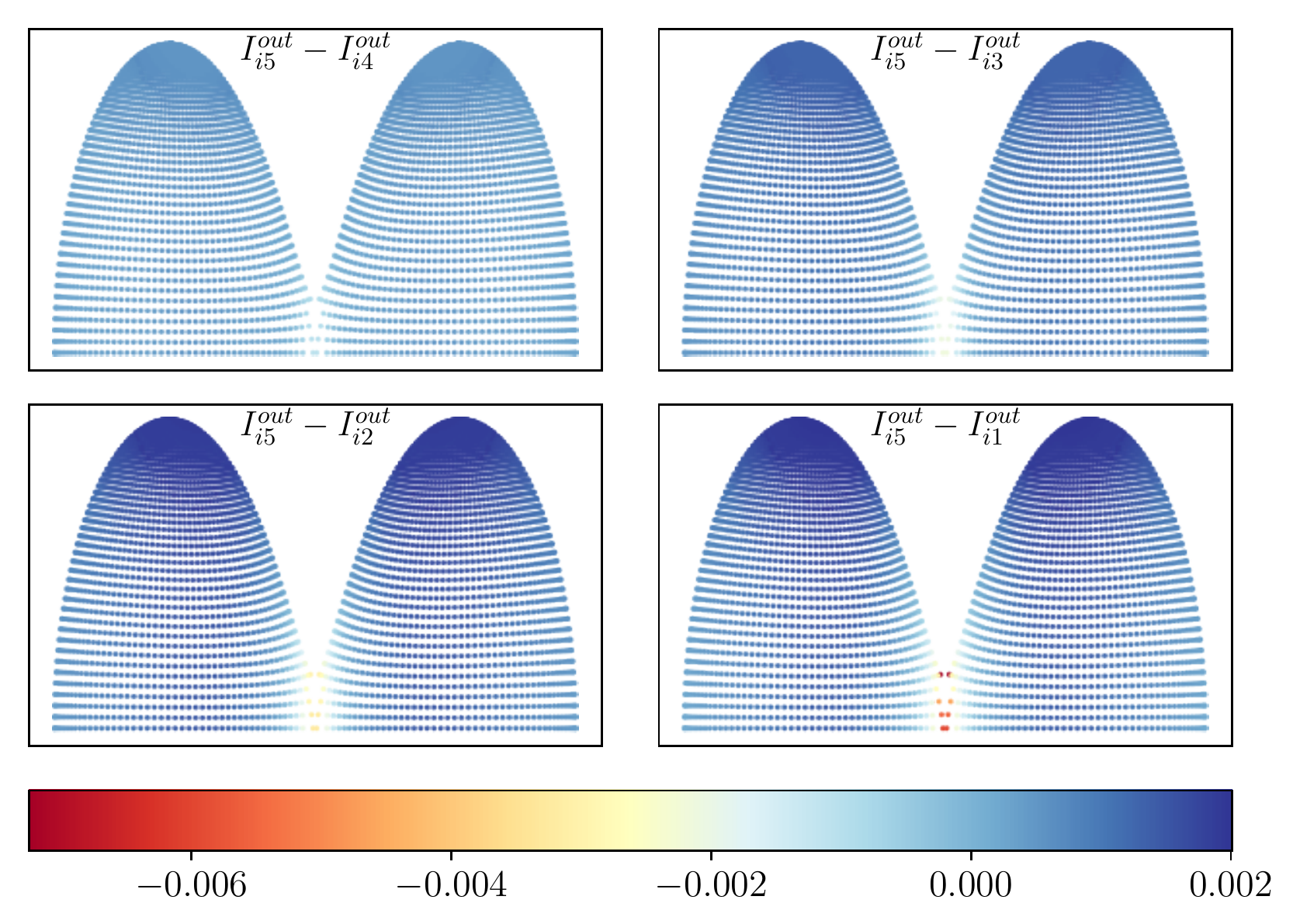}
    
        \includegraphics[width=0.495\hsize]{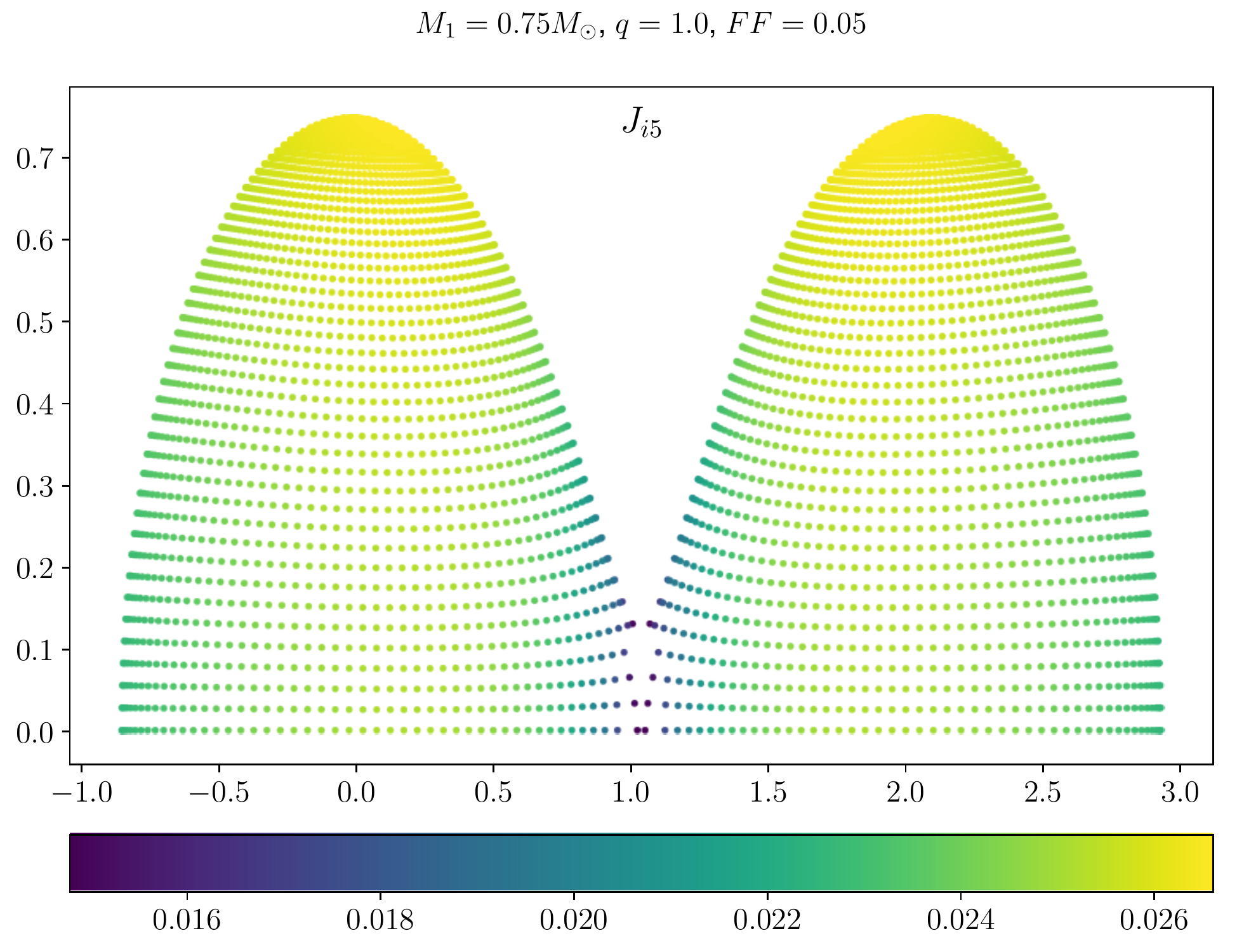}
    \includegraphics[width=0.495\hsize]{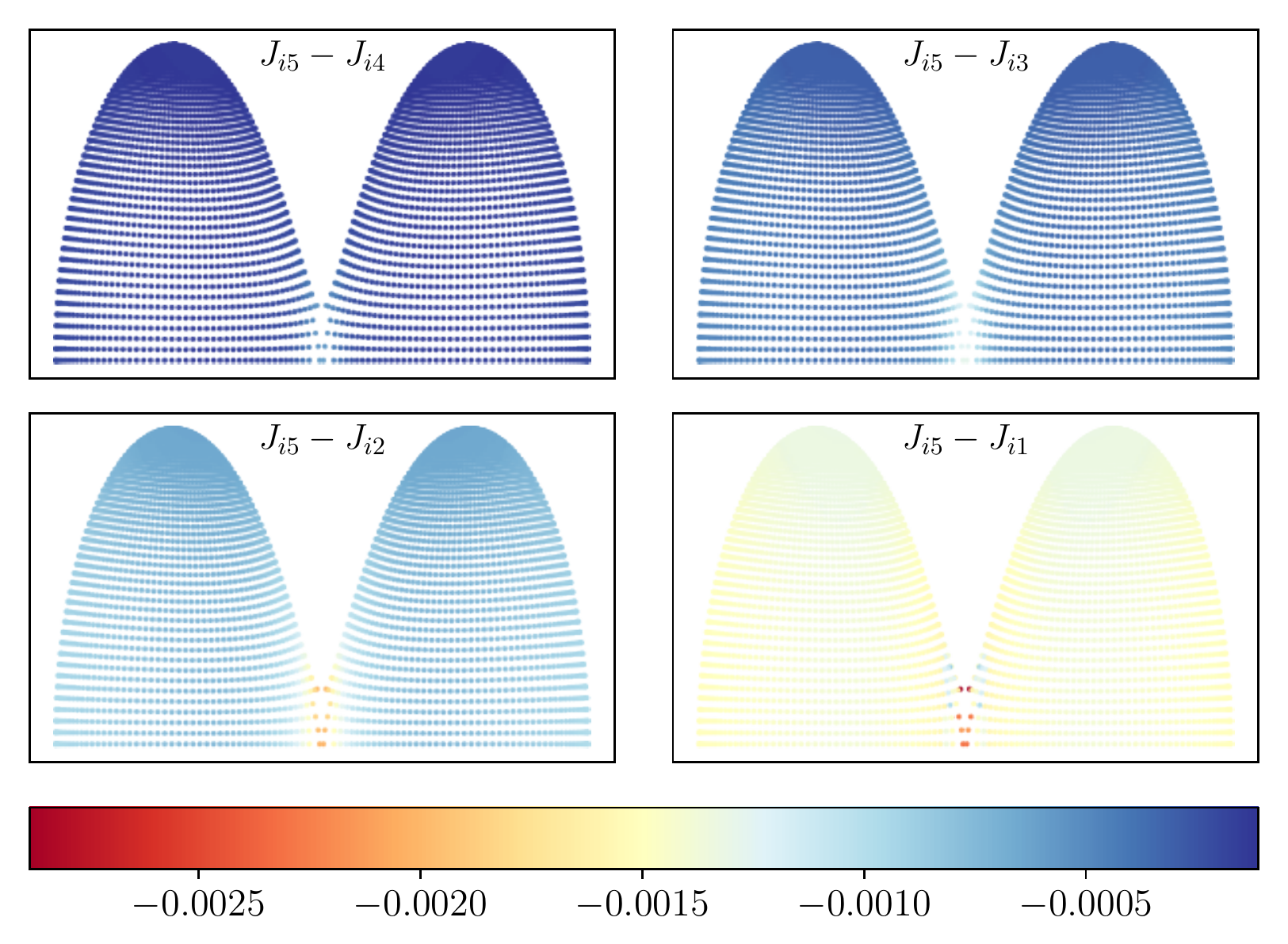}

    \caption{Surface distribution of the outward (top) and mean (bottom) intensity of a contact binary with $M_1=0.75 M_{\odot}$ after the fifth iteration. Right panels show the differences in the surface distribution between the final and each previous iteration.}
    \label{fig:m075_s}
\end{figure}

\begin{figure}[h]
    \centering
    \includegraphics[width=0.495\hsize]{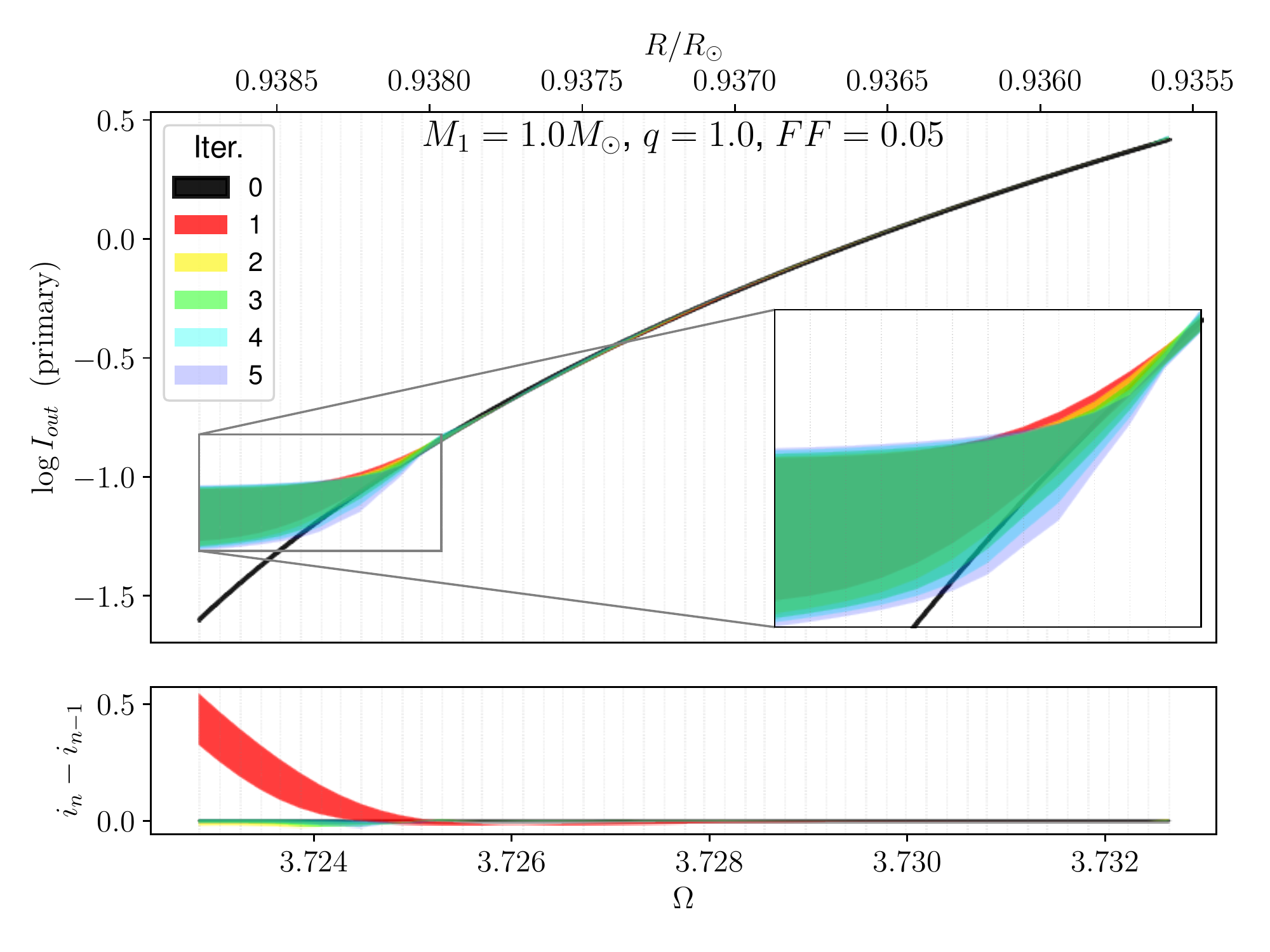}
    \includegraphics[width=0.495\hsize]{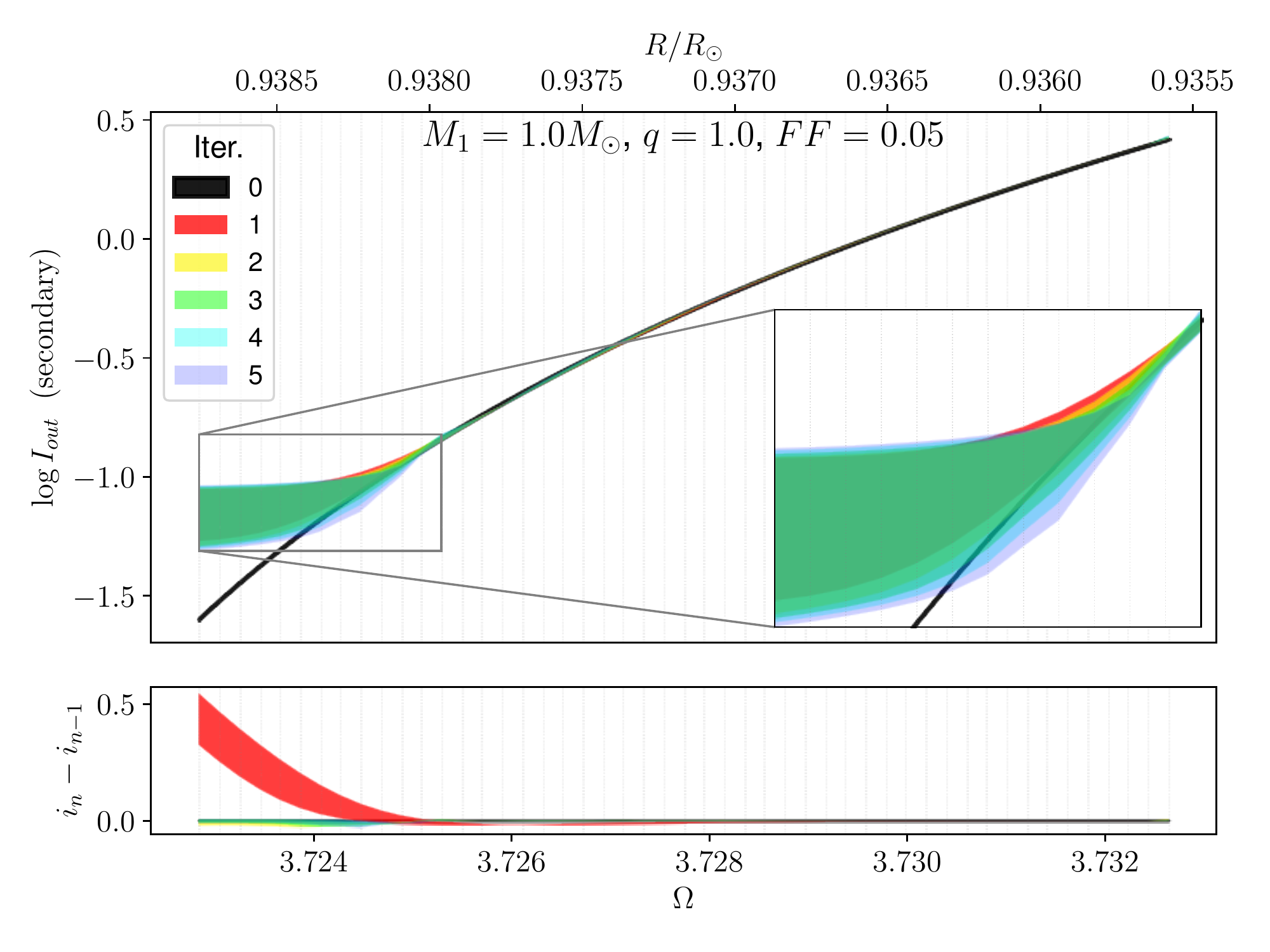}
    
    \includegraphics[width=0.495\hsize]{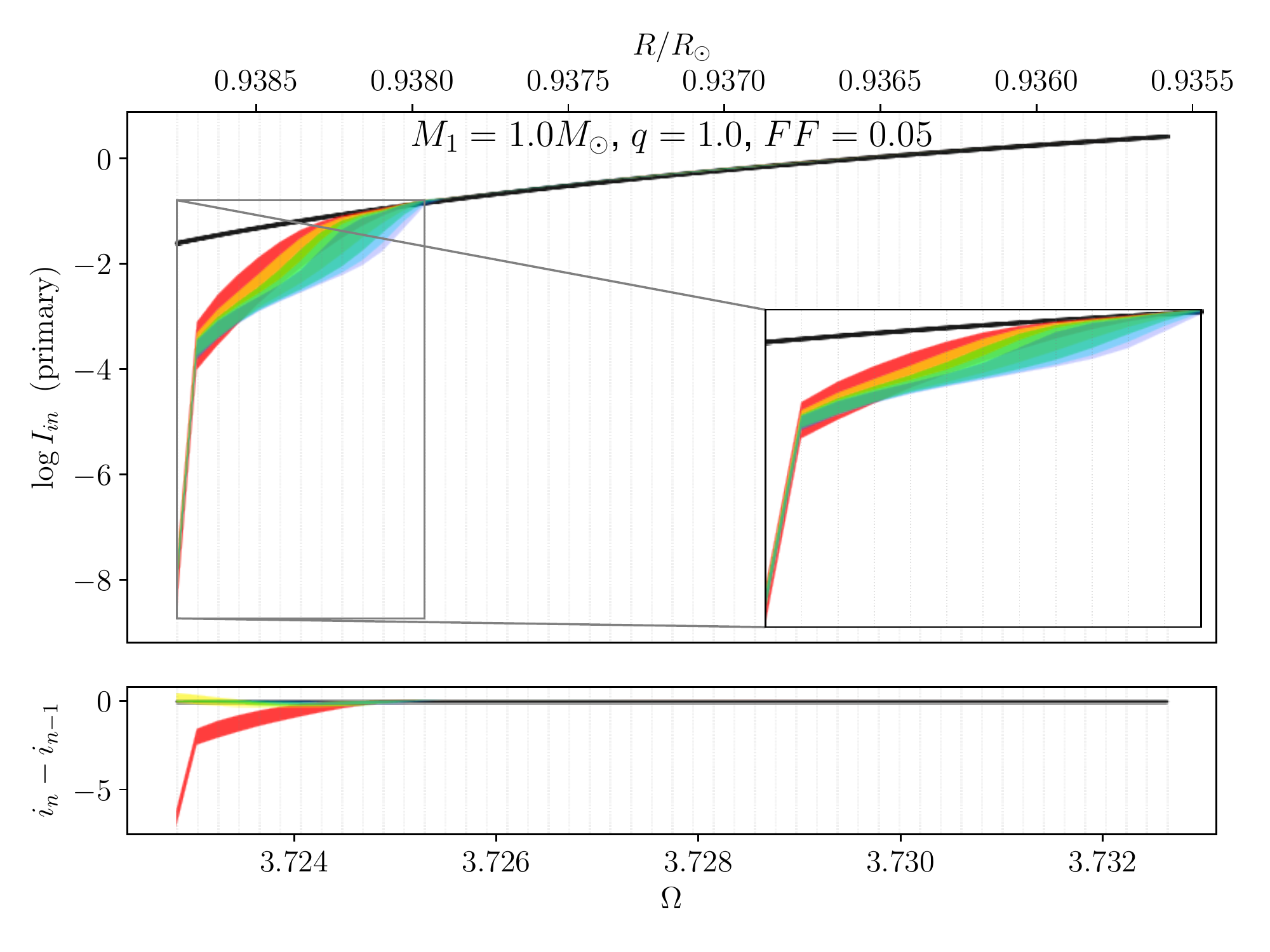}
    \includegraphics[width=0.495\hsize]{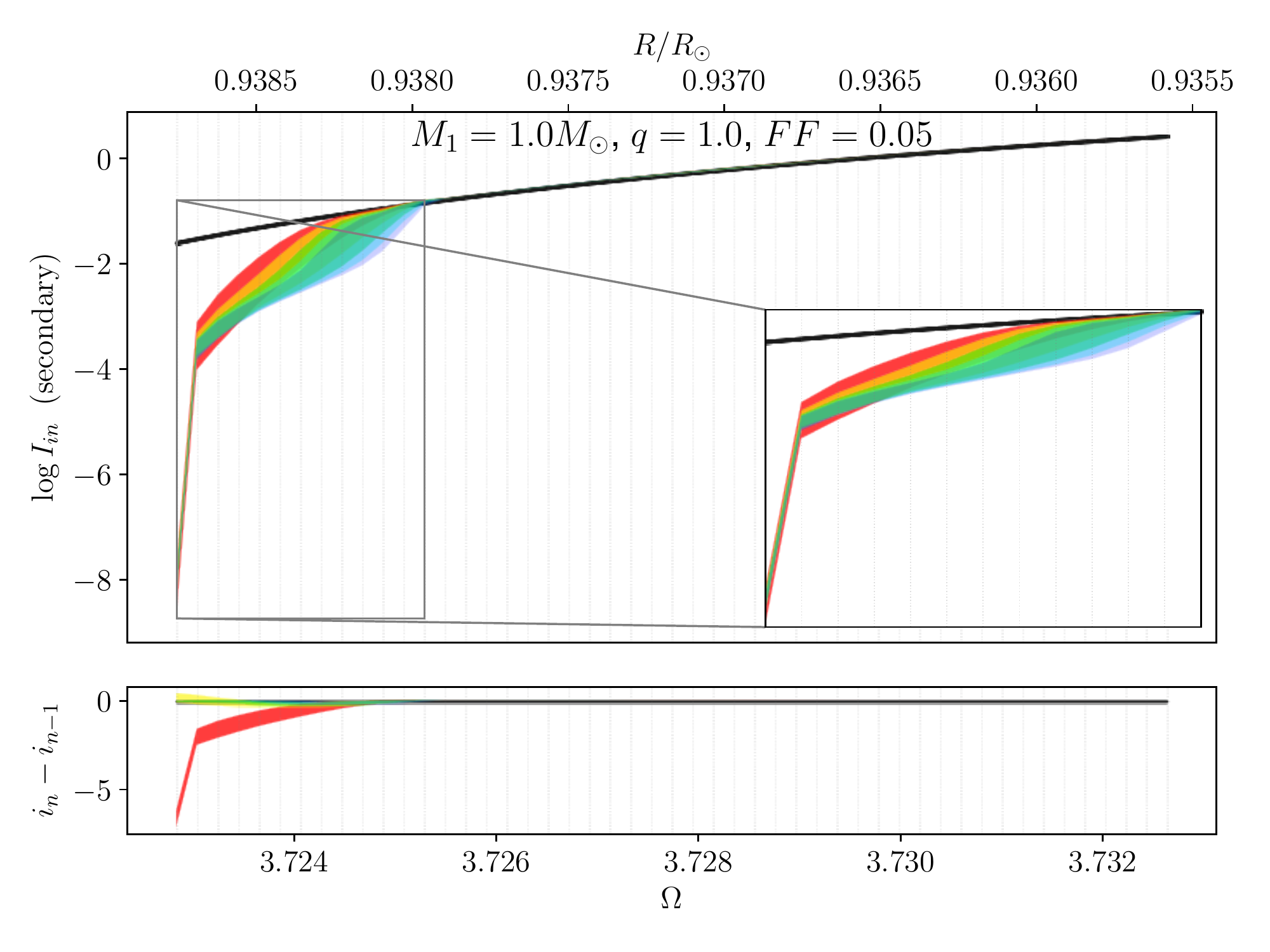}
    
    \includegraphics[width=0.495\hsize]{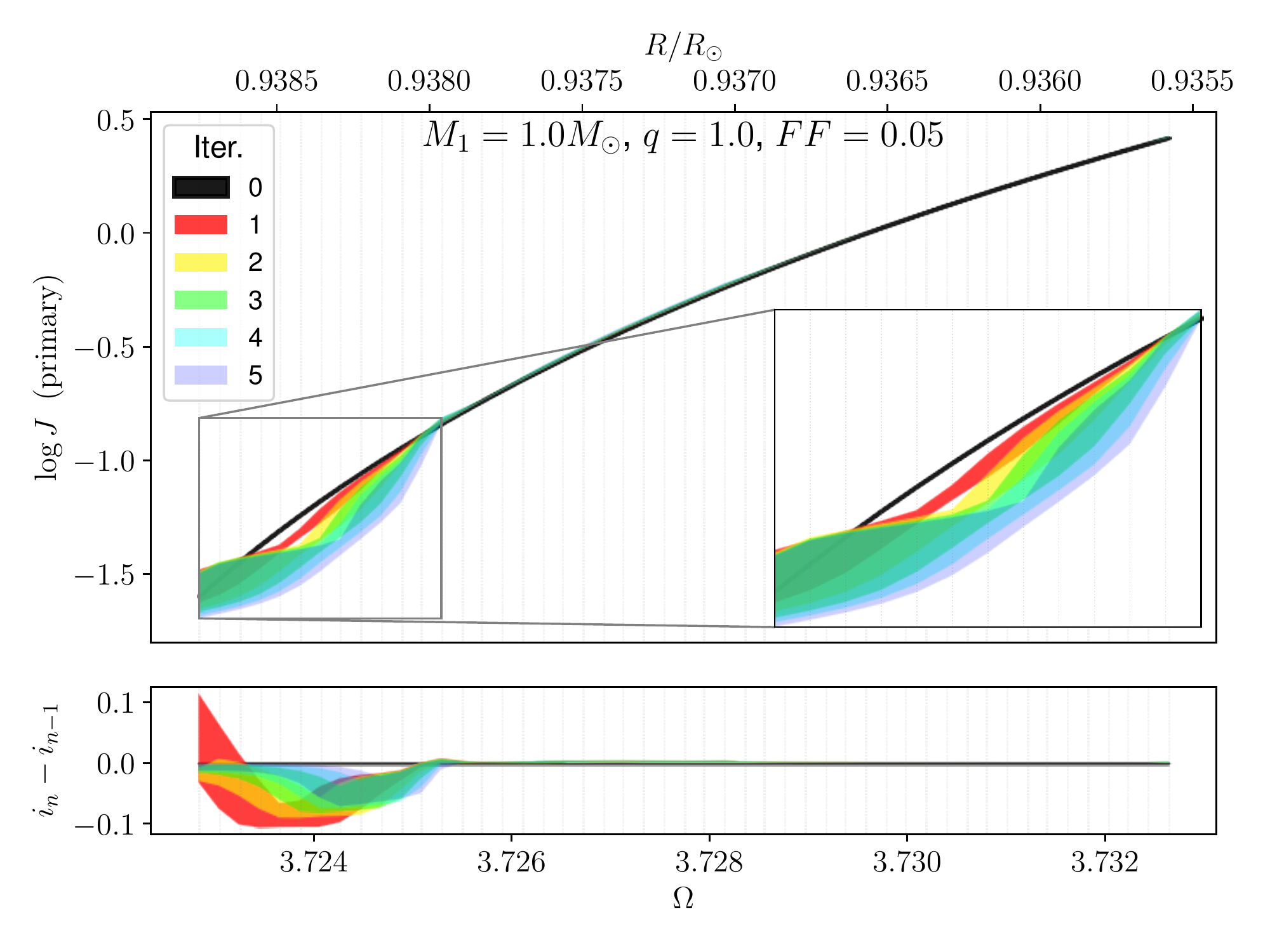}
    \includegraphics[width=0.495\hsize]{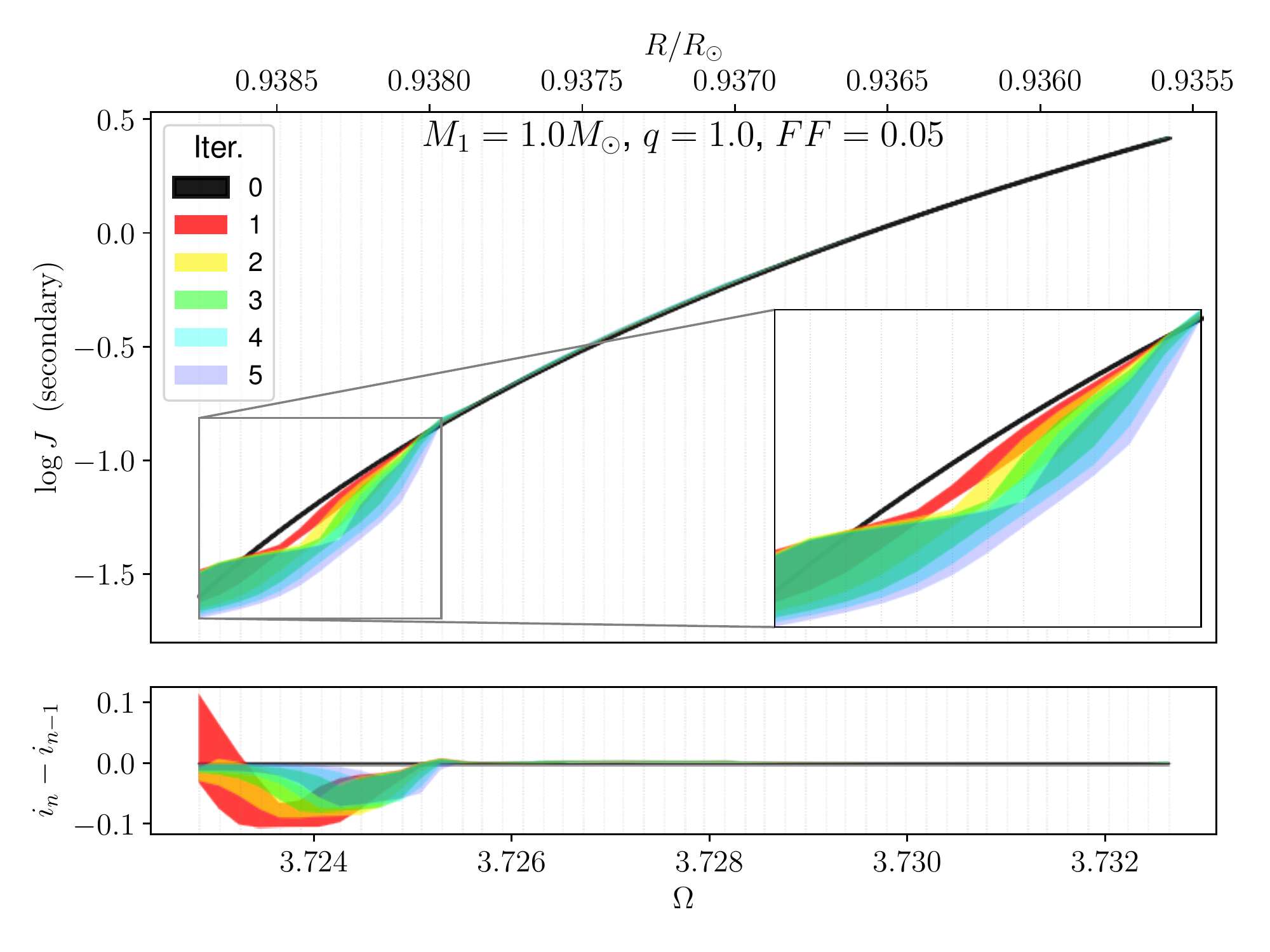}
    
    \caption{Top to bottom: outward, inward and mean intensity as a function of the potential/radius of a contact binary with $M_1=1 M_{\odot}$. Left panels: primary, right panels: secondary component. The bottom panel of each plot shows the differences between successive iterations.}
    \label{fig:m1}
\end{figure}

\begin{figure}[h]
    \centering
    \includegraphics[width=0.495\hsize]{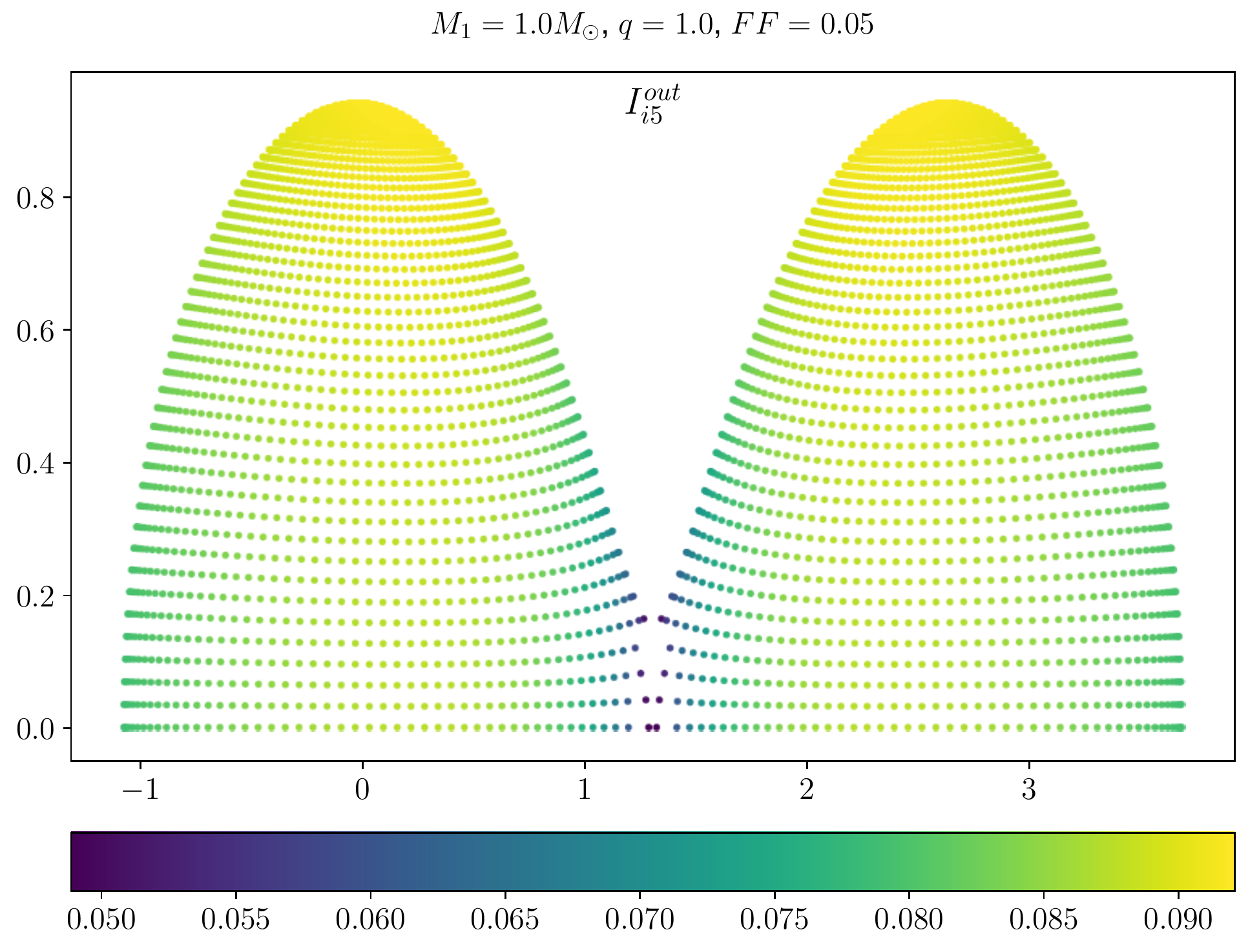}
    \includegraphics[width=0.495\hsize]{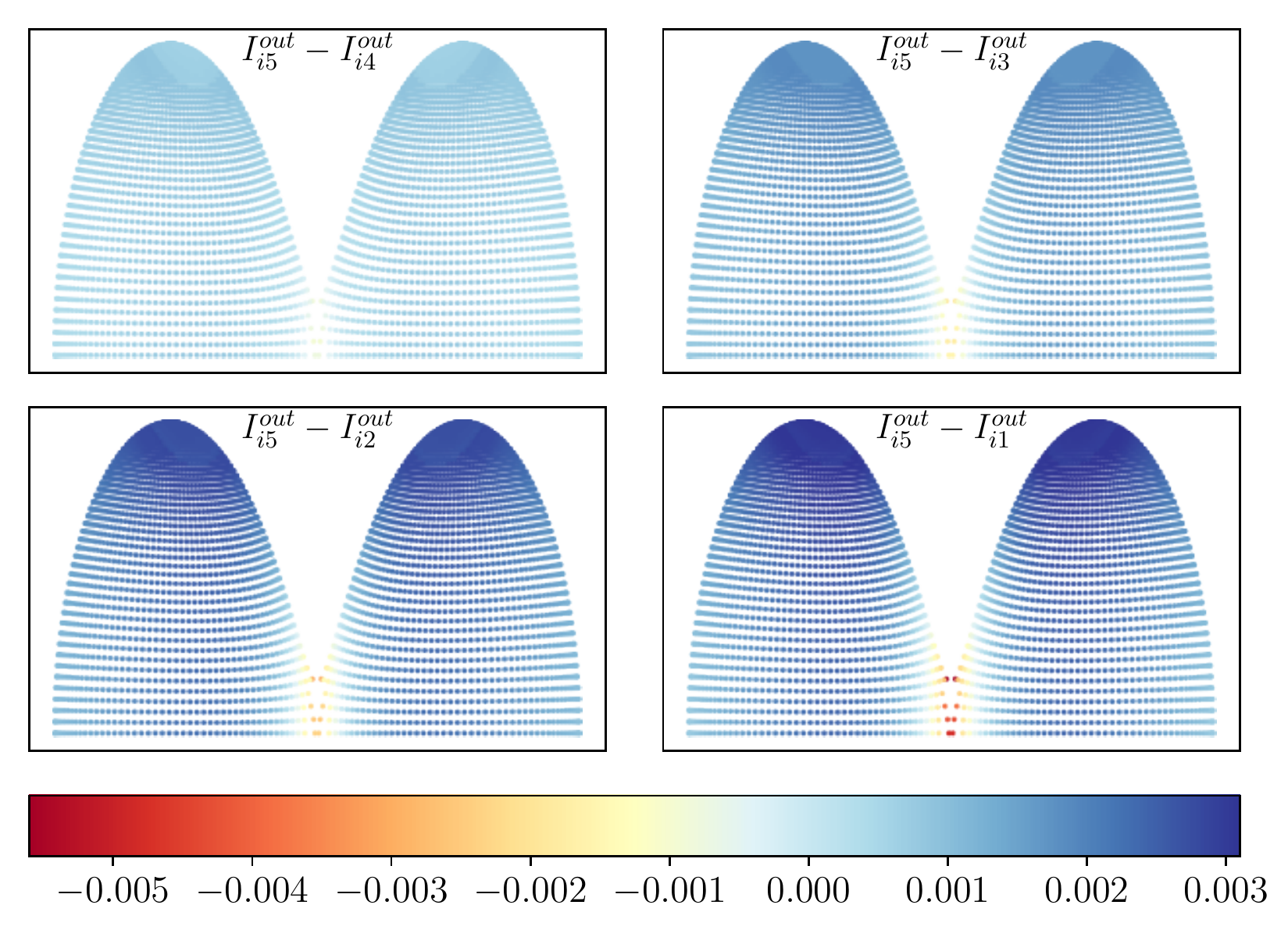}

   \includegraphics[width=0.495\hsize]{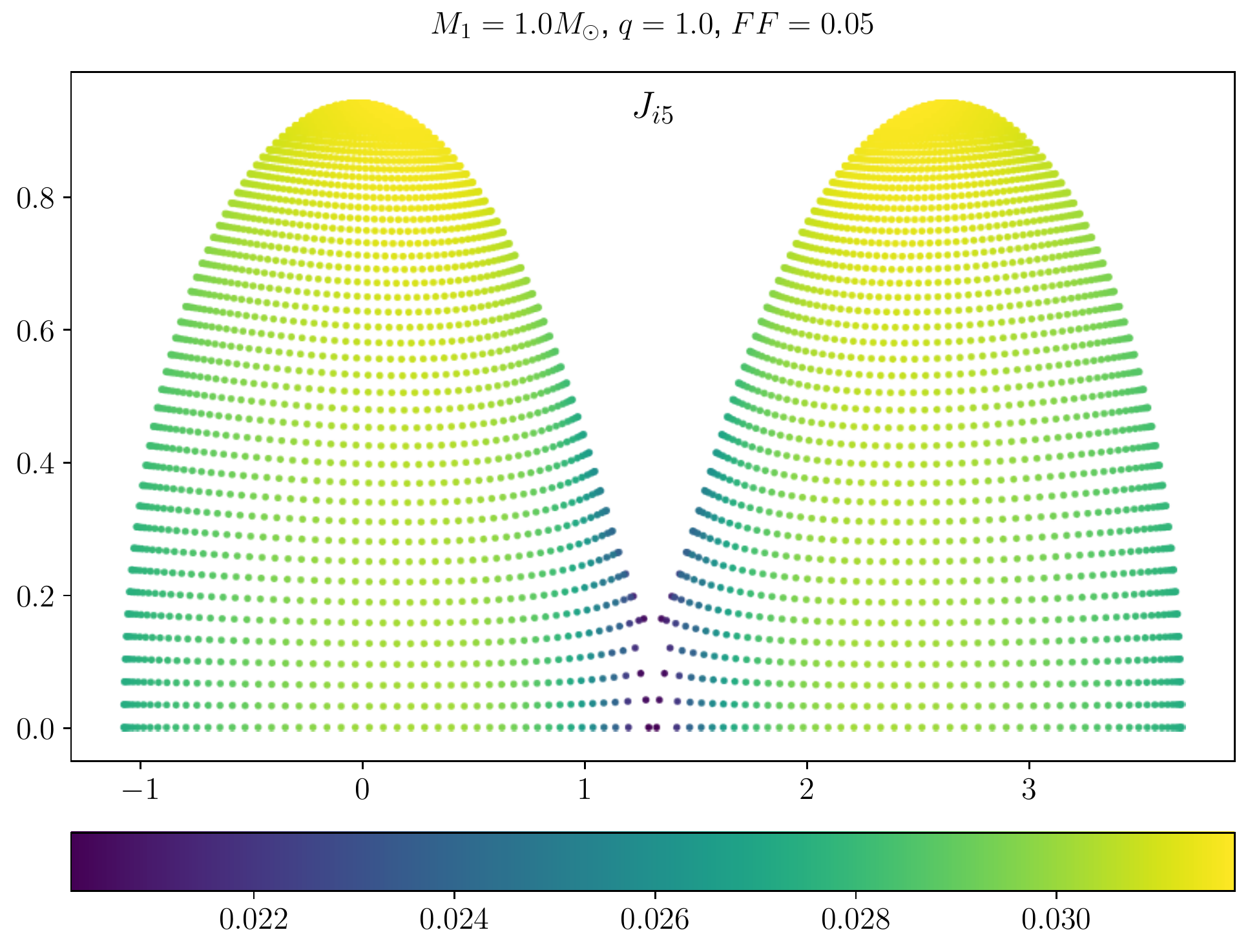}
    \includegraphics[width=0.495\hsize]{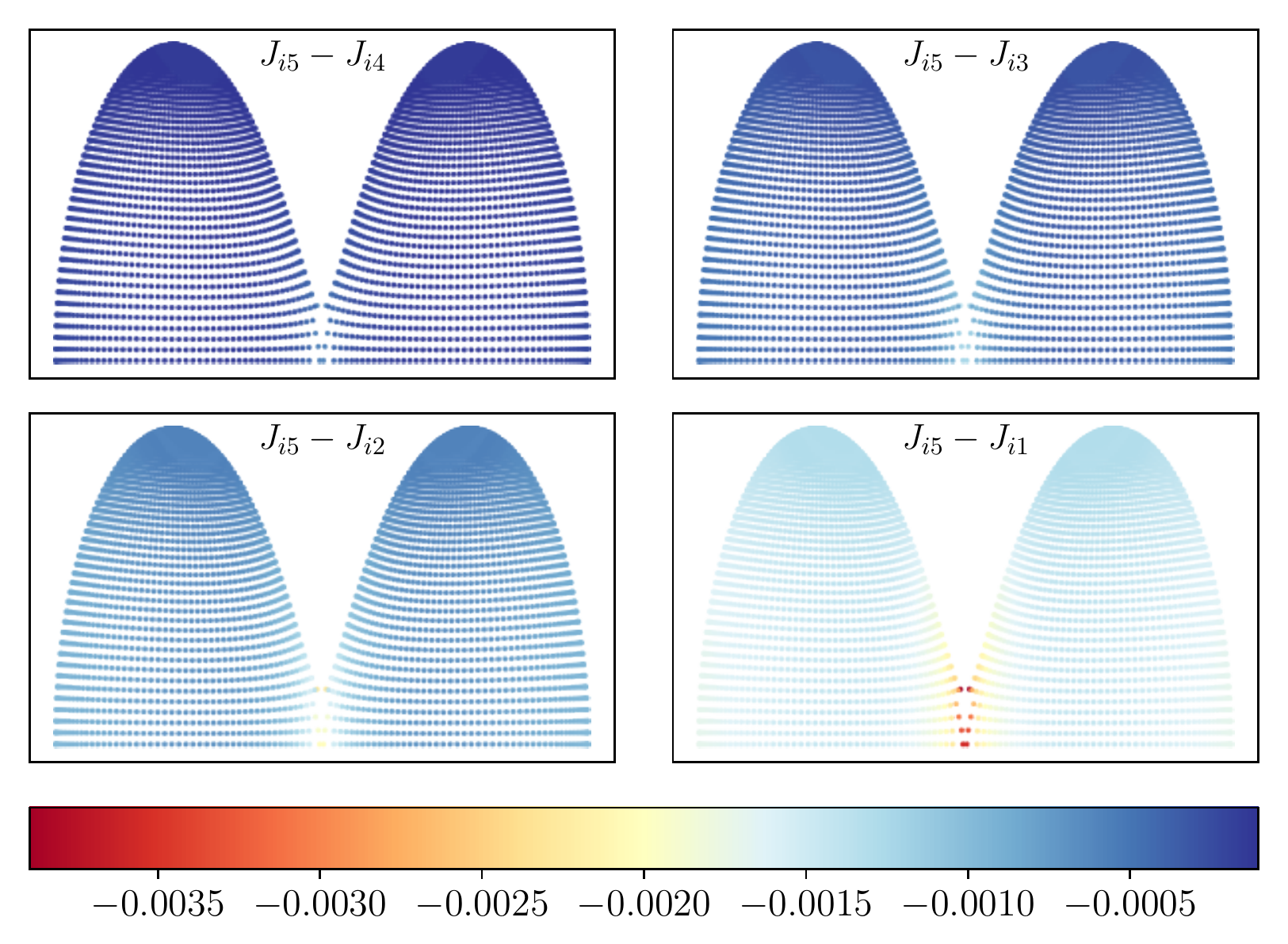}
    
    \caption{Surface distribution of the outward (top) and mean (bottom) intensity of a contact binary with $M_1=1 M_{\odot}$ after the fifth iteration. Right panels show the differences in the surface distribution between the final and each previous iteration.}
    \label{fig:m1_s}
\end{figure}

\begin{figure}[h]
    \centering
    \includegraphics[width=0.495\hsize]{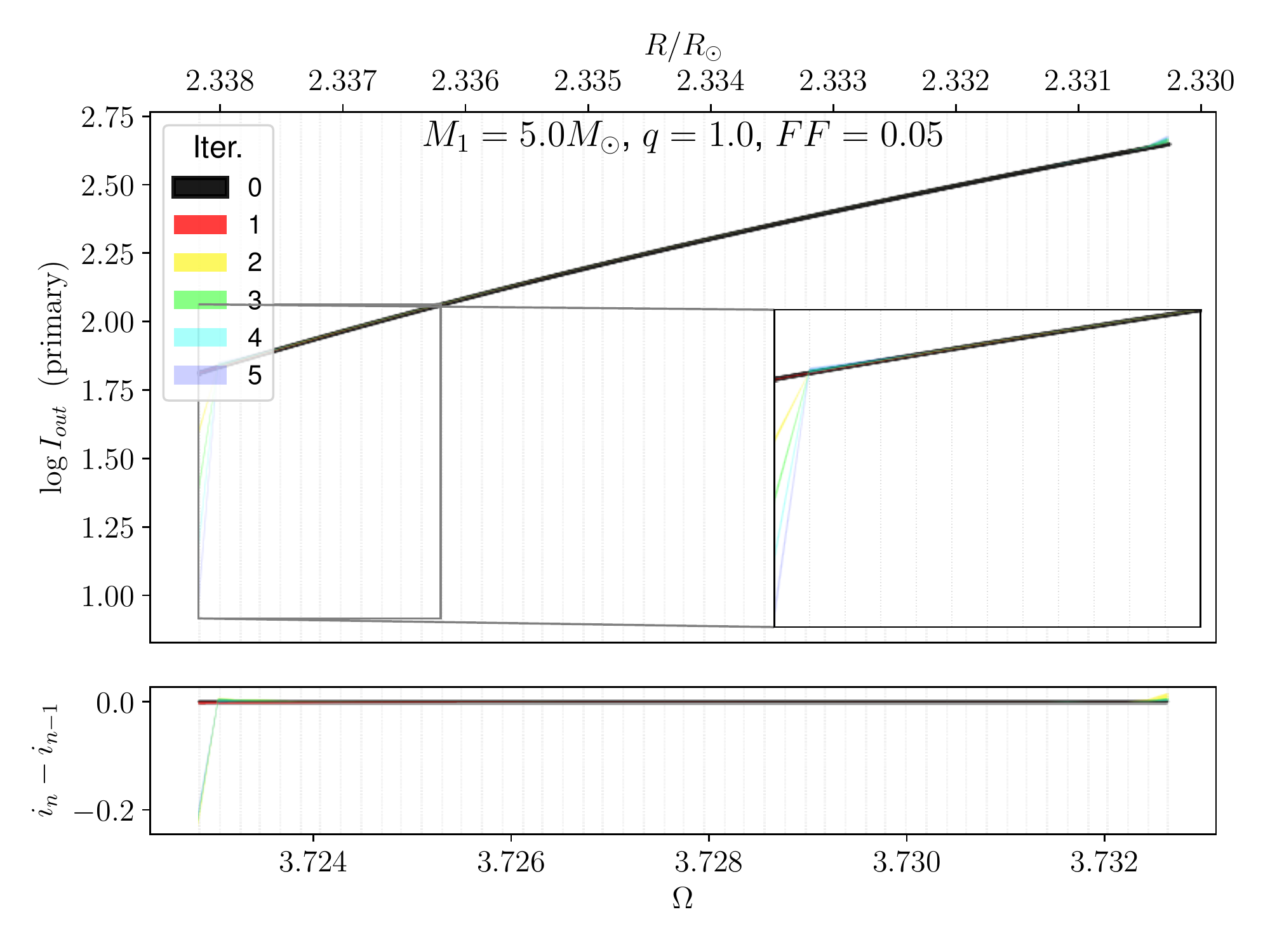}
    \includegraphics[width=0.495\hsize]{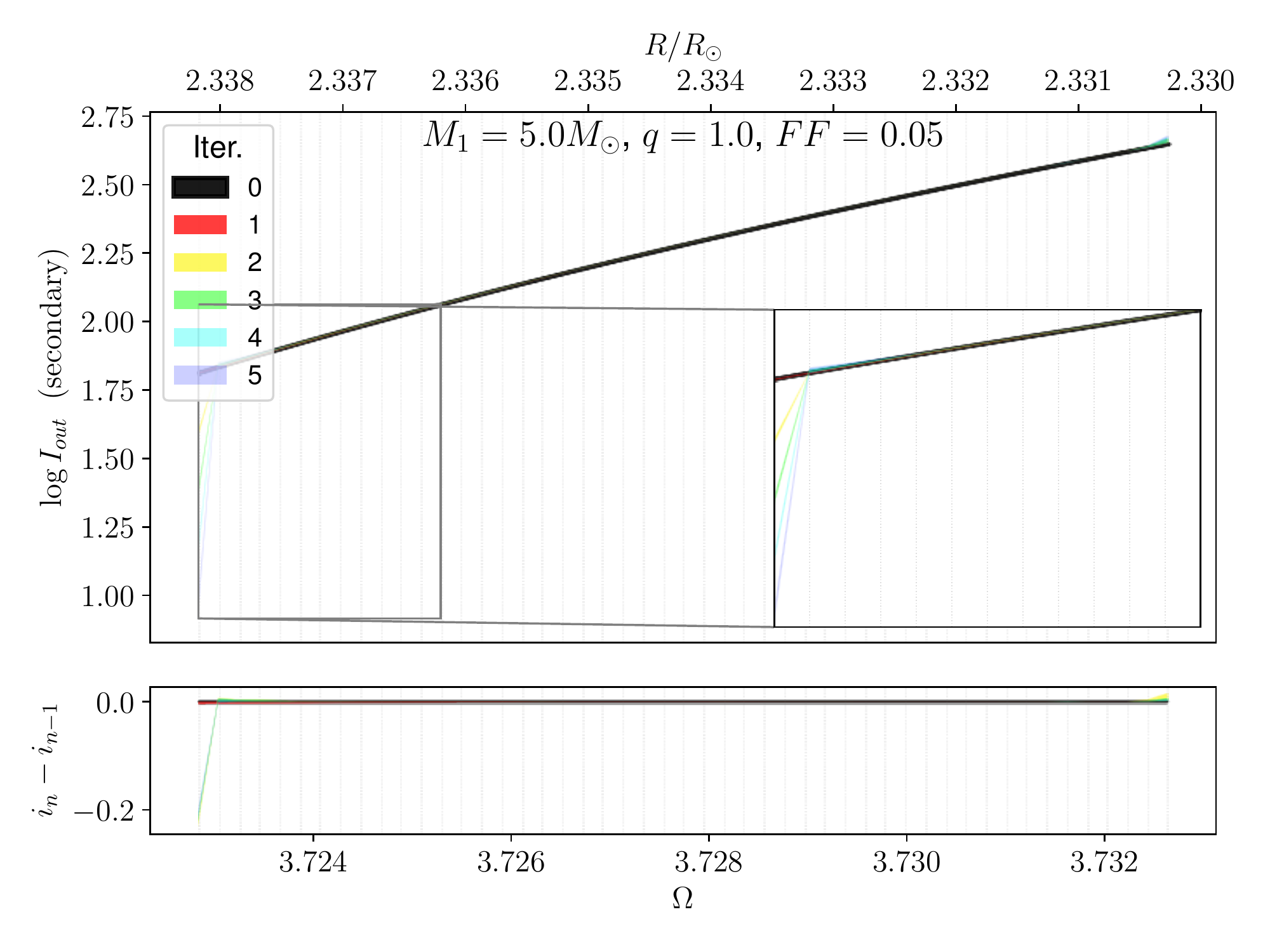}
    
    \includegraphics[width=0.495\hsize]{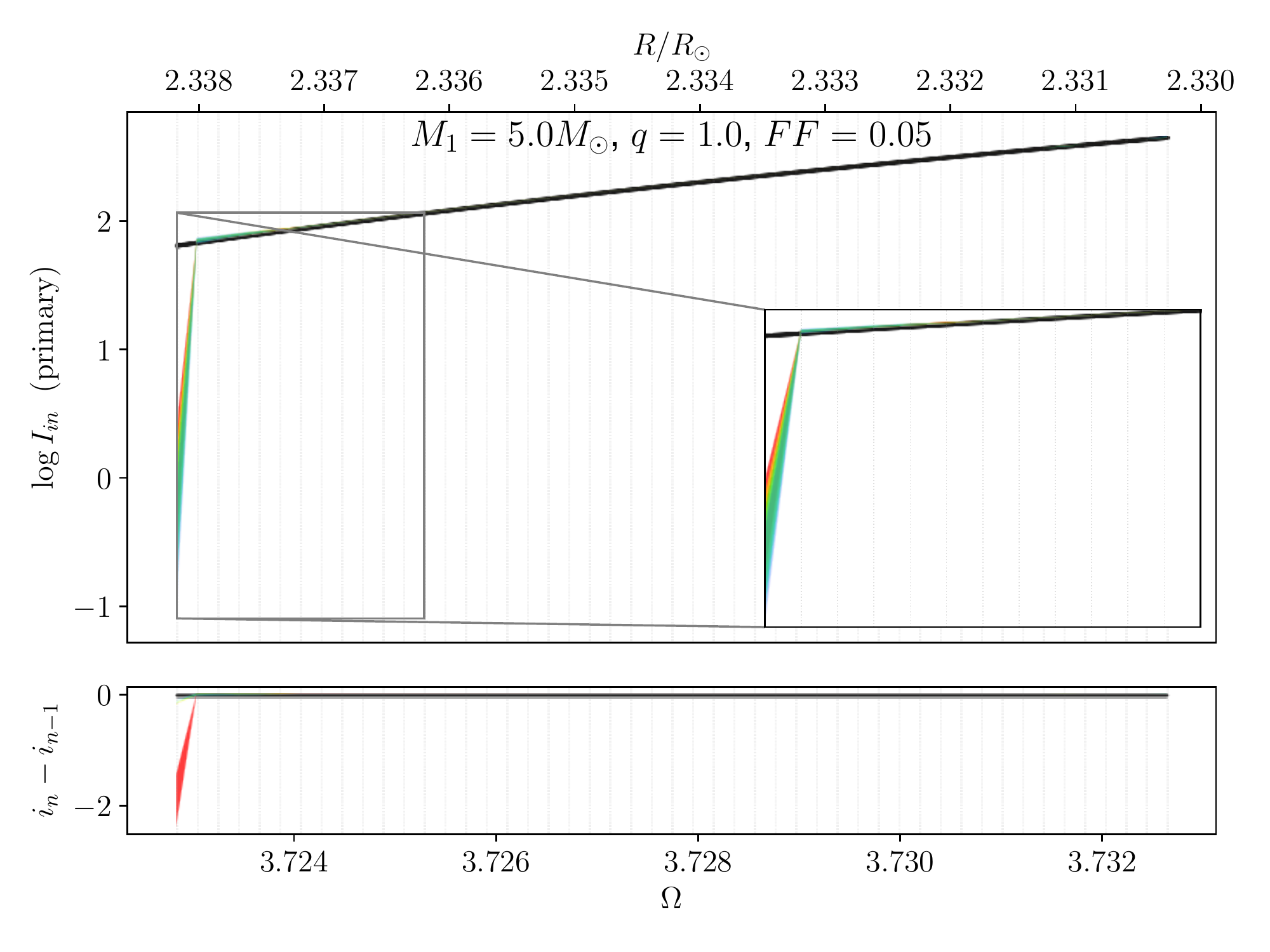}
    \includegraphics[width=0.495\hsize]{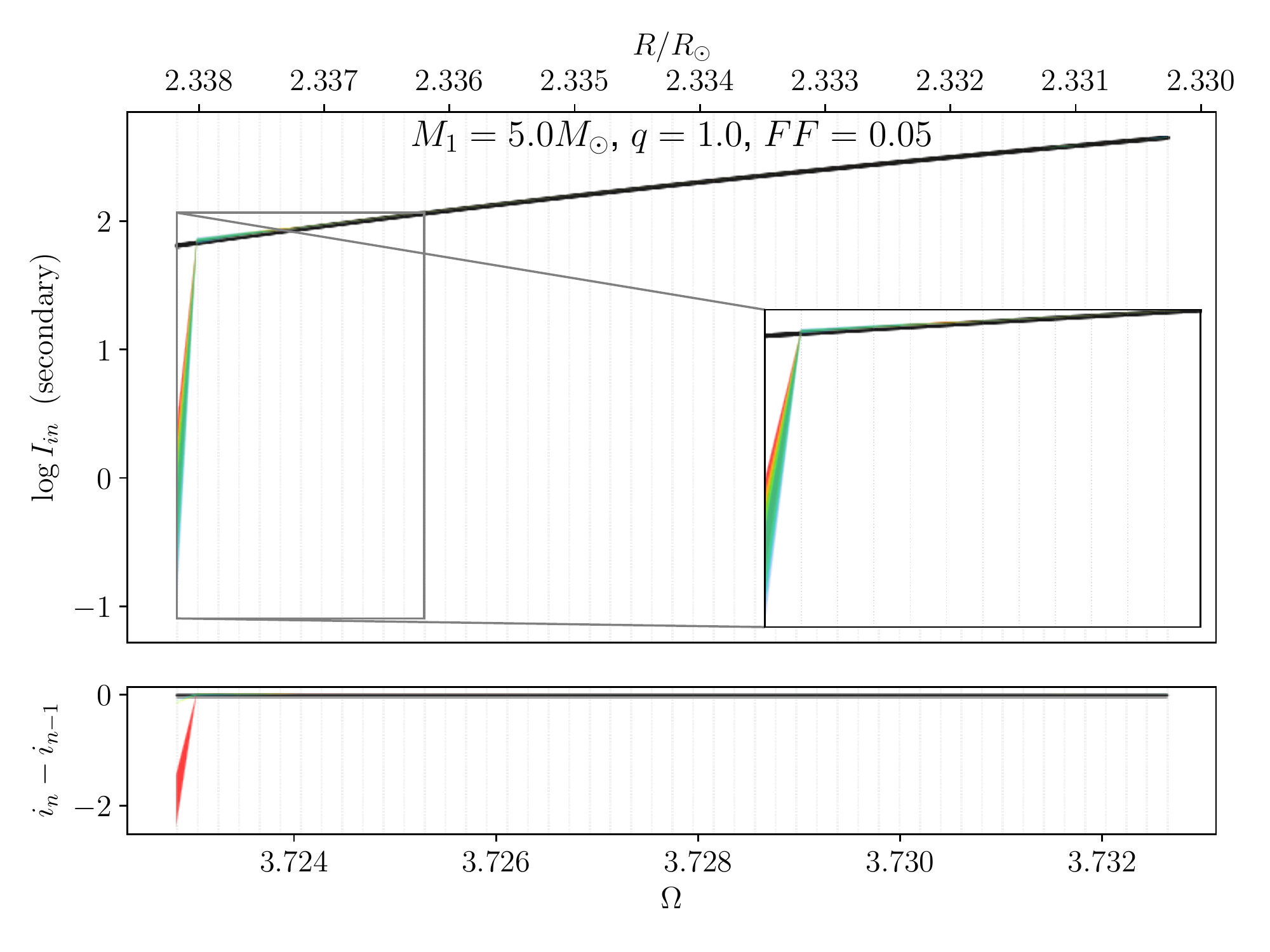}
    
    \includegraphics[width=0.495\hsize]{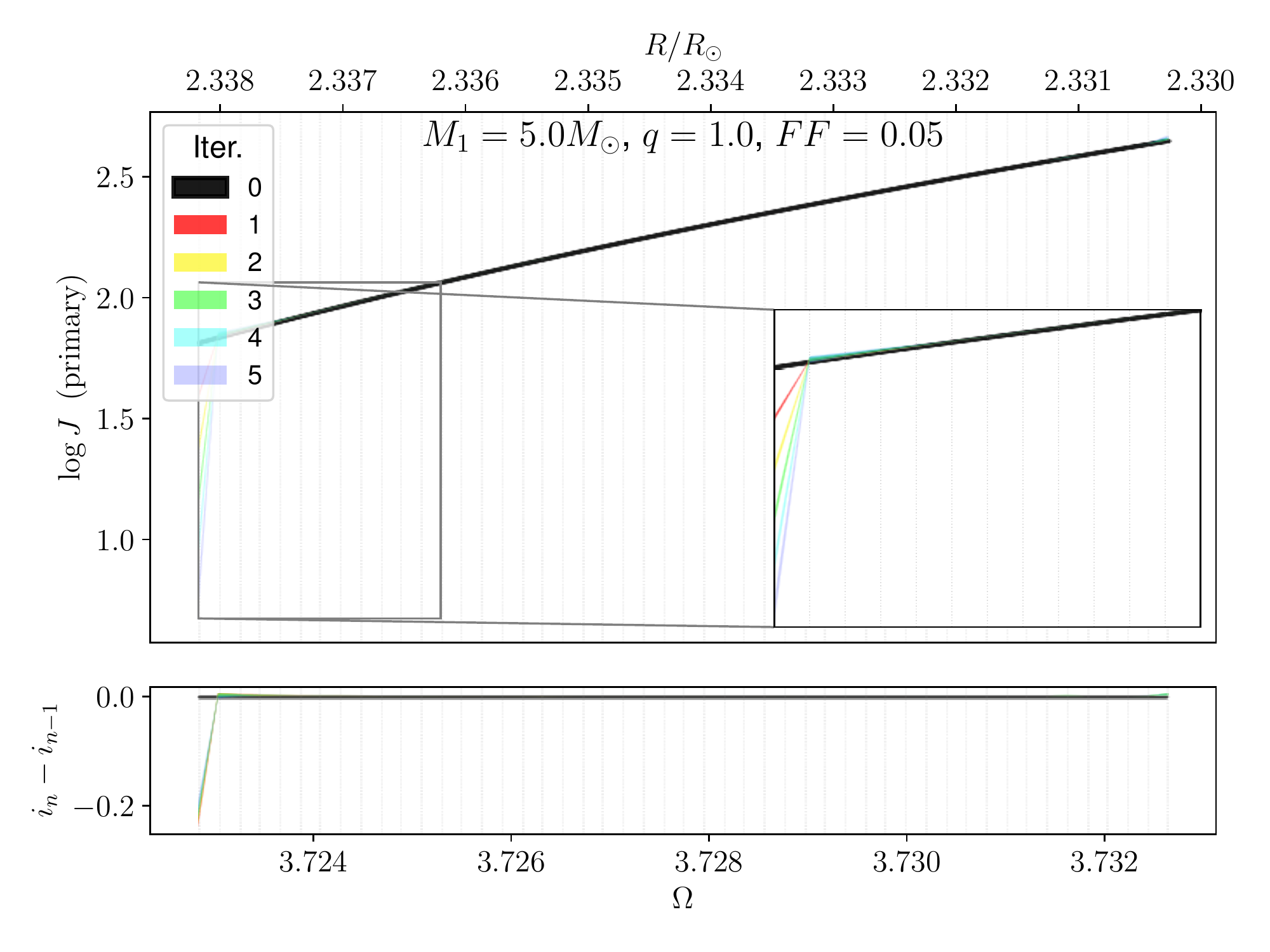}
    \includegraphics[width=0.495\hsize]{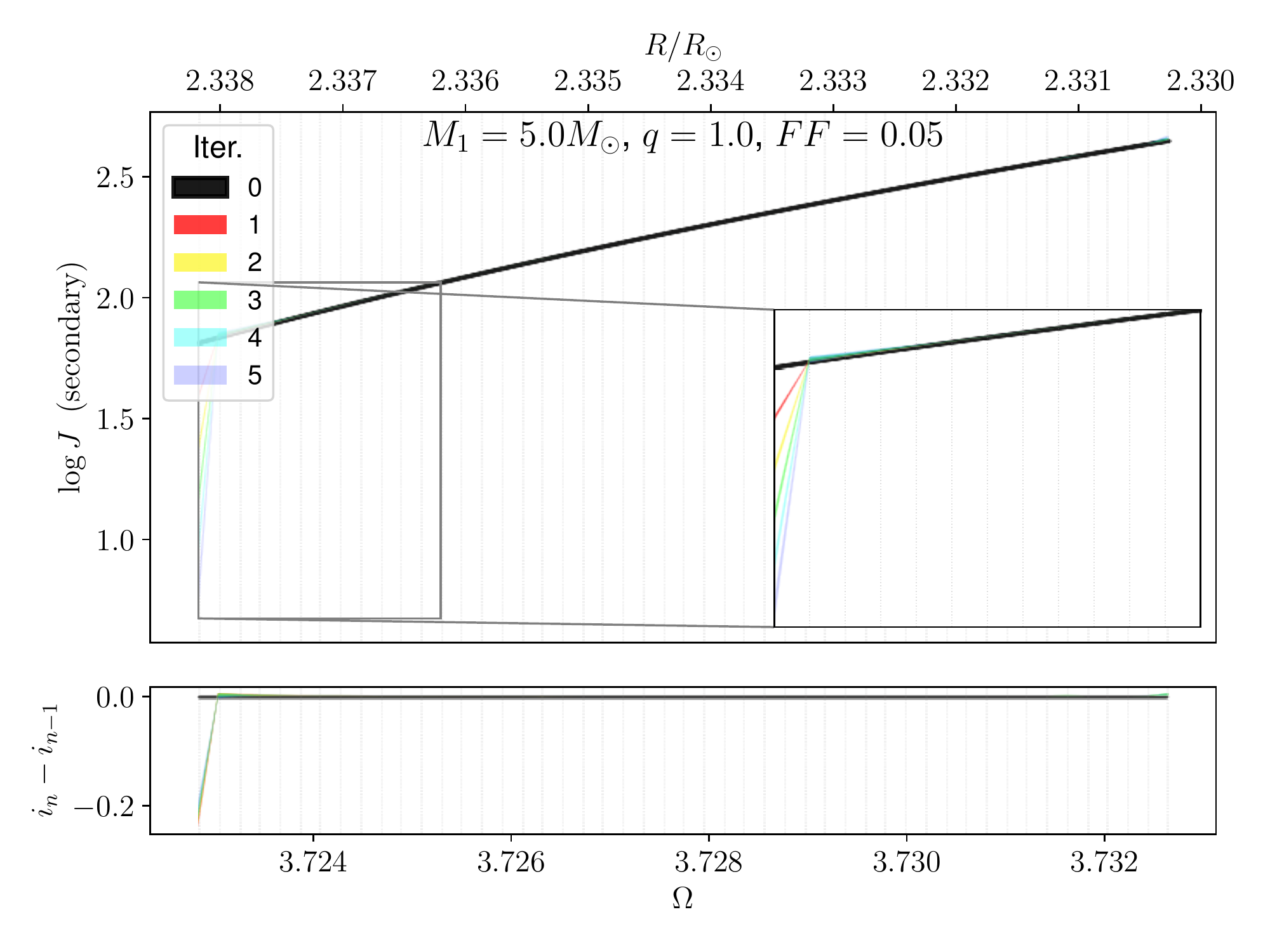}
    
    \caption{Top to bottom: outward, inward and mean intensity as a function of the potential/radius of a contact binary with $M_1=5 M_{\odot}$. Left panels: primary, right panels: secondary component. The bottom panel of each plot shows the differences between successive iterations.}
    \label{fig:m5}
\end{figure}

\begin{figure}[h]
    \centering
    \includegraphics[width=0.495\hsize]{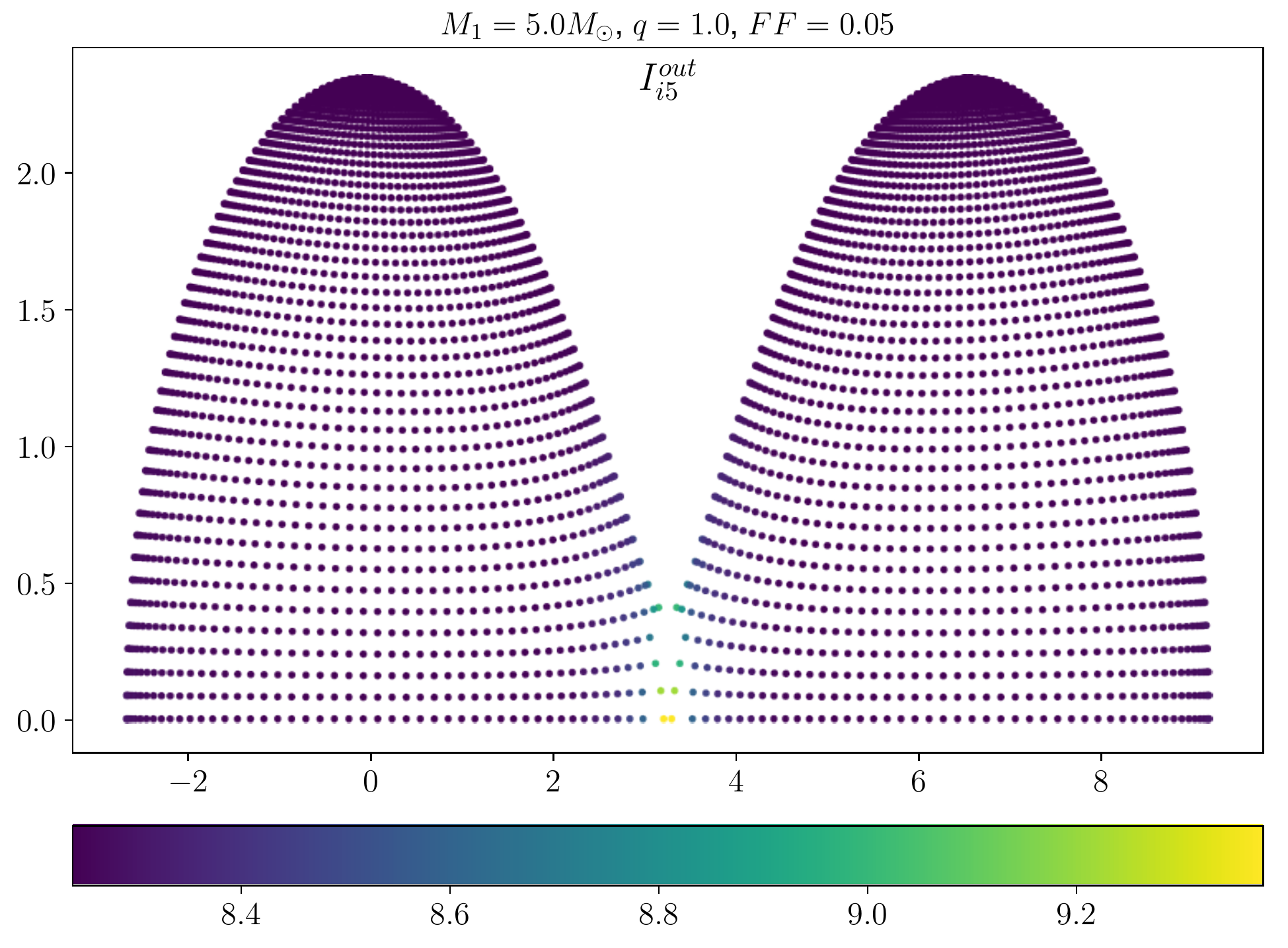}
    \includegraphics[width=0.495\hsize]{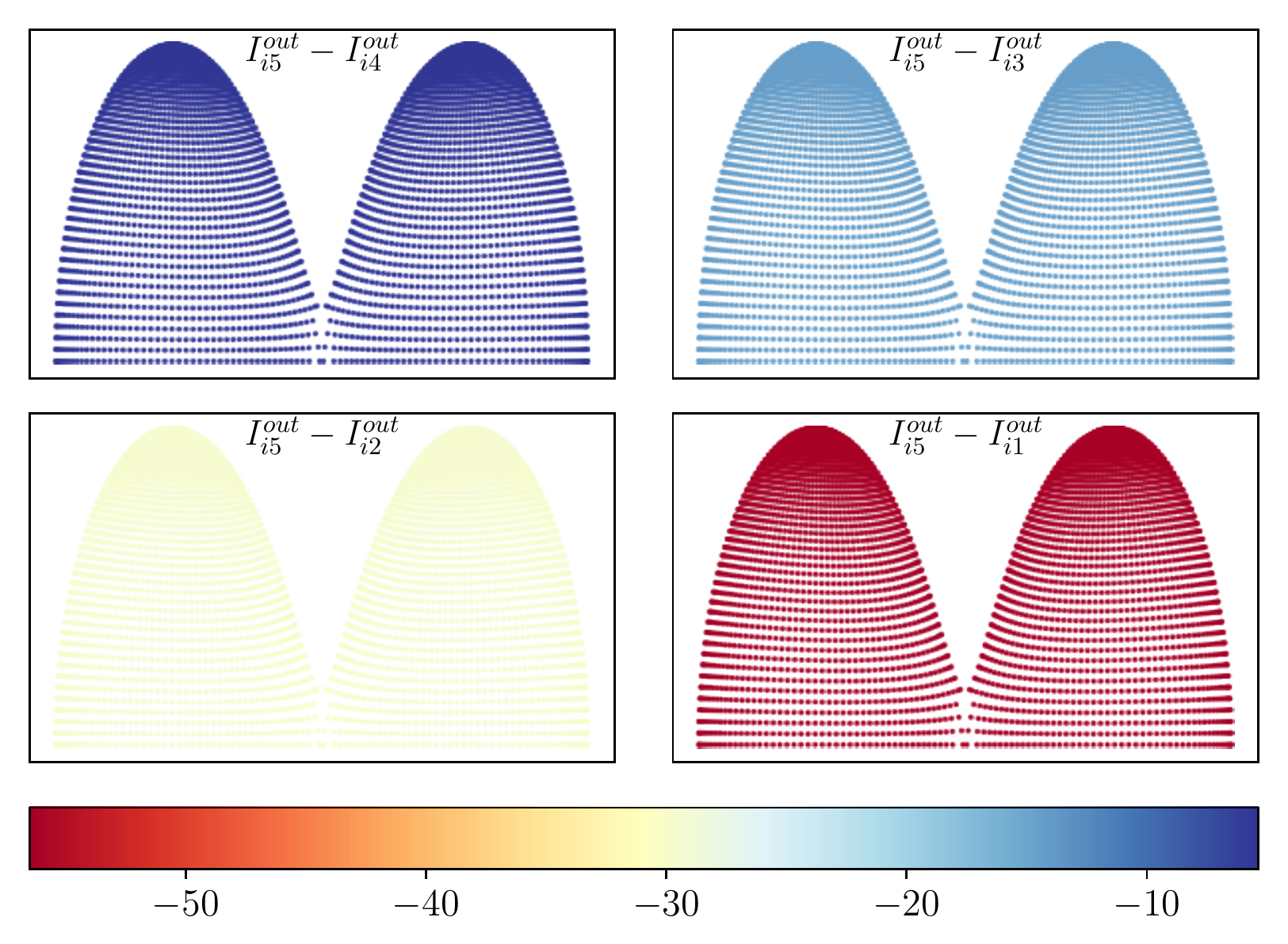}

    \includegraphics[width=0.495\hsize]{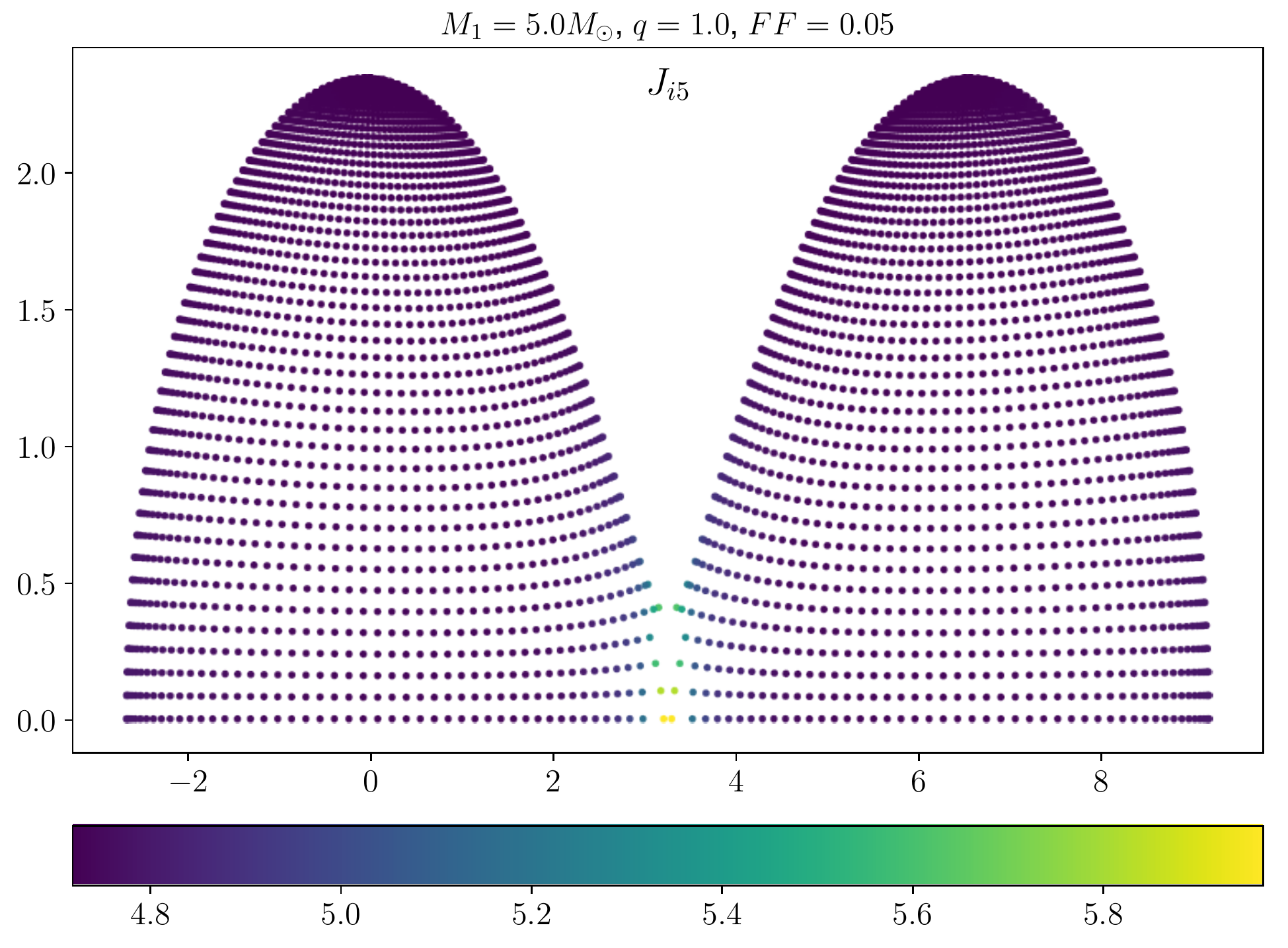}
    \includegraphics[width=0.495\hsize]{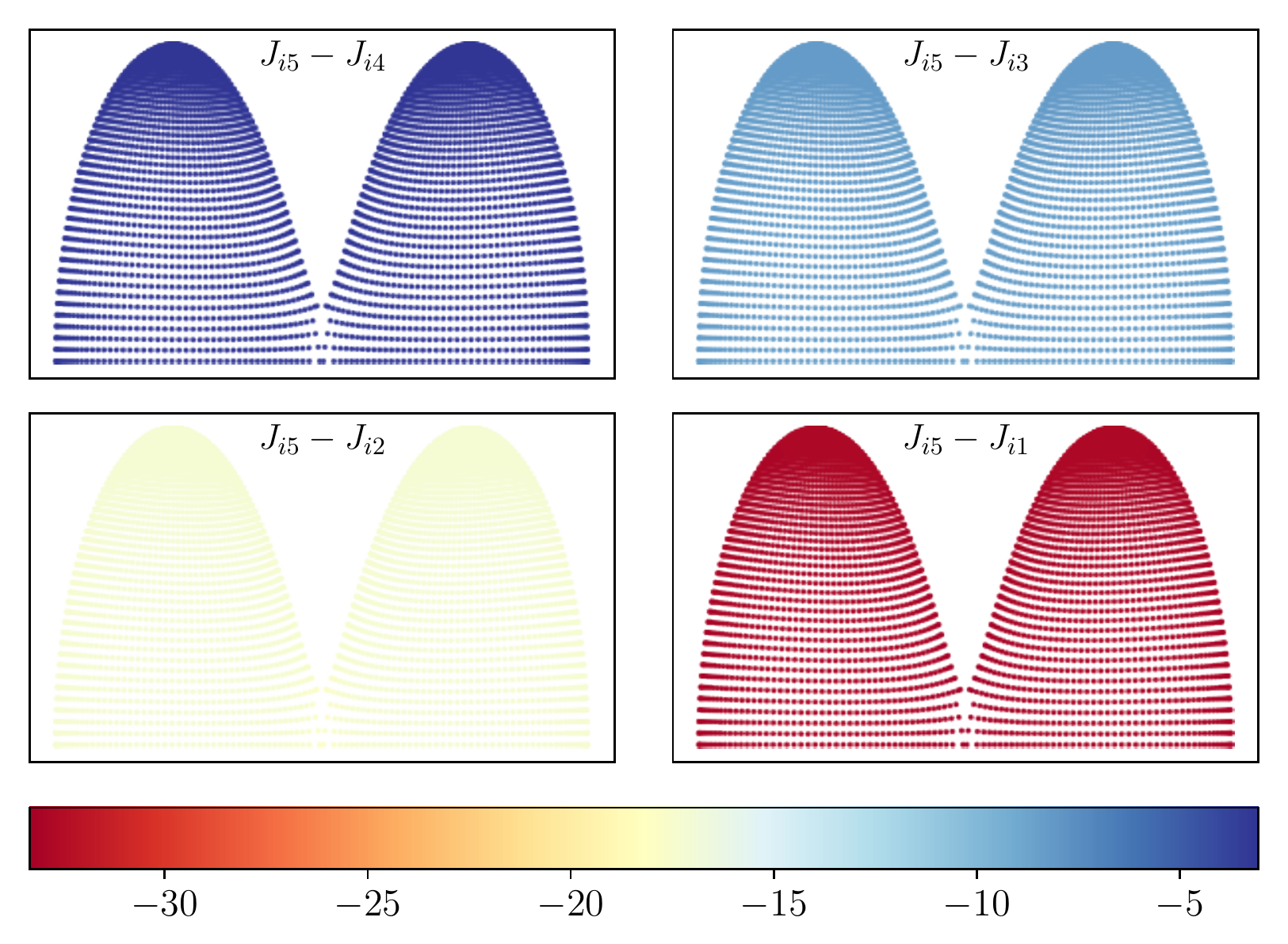}
    
    \caption{Surface distribution of the outward (top) and mean (bottom) intensity of a contact binary with $M_1=5 M_{\odot}$ after the fifth iteration. Right panels show the differences in the surface distribution between the final and each previous iteration.}
    \label{fig:m5_s}
\end{figure}

COBAIN simulation results for different stellar masses are given in Figures~\ref{fig:m075}-\ref{fig:m5_s}. In the tidally deformed polytropic approximation, the mass of the stars in a contact binary determines its internal structure, radius and surface temperature. The radiative transfer simulations of the three chosen mass values $M_1 \in \{0.75, 1.0, 5.0\} M_{\odot}$ primarily show that the default potential range of $\Delta\Omega = 0.01$ only works well for the default mass value of $M_1 = 1.0$, while for higher masses it needs to be significantly decreased in order to better sample the atmosphere layers. The unsuitable potential range also greatly affects the convergence, where it is evident that convergence within the threshold of $0.01\%$ is achieved only in the case of the default mass $1 M_{\odot}$. 

\begin{figure}[h]
    \centering
    \includegraphics[width=\hsize]{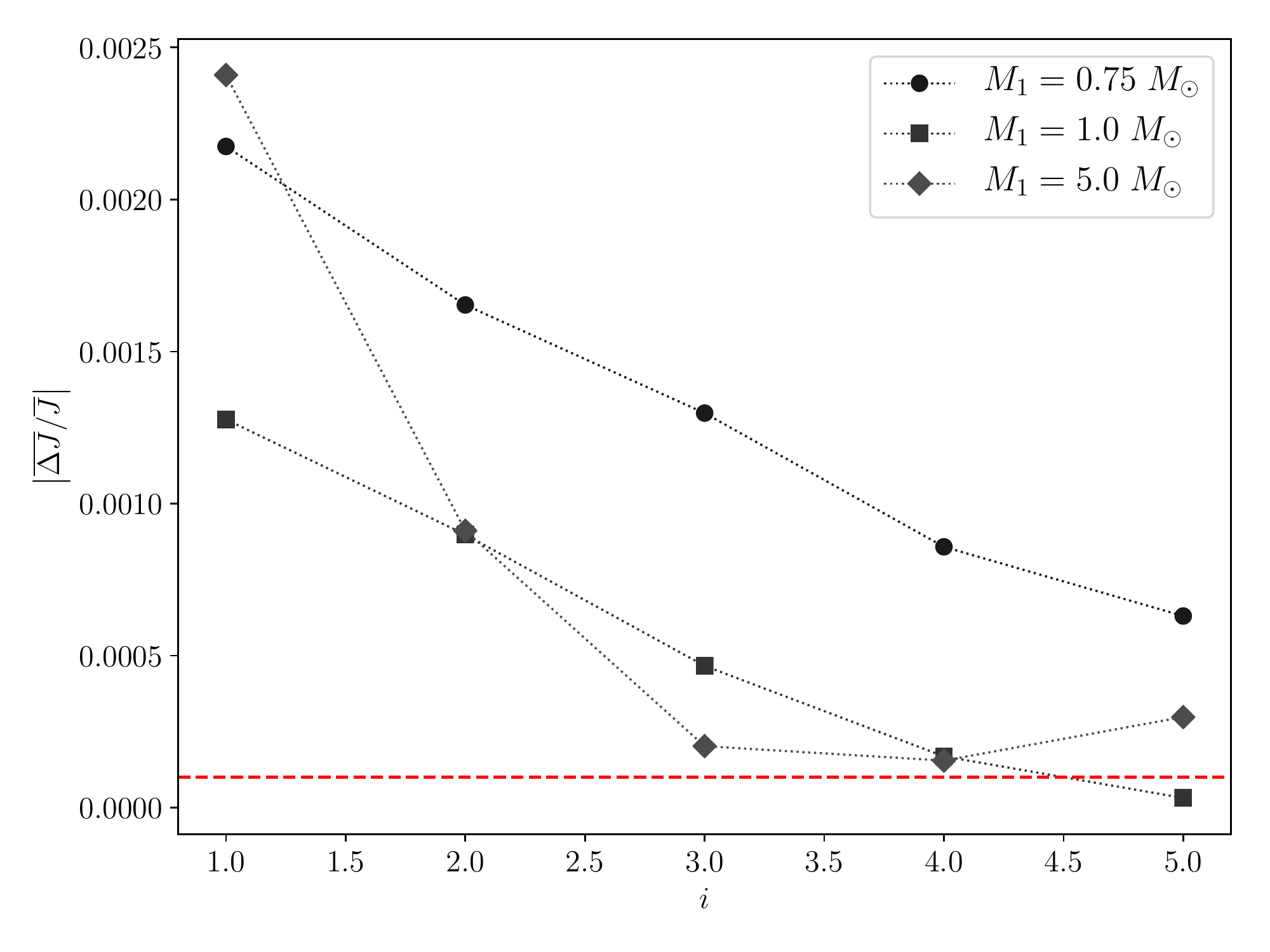}
    \caption{Convergence of the mean intensity in test systems with variable primary mass.}
    \label{fig:iters_m}
\end{figure}

\section{Discussion and future prospects}

The short orbital periods and continuous variability of contact binaries render them abundant in observational studies, with a substantial increase expected from current and future large-scale all-sky surveys. However, there is a persistent gap between the existing complex theoretical models of contact binary structure and the modern tools used for their analysis, in which the contact binary envelope is treated as a mere concatenation of two stars. In this paper, we have demonstrated that the implications on the treatment of contact binaries with these simplified models can be substantial and lead to erroneous estimates of the model parameters. The new generalized radiaitve transfer code COBAIN is being developed with the main purpose to resolve this issue. We have outlined its fundamental principles and tested its performance on a variety of models of single stars and simplified models of contact binaries. We have shown that the code works well with any geometry and is not limited to simple stellar models. 

Differential rotation models clearly indicate that the resulting intensity distribution closely follows the geometry and structure of the input model even after one iteration, and matches the expected intensity distribution in the case of stable models, like the non-rotating or a uniformly rotating star. The variety of contact binary systems tested with COBAIN shows the need for a careful selection of the grid parameters with respect to the physical parameteres of the system. We demonstrate that, in the cases where these pre-requisites are fulfilled, the radiative transfer computation converges to the chosen threshold within five iterations. A safer test of convergence would be to iterate the computation until the successive iteration differences stop decreasing, a method that will be implemented when computing atmospheres of more complex hydrodynamical models. More advanced treatment is also required to compare the results of the code simulations with other atmosphere models. This requires the computation of true physical stellar models whose interior structure cannot be described by a polytropic solution only, in particular not with one differentially rotating model. In addition, the gray atmosphere solutions do not allow for proper treatment of the effects of absorption and scattering at different wavelengths, hence monochromatic treatment is one of the key prospects for the future development of \textsc{COBAIN}. Computing the radiative transfer at a large number of wavelengths simultaneously is not a trivial task and would increase the computational expense of the code further, thus acquiring sufficient computational power is essential for the success of this project. Fortunately, the use and continuous upgrades of high-performance computing clusters has become standard practice in modern astrophysics and allows for the execution of complex and computationally expensive codes such as this one.

On the other hand, computing the atmosphere using \textsc{COBAIN} for every light curve that needs to be analyzed is completely impractical. For this purpose, the computation of contact binary atmosphere tables for different structural models and different values of the stellar properties, like mass ratio, potential, surface temperature, etc, is foreseen for the future. The atmosphere tables are intended to replace the current single star model atmospheres that are used in state-of-the-art binary star modeling software, in particular PHOEBE 2.

The full development of the code into a tool for synthesizing multi-wavelength contact binary atmospheres and contact binary spectra is an extremely ambitious undertaking that requires a lot of computational time and power - a project intended to be completed within several years, in time for the analysis of contact binary data from missions like \textit{Gaia} and \textit{LSST}, as well as the data in current eclipsing binary star catalogs, like \textit{Kepler} and \textit{OGLE-4}. This step is essential not only for uncovering the true structure and range of properties of these systems, but also for studies of their contribution to stellar populations, formation and evolution - the main scientific goal of most modern large-scale sky surveys.

\begin{acknowledgements}
This work has been supported in part by the NSF AAG grant \#1517474.

A.K. gratefully acknowledges the MSE postdoctoral fellowship of the College of Liberal Arts and Sciences at Villanova University.
\end{acknowledgements}

\bibliographystyle{aasjournal}
\bibliography{Bibliography}

\clearpage

\end{document}